\documentclass[%
 aip,
 amsmath,amssymb,
 reprint,%
]{revtex4-1}

\usepackage{bm}
\usepackage{braket}  % Allows Dirac Notation
\usepackage[caption=false]{subfig}
\usepackage{hyperref}
\usepackage{amsfonts}
\usepackage{amssymb}
\usepackage{subfig}
\usepackage{graphicx}% Include figure files
\usepackage{dcolumn}% Align table columns on decimal point
\usepackage{bm}% bold math
\usepackage{amsmath}
\usepackage[utf8]{inputenc}
\usepackage[T1]{fontenc}
\usepackage{color}
\usepackage{mathptmx}

\begin{document}

\preprint{AIP/123-QED}

\title{The Parity Operator: \\applications in quantum metrology}
% Force line breaks with \\

\author{Richard J. Birrittella}
  \email{richard.birrittella@lehman.cuny.edu}
\author{Paul M. Alsing}%
\email{paul.alsing@us.af.mil.}
\affiliation{
	Air Force Research Laboratory, Information Directorate, Rome, NY, USA, 13441
}%
\author{Christopher C. Gerry}
  \email{christopher.gerry@lehman.cuny.edu}
\affiliation{%
Department of Physics and Astronomy, Lehman College, The City University of New York, \\ $\;\;\;$ Bronx, New York, 10468-1589, USA
}%

\date{\today}% It is always \today, today,

\begin{abstract}
In this paper, we review the use of parity as a detection observable in quantum metrology as well as introduce some original findings with regards to measurement resolution in Ramsey spectroscopy and quantum non-demolition (QND) measures of atomic parity.  Parity was first introduced in the context of Ramsey spectroscopy as an alternative to atomic state detection. It was latter adapted for use in quantum optical interferometry where it has been shown to be the optimal detection observable saturating the quantum Cram\'{e}r-Rao bound for path symmetric states.  We include a brief review of the basics of phase estimation and the connection between parity-based detection and the quantum Fisher information as it applies to quantum optical interferometry. We also discuss the efforts made in experimental methods of measuring photon-number parity and close the paper with a discussion on the use of parity leading to enhanced measurement resolution in multi-atom spectroscopy.  We show how this may be of use in the construction of high-precision multi-atom atomic clocks.   
\end{abstract}

\maketitle

\begin{quotation}
	We dedicate this paper to the memory of Jonathan P. Dowling, whose body of work had a tremendous impact on the fields of quantum optical interferometry and quantum metrology.
\end{quotation}

% Comment this out to remove ToC.  
\tableofcontents
\section{\label{sec:intro} Introduction}

\noindent Observables in quantum mechanics, which are generally based on their classical counterparts, are represented as Hermitian operators in Hilbert space due to the fact that such operators possess real eigenvalues.  According to orthodox quantum mechanics, the measurement of an observable returns a corresponding eigenvalue of the operator with probabilities of obtaining specific eigenvalues determined by the details of the prepared state vector.  Usual quantum observables most commonly discussed are energy, position, momentum, angular momentum, etc., all of which have classical analogs. The spin degree of freedom of an electron is often taken as a quantum observable with no classical analog which, of course, is true in the sense that it is not possible to make sense classically out of the notion of a point particle having spin angular momentum.  Yet spin angular momentum itself surely exists in the classical world of macroscopic objects. \\ 

\noindent The same cannot be said, however, for the concept of photon-number parity.  Let $\ket{n}$, where $n \in \mathbb{Z}^{+}$, be a Fock state for a single-mode quantized electromagnetic field.  The photon-number parity of the state is defined as the evenness or oddness of the number as quantified by $\left(-1\right)^{n}$.  We define the usual boson operators $\hat{a}$ and $\hat{a}^{\dagger}$, the annihilation and creation operators of the field, respectively, which satisfy the usual commutation relation $\left[\hat{a},\hat{a}^{\dagger}\right]=1$ and the number operator $\hat{n}=\hat{a}^{\dagger}\hat{a}$ such that $\hat{n}\ket{n} = n\ket{n}$. In terms of these operators, one can introduce the photon number parity operator $\hat{\Pi} = \left(-1\right)^{\hat{a}^{\dagger}\hat{a}} = e^{i\pi\hat{a}^{\dagger}\hat{a}}$ such that $\hat{\Pi}\ket{n} = \left(-1\right)^{n}\ket{n}$.  The eigenvalues of this operator are dichotomic and thus highly degenerate.  While it is clear that $\hat{\Pi}$ is Hermitian and thus constitutes an observable, it can also be shown that there exists no classical analog to photon number parity.  This can be demonstrated by considering the energies of the quantized field: while these energies are discrete (i.e. $E_{n} = \hbar\omega\left(n+\tfrac{1}{2}\right)$), the energies of a classical field are continuous. \\

\noindent The parity operator makes frequent appearances in quantum optics and quantum mechanics.  For example it has been pointed out that the quasi-probability distribution known as the Wigner function \cite{Wigner1} \cite{Wigner3} can be expressed in terms of the expectation value of the displaced parity operator \cite{Wigner2} given for a single-mode field as

\begin{equation}
	W\left(\alpha\right) = \frac{2}{\pi}\braket{\hat{D}\left(\alpha\right)\hat{\Pi}\;\hat{D}^{\dagger}\left(\alpha\right)} 
	\label{eqn:intro1}
\end{equation} 
 
\noindent where $\hat{D}\left(\alpha\right) = e^{\alpha\hat{a}^{\dagger} - \alpha^{*}\hat{a}}$ is the usual displacement operator familiar from quantum optics, and where $\alpha$ is the displacement amplitude generally taken to be a complex number.  This relationship is the basis for reconstructing a field state through quantum state tomography \cite{Tomography}.  The parity operator has also appeared in various proposals for testing highly excited entangled two-mode field states for violations of a Bell's inequality \cite{Bell}. \\
 
\noindent Another field in which parity sees use is quantum metrology, where it serves as a suitable detection observable for reasons we will endeavor to address. Quantum metrology is the science of using quantum mechanical states of light or matter in order to perform highly resolved and sensitive measurements of weak signals like those expected by gravitational wave detectors (see for example Barsotti \textit{et al.} \cite{LIGO} and references therein) and for the precise measurements of transition frequencies in atomic or ion spectroscopy \cite{Bollinger}.  This is often done by exploiting an inherently quantum property of the state such as entanglement and/or squeezing.  The goal of quantum metrology is to obtain greater sensitivities in the measurement of phase-shifts beyond what is possible with classical resources alone, which at best can yield sensitivities at the shot noise-, or standard quantum limit (SQL).  For an interferometer operating with classical (laser) light, the sensitivity of a phase-shift measurement, $\Delta\varphi$, scales as $1/\sqrt{\bar{n}}$ where $\bar{n}$ is the average photon number of the laser field. This defines the SQL as the greatest sensitivity obtainable using classical light: $\Delta\varphi_{\text{SQL}} = 1/\sqrt{\bar{n}}$.  In cases where phase-shifts are due to linear interactions, the optimal sensitivity allowed by quantum mechanics is known as the Heisenberg limit (HL), defined as $\Delta\varphi_{\text{HL}} = 1/\bar{n}$.  Only certain states of light having no classical analog, including entangled states of light, are capable of breaching the SQL level of sensitivity.  In many cases, perhaps even most, reaching the HL requires not only a highly non-classical state of light but also a special observable to be measured. That observable turns out to be photon-number parity measured at one of the output ports of the interferometer.  This is because the usual technique of subtracting irradiances of the two output beams of the interferometers fails for many important non-classical states. Furthermore, as to be discussed in latter sections, consideration of the quantum Fisher information indicates that photon number parity serves as the optimal detection observable for path symmetric states in quantum optical interferometry.  \\

\noindent This paper is organized as follows:  Section \ref{sec:Ramsey_1} begins with a discussion on the origin of parity-based measurement in the context of atomic spectroscopy.  In section \ref{sec:Phase_est} we present a brief review of the basics of phase estimation including a concise derivation of the quantum Cram\'{e}r-Rao bound and the related quantum Fisher information and discuss how these relate to parity-based measurement, particularly in the field of quantum optical interferometry.  In Section \ref{sec: QOI} we highlight the use of several relevant states of light in quantum optical interferometry such as the N00N states and coherent light.  We show in the latter case that the use of parity does not yield sensitivity (i.e. reduced phase uncertainty) beyond the classical limit but does enhance measurement resolution.  In Section \ref{sec:exper} we discuss the experimental efforts made in performing photonic parity measurements.  Finally, in Section \ref{sec:Ramsey} we briefly return to the atomic population parity measurements, this time in the context of atomic coherent states and show that such measurements could lead to high-resolution multi-atom atomic clocks, i.e. atomic clocks of greater precision than is currently available.  We conclude the paper with some brief remarks.

\section{\label{sec:Ramsey_1} Ramsey spectroscopy with entangled and unentangled atoms}

\begin{figure*}
	\centering
	\includegraphics[width=0.9\linewidth,keepaspectratio]{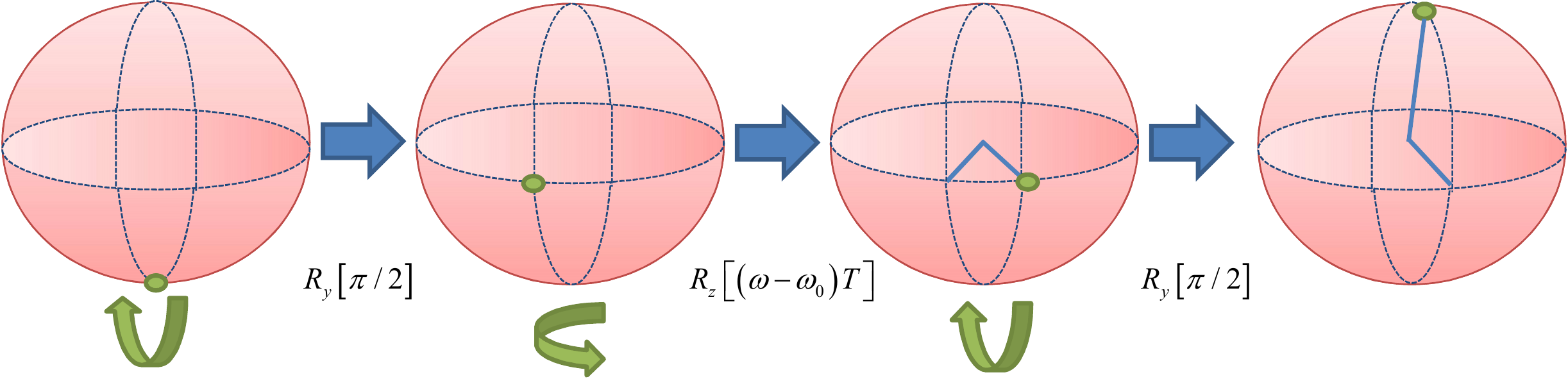}
	\caption{Bloch sphere representation for single-atom spectroscopy.  The atom is intially in its ground state. The first $\pi/2$-pulse (at frequency $\omega$) places the atom in a superposition state. The free evolution of time $T$ corresponds to a rotation in the $x-y$ plane.  The final $\pi/2$-pulse (also at frequency $\omega$) places the state of the atom near the excited state. For a frequency-spread in radiation $\Delta\omega$ the state will be nearly at the north pole of the Bloch sphere. Note $R_{i}\left[\theta\right]$ represents rotation by $\theta$ about the $i$-axis.}
	\label{fig:Bloch}
\end{figure*}

\noindent We first introduce the Dicke atomic pseudo-angular momentum operators for a collection of $N$ two-level atoms.  These are given by \cite{Yurke}

\begin{equation}
\hat{J}_{x,y,z}=\frac{1}{2}\sum_{i=1}^{N}\sigma^{\left(i\right)}_{x,y,z},
\label{eqn:ccg_1}
\end{equation}

\noindent satisfying the su(2) commutation relation of Eq.~\ref{eqn:Aa2} (see Appendix \ref{app:subsecA2}) and where $\sigma^{\left(i\right)}_{x,y,z}$ are the Pauli operators for the $i^{\text{th}}$ atom as given by $\sigma_{x}^{\left(i\right)}=\ket{e}_{i}\bra{g}+\ket{g}_{i}\bra{e}$, $\sigma_{y}^{\left(i\right)}=-i\left(\ket{e}_{i}\bra{g}-\ket{g}_{i}\bra{e}\right)$, $\sigma_{z}^{\left(i\right)}=\ket{e}_{i}\bra{e}-\ket{g}_{i}\bra{g}$ and where the ground and excited states for the $i^{\text{th}}$ atom are denoted $\ket{g}_{i},\;\ket{e}_{i}$, respectively.  Obviously the operators given in Eq.~\ref{eqn:ccg_1} are additive over all atoms.  We now introduce the corresponding the collective atomic states, the Dicke states, expressed in terms of the SU(2) angular momentum states $\ket{j,m}$ where $j=N/2$ and $m \in\{-j,-j+1,...,j\}$ which can be given as superpositions of the product states of all atoms. For $N=2j$ atoms, with $j=1/2,1,...$, the Dicke states $\ket{j,m}$ are defined in terms of the individual atomic states as  $\ket{j,j}=\ket{e_{1},e_{2},...e_{N}}=\ket{e}^{\otimes N}$ and  $\ket{j,-j}=\ket{g_{1},g_{2},...g_{N}}=\ket{g}^{\otimes N}$ with intermediate steps consisting of superpositions of  all permutations with consecutively more atoms being found in the ground state all the way down to $\ket{j,-j}$, where all atoms are in the ground state. The ladder operators given as 

\begin{align}
	\hat{J}_{+} = \hat{J}_{x}+i\hat{J}_{y} = \sum_{i=1}^{N}\ket{e}_{i}\bra{g} \nonumber \\
	\label{eqn:ccg_2} \\
	\hat{J}_{-} = \hat{J}_{x}-i\hat{J}_{y} = \sum_{i=1}^{N}\ket{g}_{i}\bra{e} \nonumber 
\end{align}

\noindent can be used repeatedly to generate expressions for all the non-extremal states in terms of the individual atoms.  We relegate further discussion of the mapping between the two sets of states to Appendix \ref{app:secA}. \\

\noindent In what follows we denote the exact transition frequency between the excited and ground states as $\omega_{0}$.  The goal of Ramsey spectroscopy is to determine the frequency with as high a sensitivity (or as low an uncertainty) as possible.  We first go through the Ramsey procedure with a single atom assumed initially in the ground state such that $\ket{\text{in}}=\ket{g}$.  The atom is subjected to a $\pi/2$ pulse described by the operator $e^{-i\tfrac{\pi}{2}\hat{J}_{y}}$ and implemented with radiation of frequency $\omega$.  For one atom, the rotation operator about the $y$-axis for an arbitrary angle is given by

\begin{equation}
	e^{-i\beta\hat{J}_{y}} = e^{-i\tfrac{\beta}{2}\sigma_{y}} = \bm{I}_{2}\cos\left(\beta/2\right)-i\sigma_{y}\sin\left(\beta/2\right),
	\label{en:ccg_3}
\end{equation}

\noindent where $\bm{I}_{2}=\ket{e}\bra{e}+\ket{g}\bra{g}$ is the identity operator in this two-dimensional subspace.  This results in the transformations 

\begin{equation}
	e^{-i\beta\hat{J}_{y}}
	\left[
		\begin{array}{c}
			\ket{e}  \\
			\ket{g}  
		\end{array}
	\right]
	=
	\left[
		\begin{array}{c}
			\cos\left(\beta/2\right)\ket{e}+\sin\left(\beta/2\right)\ket{g}  \\
			\cos\left(\beta/2\right)\ket{g}-\sin\left(\beta/2\right)\ket{e}  
		\end{array}
	\right],
	\label{eqn:ccg_4}
\end{equation}

\noindent which for $\beta=\pi/2$ we have the balanced superpositions

\begin{align}
	e^{-i\tfrac{\pi}{2}\hat{J}_{y}}\ket{e} = \frac{1}{\sqrt{2}}\left(\ket{g}+\ket{e}\right), \nonumber \\
	\label{eqn:ccg_5} \\
	e^{-i\tfrac{\pi}{2}\hat{J}_{y}}\ket{g} = \frac{1}{\sqrt{2}}\left(\ket{g}-\ket{e}\right). \nonumber
\end{align}

\begin{figure}
	\centering
	\includegraphics[width=1.0\linewidth,keepaspectratio]{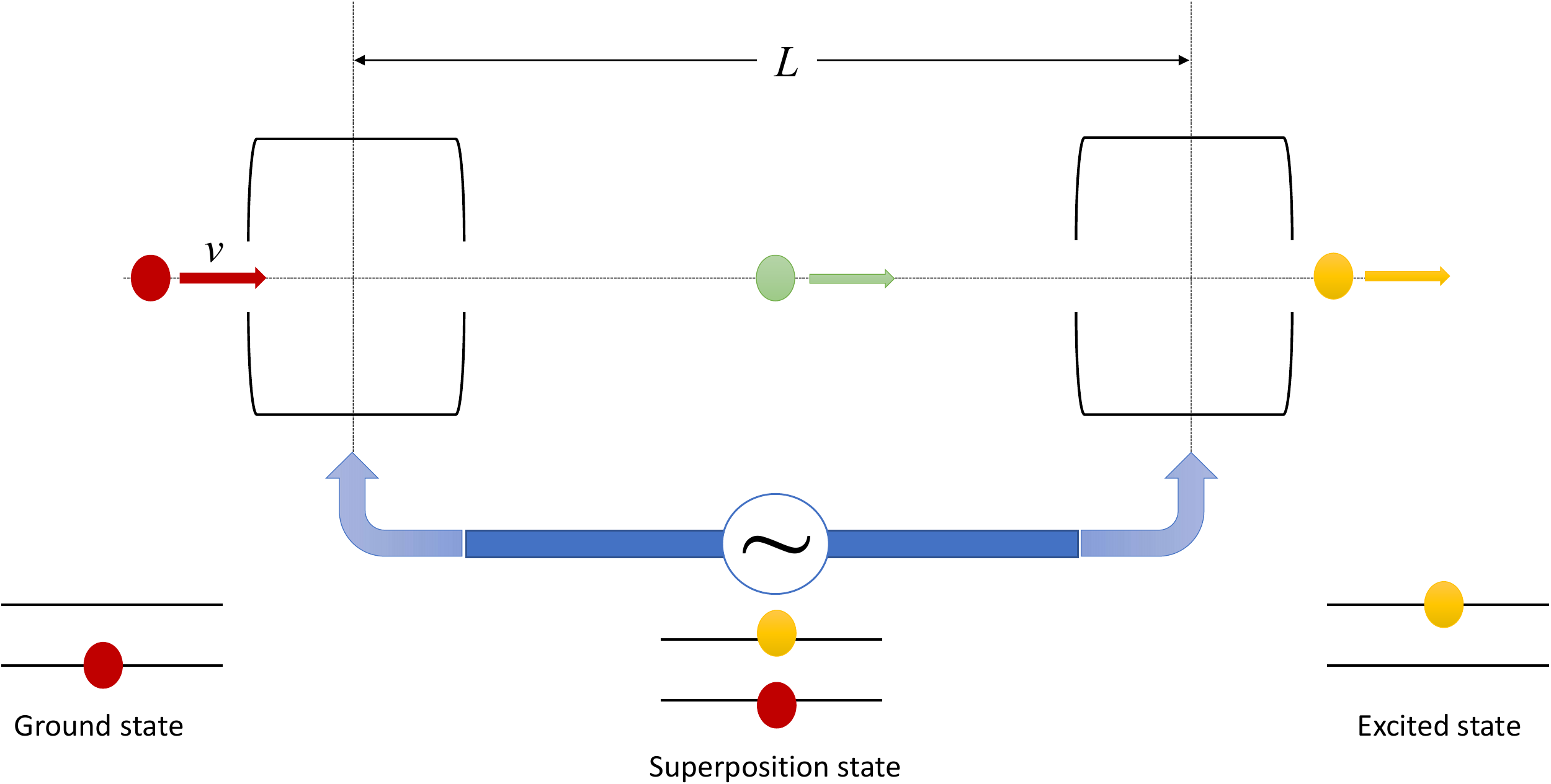}
	\caption{A sketch of the Ramsey technique employed for the case of a single atom initially prepared in the ground state.  The first $\pi/2$ pulse (frequency $\omega$) places the atom in a superposition state. The free evolution over cavity length $L$ results in a phase shift $\phi=\left(\omega_{0}-\omega\right)T$ ($T=L/v$, where $v$ is the atom velocity).  The final $\pi/2$ pulse (frequency $\omega$) places the state of the atom near the excited state.}
	\label{fig:Ramsey}
\end{figure}

\noindent Assuming the atom is initially in the ground state, the state of the atom after the first $\pi/2$ pulse becomes $\ket{\psi\left(t_{\pi/2}\right)}=e^{-i\tfrac{\pi}{2}\hat{J}_{y}}\ket{\text{in}}=\tfrac{1}{\sqrt{2}}\left(\ket{g}-\ket{e}\right)$.  This is followed by a period $T$ of free evolution (precession) governed by the operator $e^{-i\left(\omega_{0}-\omega\right)T\hat{J}_{z}}$ where once again $\omega$ is the frequency of the radiation field implementing $\pi/2$ pulses.  The state after free evolution is

\begin{equation}
	\ket{\psi\left(t_{\pi/2}+T\right)} = \frac{1}{\sqrt{2}}\left(e^{i\frac{\phi}{2}}\ket{g}-e^{-i\frac{\phi}{2}}\ket{e}\right),
	\label{eqn:ccg_6}
\end{equation}

\noindent where we have set $\phi=\left(\omega_{0}-\omega\right) T$.  After the second $\pi/2$ pulse following free evolution, we use Eq.~\ref{eqn:ccg_5} to find the final state

\begin{equation}
	\ket{\psi\left(t_{f}\right)} = i\sin\left(\phi/2\right)\ket{g} - \cos\left(\phi/2\right)\ket{e},
	\label{eqn:ccg_7}
\end{equation}

\noindent where $t_{f} = 2t_{\pi/2} + T$ and where it is assumed $T \gg 2t_{\pi/2}$ such that $t_{f}\simeq T$.  A diagram of the transformations in Bloch-sphere representation is provided in Fig.~\ref{fig:Bloch} as well as an idealized sketch of the Ramsey technique in Fig.~\ref{fig:Ramsey}.  The expectation value of $\hat{J}_{z}$ for this state is 

\begin{equation}
	\braket{\hat{J}_{z}\left(\phi\right)}_{t_{f}}^{\left(1\right)} = \frac{1}{2}\cos\phi = \frac{1}{2}\cos\left[\left(\omega_{0}-\omega\right)T\right].
	\label{eqn:ccg_8}
\end{equation}

\noindent By tuning the frequency of the driving field so as to maximize $\braket{\hat{J}_{z}\left(\phi\right)}_{t_{f}}^{\left(1\right)}$, one can estimate the transition frequency $\omega_{0}$.  \\

\noindent If we consider $N$ atoms one at a time or collectively through the Ramsey procedure \cite{RamseyBook}, then the initial state with all atoms in their ground state is the Dicke state $\ket{j,-j}$, and after the first $\pi/2$ pulse, the state generated $\ket{\psi} = e^{-i\tfrac{\pi}{2}\hat{J}_{y}}\ket{j,-j}$ is an example of an atomic coherent state (ACS) \cite{Arrechi}.  For a brief review of the Dicke states as well as a derivation of the ACS, see Appendix \ref{app:subsecA1}.  In such a state there is no entanglement among the atoms: each atom undergoes the same evolution through the Ramsey process and, because the $\hat{J}_{z}$ operator is the sum $\hat{J}_{z}=\tfrac{1}{2}\sum_{i}^{N}\sigma^{\left(i\right)_{z}}$, one easily finds that 

\begin{equation}
	\braket{\hat{J}_{z}\left(\phi\right)}_{t_{f}}^{\left(\text{ACS}\right)}  = N\braket{\hat{J}_{z}\left(\phi\right)}_{t_{f}}^{\left(1\right)} = \frac{N}{2}\cos\left[\left(\omega_{0}-\omega\right)T\right].
	\label{eqn:ccg_9}
\end{equation}

\noindent The propagation of the error in the estimation of the phase $\phi$ is given by (dropping scripts for notational convenience)

\begin{equation}
	\Delta\phi_{\text{ACS}} = \frac{\Delta\hat{J}_{z}\left(\phi\right)}{|\partial_{\phi}\braket{\hat{J}_{z}\left(\phi\right)}|} = \frac{1}{\sqrt{N}},
	\label{eqn:ccg_10}
\end{equation}

\noindent showing that the best sensitivity one can obtain with entangled atoms in the SQL.  In terms of the transition frequency, the error is given by $\Delta\omega_{0}=1/\left(T\sqrt{N}\right)$.  Before closing this section, it is worth summarizing the operator sequence required to implement Ramsey spectroscopy which we do here in terms of an arbitrary initial state $\ket{\text{in}}$.  In the Schr\"{o}inger picture, this amounts to writing the final state in terms of the initial as 

\begin{equation}
	\ket{\psi_{f}\left(\phi\right)} = e^{-i\tfrac{\pi}{2}\hat{J}_{y}}e^{-i\phi\hat{J}_{z}}e^{-i\tfrac{\pi}{2}\hat{J}_{y}}\ket{\text{in}},
	\label{eqn:ccg_11}
\end{equation}

\noindent where once again $\phi = \left(\omega_{0}-\omega\right)T$.  It is worth noting that Ramsey spectroscopy is mathematically equivalent to optical interferometry in that both can be described through the Lie algebra of SU(2) (see Appendix \ref{app:secA}).  However, the transformations of Eq.~\ref{eqn:ccg_11} is slightly different from how we will describe an interferometer in Section \ref{sec: QOI}.  \\

\noindent Now we suppose that we have $N$ atoms prepared in a maximally entangled state of the form 

\begin{equation}
	\ket{\Psi_{M}} = \frac{1}{\sqrt{2}}\left(\ket{e}^{\otimes N} + \ket{g}^{\otimes N}\right) = \frac{1}{\sqrt{2}}\left(\ket{j,j}+\ket{j,-j}\right).
	\label{eqn:ccg_12}
\end{equation}

\noindent To implement the use of this state for Ramsey spectroscopy as described by the sequence of operators in Eq.~\ref{eqn:ccg_11}, we should take the actual initial state to be $\ket{\text{in}}=e^{+i\tfrac{\pi}{2}\hat{J}_{y}}\ket{\Psi_{M}}$.  After a free evolution time $T$ our state is now

\begin{equation}
\ket{\Psi_{M}}_{T} = \frac{1}{\sqrt{2}}\left(e^{-iN\phi/2}\ket{j,j} + e^{iN\phi/2}\ket{j,-j}\right).
\label{eqn:ccg_13}
\end{equation} 

\noindent After the second $\pi/2$ pulse, we arrive at the final state 

\begin{align}
&\ket{\Psi}_{t_{f}} = \frac{1}{\sqrt{2}}\left(e^{-iN\phi/2}e^{-i\tfrac{\pi}{2}\hat{J}_{y}}\ket{j,j} + e^{iN\phi/2}e^{-i\tfrac{\pi}{2}\hat{J}_{y}}\ket{j,-j}\right), \nonumber \\
&= \frac{1}{2^{j}\sqrt{2}}\sum_{m=-j}^{j}\binom{2j}{j+m}^{1/2}\left(e^{-iN\phi/2}+\left(-1\right)^{j+m}e^{iN\phi/2}\right)\ket{j,m},
\label{eqn:ccg_14}
\end{align}

\noindent where we have used the fact that $e^{-i\tfrac{\pi}{2}\hat{J}_{y}}\ket{j,\pm j}$ are atomic coherent states generated from different extremal, or fiducial, states $\ket{j,j}\;\text{and}\;\ket{j,-j}$ (see Appendix \ref{app:ACS_A1} and \ref{app:secB} for more detail). The expression of Eq.~\ref{eqn:ccg_14} is not particularly informative when it comes to the evaluation of $\braket{\hat{J}_{z}\left(\phi\right)}_{M}$ as it can be shown that 

\begin{align}
	\braket{\hat{J}_{z}\left(\phi\right)}_{t_{f}} &= {}_{T}\braket{\Psi_{M}|e^{i\tfrac{\pi}{2}\hat{J}_{y}}\hat{J}_{z}e^{-i\tfrac{\pi}{2}\hat{J}_{y}}|\Psi_{M}}_{T} \nonumber \\
	&= {}_{T}\braket{\Psi_{M}|\left(-\hat{J}_{x}\right)|\Psi_{M}}_{T} = 0.
	\label{eqn:ccg_15}
\end{align}

\noindent Thus this expectation value furnishes no information on the phase $\phi$.  \\

\noindent To address this issue, Bollinger \textit{et al.} \cite{Bollinger} proposed measuring the quantity $\left(-1\right)^{N_{e}}$ where $N_{e}$ is the number of atoms in the excited state.  As an operator for a given $j$ value, this reads $\hat{\Pi}_{j}=\left(-1\right)^{j+\hat{J}_{z}} = e^{i\pi\left(j+\hat{J}_{z}\right)}$.  This requires us to calculate 

\begin{align}
	\braket{\hat{\Pi}_{j}}_{t_{f}} &= {}_{T}\braket{\Psi_{M}|e^{i\tfrac{\pi}{2}\hat{J}_{y}}e^{i\pi\left(j+\hat{J}_{z}\right)}e^{-i\tfrac{\pi}{2}\hat{J}_{y}}|\Psi_{M}}_{T} \nonumber \\ 
	&= \left(-1\right)^{j}{}_{T}\braket{\Psi_{M}|e^{-i\pi\hat{J}_{x}}|\Psi_{M}}_{T} \nonumber \\
	&= \left(-1\right)^{j}{}_{T}\braket{\Psi_{M}|e^{i\tfrac{\pi}{2}\hat{J}_{z}}e^{-i\pi\hat{J}_{y}}e^{-i\tfrac{\pi}{2}\hat{J}_{z}}|\Psi_{M}}_{T},
	\label{eqn:ccg_16}
\end{align}

\noindent where we have used the relations $e^{i\tfrac{\pi}{2}\hat{J}_{y}}\hat{J}_{z}e^{-i\tfrac{\pi}{2}\hat{J}_{y}}=-\hat{J}_{x}$ and $e^{i\tfrac{\pi}{2}\hat{J}_{z}}\hat{J}_{y}e^{-i\tfrac{\pi}{2}\hat{J}_{z}}=\hat{J}_{x}$. Setting $\hat{U}=e^{i\tfrac{\pi}{2}\hat{J}_{z}}e^{-i\pi\hat{J}_{y}}e^{-i\tfrac{\pi}{2}\hat{J}_{z}}$ we have  

\begin{align}
    &\braket{\hat{\Pi}_{j}}_{t_{f}} =\left(-1\right)^{j}\frac{1}{2}\left[\braket{j,j|\hat{U}|j,j} + \braket{j,-j|\hat{U}|j,-j} +   \right. \nonumber \\
	&\;\;\;\;\;\;\;\;\; \left. + e^{iN\phi}\braket{j,j|\hat{U}|j,-j} + e^{-iN\phi} \braket{j,-j|\hat{U}|j,j}       \right].
	\label{eqn:ccg_17}
\end{align}

\noindent Then with

\begin{align}
	\braket{j,j|\hat{U}|j,j} &= d_{j,j}^{j}\left(\pi\right) = 0,\nonumber\\ 
	\braket{j,-j|\hat{U}|j,-j} &= d_{-j,-j}^{j}\left(\pi\right) = 0,\nonumber \\
	\label{eqn:ccg_18} \\
	\braket{j,j|\hat{U}|j,-j} =&\left(-1\right)^{j} d_{j,-j}^{j}\left(\pi\right) = \left(-1\right)^{j},\nonumber \\
	\braket{j,-j|\hat{U}|j,j} =&\left(-1\right)^{-j} d_{-j,j}^{j}\left(\pi\right) = \left(-1\right)^{j} \nonumber,
\end{align}

\noindent where $d_{m',m}^{j}\left(\beta\right)$ are the Wigner-$d$ matrix elements discussed in Appendix \ref{app:secB}, we finally have 

\begin{equation}
	\braket{\hat{\Pi}_{j}}_{t_{f}} = \left(-1\right)^{N}\cos\left[N\left(\omega_{0}-\omega\right)T\right].
	\label{eqn:ccg_19}
\end{equation}

\noindent Note the appearance of the factor of $N$ in the argument of the cosine.  This is the consequence of the maximal entanglement between the $N$ atoms, consequently leading to an increase in the sensitivity of the frequency measurement by a factor of $1/\sqrt{N}$ over the SQL.  Again, setting $\phi=\left(\omega_{0}-\omega\right)T$, the error propagation calculus in this case results in 

\begin{equation}
	\Delta\phi_{M} = \frac{\Delta\hat{\Pi}_{j}}{|\partial_{\phi}\braket{\hat{\Pi}_{j}}_{t_{f}}|} = \frac{1}{N,}
	\label{eqn:ccg_20}
\end{equation}

\noindent or that $\Delta\omega_{0}=1/NT$.  This is considered a Heisenberg-limited uncertainty as it scales with the inverse of $N$.  A scheme for the generation of the maximally entangled state of Eq.~\ref{eqn:ccg_12} was discussed by Bollinger \textit{et al.} \cite{Bollinger} and a different scheme was discussed by Steinbach and Gerry \cite{Steinbach}.  An experimental realization of parity-measurements has been performed by Leibfried \textit{et al.} \cite{Leibfried} with three trapped ions prepared in a maximally entangled state.  \\

\noindent In the next section we will briefly review the basics of phase estimation and make the connection between parity-based measurement and the minimum phase uncertainty: the quantum Cram\'{e}r-Rao bound.

\section{\label{sec:Phase_est} Phase estimation and the quantum Cram\'{e}r-Rao bound}

\noindent Phases cannot be measured, only approximated.  This is due to the fact that there exists no Hermitian phase operator \cite{PeggBarn}.  Consequently, within the realm of quantum mechanics, the phase is treated as a classical parameter rather than a quantum observable. The general approach to interferometry, be it optical or atomic, is to encode a suitable 'probe state' with a classically-treated phase and determine the optimal detection observable for estimating its value. The art lies in determining the best combination of probe state and detection observable that yields the highest resolution and smallest phase uncertainty.  In what follows, we will discuss some of the basics of phase estimation and arrive at a relation between parity-based detection and the upper-bound on phase estimation, which determines the greatest sensitivity afforded to a given quantum state.

\subsection{\label{sec:Phase_est_CRB} The Cram\'{e}r-Rao lower bound}

\noindent In the broadest sense, an interferometer is an apparatus that can transform an input 'probe state' $\hat{\rho}$ in a manner such that the transformation can be parametrized by a real, unknown, parameter $\varphi$.  A measurement is then performed on the output state $\hat{\rho}\left(\varphi\right)$ from which an estimation of the parameter $\varphi$ takes place.  The most general formulation of a measurement in quantum theory is a positive-operator valued measure (POVM).  A POVM consists of a set of non-negative Hermitian operators satisfying the unity condition $\sum_{\epsilon}\hat{E}\left(\epsilon\right) = 1$.  Following the work of Pezz\'{e} \textit{et al.} \cite{Pezze}, the conditional probability to observe the result $\epsilon$ for a given value $\varphi$, known as the likelihood, is

\begin{equation}
	P\left(\epsilon|\varphi\right) = \text{Tr}\left[\hat{E}\left(\epsilon\right)\hat{\rho}\left(\varphi\right)\right].
	\label{eqn:2.1}
\end{equation}

\noindent If the input state is made up of $m$ independent uncorrelated subsystems such that $\hat{\rho} = \hat{\rho}^{\left(1\right)}\otimes\hat{\rho}^{\left(2\right)}\otimes\hat{\rho}^{\left(3\right)}\otimes\hat{\rho}^{\left(4\right)}\otimes....\otimes\hat{\rho}^{\left(m\right)}$ and we restrict ourselves to local operations such that the phase $\varphi$ is encoded into each subsystem and assuming independent measurements are performed on each, then the likelihood function becomes the product of the single-measurement probabilities

\begin{equation}
	P\left(\epsilon|\varphi\right) = \prod_{i = 1}^{m}P_{i}\left(\epsilon_{i}|\varphi\right)
	\label{eqn:2.3}
\end{equation}

\noindent where $P_{i}\left(\epsilon_{i}|\varphi\right) = \text{Tr}\big[\hat{E}^{\left(i\right)}\left(\epsilon_{i}\right)\hat{\rho}^{\left(i\right)}\left(\varphi\right)\big]$.  For the case of independent measurements, as described in Eq.~\ref{eqn:2.3}, often one considers the log-Likelihood function

\begin{equation}
	L\left(\epsilon|\varphi\right) \equiv \text{ln}P\left(\epsilon|\varphi\right) = \sum_{i = 1}^{m}\text{ln}P_{i}\left(\epsilon_{i}|\varphi\right).
	\label{eqn:2.4}
\end{equation}

\noindent We define the estimator $\Phi\left(\epsilon\right)$ as any mapping of a given set of outcomes, $\epsilon$, onto parameter space in which an estimation of the phase is made.  A prevalent example is the maximum-likelihood estimator (MLE) \cite{Pezze}, defined as the phase value that maximizes the likelihood function 

\begin{equation}
	\Phi_{\text{MLE}}\left(\epsilon\right) = \text{arg}\Big[\max_{\varphi} \;P\left(\epsilon|\varphi\right)\Big].
	\label{eqn:2.4.1}
\end{equation}

\noindent An estimator can be characterized by its phase dependent mean value

\begin{equation}
	\braket{\Phi}_{\varphi} = 	\sum_{\epsilon}P\left(\epsilon|\varphi\right)\Phi\left(\epsilon\right),
	\label{eqn:2.5}
\end{equation}

\noindent and its variance

\begin{equation}
	\left(\Delta\Phi\right)_{\varphi}^{2} = \sum_{\epsilon}P\left(\epsilon|\varphi\right)\left[\Phi\left(\epsilon\right) - \braket{\Phi}_{\varphi}\right]^{2}.
	\label{eqn:2.6}
\end{equation}

\noindent We will now discuss what it means for an estimator to be 'good', which in this case, refers to an estimator that provides the smallest uncertainty.  These estimators are known as \textit{unbiased} estimators, and are defined as estimators whose average value coincides with the true value of the parameter in question, that is $\braket{\Phi\left(\epsilon\right)}_{\varphi} = \varphi$ is true for all values of the parameter $\varphi$ while estimators that are unbiased in the limit of $m\to\infty$, such as the MLE, are considered asymptotically consistent. Estimators not satisfying this condition are considered \textit{biased} while estimators that are unbiased for a certain range of the parameter $\varphi$ are considered \textit{locally unbiased}. \\

\noindent We now move on to perhaps one of the most important tools in the theory of phase estimation: the Cram\'{e}r-Rao bound (CRB).  The CRB serves to set a lower bound on the variance of any arbitrary estimator and is given formally as \footnote{
	The derivation of the CRB is straightforward. First, we have
	$\tfrac{\partial\braket{\Phi}_{\varphi}}{\partial\varphi} = \sum_{\epsilon}\partial_{\varphi}\,P(\epsilon|\varphi) \Phi(\epsilon)= 
	\braket{\Phi \tfrac{\partial L}{\partial\varphi}}$, where $L(\epsilon|\varphi) \equiv \ln P(\epsilon|\varphi)$.
	Noting that  
	$\sum_{\epsilon}\partial_{\varphi}\,P(\epsilon|\varphi) 
	= \braket{\tfrac{\partial L}{\partial\varphi}} = 0$ we have
	$\left( \tfrac{\partial\braket{\Phi}_{\varphi}}{\partial\varphi} \right)^2 
	= \braket{\big( \Phi - \braket{\Phi}_\varphi \big) \tfrac{\partial L}{\partial\varphi}}^2_\varphi$ 
	$\le \braket{\big( \Phi - \braket{\Phi}_\varphi \big)^2}_\varphi \braket{ \left(\tfrac{\partial L}{\partial\varphi} \right)^2}_\varphi$
	$= \left(\Delta \Phi\right)^2_\varphi \,F(\varphi)$, where we have invoked the Cauchy-Schwarz inequality
	$\braket{A B}^2_\varphi \le \braket{A^2}_\varphi\,\braket{B^2}_\varphi$.
	Dividing by $F(\varphi)$ yields the CRB in Eq~\ref{eqn:2.13}.
} % end footnote

\begin{equation}
\left(\Delta\Phi_{\text{CRB}}\right)^{2}_{\varphi} =  \dfrac{\left(\frac{\partial\braket{\Phi}_{\varphi}}{\partial\varphi}\right)^{2}}{F\left(\varphi\right)} \leq \left(\Delta\Phi\right)^{2}_{\varphi},
\label{eqn:2.13}
\end{equation}

\noindent where the quantity $F\left(\varphi\right)$ is the classical Fisher information (FI), given by

\begin{equation}
F\left(\varphi\right) \equiv \Big\langle \left(\dfrac{\partial L\left(\epsilon|\varphi\right)}{\partial\varphi}\right)^{2} \Big\rangle = \sum_{\epsilon}\dfrac{1}{P\left(\epsilon|\varphi\right)}\left(\dfrac{\partial P\left(\epsilon|\varphi\right)}{\partial\varphi}\right)^{2},
\label{eqn:2.14}
\end{equation}

\noindent where the sum extends over all possible values of the measurement values, $\epsilon$.  While Eq.~\ref{eqn:2.13} is the most general form the CRB, it is most useful for the cases of unbiased estimators where the numerator on the right-hand side, $\partial_{\varphi}\braket{\Phi}_{\varphi} = 1$.  For this case, the CRB is simply given as the inverse of the Fisher information $F\left(\varphi\right)$.  An estimator that saturates the CRB is said to be efficient.  The existence, however, of an efficient estimator depends on the properties of the probability distribution.  It is also worth noting that the derivation of the Fisher information of Eq.~\ref{eqn:2.14} assumed an initial state $\rho$ comprised of $m$-independent subsystems.  It is straight forward to show the additivity of the Fisher information $F\left(\varphi\right) = \sum_{i=1}^{m}F^{\left(i\right)}\left(\varphi\right)$ using Eq.~\ref{eqn:2.14} and plugging in Eq.~\ref{eqn:2.3} and \ref{eqn:2.4}, where $F^{\left(i\right)}\left(\varphi\right)$ is the Fisher information of the $i^{\text{th}}$ subsystem. For $m$ identical subsystems and measurements, this yields $F\left(\varphi\right) = mF^{\left(1\right)}\left(\varphi\right)$, where $F^{\left(1\right)}\left(\varphi\right)$ is the Fisher information for a single-measurement and $m$ is the total number of measurements. This is the form of the Fisher information most often used in the literature.  

\subsection{\label{sec:Phase_est_qCRB} Quantum Fisher information and the upper bound}

\noindent We now turn our attention towards discussing an upper bound \footnote{We follow the language found in the literature with regards to defining the upper and lower bounds on phase estimation; i.e. the CRB is the upper bound and the qCRB is the lower bound.} on phase estimation, known as the quantum Cram\'{e}r-Rao bound (qCRB), which in turn will be dependent on the quantum Fisher information $F_{Q}$ (QFI).  We obtain this upper bound by maximizing the FI over all possible POVMs \cite{QFI}\cite{QFI2},

\begin{equation}
F_{\text{Q}}\left[\hat{\rho}\left(\varphi\right)\right] \equiv \max_{\{\hat{E}\left(\epsilon\right)\}}F\left[\hat{\rho}\left(\varphi\right),\{\hat{E}\left(\epsilon\right)\}\right],
\label{eqn:2.16}
\end{equation}

\noindent where this quantity is known as the quantum Fisher information.  It is important to note that the quantum Fisher information is independent of the POVM used. This quantity can be expressed as \footnote{
	The QFI in Eq.~\ref{eqn:2.17} has a pleasing geometrical interpretation: it is the infinitesimal version of the quantum fidelity 
	$\mathcal{F}_Q(\hat{\rho}_1,\hat{\rho}_2) = \textrm{Tr}\left[\sqrt{\sqrt{\hat{\rho}_2}\, \hat{\rho}_1\sqrt{\hat{\rho}_2}}\right] $
	between two density matrices in the sense that 
	$\int_0^\zeta\sqrt{F_Q}(\zeta') d\zeta' = \sqrt{\mathcal{F}_Q}(\zeta)$ along the geodesic curve connecting $\hat{\rho}_1$ and  $\hat{\rho}_2$ parameterized by $\zeta$.
	This can be seen a follows: $\mathcal{F}_Q(\hat{\rho}_1,\hat{\rho}_2) = \textrm{max}_U \textrm{Tr}[\sqrt{\hat{\rho}_1} \sqrt{\hat{\rho}_2}] $
	$ =\textrm{Tr}[|\sqrt{\hat{\rho}_1} \sqrt{\hat{\rho}_2}\,|] $ = (where $|A| = \sqrt{A A^\dagger}$)
	over all purifications of the density matrices.
	A purification of a density matrix $\hat{\rho}_i$  is a pure state $\ket{\psi_i} =(U^{(f)}\otimes\sqrt{\hat{\rho}_i}) \ket{\Gamma}$ where $\ket{\Gamma}=\sum_k \ket{k}\otimes\ket{k}$ \cite{Wilde:2017} such that $\textrm{Tr}_f[\ket{\psi_i}\bra{\psi_i} ]= \hat{\rho}_i$. We can think of this as a fibre bundle where the base space is the space of all positive Hermitian operators, not necessarily of unit trace (the positive cone), and sitting above each (un-normalized) quantum state $\hat{\rho}_i$ is the vector space (fibre, $f$) of its purifications, which as operators can be represented the vector $A_i=\sqrt{\hat{\rho}_i}$. The arbitrary unitary $U^{(f)}$ is the freedom to move the vector $A_i$ around in the fibre. Now the fidelity is given as 
	$\sqrt{\mathcal{F}_Q}(\zeta) = \textrm{max}|\textrm{Tr}[A_1 A^\dagger_2]|$ over all purifications 
	(i.e. over all $U^{(f)}_1 U^{\dagger(f)}_2)$.  
	The \textit{Bures angle} $d_B$ is given by 
	$\cos\left[d_B(\hat{\rho}_1,\hat{\rho}_2)\right] = \sqrt{\mathcal{F}_Q}(\hat{\rho}_1,\hat{\rho}_2)$
	is the length of the geodesic curve within the subspace of (unit trace) density matrices connecting $\hat{\rho}_1$ and $\hat{\rho}_2$.
	The infinitesimal version of this is given by the \textit{Bures metric} \cite{GeomQStates:2006} 
	$ds^2_B =  \textrm{Tr}[dA_1 dA^\dagger_2] = \tfrac{1}{4} \textrm{Tr}[d\hat{\rho}(\zeta) L_\zeta] =
	\tfrac{1}{4} \textrm{Tr}[\hat{\rho}(\zeta) L^2_\zeta] = F_Q(\zeta) d\zeta^2$, where the last expression is just Eq.~\ref{eqn:2.17}.
	This last expression could also be interpreted as the speed $ds_B/d\zeta = \sqrt{F_Q}(\zeta)$ along the geodesic connecting the two quantum states 
	$\hat{\rho}_1$ and $\hat{\rho}_2$ (our \textit{input} and \textit{output} states along which $\zeta$ varies) is governed by the (square root) of the QFI. Note further that for pure states 
	$\tfrac{1}{4}F_Q(\zeta) =\left[ \braket{\partial_\zeta\psi|\partial_\zeta\psi} - |\braket{\psi|\partial_\zeta\psi}|^2\right] =
	\textrm{Tr}[\ket{\partial_\zeta\psi}\bra{\partial_\zeta\psi}\, P_\perp]$ where $P_\perp = I - \ket{\psi}\bra{\psi}$ is the projector onto states perpendicular  
	to $\ket{\psi}$. 
	Thus $\ket{\nabla_\zeta\,\psi}\equiv P_\perp\,\ket{\partial_\zeta\psi}$ $=\ket{\partial_\zeta\psi} - \ket{\psi} \braket{\psi|\partial_\zeta \psi} $   is the intrinsic
	\emph{covariant derivative} \cite{Provost_Vallee:1980} pointing across fibres that is horizontal (tangent) to the base (parameter, $\zeta$) space, 
	with $\braket{\psi|\partial_\zeta \psi}$ the U(1) connection (in the complex Hermitian line bundle) \cite{Ben-Aryeh:2004,GeomQStates:2006,Frankel_3rdEd:2012}. 
	The QFI is just the norm of this covariant derivative, $\tfrac{1}{4}F_Q(\zeta) = ||\ket{\nabla_\zeta\,\psi}||^2 = \braket{\nabla_\zeta\,\psi|\nabla_\zeta\,\psi}$. 
	From the discussion after 
	Eq.~\ref{eqn:2.50} with the parity operator considered as a unitary evolution operator 
	$\ket{\psi(\phi)}=\hat{\Pi}_b^{(\phi)}\ket{\psi(0)} = e^{-i\phi\,\hat{n}_b}\ket{\psi(0)}$ we obtain 
	$\ket{\nabla_\phi\psi} = -i(\hat{n}_b - \bar{n}_b)\ket{\psi(\phi)}$, with $ \bar{n}_b = \bra{\psi(\phi)} \hat{n}_b\ket{\psi(\phi)}$, and 
	$F_{\Pi_b}(\phi) = 4\,||\ket{\nabla_\phi\,\psi}||^2=4\,\braket{(\Delta \hat{n}_b)^2}_\phi$ as before.
	Note, these geometric concepts can be extended to multiparameter estimation \cite{Guo:2016}, 
	where $H = H(\zeta_1,\zeta_2,\ldots)$ and
	the Quantum Geometric Tensor $Q_{ij} = \braket{\partial_i \psi|P_\perp|\partial_i \psi}$  
	$=\braket{\partial_i H\,\partial_j H} - \braket{\partial_i H}\braket{\partial_j H}$ (where $\partial_i = \partial_{\zeta_i}$)
	takes central role \cite{QFI}, with the unifying properties that 
	$\textrm{Re}(Q_{ij}) = \tfrac{1}{4} \mathcal{F}_{QFI}= Cov(H_i,H_j) =  \braket{\tfrac{1}{2}\{H_i\,H_j\}} - \braket{H_i}\braket{H_j}$ 
	are the elements of 	the QFI Matrix \cite{QFI}, and 
	$\textrm{Im}(Q_{ij})=-\tfrac{1}{2} \Omega_{ij} =
	-\tfrac{1}{4} \left(\braket{\partial_i \psi | \partial_j \psi } - \braket{ \partial_j \psi  |  \partial_i \psi }\right)$
	are the Berry (Phase) Curvatures \cite{Samuel_Bhandari:1988,BerryBook,Ben-Aryeh:2004}.
}% end footnote

\begin{equation}
F_{\text{Q}}\left[\hat{\rho}\left(\varphi\right)\right] = \text{Tr}\left[\hat{\rho}\left(\varphi\right)\hat{L}_{\varphi}^{2}\right],
\label{eqn:2.17}
\end{equation}

\noindent where $\hat{L}_{\varphi}$ is known as the symmetric logarithmic derivative (SLD) \cite{SLD} defined as the solution to the equation

\begin{equation}
\frac{\partial\hat{\rho}\left(\varphi\right)}{\partial\varphi} = \frac{\hat{\rho}\left(\varphi\right)\hat{L}_{\varphi} + \hat{L}_{\varphi}\hat{\rho}\left(\varphi\right)}{2}.
\label{eqn:2.18}
\end{equation}

\noindent The chain of inequalities is now

\begin{equation}
\left(\Delta\Phi\right)^{2}_{\varphi} \geq \left(\Delta\Phi\right)^{2}_{\text{CRB}} \geq \left(\Delta\Phi\right)^{2}_{\text{qCRB}},
\label{eqn:2.19}
\end{equation}

\noindent where it follows that the quantum Cram\'{e}r-Rao bound (also known as the Helstrom bound \cite{Helstrom})  is given by

\begin{equation}
\left(\Delta\Phi\right)^{2}_{\text{qCRB}} \equiv \frac{\left(\frac{\partial\braket{\Phi}}{\partial\varphi}\right)^{2}}{F_{\text{Q}}\left[\hat{\rho}\left(\varphi\right)\right]}.
\label{eqn:2.20}
\end{equation}

\noindent Since the qCRB is inversely proportional to the QFI and the QFI itself is a maximization over all possible POVMs, it is clear to see how the qCRB serves as an upper bound on phase estimation.

\subsubsection{\label{sec:Phase_est_qCRB_Fish} Calculating the QFI for pure and mixed states}

\noindent Here we work through a suitable expression for the QFI, using our definition of the SLD given in Eq.~\ref{eqn:2.18}, in terms of the complete basis $\{\ket{n}\}$, where our density operator is now given generally as $\hat{\rho}\left(\varphi\right) = \sum_{n}p_{n}\ket{n}\bra{n}$.   Following the work of Pezz\'{e} \textit{et al.} \cite{Pezze}, the quantum Fisher information can be written in this basis as

\begin{align}
F_{\text{Q}}\left[\hat{\rho}\left(\varphi\right)\right] &= \text{Tr}\big[\hat{\rho}\left(\varphi\right)\hat{L}^{2}_{\varphi}\big] = \sum_{k,\;k'}p_{k}|\langle k|\hat{L}_{\varphi}|k' \rangle|^{2}\nonumber \\
&= \sum_{k,\;k'}\frac{p_{k} + p_{k'}}{2}\times|\langle k|\hat{L}_{\varphi}|k' \rangle|^{2}.
\label{eqn:2.30}
\end{align}

\noindent Thus it is sufficient to know the matrix elements of the SLD, $\langle k|\hat{L}_{\varphi}|k'\rangle$ in order to calculate the QFI.  Using Eq.~\ref{eqn:2.18} and our general density operator, it is easy to show

\begin{equation}
\langle k|\hat{L}_{\varphi}|k'\rangle = \left(\frac{2}{p_{k} + p_{k'}}\right)\times\langle k|\partial_{\varphi}\hat{\rho}\left(\varphi\right)|k'\rangle,
\label{eqn:2.31}
\end{equation}

\noindent which makes Eq.~\ref{eqn:2.30} \cite{Hubner}

\begin{equation}
F_{\text{Q}}\left[\hat{\rho}\left(\varphi\right)\right] = \sum_{k,\;k'}\frac{2}{p_{k} + p_{k'}}\times |\langle k|\partial_{\varphi}\hat{\rho}\left(\varphi\right)|k'\rangle|^{2}.
\label{eqn:2.32}
\end{equation}

\noindent We proceed further through the use of the definition

\begin{equation}
\partial_{\varphi}\hat{\rho}\left(\varphi\right) = \sum_{k}\left(\partial_{\varphi}p_{k}\right)\ket{k}\bra{k} + \sum_{k}p_{k}\ket{\partial_{\varphi}k}\bra{k} + \sum_{k}p_{k}\ket{k}\bra{\partial_{\varphi}k},
\label{eqn:2.33}
\end{equation}

\noindent which is a simple application of the chain rule for derivatives.  Using the identity $\partial_{\varphi}\langle k|k'\rangle = \langle \partial_{\varphi}k|k'\rangle + \langle k|\partial_{\varphi}k'\rangle \equiv 0$, the matrix elements in Eq.~\ref{eqn:2.32} become

\begin{equation}
\langle k|\partial_{\varphi}\hat{\rho}\left(\varphi\right)|k'\rangle = \left(\partial_{\varphi}\hat{\rho}\left(\varphi\right)\right)\delta_{k,\;k'} + \left(p_{k} - p_{k'}\right)\langle\partial_{\varphi}k|k'\rangle.
\label{eqn:2.34}
\end{equation}

\noindent The SLD and QFI then become

\begin{eqnarray}	
&\hat{L}_{\varphi} = \sum_{k}\frac{\partial_{\varphi}p_{k}}{p_{k}}\times\ket{k}\bra{k} + 2\sum_{k,\;k'}\frac{p_{k} - p_{k'}}{p_{k} + p_{k'}}\times\langle\partial_{\varphi}k|k'\rangle\ket{k'}\bra{k},\nonumber\\
\\
&F_{\text{Q}}\left[\hat{\rho}\left(\varphi\right)\right] = \sum_{k}\frac{\left(\partial_{\varphi}p_{k}\right)^{2}}{p_{k}} + 2\sum_{k,\;k'}\frac{\left(p_{k} - p_{k'}\right)^{2}}{p_{k} + p_{k'}}\times|\langle\partial_{\varphi}k|k'\rangle|^{2},\nonumber
\label{eqn:2.35}
\end{eqnarray}

\noindent respectively.  These results, we show next, simplify in the case of pure states where we can write $\hat{\rho}\left(\varphi\right) = \ket{\psi\left(\varphi\right)}\bra{\psi\left(\varphi\right)}$ and consequently $\partial_{\varphi}\hat{\rho}\left(\varphi\right) = \partial_{\varphi}\hat{\rho}^{2}\left(\varphi\right) = \hat{\rho}\left(\varphi\right)\big[\partial_{\varphi}\hat{\rho}\left(\varphi\right)\big] + \big[\partial_{\varphi}\hat{\rho}\left(\varphi\right)\big]\hat{\rho}\left(\varphi\right)$.  Using this, and a cursory glance at Eq.~\ref{eqn:2.18}, it is clear the SLD becomes

\begin{align}
\hat{L}_{\varphi} &= 2\big[\partial_{\varphi}\hat{\rho}\left(\varphi\right)\big] = 2\big[\partial_{\varphi}\ket{\psi\left(\varphi\right)}\bra{\psi\left(\varphi\right)}\big]\nonumber\\
&= 2\ket{\partial_{\varphi}\psi}\bra{\psi} + 2\ket{\psi}\bra{\partial_{\varphi}\psi},
\label{eqn:2.36}
\end{align} 

\noindent where in the last step, the $\varphi$-dependency of $\ket{\psi}$ is implicit for notational convenience.  Plugging this directly into the first line of Eq.~\ref{eqn:2.30} yields

\begin{align}
F_{\text{Q}}\left[\hat{\rho}\left(\varphi\right)\right] &= \text{Tr}\big[\hat{\rho}\left(\varphi\right)\hat{L}^{2}_{\varphi}\big] = \langle\psi|\hat{L}^{2}_{\varphi}|\psi\rangle \nonumber \\
&= 4\{\langle\partial_{\varphi}\psi|\partial_{\varphi}\psi\rangle - |\langle\partial_{\varphi}\psi|\psi\rangle|^{2}\}, 
\label{eqn:2.37}
\end{align}

\noindent  which is the form of the QFI most often used in quantum metrology literature.  Next we move on to discussing a specific detection observable: photon number parity.  

\subsubsection{\label{sec:Phase_est_QCRB_parity} Connection to parity-based detection}

\noindent The central theme discussed throughout this paper is the use of the quantum mechanical parity operator as a detection observable in quantum optical interferometry.  The use of parity as a detection observable first came about in conjunction with high precision spectroscopy, by Bollinger \textit{et al.} \cite{Bollinger}.  It was first adapted and formally introduced for use in quantum optical interferometry by C. C. Gerry \textit{et al.} \cite{GerryParityAgain}\cite{GerryParity2}. A detection observable is said to be optimal if for a given state, the CRB achieves the qCRB, that is,

\begin{equation}
\left(\Delta\Phi\right)_{\varphi}^{2} \geq \left(\Delta\Phi\right)_{\text{CRB}}^{2} = \left(\Delta\Phi\right)_{\text{qCRB}}^{2}.
\label{eqn:2.38}
\end{equation}

\noindent Furthermore, parity detection achieves maximal phase sensitivity at the qCRB for all pure states that are path symmetric \cite{PathSym1} \cite{PathSym2}. For the purposes of this paper, it is sufficient to derive the classical Fisher information. We start from the expression for the classical Fisher information, assuming a single measurement is performed, given in Eq.~\ref{eqn:2.14}.  For parity \cite{PathSym}, the measurement outcome $\epsilon$ can either be positive $+$ or negative $-$ and satisfies $P\left(+|\varphi\right) + P\left(-|\varphi\right) \equiv 1$.  The expectation value of the parity operator can then be expressed as a sum over the possible eigenvalues weighted with the probability of that particular outcome leading to 
\begin{align}
\langle\hat{\Pi}\rangle &= \sum_{i}P\left(i|\varphi\right)\lambda_{i} = P\left(+|\varphi\right) - P\left(-|\varphi\right)\nonumber\\
&= 2P\left(+|\varphi\right) - 1 = 1 - 2P\left(-|\varphi\right).
\label{eqn:2.40}
\end{align}

\noindent Likewise, we can calculate the variance

\begin{align}
\left(\Delta\hat{\Pi}\right)^{2} &= \langle\hat{\Pi}^{2}\rangle - \langle\hat{\Pi}\rangle^{2} = 1 - \langle\hat{\Pi}\rangle^{2} \nonumber\\
&= 1 - \left(P\left(+|\varphi\right) - P\left(-|\varphi\right)\right)^{2} \nonumber \\
&= 1 - \left(P\left(+|\varphi\right) + P\left(-|\varphi\right)\right)^{2} + 4P\left(+|\varphi\right)P\left(-|\varphi\right)\nonumber \\
&= 4P\left(+|\varphi\right)P\left(-|\varphi\right).
\label{eqn:2.41}
\end{align}

\noindent Finally, from Eq.~\ref{eqn:2.40}, it follows that

\begin{equation}
\frac{\partial P\left(+|\varphi\right)}{\partial\varphi} = \frac{1}{2}\frac{\partial\langle\hat{\Pi}\rangle}{\partial\varphi} = -\frac{\partial P\left(-|\varphi\right)}{\partial\varphi}.
\label{eqn:2.42}
\end{equation}

\noindent Combining Eq.~\ref{eqn:2.41} and \ref{eqn:2.42} we find

\begin{align}
F^{\left(1\right)}\left(\varphi\right) &= \sum_{\epsilon_{1}}\dfrac{1}{P\left(\epsilon_{1}|\varphi\right)}\left(\dfrac{\partial P\left(\epsilon_{1}|\varphi\right)}{\partial\varphi}\right)^{2} \nonumber \\
&= \frac{1}{\left(\Delta\hat{\Pi}\right)^{2}}\Bigl\lvert\frac{\partial\langle\hat{\Pi}\rangle}{\partial\varphi}\Bigr\rvert^{2},
\label{eqn:2.43}
\end{align} 

\noindent making the CRB / qCRB 

\begin{equation}
\Delta\Phi_{\text{CRB/qCRB}} =  \frac{\sqrt{1 - \langle\hat{\Pi}\rangle^{2}}}{\bigl\lvert \partial_{\varphi}\langle\hat{\Pi}\rangle \bigr\rvert} = \frac{1}{\sqrt{F^{\left(1\right)}\left(\varphi\right)}} .
\label{eqn:2.44}
\end{equation}

\noindent Eq.~\ref{eqn:2.44} is simply the phase uncertainty obtained through the error propagation calculus. This is advantageous over other means of detection, such as photon-number counting, because the use of parity does not require any pre- or post- data processing.  By comparison, photon number counting typically works by construction of a phase probability distribution conditioned on the outcome of a sequence of $m$ measurements \cite{Uys}.  After a sequence of detection events, the error in the phase estimate is determined from this distribution.  While this provides phase estimation at the qCRB, it lacks the advantage of being directly determined from the signal, unlike the use of parity \cite{PathSym}. There are disadvantages to using parity, however.  Performance of photon number parity is highly susceptible to losses.  Parity also achieves maximal phase sensitivities at particular values of the phase, restricting its use to estimating local phases \footnote{The exception being the optical $N00N$ state, for which parity is optimal for all values of the phase.}. Restricting its use to local parameter estimation, however, is not terribly problematic in interferometry, as one is interested in measuring small changes to parameters that are more-or-less known.    \\

\noindent It is worth pointing out that the optimal POVM depends, in general, on $\varphi$.  This is somewhat problematic as it requires one to \textit{already know} the value of the parameter $\varphi$ in order to choose an optimal estimator.  Some work has been done to overcome this obstacle \cite{QFI3} which concludes the QFI can be asymptotically obtained in a number of measurements without any knowledge of the parameter.  For all cases considered throughout this paper, we will use parity as our detection observable (except in cases where we wise to draw comparisons between observables), which we know saturates the qCRB.  We will now move on to discuss how one calculates the QFI in quantum optical interferometry.

\subsection{\label{sec:Phase_est_QCRB_QOI} Calculating the QFI in quantum optical interferometry}

\noindent We use the Schwinger realization of the su(2) algebra with two sets of boson operators, discussed in detail in Appendix \ref{app:subsecA2}, to describe a standard Mach-Zehnder interferometer \cite{Yurke}.  In this realization, the quantum mechanical beam splitter can be viewed as a rotation about a given (fictitious) axis determined by the choice of angular momentum operator, i.e. the choice of a $\hat{J}_{x}$-operator performs a rotation about the $x$-axis while the choice of a $\hat{J}_{y}$-operator performs a rotation about the $y$-axis.  An induced phase shift, assumed to be in the $b$-mode, is described by a rotation about the $z$-axis described by the use of the $\hat{J}_{z}$-operator.  The state just before the second beam splitter is given as

\begin{equation}
\ket{\psi\left(\varphi\right)} = e^{-i\varphi\hat{J}_{z}}e^{-i\frac{\pi}{2}\hat{J}_{x}}\ket{\text{in}},
\label{eqn:2.45}
\end{equation}

\noindent where we are assuming the beam splitters to be 50:50.  This in turn makes the derivative 

\begin{equation}
\ket{\psi'\left(\varphi\right)} = -i e^{-i\varphi\hat{J}_{z}}\hat{J}_{z}e^{-i\frac{\pi}{2}\hat{J}_{x}}\ket{\text{in}},
\label{eqn:2.46}
\end{equation}

\noindent leading to

\begin{equation}
\big\langle \psi'\left(\varphi\right)|\psi\left(\varphi\right) \big\rangle = i\braket{\text{in}|\hat{J}_{y}|\text{in}},
\label{eqn:2.47}
\end{equation}

\noindent and

\begin{equation}
\big\langle \psi'\left(\varphi\right)|\psi'\left(\varphi\right) \big\rangle = \braket{\text{in}|\hat{J}^{\;2}_{y}|\text{in}},
\label{eqn:2.48}
\end{equation}

\noindent where we have made use of the Baker-Hausdorf identity in simplifying

\begin{equation}
e^{i\frac{\pi}{2}\hat{J}_{x}}\hat{J}_{z}e^{-i\frac{\pi}{2}\hat{J}_{x}} = \hat{J}_{y}.
\label{eqn:2.49}
\end{equation}

\noindent Combining Eq.~\ref{eqn:2.47} and \ref{eqn:2.48} into Eq.~\ref{eqn:2.37} yields for the QFI

\begin{align}
F_{\text{Q}}\left[\hat{\rho}\left(\varphi\right)\right] &= 4\{\langle\partial_{\varphi}\psi|\partial_{\varphi}\psi\rangle - |\langle\partial_{\varphi}\psi|\psi\rangle|^{2}\}\nonumber \\
&= 4\big\langle \big( \Delta\hat{J}_{y} \big)^{2} \big\rangle_{\text{in}},
\label{eqn:2.50}
\end{align}

\noindent which is simply the variance of the $\hat{J}_{y}$-operator with respect to the initial input state $\ket{\text{in}}$.  This is the form of the QFI used in all of the following interferometric calculations.  One important thing to notice is that the quantum Fisher information depends solely on the initial state and not on the value of the phase $\varphi$ to be measured.  \\

\noindent Note that Eq.~\ref{eqn:2.50} is a general result. Let $\mathcal{\hat{O}}$ be a generator of a \textit{flow} parameterized by $\zeta$ 
such that the wave function evolves according to the Schr\"{o}dinger equation  $i\partial_\zeta\ket{\psi} = \mathcal{\hat{O}}\ket{\psi}$ with solution
$\ket{\psi(\zeta)} = e^{-i \hat{\mathcal{O}} \zeta}\ket{\psi(0)}$. Then $\ket{\partial_\zeta\psi} = -i \hat{\mathcal{O}} \ket{\psi}$
so that $\braket{\partial_\zeta\psi|\partial_\zeta\psi} =  \braket{\hat{\mathcal{O}}^2}$, 
and $|\braket{\psi|\partial_\zeta\psi}|^2 =  \braket{\hat{\mathcal{O}}}^2$.
Hence $F_Q(\zeta) =4 \left[ \braket{\partial_\zeta\psi|\partial_\zeta\psi} - |\braket{\psi|\partial_\zeta\psi}|^2\right] 
= 4 \braket{\left( \Delta\hat{\mathcal{O}} \right)^2}_\zeta$.
Thus, both Eq.~\ref{eqn:2.13} and Eq.~\ref{eqn:2.20} yield a Fourier-like uncertainty relation for unbiased estimators of the form
$\left( \Delta\Phi \right)_\zeta \,  \braket{\left( \Delta\hat{\mathcal{O}} \right)}_\zeta \ge 1/2$ (if we do not use the maximum FI)  
reminiscent of $\Delta x\, \Delta k \ge 1/2$, so that a precise measurement $\left( \Delta\Phi \right)_\zeta\ll 1/2$ of $\zeta$ requires a large uncertainty (variance)  $\braket{\left( \Delta\hat{\mathcal{O}} \right)}_\zeta\gg 1/2$ in its generation.
Noting that $\hat{\Pi}_b = (-1)^{\hat{b}^\dagger\hat{b}} = e^{i\pi\,\hat{n}_b}$ is both unitary as well as Hermitian, 
define $\hat{\Pi}_b^{(\phi)} \equiv e^{-i\phi\,\hat{n}_b}$ with $\hat{\mathcal{O}}\to\hat{n}_b$ and $\zeta\to\phi$ (which we take to be $\phi=-\pi$ for the parity operator). 
Thus, a measurement $\hat{\Pi}_b^{(\phi)}\ket{\psi(0)} = e^{-i\phi\,\hat{n}_b}\ket{\psi(0)}$  can be thought of as a Schr\"{o}dinger-like evolution in $\phi$ with 
generator $\hat{n}_b$.
Then, from above 
$F_{\Pi_b}(\phi) =  4 \braket{\left( \Delta\hat{n}_b\right)^2}_\phi$. This result is independent of the parameter $\phi$.
This leads to the insightful interpretation of Eq.~\ref{eqn:2.44}
as $\Delta\Phi_{\text{CRB/qCRB}}\,\Delta \hat{n}_b = \tfrac{1}{2}$, i.e. a more formal statement of the classical 
number-phase uncertainty relationship $\Delta\phi\,\Delta N\ge \tfrac{1}{2}$. In fact, we now see that the parity operator $\hat{\Pi}_b$
leads to the \textit{minimal} phase uncertainty relationship possible, since it saturates the inequality. \\

\noindent Next we will discuss the phase uncertainty and measurement resolutions obtained using parity-based detection in interferometry for a number of cases in which input classical and/or quantum mechanical states of light are considered.

\section{\label{sec: QOI} Quantum Optical Interferometry}

\begin{figure}
	\centering
	\includegraphics[width=1.0\linewidth,keepaspectratio]{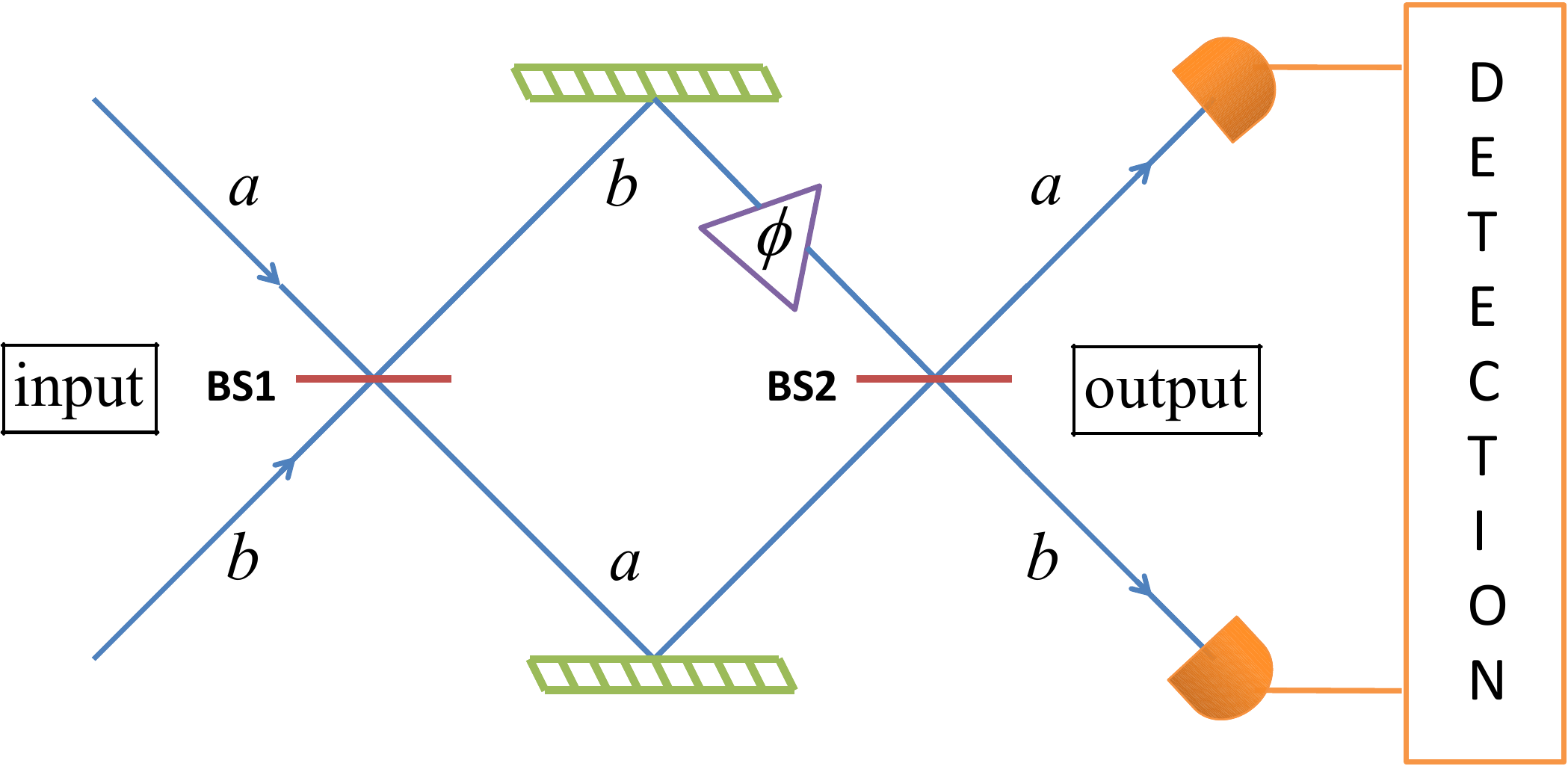}
	\caption{A sketch of a standard Mach-Zehnder interferometer (MZI). Light propagates through one (or both) input ports of the MZI. A phase shift occurs after the first beam splitter due to a relative path length difference between the two arms of the MZI. Measurement occurs on one (or both) modes of the output.}
	\label{fig:MZI}
\end{figure}

\noindent In this section we highlight several different interferometric schemes involving both classical and non-classical states of light comparing the use of several different detection observables. Here we show that in cases where the bound on phase sensitivity is not saturated, parity-based detection yields sub-SQL limited phase sensitivity and can, in certain cases, approach or out-perform the HL of phase sensitivity.  We also draw attention to the correlation between parity-based detection and the saturation of the qCRB.  Before we get into certain cases, we will provide the reader with a concise derivation of the output state of an interferometer for arbitrary initial states as well as the average value of an arbitrary detection observable and the subsequent phase uncertainty. \\

\noindent Consider an interferometer transforming an initial state $\ket{\text{in}}$ according to 

\begin{equation}
	\ket{\text{out}} = e^{i\tfrac{\pi}{2}\hat{J}_{x}}e^{-i\phi\hat{J}_{z}}e^{-i\tfrac{\pi}{2}\hat{J}_{x}}\ket{\text{in}} = e^{-i\phi\hat{J}_{y}}\ket{\text{in}},
	\label{eqn:qoi_1}
\end{equation}

\noindent where we employ the Schwinger realization of the SU(2) Lie algebra (see Appendix \ref{app:subsecA2}). It is worth comparing the expression for the output state of Eq.~\ref{eqn:qoi_1} with that of the final state obtained when one performs Ramsey spectroscopy, as shown in Eq.~\ref{eqn:ccg_11}. The $\pi/2$ pulses performed on atomic states is analogous to a beam splitter transformation affecting two boson modes in that both can be described in terms of the su(2) Lie algebra. The same can be said of how the phase $\phi$ is encoded in both procedures, though they have very different physical interpretations.  For atomic systems, the phase shift arises during the period of free evolution while in interferometry it stems from a relative path length difference between the two arms of the interferometer.  \\

\noindent For the most general of separable initial states, $\ket{\text{in}} = \ket{\psi_{0}^{\left(1\right)}}_{a}\otimes\ket{\psi_{0}^{\left(2\right)}}_{b}$, where in the photon number basis $\ket{\psi_{0}^{\left(i\right)}}=\sum_{n=0}^{\infty}C_{n}^{\left(i\right)}\ket{n}$, $n \in \mathbb{Z}^{+}$, the input state can be expressed as

\begin{align}
	\ket{\text{in}} &= \sum_{n=0}^{\infty}\;\;\sum_{q=0}^{\infty}C_{n}^{\left(1\right)}C_{q}^{\left(2\right)}\ket{n}_{a}\otimes\ket{q}_{b} \nonumber \\
	&= \sum_{j=0,1/2,..}^{\infty}\sum_{m=-j}^{j}C_{j+m}^{\left(1\right)}C_{j-m}^{\left(2\right)}\ket{j,m}.
	\label{eqn:initial}
\end{align}

\noindent Working in the Schr\"{o}dinger picture, the transformation of Eq.~\ref{eqn:qoi_1} acting on this initial state yields the output state

\begin{align}
	\ket{\text{out}} &= \sum_{j'=0,1/2,..}^{\infty}\sum_{m'=-j'}^{j'}\braket{j',m'|e^{-i\phi\hat{J}_{y}}|\text{in}}\;\ket{j',m'} \nonumber \\
	&= \sum_{j=0,1/2,..}^{\infty}\sum_{m'=-j}^{j} \Gamma_{j,m'}\left(\phi\right) \ket{j+m'}_{a}\otimes\ket{j-m'}_{b},
	\label{eqn: qoi_2}
\end{align}

\noindent where we have inserted a complete set of states in the angular momentum basis and where the phase-dependent state coefficients are given by 

\begin{equation}
	\Gamma_{j,m'}\left(\phi\right) = \sum_{m=-j}^{j}C_{j+m}^{\left(1\right)}C_{j-m}^{\left(2\right)} d_{m',m}^{j}\left(\phi\right).
	\label{eqn:qoi_3}
\end{equation}

\noindent The phase-dependent term in Eq.~\ref{eqn:qoi_3} are the well known Wigner-$d$ matrix elements discussed in some detail in Appendix \ref{app:subsecB1}.  Note that if our initial state is an entangled two-mode state, the corresponding coefficients would be of the form $A_{n,p}$ where $A_{n,p} \neq C_{n}^{\left(1\right)}C_{p}^{\left(2\right)}$.  For an arbitrary detection observable $\hat{\mathcal{O}}$, the expectation value can be calculated directly as 

\begin{equation}
	\braket{\hat{\mathcal{O}}}_{\text{out}} = \sum_{j,m'}\sum_{J,M} \Gamma_{J,M}^{*}\left(\phi\right)\Gamma_{j,m'}\left(\phi\right) \braket{J,M|\hat{\mathcal{O}}|j,m'}.
	\label{eqn:qoi_4}
\end{equation}

\noindent From this the phase uncertainty can be found through use of the usual error propagation calculus to be 

\begin{equation}
	\Delta\phi_{\hat{\mathcal{O}}} = \frac{\Delta\mathcal{\hat{O}}_{\text{out}}}{|\partial_{\phi} \braket{\hat{\mathcal{O}}}_{\text{out}} |}.
	\label{eqn:qoi_5}
\end{equation}

\noindent From the perspective of the Heisenberg picture, the transformed observable $\mathcal{\hat{O}}'$ is given by $\mathcal{\hat{O}}'=e^{i\phi\hat{J}_{y}}\mathcal{\hat{O}}e^{-i\phi\hat{J}_{y}}$ and the derivative of its expectation value 

\begin{align}
	\partial_{\phi}\braket{\hat{\mathcal{O}}}_{\text{out}}=\partial_{\phi}\braket{\hat{\mathcal{O}}'}_{\text{in}} = i\braket{\left[\hat{J}_{y},\mathcal{\hat{O}}'\right]}_{\text{in}}.
	\label{eqn:qoi_5.1}
\end{align}

\noindent We remind the reader that the greatest phase sensitivity afforded by classical states is the standard quantum limit (SQL),  $\Delta\phi_{\text{SQL}} = 1/\sqrt{\bar{n}}$ while for quantum states the phase sensitivity is bounded by the Heisenberg limit (HL) $\Delta\phi_{\text{HL}} = 1/\bar{n}$.  We note that while the HL serves as a bound on phase sensitivity, it \textit{has} been demonstrated to be beaten by some quantum states for low (but still $>1$) average photon numbers \cite{AnisimovTMSVS}.  The goal of this section is to highlight the effectiveness of parity-based measurement performed on one of the output ports.  Unless otherwise stated, we assume measurement is performed on the output $b$-mode without loss of generality \footnote{Generally speaking, the expectation value of the parity operator for each output mode is related by a phase shift.  Often the output mode for which the parity expectation value peaks at $\phi=0$ is considered.  However, either output mode is suitable with the right choice of  phase-shifter.}.  We are now ready to discuss several different cases.  \\

\subsection{\label{sec:QOI_N00N} The N00N states}

\noindent The optical $N00N$ state has been extensively studied for use in high-precision quantum metrology \cite{GerryParity2}  \cite{Dowling1} \cite{Dowling2} \cite{Dowling3}.  It is defined as the superposition state in which $N$ photons are in one mode (labeled the $a$-mode) while none are in the either ($b$-mode) $\ket{N,0}_{a,b}$ and where no photons are in the $a$-mode and $N$ photons are in the $b$-mode $\ket{0,N}_{a,b}$, and can be written generally as  

\begin{equation}
	\ket{\psi}_{N00N} = \frac{1}{\sqrt{2}}\left(\ket{N,0} + e^{i\Phi_{N}}\ket{0,N}\right),
	\label{eqn:N00N_1}
\end{equation}

\noindent where $\Phi_{N}$ is a relative phase factor that may depend on $N$ and whose value will generally depend on the method of state generation. The origin of the moniker "$N00N$ state" is obvious, though such states are also known as maximally path-entangled number states as the path of the definite number of photons $N$ in the superposition state of Eq.~\ref{eqn:N00N_1} can be interpreted as being objectively indefinite. \\

\noindent Let us consider an interferometric scheme in which the initial state is given by $\ket{\text{in}} = \ket{N,0}_{a,b}$ and the first beam splitter is replaced by an optical device that transforms the initial state into the $N00N$ state of Eq.~\ref{eqn:N00N_1} (a "magic" beam splitter, so to speak).  The state after the phase shift (taken to be in the $b$-mode) is 

\begin{equation}
	\ket{\psi_{N}\left(\phi\right)} = \frac{1}{\sqrt{2}}\left(\ket{N,0}_{a,b}+e^{iN\phi+\Phi_{N}}\ket{0,N}_{a,b}\right),
	\label{eqn:N00N_2}
\end{equation}

\noindent amounting to an additional relative phase shift of $N\phi$.  Finally, the state after the second beam splitter is \cite{GerryParity2}

\begin{equation}
	\ket{\text{out}} = \frac{1}{\sqrt{2^{N+1}N!}}\left[\left(\hat{a}^{\dagger}+i\hat{b}^{\dagger}\right) + e^{i\left(N\phi+ \Phi_{N}\right)}\left(\hat{b}^{\dagger} + i\hat{a}^{\dagger}\right)\right]\ket{0,0}_{a,b},
	\label{eqn:N00N_3}
\end{equation}

\noindent which for the case of $N=1$ yields the same phase sensitivity as the $\ket{\text{in}}=\ket{1,0}_{a,b}$ state when implementing intensity-difference measurements in a regular MZI scheme. For the case of $N=2$, the output state can be written as  

\begin{align}
\ket{\text{out,2}} = &\frac{1}{2\sqrt{2}} \left[\left(1-e^{i2\phi+ \Phi_{2}}\right)\ket{2,0}_{a,b} + \right. \nonumber \\
& \left. + \sqrt{2}i\left(1+e^{i2\phi+ \Phi_{2}}\right)\ket{1,1}_{a,b} +\left(e^{i2\phi+ \Phi_{2}}-1\right)\ket{0,2}_{a,b}  \right].
\label{eqn:N00N_4}
\end{align}

\noindent Interestingly, for all values $N > 1$, intensity-difference measurements fails to capture any phase-shift dependence; this results in a measurement of zero irrespective of the value of $\phi$, making it an unsuitable detection observable for this choice of input state.   \\

\noindent A Hermitian operator was introduced by Dowling \textit{et al.} \cite{Dowling1} \cite{Dowling2} of the form $\hat{\Sigma}_{N} = \ket{N,0}_{a,b}\bra{0,N} + \ket{0,N}_{a,b}\bra{N,0}$ whose expectation value with respect to the state Eq.~\ref{eqn:N00N_1}, $\braket{\hat{\Sigma}_{N}} = \cos N\phi$, depends on the phase $N\phi$ with interference fringes whose oscillation period is $N$ times shorter than that of the single-photon case. The phase uncertainty obtained through the error propagation calculus can be found easily to be 

\begin{equation}
	\left(\Delta\phi\right)^{2} = \frac{\braket{\hat{\Sigma}_{N}^{2}} - \braket{\hat{\Sigma_{N}}}^{2}}{\left(\partial_{\phi}\braket{\hat{\Sigma_{N}}}\right)^{2}} = \frac{1 - \cos^{2}N\phi}{N^{2}\sin^{2}N\phi} = \frac{1}{N^{2}},
	\label{eqn:N00N_5}
\end{equation}   

\noindent which is an improvement over the classical (SQL) limit by a factor of $1/\sqrt{N}$.  The result of Eq.~\ref{eqn:N00N_5} follows from the heuristic number-phase relation $\Delta\phi\Delta N \geq 1$.  For the $N00N$ state of Eq.~\ref{eqn:N00N_1}, the uncertainty in photon number is $N$, the total average photon number, immediately giving the equality $\Delta\phi = 1/N$. Extrapolating for arbitrary states, we define the Heisenberg limit (HL) $\Delta\phi_{\text{HL}} = 1/\bar{n}_{\text{tot}}$ where $\bar{n}_{\text{tot}}$ is the total average photon number inside the interferometer. \\

\noindent It was found that the results of the projection operator employed by Dowling \textit{et al.} can be realized through implementation of photon-number parity measurements performed on one mode \cite{GerryParity2}.  As a demonstration, consider the $N=2$ case described by the output state given by Eq.~\ref{eqn:N00N_4}.  The action of the parity operator on the $b$-mode, $\hat{\Pi}_{b}=\left(-1\right)^{\hat{b}^{\dagger}\hat{b}}$, results in a sign flip on the center term

\begin{align}
	&\hat{\Pi}_{b}\ket{\text{out,2}} =\frac{1}{2\sqrt{2}} \left[\left(1-e^{i2\phi+ \Phi_{2}}\right)\ket{2,0}_{a,b} + \right. \nonumber \\
	& \left. - \sqrt{2}i\left(1+e^{i2\phi+ \Phi_{2}}\right)\ket{1,1}_{a,b} +\left(e^{i2\phi+ \Phi_{2}}-1\right)\ket{0,2}_{a,b}  \right],
	\label{eqn:N00N_6}
\end{align}

\noindent leading to the expectation value $\braket{\hat{\Pi}_{b}}=\cos \left(2\phi + \Phi_{2}\right)$, similar to the result obtained through use of the projection operator employed by Dowling \textit{et al.}  Unlike the method of measuring the intensity-difference between modes, this carries relevant phase information from which an estimation can be made. For the arbitrary $N$ case, it was found

\begin{align}
	\braket{\hat{\Pi}_{b}} &= \frac{i^{N}}{2}\left[e^{i\left(N\phi+ \Phi_{N}\right)} + \left(-1\right)^{N}e^{-i\left(N\phi+ \Phi_{N}\right)}\right] \nonumber \\
	&= 
	\begin{cases}
	\left(-1\right)^{N/2}\cos\left(N\phi+ \Phi_{N}\right) & N\;\text{even}, \\
	\left(-1\right)^{\left(N+1\right)/2}\sin\left(N\phi+ \Phi_{N}\right) & N\;\text{odd},
	\end{cases}
	\label{eqn:N00N_7}
\end{align}

\noindent and consequently $\Delta\phi = 1/N$.  It is important to note that while photon-number parity functions similarly to the projection operator put forth by Dowling \textit{et al.}, they are not equivalent.  This is clear as $\hat{\Sigma}_{N}$ is not directly connected to an observable. Furthermore, there is presently no physical realization of the projector $\hat{\Sigma}_{N}$ capable of being utilized experimentally.  \\

\noindent Clearly states similar in form to Eq.~\ref{eqn:N00N_1} after the first beam splitter are favorable in interferometry.  Let us begin by considering a case in which the state after beam splitting can be written in terms of a superposition of $N00N$ states.

\subsection{\label{sec:QOI_cohEnt} Entangled coherent states}

\noindent One such case known to yield Heisenberg-limited phase sensitivity is the case in which the action of the first beam splitter results in an entangled coherent state (ECS) \cite{ECS3}.  It has been shown by Israel \textit{et al.} \cite{Israel} that such a state can be generated through the mixing of coherent light with a squeezed vacuum at the first beam splitter (a case to be discussed in greater detail in a latter section).  Another method involves the mixing of coherent- and cat- states.  The coherent state is given by $\ket{\alpha} = e^{-|\alpha|^{2}/2}\sum_{n=0}^{\infty}\frac{\alpha^{n}}{\sqrt{n!}}\ket{n}$ and constitutes the most classical of quantized fields states, characterized as light from a well phase-stabilized laser.  It is important to point out that while coherent light maintains classical properties, it is still a quantum state of light as it is defined in terms of a quantized electromagnetic field. The generalized cat state \cite{Yurke2} is expressed as a superposition of equal-amplitude coherent states differing by a $\pi$-phase shift. Such states have been studied extensively in the context of phase-shift measurements \cite{ECS1}.  The initial state can be written as 

\begin{equation}
	\ket{\text{in}} = \ket{\beta}_{a}\otimes \mathcal{N}\left(\ket{\gamma}_{b}+e^{i\theta}\ket{-\gamma}_{b}\right),
	\label{eqn:ecs_1}
\end{equation}

\noindent where $\mathcal{N}$ is the cat state normalization factor given by $\mathcal{N} = 1/\sqrt{2\left(1+e^{-2|\gamma|^{2}}\cos\theta\right)}$. The QFI can be calculated immediately for this input state using Eq.~\ref{eqn:2.50}.  For large $\gamma$, $\braket{\hat{J_{+}}} = \braket{\hat{J_{-}}} = \braket{\hat{J_{y}}} \simeq 0$ and 

\begin{align}
	\braket{J_{+}^{2}} &= \beta^{*\;2}\gamma^{2},\;\;\;\;\;\;\;\;\;\;\;\;\;\;\;\;\;\;\;\;\;\;\;\;\; \braket{J_{-}^{2}} = \beta^{2}\gamma^{*\;2} \\
	\braket{J_{+}\hat{J}_{-}} &= |\beta|^{2}\left(1+|\gamma|^{2}\right),\;\;\;\;\;\; \braket{J_{-}\hat{J}_{+}} = |\gamma|^{2}\left(1+|\beta|^{2}\right),
	\label{eqn:ecs_1_fix}
\end{align}  

\noindent resulting in 

\begin{align}
	F_{Q} &= 4\braket{\left(\Delta\hat{J}_{y}\right)^{2}}_{\text{in}} = 4\braket{\hat{J}_{y}^{2}} \nonumber \\
	&= |\beta|^{2} + |\gamma|^{2} + 2|\beta|^{2}|\gamma|^{2}\left(1-\cos\left[2\left(\theta_{\beta} - \theta_{\gamma}\right)\right]\right),
	\label{eqn:ecs_2_fix}
\end{align}

\noindent where $\left(\theta_{\beta}-\theta_{\gamma}\right)$ is the phase difference between coherent states $\ket{\beta}$ and $\ket{\gamma}$. Setting $\beta=\alpha/\sqrt{2}$ and $\gamma=-i\beta=-i\alpha/\sqrt{2}$, the total average photon number in the interferometer in the limit of large coherent state amplitude is $|\beta|^{2} + |\gamma|^{2} = |\alpha|^{2}$. Plugging these into Eq.~\ref{eqn:ecs_2_fix} and noting now that $\theta_{\beta}-\theta_{\gamma}=\pi/2$, the minimum phase uncertainty is given by

\begin{equation}
\Delta\phi_{\text{min}}  \xrightarrow{\text{large}\; \alpha}\frac{1}{|\alpha|^{2}} = \Delta\phi_{\text{HL}}.
\label{eqn:esc_6}
\end{equation}  

\noindent Let us calculate the state after the first beam splitter for this choice of input state.  Through the usual beam splitter transformations for coherent states, we have

\begin{align}
	\ket{\beta}_{a}\ket{\gamma}_{b}&\xrightarrow{\text{BS 1}} \ket{\tfrac{1}{\sqrt{2}}\left(\beta+i\gamma\right)}_{a}\ket{\tfrac{1}{\sqrt{2}}\left(\gamma+i\beta\right)}_{b}, \nonumber \\
	\label{eqn:esc_2} \\
	\ket{\beta}_{a}\ket{-\gamma}_{b}&\xrightarrow{\text{BS 1}} \ket{\tfrac{1}{\sqrt{2}}\left(\beta-i\gamma\right)}_{a}\ket{\tfrac{1}{\sqrt{2}}\left(-\gamma+i\beta\right)}_{b}, \nonumber
\end{align}

\noindent making the state after the first beam splitter 

\begin{align}
	\ket{\text{out, BS 1}} &= \mathcal{N} \left( \ket{\tfrac{1}{\sqrt{2}}\left(\beta+i\gamma\right)}_{a}\ket{\tfrac{1}{\sqrt{2}}\left(\gamma+i\beta\right)}_{b} +    \right. \nonumber \\
	&\;\;\;\;\;\;\;\;  \left.  + e^{i\theta} \ket{\tfrac{1}{\sqrt{2}}\left(\beta-i\gamma\right)}_{a}\ket{\tfrac{1}{\sqrt{2}}\left(-\gamma+i\beta\right)}_{b}  \right].
	\label{eqn:esc_3}
\end{align}

\begin{figure}
	\centering
	\includegraphics[width=1.0\linewidth,keepaspectratio]{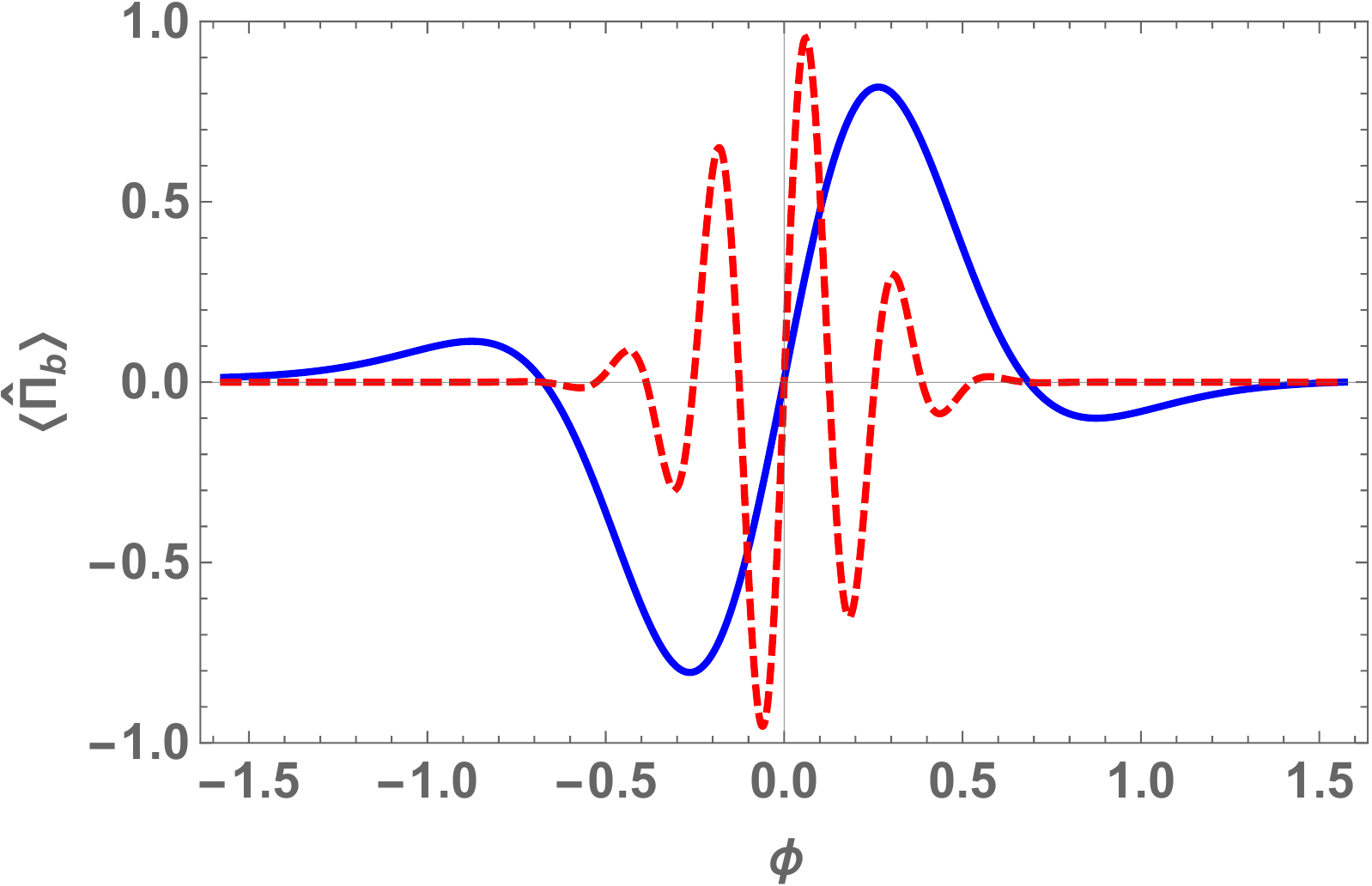}
	\caption{Expectation value of the parity operator $\braket{\hat{\Pi}_{b}}$ for $\bar{n}_{\text{tot}}=5$ (blue, solid) and  $\bar{n}_{\text{tot}}=25$ (red, dashed).  The peaks of the curves narrows and migrate towards the phase origin, which serves as an inflection point. Note for these curves, the relative phase shift in Eq.~\ref{eqn:ecs_1} is chosen as $\theta=\pi/2$ given our interferometric scheme of Eq.~\ref{eqn:qoi_1} to align with the results found by Gerry \textit{et al} \cite{ECS}. }
	\label{fig:ECS_parity}
\end{figure}

\begin{figure}
	\centering
	\includegraphics[width=0.95\linewidth,keepaspectratio]{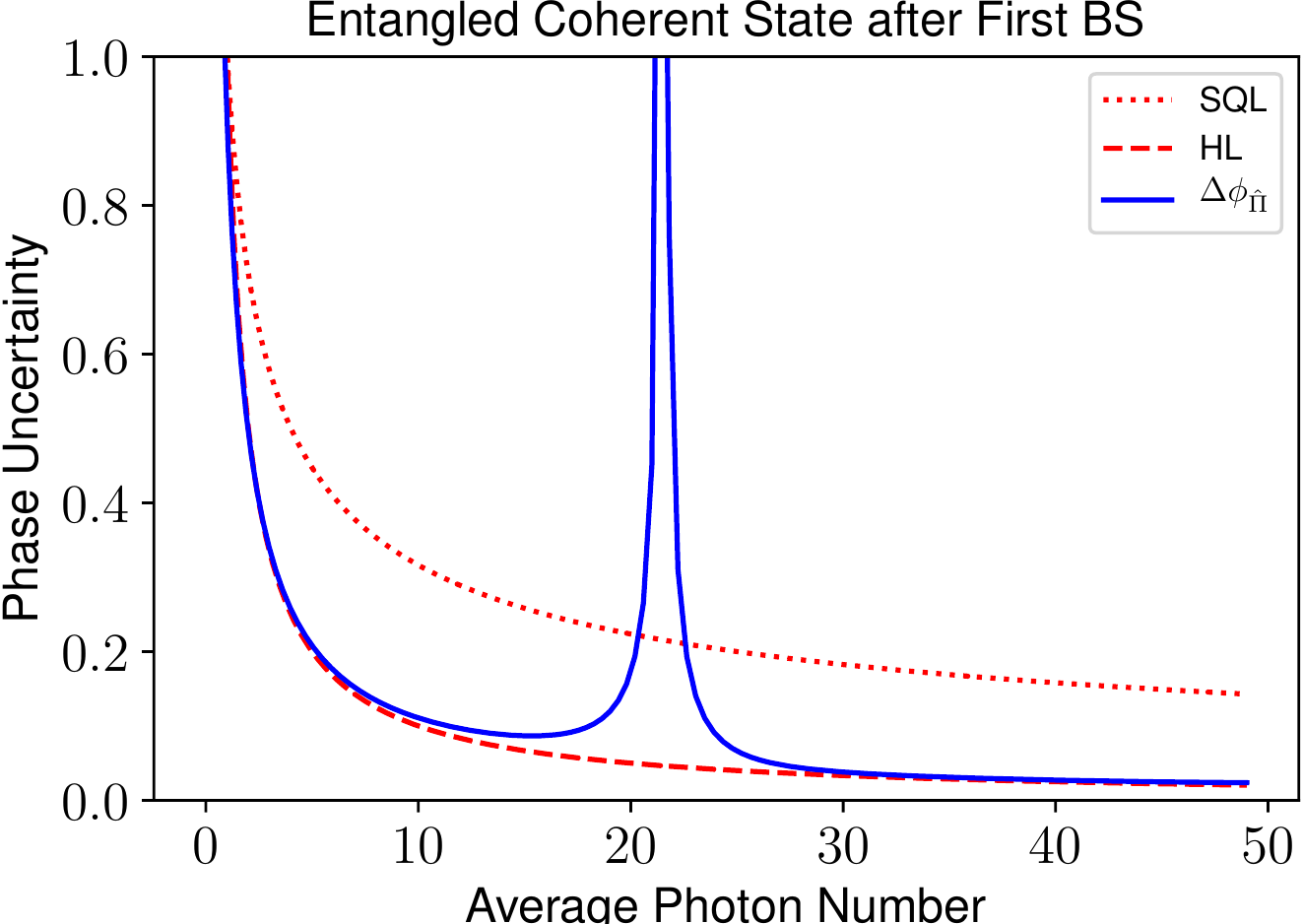}
	\caption{Phase uncertainty obtained when the state after the first beam splitter is an entangled coherent state of the form Eq.~\ref{eqn:esc_4} for a relative phase $\theta=\pi/2$ and phase value $\phi=\pi/45$. Note that aside from the spike at $\bar{n}_{\text{tot}}\sim 21$, the phase uncertainty follows closely along the HL.}
	\label{fig:ECS_PU}
\end{figure}

\noindent Once again setting $\beta=\alpha/\sqrt{2}$ and $\gamma = -i\beta = -i\alpha/\sqrt{2}$, this becomes

\begin{equation}
	\ket{\text{out, BS 1}} = \mathcal{N}\left(\ket{\alpha}_{a}\ket{0}_{b} + e^{i\theta}\ket{0}_{a}\ket{i\alpha}_{b}\right),
	\label{eqn:esc_4}
\end{equation}

\noindent the entangled coherent state.  This state was studied by Gerry \textit{et al.} \cite{ECS} where the first beam splitter was replaced by an asymmetric non-linear interferometer (ANLI) yielding the state  $\ket{\text{out, ANLI}}\propto \big(\ket{0,i\alpha}_{a,b}+i\ket{-\alpha,0}_{a,b}\big)$, similar to Eq.~\ref{eqn:esc_4}. This state can be written as a superposition of $N00N$ states as per 

\begin{align}
	\ket{\text{out, ANLI}} = e^{-\frac{1}{2}|\alpha|^{2}}&\sum_{N=0}^{\infty} \frac{\alpha^{N}}{\sqrt{N!}}\left(-1\right)^{N}\times \nonumber \\
	&\;\;\;\;\times\frac{1}{\sqrt{2}}\left(\ket{N,0}_{a,b}+e^{-i\Phi_{N}}\ket{0,N}_{a,b}\right),
	\label{eqn:esc_7}
\end{align}  

\noindent where $\Phi_{N} = \tfrac{\pi}{2}\left(N+1\right)$.  Parity-based detection was considered on the output $b$-mode after acquiring a phase shift $\phi$ and passing through the second beam splitter to find

\begin{align}
	\braket{\hat{\Pi}}_{b}&= e^{-\bar{n}}\left(1+e^{\bar{n}\cos\phi}\sin\left(\bar{n}\sin\phi\right)\right),\label{eqn:esc_8} \\
	\nonumber \\
	\partial_{\phi}\braket{\hat{\Pi}}_{b}&= \bar{n}e^{-\bar{n}\left(1-\cos\phi\right)}\cos\left(\phi+\bar{n}\sin\phi\right),\label{eqn:esc_8b}
\end{align}

\noindent and from the error propagation calculus, Eq.~\ref{eqn:qoi_5}, the phase uncertainty for $\phi\to 0$ is

\begin{equation}
	\Delta\phi_{\hat{\Pi}_{b}} = \frac{\sqrt{1-e^{-2\bar{n}}}}{\bar{n}},
	\label{eqn:esc_9}
\end{equation}

\noindent which subsequently yields the HL in the limit of large $|\alpha|$. Plots of the expectation value of the parity operator, Eq.~\ref{eqn:esc_8}, and corresponding phase uncertainty, Eq.~\ref{eqn:esc_9}, found by Gerry \textit{et al.} \cite{ECS} are given in Figs.~\ref{fig:ECS_parity} and \ref{fig:ECS_PU}, respectively.  Clearly distributions after the first beam splitter that are reminiscent in form to the $N00N$ state tends towards providing greater phase sensitivity.  \\

\noindent Another such case to consider is the case of the input twin-Fock states.  As per the very well known Hong-Ou-Mandel (HOM) effect, when the initial state $\ket{1,1}_{a,b}$ is incident upon a 50:50 beam splitter, the resulting state is the two-photon $N00N$ state $\ket{2,0}_{a,b}+\ket{0,2}_{a,b}$.  Let us now consider the case in which the arbitrary state $\ket{N,N}_{a,b}$ is taken as the initial state.   

\subsection{\label{sec:QOI_TFS} Twin-Fock state input}

\begin{figure}
	\centering
	\includegraphics[width=0.90\linewidth,keepaspectratio]{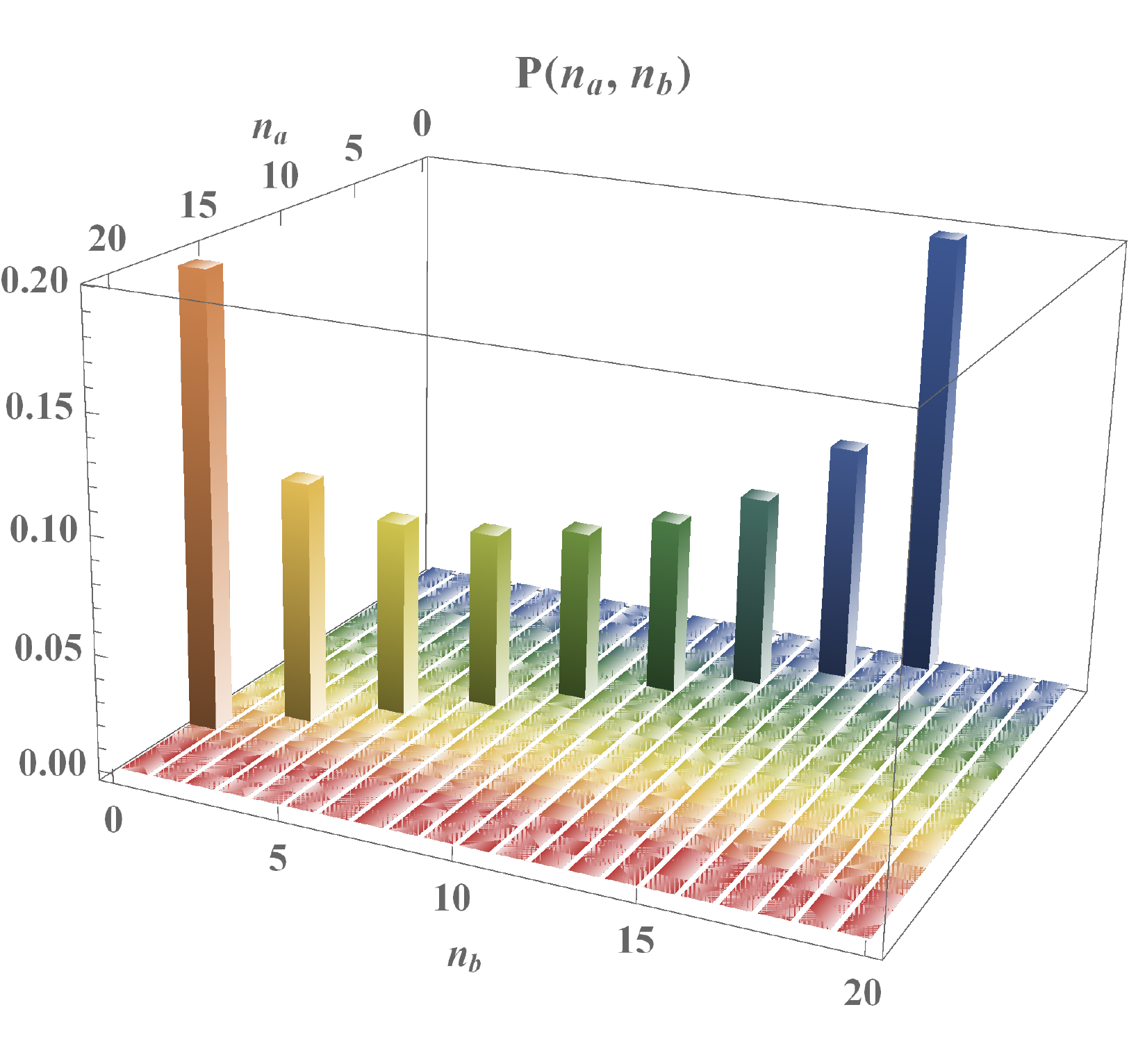}
	\caption{Two-mode joint photon number distribution resulting from beamsplitting the state $\ket{\text{in}}=\ket{N,N}_{a,b}$. The "bathtub" shape of the arcsine distribution is peaked at the $\ket{N,0}_{a,b}$ and $\ket{0,N}_{a,b}$ states.  This is plotted for $N=8$.}
	\label{fig:TFS_prob}
\end{figure}

\noindent It was first pointed out by Holland and Burnett \cite{TwinFock} that interferometric phase measurements when considering input twin-Fock states $\ket{N,N}_{a,b}$  asymptotically approaches the HL. They found this by studying the phase-difference distribution for the states inside the MZI just prior to the second beam splitter.  On the other hand, Bollinger \textit{et al.} \cite{Bollinger} \footnote{Their discussion is in the context of spectroscopy using maximally entangled states of a system of $N$ two-level trapped ions.} showed that if the state just after the first beam splitter is somehow a maximally entangled state (MES) of the form $\propto \ket{2N,0}_{a,b} + e^{-i\Phi_{N}}\ket{0,2N}_{a,b}$, $N\in\mathbb{Z}^{+}$, then the phase uncertainty is \textit{exactly} the HL:  $\Delta\phi_{\text{HL}} = 1/2N$.  The problem is such a state is incredibly difficult to produce; in fact, it cannot be done with an ordinary beam splitter.  Schemes for generating such states using both nonlinear devices and linear devices used in conjunction with conditional measurements have been proposed \cite{GerryParityAgain} \cite{Campos1} \cite{Benmoussa} \cite{Kok} \cite{Kok2} \cite{Fiurasek} \cite{Benmoussa2}.  For an initial state described by state coefficients  $C_{n}^{\left(1\right)}C_{q}^{\left(2\right)} = \delta_{N,n}\delta_{N,q}$ as per Eq.~\ref{eqn:initial}, the state after the first beam splitter is the well-known arcsine (AS) state \cite{Campos}, given by 

\begin{align}
	\ket{\text{out, BS1}} &= \sum_{k=0}^{N} \left(-1\right)^{N-k}\Bigg[\binom{2k}{k}\binom{2N-2k}{N-k}\left(\frac{1}{2}\right)^{2N}\Bigg]^{1/2}\times, \nonumber\\
	& \times \ket{2k}_{a}\otimes\ket{2N-2k}_{b},
	\label{eqn:tfs_1}
\end{align}

\noindent where in this case a $\hat{J}_{y}$-type beam splitter was considered rather than a $\hat{J}_{x}$-type of the previous sections.  This simply amounts to a relative phase difference between terms in the sum in Eq.~\ref{eqn:tfs_1}. Clear for the $N=1$ case, we recover the well known two-photon $N00N$ state that has long been available in the laboratory \cite{HOM}. For $N>1$, the state of Eq.~\ref{eqn:tfs_1} does not cleanly result in the $N$-photon $N00N$ state, but instead a superposition of the $N$-photon $N00N$ state and other (but not all) permutations of the state $\ket{p,q}$ where $p+q=2N$.  Due to the strong correlations between photon number states of the two modes, the only nonzero elements of the joint-photon number probability distribution are the joint probabilities for finding $2k$ photons in the $a$-mode and $2N-2k$ photons in the $b$-mode, given by

\begin{align}
	P_{\text{AS}}\left(2k,2N-2k\right) &= |{}_{a,b}\braket{2k,2N-2k|\text{out, BS1}}|^{2} \nonumber \\
	&= \binom{2k}{k}\binom{2N-2k}{N-k}\left(\frac{1}{2}\right)^{2N},\;\;\;\;k\in\left[0,N\right]
	\label{eqn:tsf_2}
\end{align}

\noindent forming a distribution known as the fixed-multiplicative discrete arcsine law of order $N$ \cite{MathBook}, deriving the name of the state. Such a distribution is characterized by the "bathtub" shape of an arcsine distribution, with peaks occuring for the $\ket{N,0}_{a,b}$ and $\ket{0,N}_{a,b}$  states, as shown in Fig.~\ref{fig:TFS_prob}. The phase properties of this state were studied by Campos \textit{et al.} \cite{Campos}.  \\

\noindent Next we will compare the qCRB for this choice of initial state against the phase uncertainty obtained via parity-based measurements.  Once again, the qCRB can be calculated directly from the initial state through Eq.~\ref{eqn:2.50}. It is plainly evident that $\braket{J}_{+}=\braket{J}_{-}=\braket{J}_{y}=0$,  $\braket{\hat{J}_{+}\hat{J}_{-}} = \braket{\hat{J}_{-}\hat{J}_{+}} = N\left(N+1\right)$ and consequently

\begin{equation}
	\Delta^{2}\hat{J}_{y} = \frac{N}{2}\left(1+N\right)\to \Delta\phi_{\text{min}} = \frac{1}{\sqrt{2N\left(1+N\right)}}.
	\label{eqn:tfs_3}
\end{equation}

\noindent Right away it is clear that for the case of $N=1$, the minimum phase uncertainty provides the HL: $\Delta\phi_{\text{HL}} = 1/2N = 1/2$. Next we consider the use of parity performed on the output $b$-mode as a detection observable. The expectation value of the parity operator can be calculated directly with respect to the state of Eq.~\ref{eqn:tfs_1}, accounting for the phase shift and assuming the second beam splitter is of the $\hat{J}_{x}$-type, to be

\begin{align}
	\braket{\hat{\Pi}_{b}}_{\text{AS}} &= \braket{\text{in}|\hat{\Pi}_{b}|\text{in}}\nonumber \\
	 &=\braket{\text{out, BS 1}|e^{-i\phi\hat{J}_{z}}e^{-i\tfrac{\pi}{2}\hat{J}_{x}}\hat{\Pi}_{b}e^{i\tfrac{\pi}{2}\hat{J}_{x}}e^{i\phi\hat{J}_{z}}|\text{out, BS 1}} \nonumber \\
	&= \sum_{k=0}^{N}e^{i2\phi\left(N-2k\right)}\binom{2k}{k}\binom{2N-2k}{N-k}\left(\frac{1}{2}\right)^{2N},
	\label{eqn:tsf_4}
\end{align}

\noindent The imaginary part of Eq.~\ref{eqn:tsf_4} sums identically to zero as it is the product of an even times and odd function of $k$.  The real part is identically a Legendre polynomial, making

\begin{equation}
	\braket{\hat{\Pi}_{b}}_{\text{AS}} = P_{N}\left[\cos\left(2\phi\right)\right].
	\label{eqn:tfs_5}
\end{equation}

\begin{figure}
	\centering
	\includegraphics[width=0.97\linewidth,keepaspectratio]{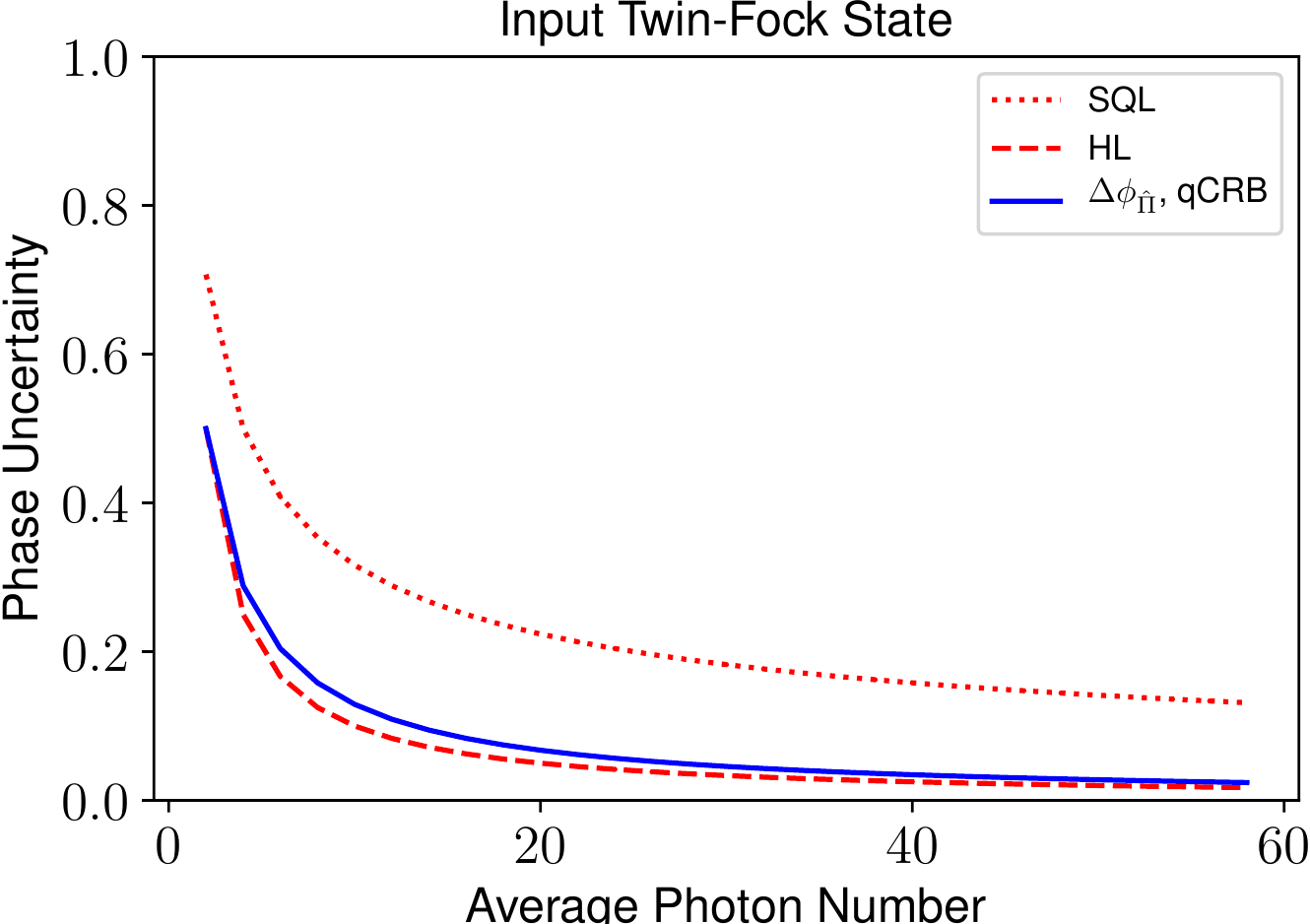}
	\caption{Parity-based phase uncertainty with phase $\phi=10^{-4}$ for the case of an input twin-Fock state.  Note the phase uncertainty obtained corresponds with the qCRB.}
	\label{fig:TFS_PU}
\end{figure}

\noindent With this, the phase uncertainty can be computed directly from the error propagation calculus given by Eq.~\ref{eqn:qoi_5}.  For $N=1$, the expectation value of the parity operator is $\braket{\hat{\Pi}_{b}}_{\text{AS}} =\cos\left(2\phi\right)$ leading to $\Delta\phi = 1/2$, the HL. For $N=2$, it can be shown $\braket{\hat{\Pi}_{b}}_{\text{AS}} = 1/4 + 3/4\cos\left(4\phi\right)$, which in the limit of $\phi\to 0$ yields $\Delta\phi = 1/\sqrt{12}=0.2886$; close to the HL of $1/4=0.25$.  These results are in agreement with the minimum phase uncertainty obtained through calculation of the qCRB for this state, Eq.~\ref{eqn:tfs_3}. A plot of the parity-based phase uncertainty and corresponding qCRB can be found in Fig.~\ref{fig:TFS_PU}.  Another interesting feature of Eq.~\ref{eqn:tfs_5} occurs when considering measurements around $\phi=\pi/2$.  Through use of standard identities involving Legendre polynomials, it can be shown that Eq.~\ref{eqn:tfs_5} becomes $\braket{\hat{\Pi}_{b}}^{\left(\phi=\pi/2\right)}_{\text{AS}} = \left(-1\right)^{N}$, corresponding to a peak in the curve yielding the same degree of resolution as $\phi=0$, but takes the maximal (minimal) value of $\pm 1$ depending on the value of $N$.  This may prove to be of use in verifying one has lossless conditions within the interferometer as typically the experimenter would have foreknowledge of the value $N$ and therefore know what the value in a $\pi/2$-shifted interferometer should be. Measurement to the contrary can point towards the presence of losses in the system. \\

\noindent In the next section, we will briefly consider states comprised of superpositions of twin-Fock states.

\subsubsection{\label{{sec:QOI_Super_TFS}} Superpositions of twin-Fock states}

\noindent Highly photon-number correlated continuous variable two-mode states have been investigated for use in quantum optical interferometry.  It is immediately clear that such states are entangled as they are already in Schmidt form $\ket{\Psi} = \sum_{n=0}^{\infty}B_{n}\ket{n,n}_{a,b}$ with total average photon number $\bar{n}_{\text{tot}} = 2\sum_{n=0}^{\infty}|B_{n}|^{2}n$. In terms of the initial state of Eq.~\ref{eqn:initial}, the coefficients for such a state are given by $C_{n}^{\left(1\right)}C_{q}^{\left(2\right)} = B_{n}\delta_{q,n}$.  The expectation value of the parity operator is readily calculable for arbitrary correlated two-mode states of this form using the results of the twin-Fock state input of Eq.~\ref{eqn:tfs_5}:

\begin{equation}
	\braket{\hat{\Pi}_{b}}_{\text{corr}} = \sum_{n=0}^{\infty}|B_{n}|^{2}P_{n}\left[\cos\left(2\phi\right)\right],
	\label{eqn:stfs_1}
\end{equation} 

\noindent where once again $P_{n}\left(x\right)$ are the Legendre polynomial.  The most well-known and studied correlated two-mode state is the two-mode squeezed vacuum state (TMSVS). The TMSVS is a laboratory standard, routinely produced through parametric down conversion \cite{BoydBook}: a second order nonlinear effect in which a pump photon of frequency $2\omega$ is annihilated and two photons, each of frequency $\omega$, are produced. The correlation between modes is due to the pair-creation of photons resulting from the down-conversion process. Note that under the parametric approximation, the pump is treated as a classical and non-depleting field.  Consequently the pump is not treated as a quantized mode. The TMSVS state coefficients are given by $B_{n}^{\left(\text{TMSVS}\right)}=\left(1-|z|^{2}\right)^{1/2}z^{n}$.  The parameter $z$ is a complex number constrained to be $|z|<1$ and can be expressed in terms of the pump field parameters, $\gamma$ and $2\phi$ being the pump amplitude and phase respectively,  as $z=e^{i 2\phi}\tanh r$ where $r=|\gamma|t$ is the squeeze parameter. Interestingly, coupling the TMSVS coefficients with Eq.~\ref{eqn:stfs_1} yields for the phase value $\phi=\pi/2$

\begin{equation}
	\braket{\hat{\Pi}_{b}}_{\text{tmsvs}}^{\left(\phi=\pi/2\right)} = \frac{1}{\cosh\left(2r\right)} = 
	\begin{cases}
		 1, & r \to 0 \\
		\\
		 2e^{-2r}. & r \gg 1 
	\end{cases}
	\label{eqn:stfs_1a}
\end{equation}

\begin{figure}
	\centering
	\includegraphics[width=0.97\linewidth,keepaspectratio]{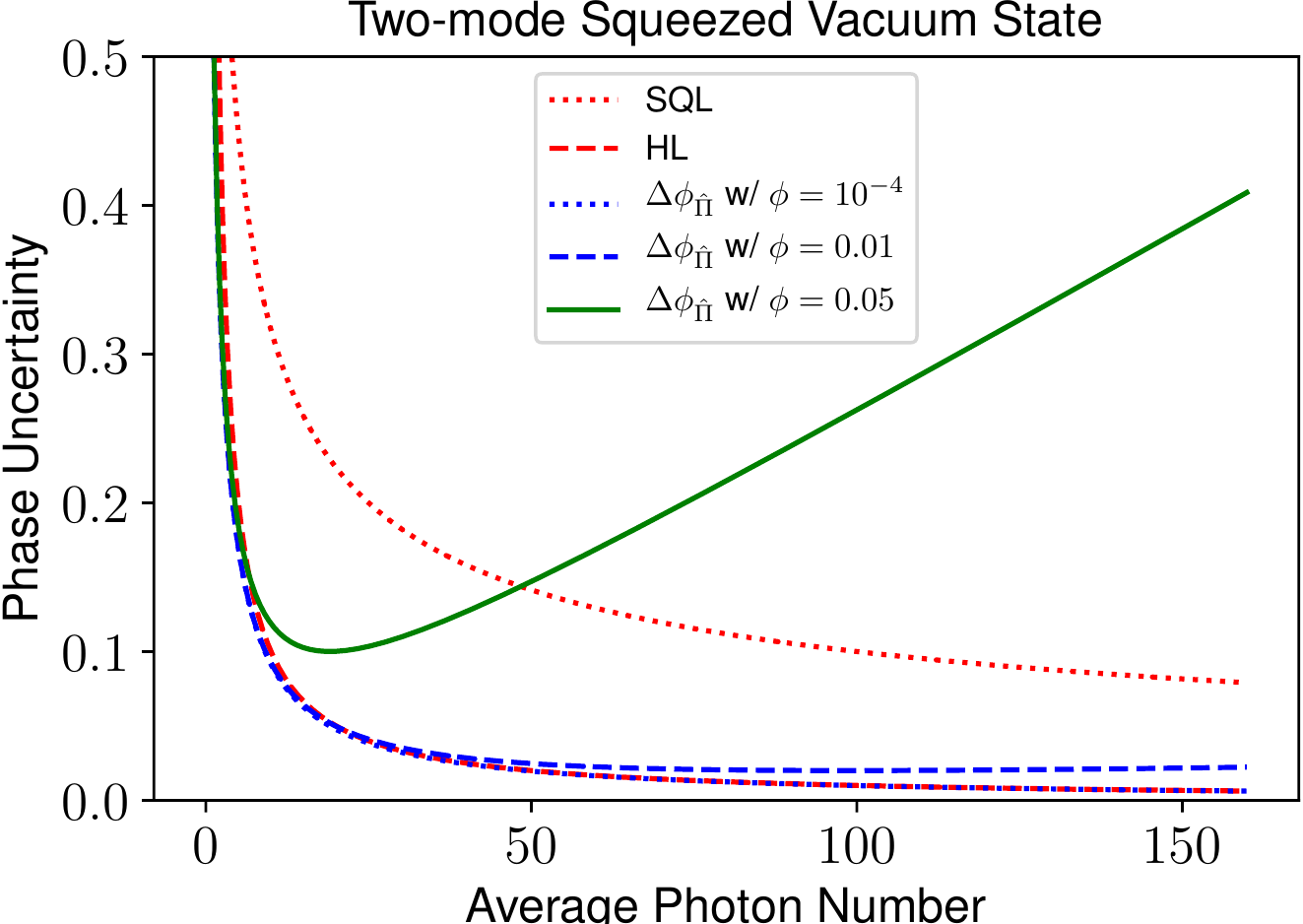}
	\caption{Phase uncertainty using parity-based detection for the input two-mode squeezed vacuum state.  For phases close to zero, the phase uncertainty yields sub-HL sensitivity, however rapidly degrades as the phase deviates from zero.}
	\label{fig:TMSVS_PU}
\end{figure}

\noindent In other words, measurement of photon-number parity on one of the output modes provides a direct measure of the degree of squeezing i.e., determination of the squeeze parameter $r$.  For the TMSVS, parity-based detection yields a phase uncertainty that falls below the HL for $\phi=0$; an effect that has been pointed out in the past by Anisimov \textit{et al.} \cite{AnisimovTMSVS}, explained in terms of the Fisher information. Once again using Eq.~\ref{eqn:2.50} and noting for such a correlated state that $F_{Q} = 2\braket{\hat{J}_{+}\hat{J}_{-}}_{\text{in}}$, it can be shown that the TMSVS minimum phase uncertainty is

\begin{equation}
	\Delta\phi_{\text{min}} = \frac{1}{\sqrt{\bar{n}_{\text{tot}}\left(2+\bar{n}_{\text{tot}}\right)}},
	\label{eqn:stfs_2}
\end{equation}   

\noindent where for the TMSVS $\bar{n}_{\text{tot}} = 2\sinh^{2}r$. This means that the TMSVS has the potential for super sensitive phase estimation; clearly the phase estimate of Eq.~\ref{eqn:stfs_2} is sub-HL. This is seemingly a violation of the bound set for quantum states.  It has been argued that such a limitation, based on the heuristic photon number-phase relation $\Delta\phi\Delta N \geq 1$, is reasonable in the case of definite photon number (finite energy) but proves to be an incomplete analysis when considering the effect of photon-number fluctuations.  Hoffman \cite{PathSym1} suggests a more direct definition of the limit on phase estimation for quantum states in terms of the second moment of $\hat{n}$, $\Delta\varphi = 1/\braket{\hat{n}^{2}}$.  This provides better sensitivity in phase measurements than the HL as $\braket{\hat{n}^{2}}$ contains direct information about fluctuations that $\braket{\hat{n}}^{2}$ does not. In fact this is why parity-based measurement yields greater sensitivity: it contains all moments of $\hat{n}$.  For the case of the TMSVS, the Hoffman limit is given by $\Delta\phi = 1/\sqrt{2\bar{n}\left(\bar{n}+1\right)}$.  With parity-based detection, sensitivity of the phase estimate is better than allowed by the HL but is never better than the Hoffman limit (the sub-HL sensitivity is prominent for low (but still $>1$) average photon numbers but asymptotically converges to the HL for $\phi=0$). It was pointed out by Anisimov \textit{et al.} \cite{AnisimovTMSVS} as well as Gerry \textit{et al.} \cite{GerryPCS} that the TMSVS, using parity measurements, has superior phase sensitivity near $\phi=0$ but degrades rapidly as the phase difference deviates from zero as shown in Fig.~\ref{fig:TMSVS_PU}, making the state sub-optimal for interferometry. On the other hand, the case in which one has parametric down-conversion with coherent states seeding the signal and idler modes, or two-mode squeezed coherent states (TMSCS), was considered by Birrittella \textit{et al.} \cite{TMSCS} who found a measurement resolution and phase sensitivity dependent on a so-called cumulative phase (the sum of the initial field phases) that yields, for low squeezing, sub-SQL phase sensitivity that does not degrade like the TMSVS does as the phase deviates from zero for the optimal choice of cumulative phase. The cumulative-phase-dependent state statistics and entanglement properties of the state resulting from coherently-stimulated down-conversion with a quantized pump field has been studied by Birrittella \textit{et al.} \cite{Birrittella3}, however the state has not yet been considered in the context of interferometry.\\

\noindent Another correlated two-mode state that has been extensively studied in the context of interferometry and metrology are the pair coherent states (PCS) \cite{PCS}, or circle states, which have the form

\begin{equation}
	\ket{\zeta} = \mathcal{N}\int_{0}^{2\pi}\ket{\sqrt{\zeta}e^{i\theta}}\ket{\sqrt{\zeta}e^{-i\theta}}d\theta,
	\label{eqn:stfs_3}
\end{equation}  

\noindent where $\ket{\sqrt{\zeta}e^{\pm i\theta}}$ are the Glauber coherent states.  In terms of the number state basis, the PCS can be written as

\begin{equation}
	\ket{\zeta} = \frac{1}{\sqrt{I_{0}\left(2|\zeta|\right)}}\sum_{n=0}^{\infty}\frac{\zeta^{n}}{n!}\ket{n,n}_{a,b},
	\label{eqn:stfs_4}
\end{equation}

\noindent where $I_{0}\left(2|\zeta|\right)$ is the modified Bessel function of order zero and $\zeta$ is a complex number defined such that $\ket{\zeta}$ is a right-eigenstate of the joint photon-annihilation operators $\hat{a}\hat{b}\ket{\zeta} = \zeta\ket{\zeta}$ and $\left(\hat{a}^{\dagger}\hat{a}-\hat{b}^{\dagger}\hat{b}\right)\ket{\zeta}=0$ \footnote{The generalized pair coherent state is defined such that $\left(\hat{a}^{\dagger}\hat{a}-\hat{b}^{\dagger}\hat{b}\right)\ket{\zeta,q}=q\ket{\zeta,q}$.  For the purposes of this review article, we assume $q=0$ without loss of generality.}. Such a state has been shown to exhibit sub-Poissonian statistics which results in sub-SQL phase uncertainty, enhanced measurement resolution and a high signal-to-noise ratio: results very close to those obtained from input twin-Fock states \cite{GerryPCS}. Furthermore, the PCS does not display the degradation of phase sensitivity as $\phi$ deviates from zero as the TMSVS does, making it more stable for interferometric measurements.  The problem, however, lies in generating such a state.  Several schemes have been proposed, most notably a scheme involving the use of third order cross-Kerr coupling between coherent states and the implementation of a state-reductive measurement \cite{PCSgen}, resulting in the projection of the PCS in bursts.  Currently, the PCS has yet to be experimentally realized.

\subsection{\label{sec:QOI_coh_SqVac} Coherent light mixed with a squeezed vacuum state}

\noindent The original interferometric scheme for reducing measurement error was proposed by Caves \cite{Caves} in the context of gravitational wave detection.  Here we consider the case in which the input state is a product of coherent light in one port and a single-mode squeezed vacuum state (SVS) in the other, given by $\ket{\text{SVS}}=\sum_{N=0}^{\infty}S_{N}\ket{N}$ with average photon number $\bar{n}_{\text{SVS}}=\sinh^{2}r$, and with state coefficients

\begin{equation}
S_{N} = \left(-1\right)^{N/2}\Bigg[\frac{N!}{2^{N}\left(\tfrac{N}{2}!\right)^{2}}\frac{\tanh^{N}r}{\cosh r}\Bigg]^{1/2}\cos^{2}\left(N\pi/2\right),
\label{eqn:cohsq_1}
\end{equation}

\noindent where once again the parameter $r$ is the squeeze parameter.  Such a state can be generated by beam splitting a TMSVS; the resulting two-mode state is a product of two single-mode SVS offset from each other by a $\pi$-phase shift \cite{Buzek}. The initial state is then $\ket{\text{\text{in}}}=\ket{\alpha}_{a}\otimes\ket{\text{SVS}}_{b}$ with state coefficients as per Eq.~\ref{eqn:initial} given by $C_{n}^{\left(1\right)}C_{p}^{\left(2\right)}=e^{-|\alpha|^{2}/2}\tfrac{\alpha^{n}}{\sqrt{N!}}\times S_{p}$.  This choice of input state has been shown to produce the ECS after the first beam splitter \cite{Israel}  and has been studied extensively by Pezz\'{e} \textit{et al.} \cite{Pezze2} who considered a Bayesian phase inference protocol and showed that the phase sensitivity saturates the CRB and go on to demonstrate that the phase sensitivity can reach the HL $\Delta\phi \sim 1/\bar{n}_{\text{tot}}$ independently of the true value of the phase shift.  As the input state is path-symmetric \cite{PathSym2}, it is sufficient to determine the CRB through calculation of the classical Fisher information, which they found for a single measurement to be

\begin{align}
	F^{\left(1\right)}\left(\phi\right) &= \sum_{N_{c},N_{d}}^{\infty}\frac{1}{P\left(N_{c},N_{d}|\phi\right)}\left(\frac{\partial P\left(N_{c},N_{d}|\phi\right)}{\partial \phi}\right)^{2} \nonumber \\
	&= |\alpha|^{2}e^{2r} + \sinh^{2}r,
	\label{eqn:cohsq_2}
\end{align}

\noindent where $P\left(N_{c},N_{d}|\phi\right)$ is the conditional probability of measuring $N_{c}$ photons in one output mode and $N_{d}$ in the other and is given in terms of the Wigner-$d$ rotation elements (see Appendix \ref{app:secB}) by 

\begin{equation}
P\left(N_{c},N_{d}|\phi\right) = e^{-|\alpha|^{2}}\Bigg|\sum_{n=0}^{2j}\frac{\alpha^{2j-n}}{\left(2j-n\right)!}S_{n}\;d_{m,j-n}^{j}\left(\phi\right)\Bigg|^{2},
\label{eqn:cohsq_3}
\end{equation}

\noindent with $j=\left(N_{c} + N_{d}\right)/2$ and $m=\left(N_{c}-N_{d}\right)/2$.  For the regime in which both input ports are of roughly equal intensity $|\alpha|^{2}\simeq \sinh^{2}r=\bar{n}_{\text{tot}}/2$,and assuming large $r$ such that $\bar{n}_{\text{tot}}\simeq e^{2r}/2$ it can be shown 

\begin{equation}
	\Delta\phi_{\text{min}} = \frac{1}{\sqrt{\bar{n}_{\text{tot}}\left(\bar{n}_{\text{tot}}+\tfrac{1}{2}\right)}}\xrightarrow{\text{large } \bar{n}_{\text{tot}}} \frac{1}{\bar{n}_{\text{tot}}} = \Delta\phi_{\text{HL}}.
	\label{eqn:cohsq_4}
\end{equation}
  
\noindent The same scheme was considered by Birrittella \textit{et al.} \cite{Birrittella1} using photon-number parity performed in the output $b$-mode as the detection observable. In particular, they studied a regime in which the two-mode joint-photon number distribution was parameterized such that the it was symmetrically populated along the borders with no population in the interior, mimicking the case of the $N00N$ state generated after the first beam splitter.  The parameters $\alpha$ and $r$ were chosen to be relevant to an experiment performed by Afek \textit{et al.} \cite{Afek} based on the $N00N$ state within the superposition of $N00N$ states (found in the output state of the first beam splitter) in which they obtained high phase sensitivity and super-resolution.  They achieved this measurement scheme by counting only the coincident counts where the total photon numbers counted added up to the selected value of $N$.  In other words, they measured $\braket{\hat{a}^{\dagger}\hat{a}\hat{b}^{\dagger}\hat{b}}_{\text{out}}$ but retained only the counts where if one detector detects $m$ photons, the other detects $N-m$, and where all other counts $\neq N$ are discarded. They reported sub-SQL phase sensitivity with this scheme as well as super-resolved measurement.  It is worth pointing out, however, that this tends to work better for low average photon numbers as the large-photon-number case cannot be reasonably expressed as a superposition of $N00N$ states.\\ 

\noindent With parity measurements performed on one of the output beams, it is not necessary nor possible to restrict oneself to a definite $N$-photon $N00N$ state, which can be advantageous.  The total number of photons inside the interferometer for this input state is indeterminate, but the Heisenberg limit is approached in terms of the average of the total photon number.  It has been shown by Seshadreesan \textit{et al.} \cite{Seshadreesan} that photon-number parity-based interferometry reaches the HL in the case of equal-intensity light incident upon a 50:50 beam splitter.  The use of photon-subtracted squeezed light in one of the input ports has also been studied by Birrittella \textit{et al.} \cite{Birrittella2} who showed the state after the first beam splitter resembled an ECS of higher average photon number than that generated through the use of a squeezed vacuum state.\\

\noindent Next we will consider the case in which purely classical light is initially in one mode and the most quantum of quantized field states, a Fock (or number state), is initially in the other.      

\subsection{\label{sec:QOI_cohN} Coherent light mixed with $N$ photons}

\noindent Next we consider the choice of input state $\ket{\text{in}}=\ket{\alpha}_{a}\otimes\ket{N}_{b}$, that is, coherent light in the input $a$-mode and a Fock state of $N$ photons in the input $b$-mode.  This was studied by Birrittella \textit{et al.} \cite{Birrittella1} in the context of parity-based phase estimation.  In terms of Eq.~\ref{eqn:initial}, the two-mode state coefficients are $C_{n}^{\left(1\right)}C_{q}^{\left(2\right)} = e^{-|\alpha|^{2}/2}\alpha^{n}/\sqrt{n!}\times\delta_{q,N}$, which in terms of the angular momentum basis states can be written as

\begin{equation}
\ket{\text{in}} = e^{-\frac{1}{2}|\alpha|^{2}}\sum_{j=N/2,..}^{\infty}\frac{\alpha^{2j-N}}{\sqrt{\left(2j-N\right)!}}\ket{j,j-N},
\label{eqn:cohN_1}
\end{equation}

\noindent where the sum over $j$ includes all half-odd integers.  It is worth pointing out a characteristic of this particular state upon beam splitting.  Consider the state after the first beam splitter, given by (see Appendix \ref{app:subsecA2}) $\ket{\text{out, BS 1}} = e^{-i\tfrac{\pi}{2}\hat{J}_{x}}\ket{\text{in}}$.  For the special case of $N=1$, this state can be written as 

\begin{align}
\ket{\text{out, BS 1}}_{N=1} &= e^{-\frac{1}{2}|\alpha|^{2}}\sum_{n=0}^{\infty}\sum_{q=0}^{\infty}\frac{\big(\tfrac{\alpha}{\sqrt{2}}\big)^{n}\big(\tfrac{i\alpha}{\sqrt{2}}\big)^{q}}{\sqrt{n!p!}} \times\nonumber \\
&\;\;\;\;\;\;\;\;\;\;\;\;\;\; \times \left(1-\delta_{n,0}\delta_{q,0}\right)\;\gamma_{n,q}\ket{n,q}_{a,b},
\label{eqn:cohN_2}
\end{align}

\noindent where the factor $\gamma_{n,q}$ is given by 

\begin{align}
\gamma_{n,q} &= \frac{4i\left(-1\right)^{1-q}2^{n+q+1}n!}{\alpha \left(n+q+1\right)!}\;\; {}_{2}\tilde{F}_{1}\left(2,n+1;2-q;-1\right),
\label{eqn:cohN_3}
\end{align}

\noindent and where ${}_{2}\tilde{F}_{1}\left(a,b;c;z\right)$ is a regularized hypergeometric function (see Appendix \ref{app:subsecB1}).  For $n=q$, this function is identically zero $\forall q \neq 0$.  Note that the term $n=q=0$ is a term that does not appear in the sum in Eq.~\ref{eqn:cohN_2} (due to the presence of the $N=1$ initial state). This coincides with a line of destructive interference along the diagonal line $n=q$ resulting in a bimodal distribution.  This effect persists for odd values of $N$ resulting in a symmetric (assuming 50:50 beam splitter) $\left(N+1\right)$-modal distribution with the peaks of the distribution migrating towards the respective axes. The same structure of distribution occurs for even $N$ as well, however lines of contiguous zeroes do not occur.  This is worth noting since a distribution like this is reminiscent of the well known twin-Fock state input case, discussed earlier, in which the state after beam splitting are the so-called arcsine, or "bat", states \cite{Campos}.  It has long been known that the input twin-Fock state case, and states with similar distributions post-beam splitter, leads to sub-SQL sensitivity \cite{GerryParity2} \cite{Dowling1} \cite{Dowling2} \cite{Campos}. 

\begin{figure}[h]
	\centering
	\hspace*{0.0cm}
	\subfloat[][]{\includegraphics[width=0.97\linewidth,keepaspectratio]{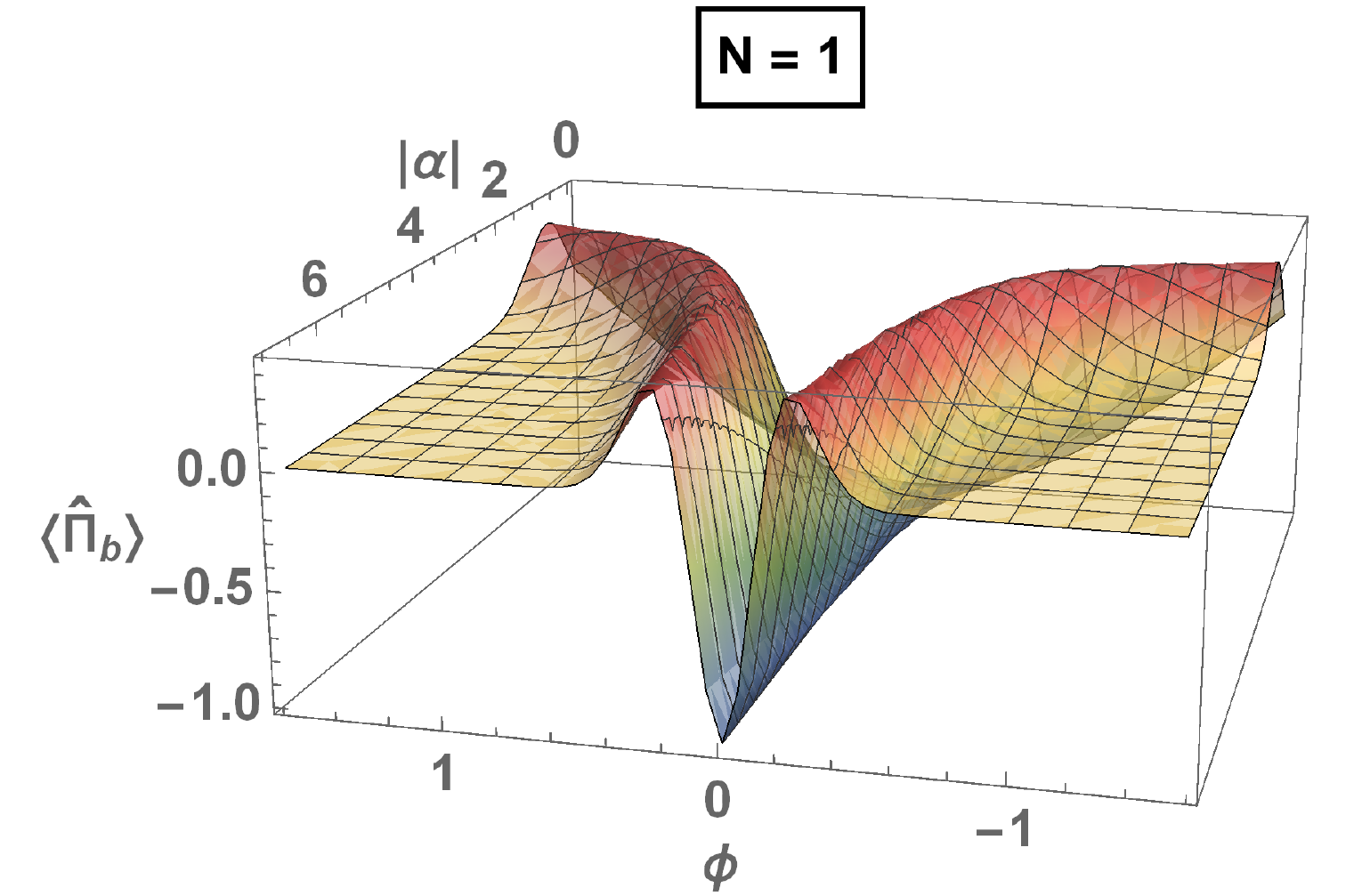}
		\label{fig:AlphaN_1}}
	\\
	\hspace*{0.0cm}
	\subfloat[][]{\includegraphics[width=0.92\linewidth,keepaspectratio]{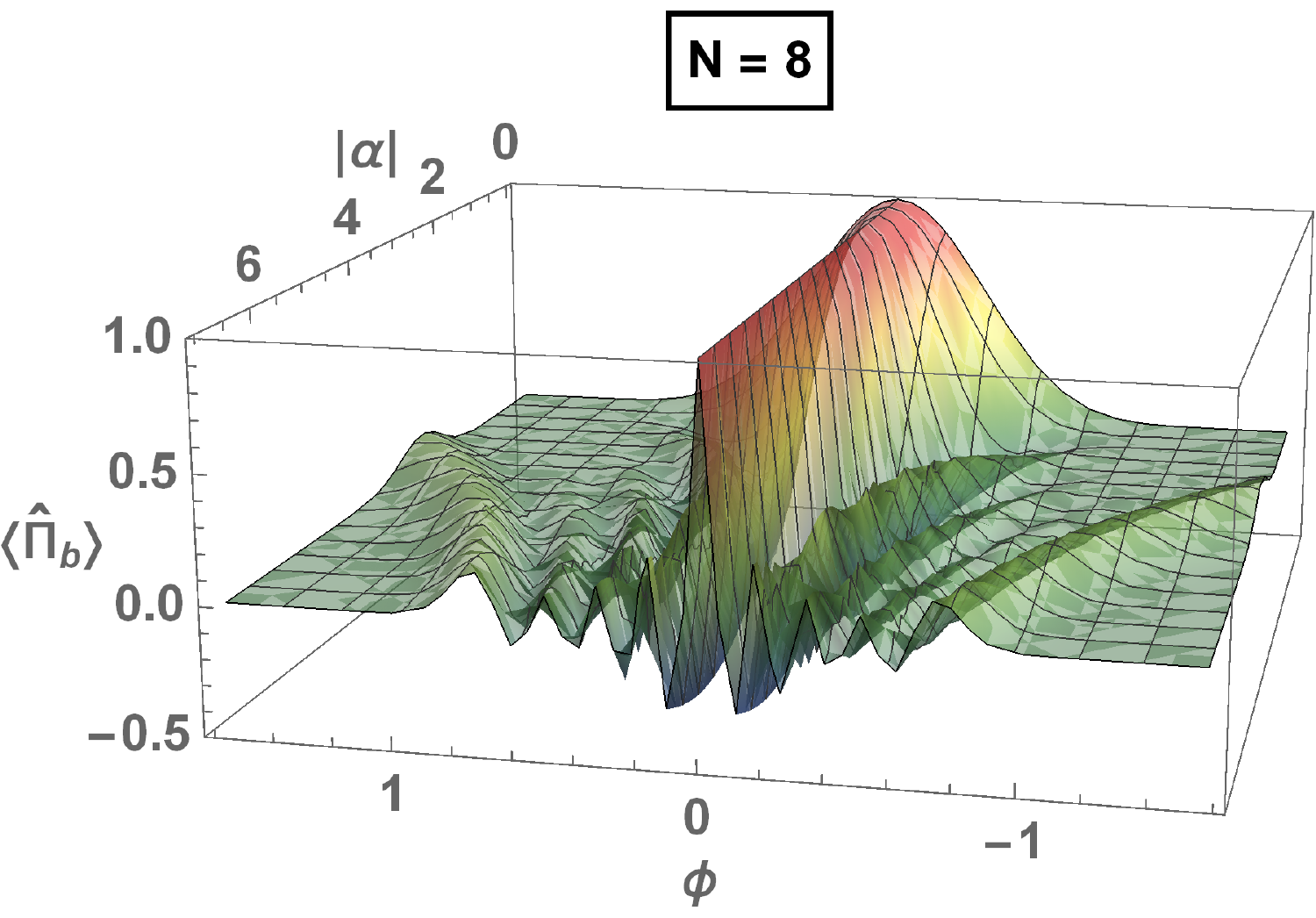}
		\label{fig:AlphaN_8}}
	\caption{Expectation value of the parity operator taken with respect to the output $b$-mode for an input state $\ket{\text{in}} = \ket{\alpha}_{a}\ket{N}_{b}$ for \protect\subref{fig:AlphaN_1} $N=1$ and \protect\subref{fig:AlphaN_8} $N=8$. Note the narrowness of the peak for increasing $|\alpha|$ and $N$.}
	\label{fig:AlphaN_parity}
\end{figure}

\subsubsection{\label{sec:QOI_cohN_res} Measurement resolution using parity-based detection}

\noindent Once again we consider the use of photon-number parity performed on the output $b$-mode as our detection observable.  Using in input state of Eq.~\ref{eqn:cohN_1}, the expectation value of the parity operator can be computed from Eqs.~\ref{eqn:qoi_3} and \ref{eqn:qoi_4} to find

\begin{equation}
\braket{\hat{\Pi}_{b}\left(\phi\right)}_{N} =\left(-1\right)^{N}e^{-|\alpha|^{2}}\sum_{j=N/2,..}^{\infty}\frac{|\alpha|^{2\left(2j-N\right)}}{\left(2j-N\right)!}d_{j-N,j-N}^{j}\left(2\phi\right).
\label{eqn:cohN_4}
\end{equation}

\begin{figure}
	\centering
	\includegraphics[width=1\linewidth,keepaspectratio]{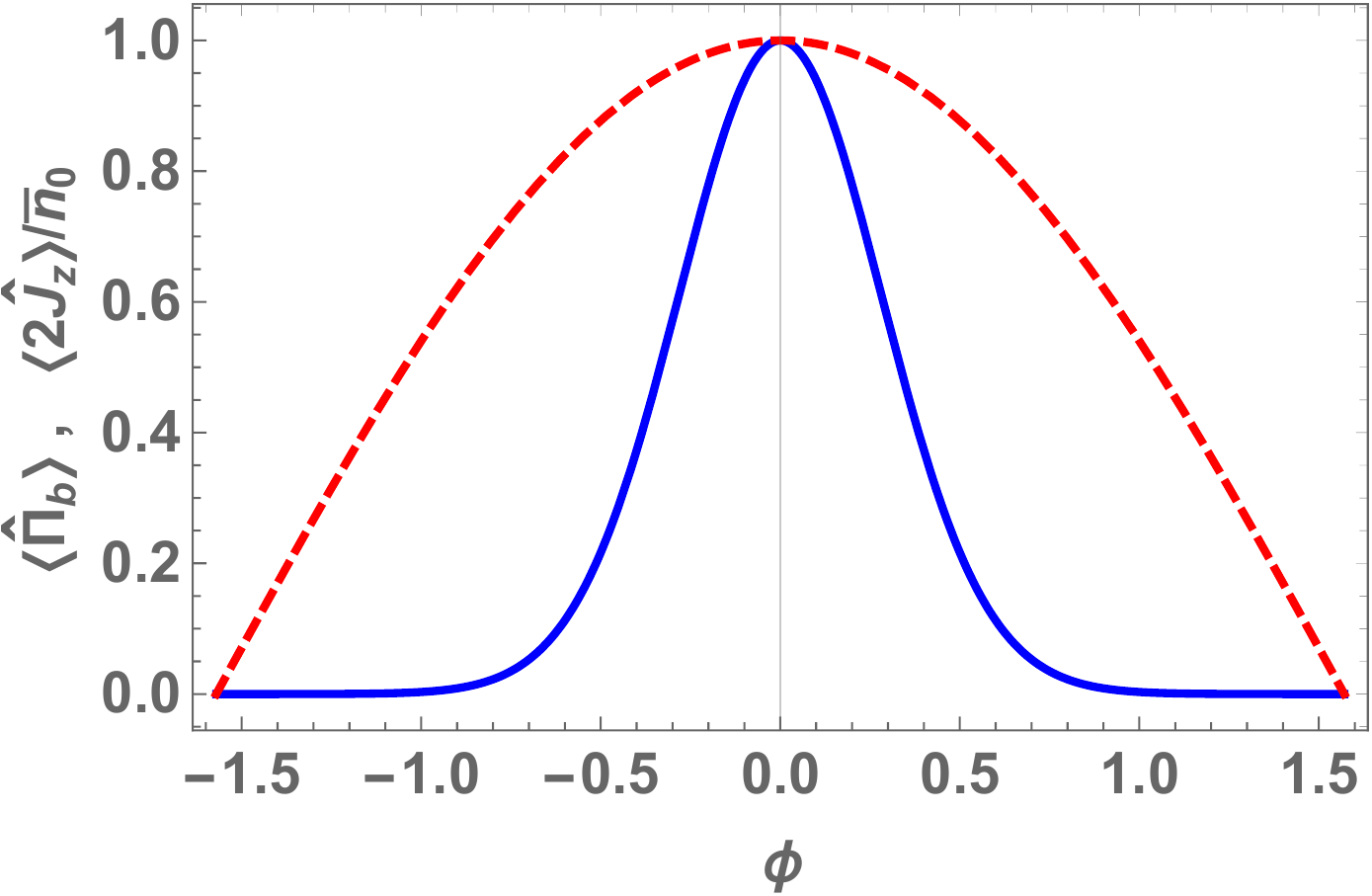}
	\caption{Comparison of measurement resolution obtained for an input state $\ket{\text{in}} = \ket{\alpha}_{a}\otimes\ket{0}_{b}$ between scaled intensity-difference $\braket{2\hat{J}_{z}}/\bar{n}_{0}$ (red, dashed) and parity $\braket{\hat{\Pi}_{b}}$ (blue, solid). The coherent state amplitude is taken to be  $|\alpha|=5$.}
	\label{fig:coh_vac}
\end{figure}

\noindent In the special case of $N=0$, we obtain the result found by Chiruvelli and Lee \cite{Chiruvelli} and discussed by Gao \textit{et al.} \cite{SupRes}, which will be expanded upon in greater detail in the following section. For small angles, $\phi\to 0$, $\braket{\hat{\Pi}_{b}\left(0\right)}_{0}\to 1$ but becomes narrower about $\phi=0$ for increasing values of $|\alpha|$, as seen in Fig.~\ref{fig:AlphaN_parity}.  The signal is not super-resolved in the usual sense of having oscillation frequencies scaling as $M\phi$ for integer $M>1$.  However, compared with the corresponding result for output subtraction, we can see the signal for parity measurement is much narrower, seen in Fig.~\ref{fig:coh_vac}.  It is in this sense that Gao \textit{et al.} \cite{SupRes} interpret parity-based measurement to be super-resolved.  Furthermore, for arbitrary $N$, the parity of the state is reflected by the expectation value of the parity operator evaluated at $\phi=0$:  $\braket{\hat{\Pi}_{b}\left(0\right)}_{N} = \left(-1\right)^{N}$.  The peak (or trough) centered at $\phi=0$ also narrows for increasing values of $N$. This can also be seen in Fig.~\ref{fig:AlphaN_parity}.

\subsubsection{\label{sec:QOI_cohN_uncertain} Phase uncertainty: approaching the Heisenberg limit}

\noindent We begin with an analysis of the phase uncertainty obtained by computation of Eq.~\ref{eqn:qoi_5} with $\hat{J}_{z}$ as the detection observable.  Plugging in the coefficients obtained using Eq.~\ref{eqn:qoi_3} for this choice of initial state and using Eq.~\ref{eqn:qoi_4} yields a phase uncertainty

\begin{equation}
\Delta\phi_{\hat{J}_{z}} = \frac{\sqrt{|\alpha|^{2}+N\left(1+2|\alpha|^{2}\sin^{2}\phi\right)}}{|\left(|\alpha|^{2}-N\right)\sin\phi|}.
\label{eqn:cohN_5}
\end{equation}

\noindent For $N=0$, we recover the well known SQL.  What is important to note about Eq.~\ref{eqn:cohN_5} is that for fixed $|\alpha|$ the optimal noise reduction achievable is the SQL and occurs only for the case of $N=0$.  For other values of $N$, the noise level rises to above the SQL.  In particular, if the initial average photon numbers of the two modes are around the same value, i.e. $|\alpha|^{2} \simeq N$, the noise level becomes very high. \\

\begin{figure}[h]
	\centering
	\hspace*{0.0cm}
	\subfloat[][]{\includegraphics[width=0.98\linewidth,keepaspectratio]{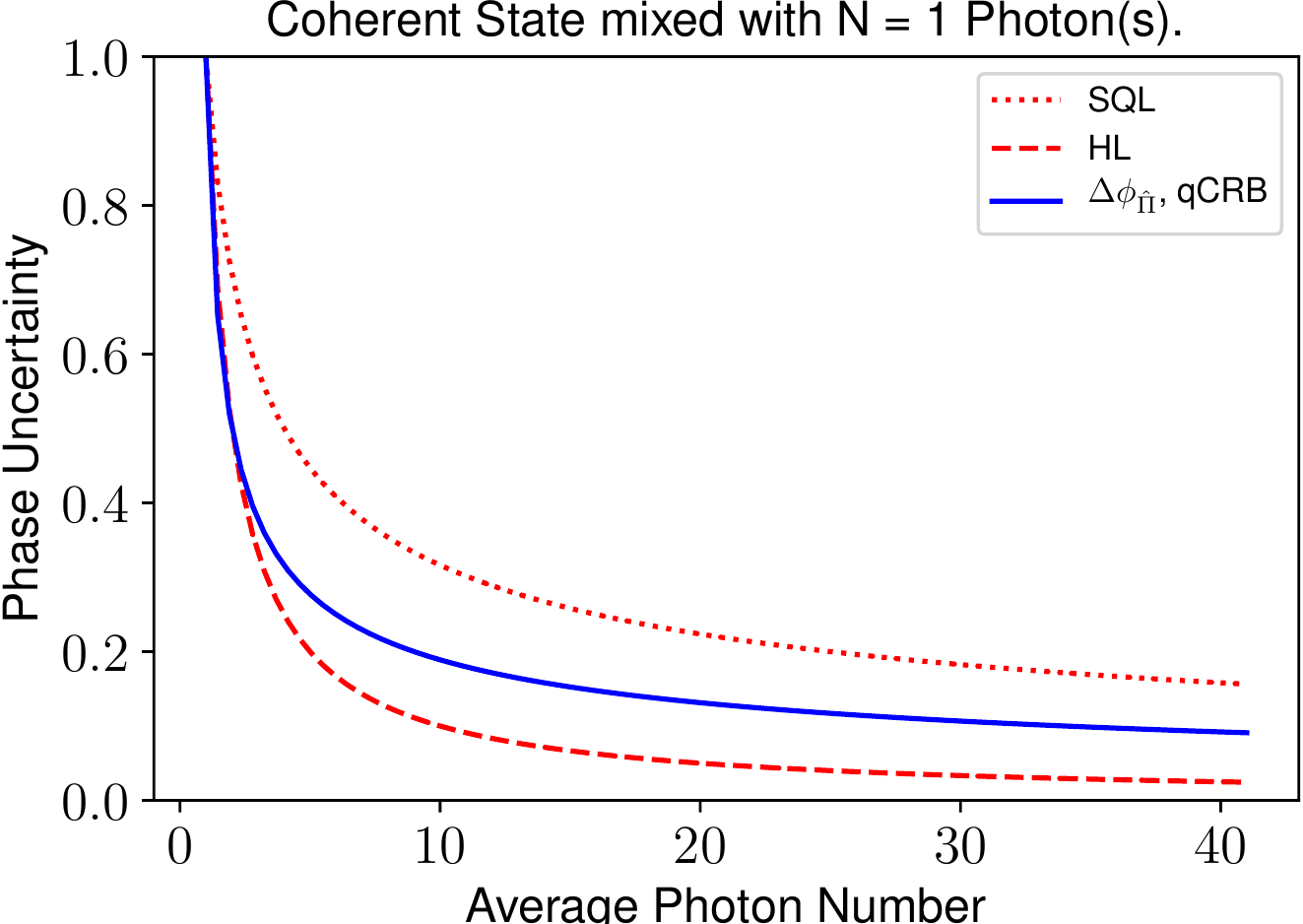}
		\label{fig:PU_N1}}
	\\
	\hspace*{0.0cm}
	\subfloat[][]{\includegraphics[width=1.0\linewidth,keepaspectratio]{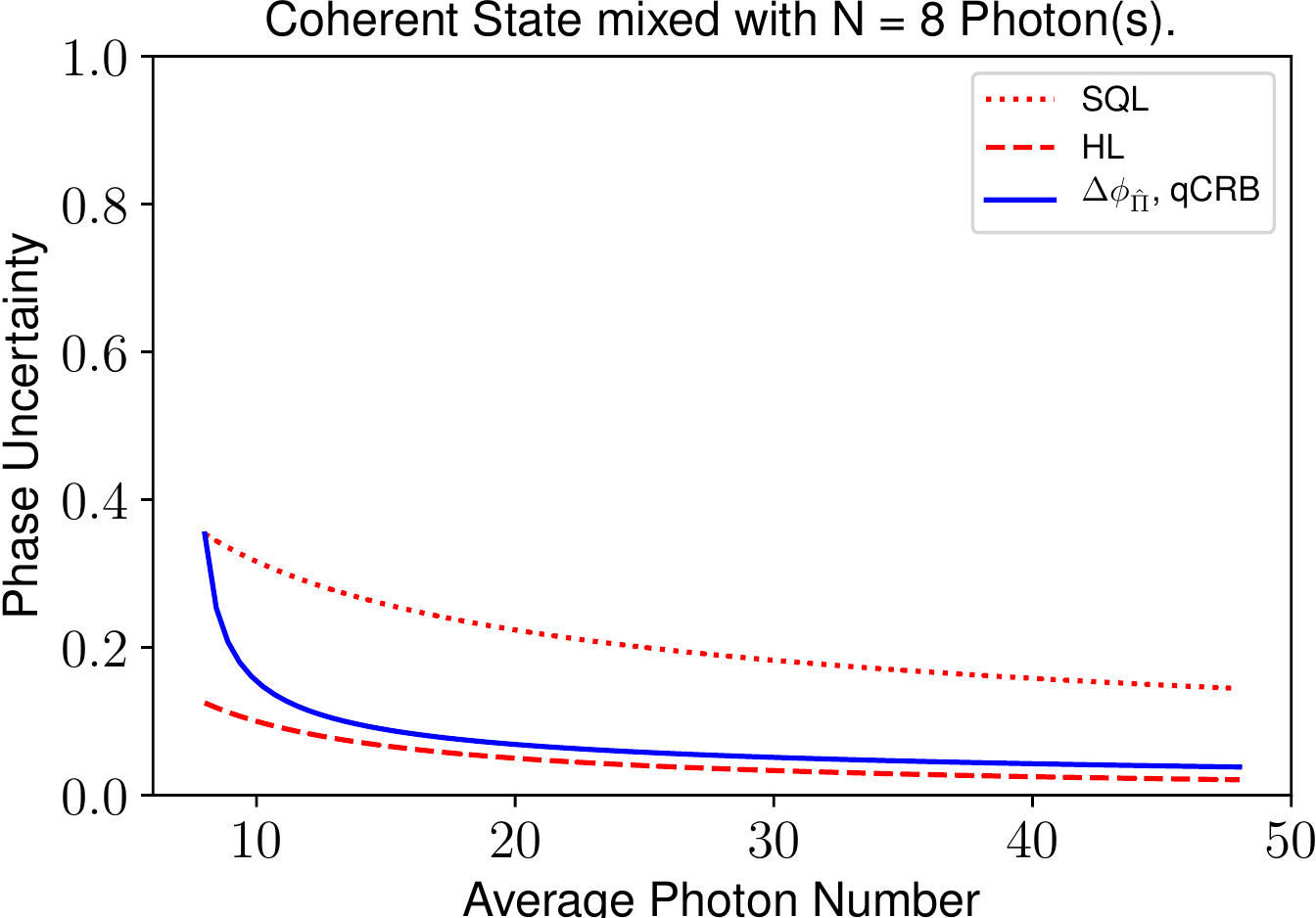}
		\label{fig:PU_N8}}
	\caption{Phase uncertainty computed with phase $\phi=10^{-4}$ for the input state $\ket{\text{in}} = \ket{\alpha}_{a}\ket{N}_{b}$, with accompanying SQL and HL, for  \protect\subref{fig:PU_N1} $N=1$ and \protect\subref{fig:PU_N8} $N=8$.  In the limit of $N\to\infty$, the phase uncertainty approaches the HL. The blue line in each figure also corresponds to the qCRB for this choice of initial state.}
	\label{fig:phase_uncert}
\end{figure}

\noindent Parity-based detection fares quite a bit better. The phase uncertainty in this case is computed numerically using Eq.~\ref{eqn:qoi_5} with $\hat{\Pi}_{b}$ as the detection observable.  The phase uncertainty, along with corresponding SQL and HL, $\Delta\phi_{\text{SQL}}=1/\sqrt{|\alpha|^{2}+N}$ and $\Delta\phi_{\text{HL}} = 1/\left(|\alpha|^{2} + N\right)$ respectively, are plotted for several different values of $N$ in Fig.~\ref{fig:phase_uncert}. The largest gain in sensitivity is achieved in the case of going from $N=0$, wherein the phase uncertainty falls along the SQL, to $N=1$, where the phase uncertainty is sub-SQL.  The noise reduction approaches the HL for increasing values of $N$.  \\

\noindent One can also calculate the QFI using Eq.~\ref{eqn:2.50} and subsequent qCRB, which we remind the reader is independent of detection observable and depends solely on the choice of initial state, to find the minimum phase uncertainty attainable \cite{Birrittella1} 

\begin{equation}
\Delta\phi_{\text{min}} = \frac{1}{\sqrt{|\alpha|^{2} + N\left(1+2|\alpha|^{2}\right)}}.
\label{eqn:cohN_6}
\end{equation} 

\noindent This result follows the blue curve corresponding to parity-based phase sensitivity in Fig.~\ref{fig:phase_uncert} precisely for small values of the phase. It is worth pointing out that for small $\alpha$, the phase uncertainty scales as $1/\sqrt{N}$, which follows the SQL.   

\subsection{\label{sec:QOI_coh} Interferometry with strictly classical (coherent) light}

\noindent So far we have investigated the use of several different quantum states of light as our initial state to the MZI and have shown sub-SQL phase sensitivity in all cases when considering parity-based detection.  Now we will turn our attention to the case where strictly classical light is used.  It is well known that this case will yield phase sensitivities at the SQL, the greatest sensitivity attainable using classical light, for the optimal choice of phase when considering intensity-difference measurements.  So the question is:  what is to be gained by considering the use of photon-number parity?  Here we will endeavor to shed some light on this question. \\

\noindent We start by considering coherent light in one of the input ports of the interferometer, making our initial state $\ket{\text{in}} = \ket{\alpha}_{a}\otimes\ket{0}_{b}$, where once again the coherent state is given by $\ket{\alpha} = e^{-|\alpha|^{2}/2}\sum_{n=0}^{\infty}\frac{\alpha^{n}}{\sqrt{n!}}\ket{n}$. Considering the transformation of Eq.~\ref{eqn:qoi_1} in which the beam splitters are of $\hat{J}_{x}$-type, the state after the first beam splitter is given by (see Appendix \ref{app:subsecA3}) 

\begin{equation}
\ket{\alpha}_{a}\ket{0}_{b} \xrightarrow{\text{BS 1}} \ket{\tfrac{\alpha}{\sqrt{2}}}_{a}\ket{-\tfrac{i\alpha}{\sqrt{2}}}_{b},
\label{eqn:coh_1}
\end{equation}

\noindent and the state after the accumulated phase shift in the $b$-mode (the $\hat{J}_{z}$ operation introduces an anti-symmetric phase shift of $\pm\phi/2$ in each mode, which can be treated as a phase shift of $\phi$ in one arm of the MZI without loss of generality) is given by

\begin{equation}
\ket{\tfrac{\alpha}{\sqrt{2}}}_{a}\ket{\tfrac{i\alpha}{\sqrt{2}}}_{b} \xrightarrow{\phi\text{-shift}} \ket{\tfrac{\alpha}{\sqrt{2}}}_{a}\ket{-\tfrac{i\alpha e^{i\phi}}{\sqrt{2}}}_{b}.
\label{eqn:coh_2}
\end{equation}

\noindent Finally, the state after the second beam splitter is 

\begin{equation}
\ket{\tfrac{\alpha}{\sqrt{2}}}_{a}\ket{-\tfrac{i\alpha e^{i\phi}}{\sqrt{2}}}_{b} \xrightarrow{\text{BS 2}} \ket{\tfrac{\alpha}{2}\left(1+e^{i\phi}\right)}_{a}\ket{\tfrac{i\alpha}{2}\left(1-e^{i\phi}\right)}_{b}.
\label{eqn:coh_3}
\end{equation}

\noindent Next we compare the phase uncertainty obtained using two different detection observables: taking the intensity-difference between modes and performing photon number parity on one of the output modes.

\subsubsection{\label{sec:QOI_coh_diff} Difference in output mode intensities}

\noindent The intensity of a quantized field is proportional to the average photon number of the quantum state of the field \cite{GerryBook}, $I \propto \braket{\hat{n}}$.  Consequently, the difference between mode intensities at the output of the second beam splitter can be written as $\delta I \propto \braket{\hat{a}^{\dagger}\hat{a} - \hat{b}^{\dagger}\hat{b}}$.  In terms of the su(2) Lie algebra, this amounts to the expectation value of the operator $2\hat{J}_{z}$.  Using Eq.~\ref{eqn:coh_3}, the mode intensities are given given by $\braket{\hat{n}_{a\left(b\right)}} = \tfrac{\bar{n}_{0}}{2}\left(1 \pm \cos\phi\right)$ where $\bar{n}_{0} = |\alpha|^{2}$ (note we are assuming a lossless interferometer such that $\bar{n}_{0} = \braket{\hat{n}_{a}} + \braket{\hat{n}_{b}}$ is conserved) leading to the average value $\braket{2\hat{J}_{z}} = \bar{n}_{0}\cos\phi$.  It is also straight-forward to show that $\braket{\left(2\hat{J}_{z}\right)^{2}} = \bar{n}_{0}\left(1 + \braket{2\hat{J}_{z}}\cos\phi\right)$.  Combining these expressions and Eq.~\ref{eqn:qoi_5} yields a phase uncertainty

\begin{equation}
\Delta\phi_{2\hat{J}_{z}} = \frac{\sqrt{\braket{\hat{J}_{z}^{\;2}}-\braket{\hat{J}_{z}}^{2}}}{|\partial_{\phi}\braket{\hat{J}_{z}}|} = \frac{1}{\sqrt{\bar{n}_{0}}|\sin\phi|},
\label{eqn:coh_4}
\end{equation}

\noindent which yields the SQL of phase sensitivity for the value of the phase $\phi=\pi/2$, which means that the detection of small phase shifts, such as what would be expected in gravitational wave detectors, would have a high degree of uncertainty.  Of course, one could compensate for this by inserting a $\pi/2$-phase-shifting element which would have the effect of replacing the $\sin\phi$ with a $\cos\phi$ in Eq.~\ref{eqn:coh_4}.  It is worth pointing out that the fact that $\Delta\hat{J}_{z}$ does not vanish is an indication that the quantum fluctuations of the vacuum (note the coherent state has the same quantum fluctuations as the vacuum) has the effect of limiting the precision of the phase-shift measurement.  Next we consider the use of parity, performed on the output $b$-mode, as our detection observable.  

\subsubsection{\label{sec:QOI_coh_parity} Parity-based detection}

\noindent We define parity with respect to the $b$-mode as $\hat{\Pi}_{b} = \left(-1\right)^{\hat{b}^{\dagger}\hat{b}}$ and likewise for the $a$-mode $\hat{\Pi}_{a} = \left(-1\right)^{\hat{a}^{\dagger}\hat{a}}$. From this the corresponding expectation values and their first derivatives are found to be \cite{Chiruvelli}   

\begin{align}
\braket{\hat{\Pi}_{a\left(b\right)}} &= e^{-\bar{n}_{0}\left(1 \pm \cos\phi\right)}, \label{eqn:coh_5a} \\
\nonumber\\
\partial_{\phi} \braket{\hat{\Pi}_{a\left(b\right)}} &= \pm\bar{n}_{0}\sin\phi\braket{\hat{\Pi}_{a\left(b\right)}},\label{eqn:coh_5b}
\end{align}

\noindent Noting that $\braket{\hat{\Pi}_{a\left(b\right)}^{2}}\equiv 1$, the phase uncertainty can immediately be found from the error propagation calculus  

\begin{align}
\Delta\phi_{\hat{\Pi}_{a}} &= \frac{\sqrt{1-\braket{\hat{\Pi}_{a}}^{2}}}{|\partial_{\phi}\braket{\hat{\Pi}_{a}}|} = \frac{\sqrt{e^{4\bar{n}_{0}\cos^{2}\phi/2}-1}}{\bar{n}_{0}|\sin\phi|}, \label{eqn:coh_6a}\\
& \nonumber \\
\Delta\phi_{\hat{\Pi}_{b}} &= \frac{\sqrt{1-\braket{\hat{\Pi}_{b}}^{2}}}{|\partial_{\phi}\braket{\hat{\Pi}_{b}}|} = \frac{\sqrt{e^{4\bar{n}_{0}\sin^{2}\phi/2}-1}}{\bar{n}_{0}|\sin\phi|}.
\label{eqn:coh_6}
\end{align}

\noindent The curves for both $\braket{\hat{\Pi}_{a\left(b\right)}}$ are displaced from one another by a $\pi$-phase shift, as evident by Eq.~\ref{eqn:coh_5a}.  This implies that while the peak for $\braket{\hat{\Pi}_{b}}$ occurs at $\phi = 0$, the peak for $\braket{\hat{\Pi}_{a}}$ occurs at $\phi=\pi$. We can expand Eqs.~\ref{eqn:coh_6a} and \ref{eqn:coh_6} about their respective optimal phase values to find

\begin{align}
\Delta\phi_{\hat{\Pi}_{a}} &\stackrel{\phi\to \pi}{\simeq} \frac{1}{\sqrt{\bar{n}_{0}|\cos\phi|}},  \label{eqn:coh_7a}\\
\nonumber \\
\Delta\phi_{\hat{\Pi}_{b}} &\stackrel{\phi\to 0}{\simeq} \frac{1}{\sqrt{\bar{n}_{0}|\cos\phi|}}, \label{eqn:coh_7}
\end{align}

\noindent which is in agreement with the minimum phase uncertainty attainable for this choice of initial state, the qCRB.  This can be quickly verified by calculating the QFI using Eq.~\ref{eqn:2.50} to immediately give the SQL $\Delta\phi_{\text{min}} = 1/\sqrt{\bar{n}_{0}} = \Delta\phi_{\text{SQL}}$. \\

\noindent The output signal itself is not super-resolved in the usual sense of having $N\phi$ ($N>1$) oscillation frequency scaling.  However, the narrowing of the peak when considering parity-based measurement as opposed to the usual method of taking the intensity-difference has been defined as a form of super-resolution by Gao \textit{et al} \cite{SupRes}. It is worth pointing out that for input states displaying quantum properties, such as all of the cases considered in the previous sections, super-resolution tends towards providing a greater degree of phase sensitivity. On the other hand, super-resolution has also been demonstrated in the absence of entangled states using light exhibiting strictly classical interference \cite{Resch}. We also note the method of photon-number parity detection has recently been discussed in the context of SU(1,1) interferometry \cite{Li}.\\ 

\section{\label{sec:exper} Experimental realizations of parity-based detection}

\noindent As emphasized in the Introduction, the parity operator, whether in the context of atomic (or spin) systems or in the context of photon number, is a Hermitian operator and is therefore an observable; but one that does not have a classical analog.  The being the case, the question becomes how can parity be measured or least determined through some measurement process?  The obvious way to do that is through counting the number of atoms \cite{Hume} in the excited (or ground), something that can be done through a process known as electron shelving [], or counting the number of photons in an optical field and raising $-1$ to that power.  Of course, this means that the counting itself must be possible with a resolution at the level of a single atom or photon: a challenging prospect for the cases where the number or atoms or photons is large.  Ideally one would like to be able to determine parity directly, by which we mean through a technique with a readout of $\pm 1$ without directly measuring the number of atoms or photons.  This could possibly be done with quantum non-demolition (QND) measurements.  It is worth noting that QND measurements can be used to measure the number of atoms or photons as well, but what we have in mind is the measurement of parity wherein the detector does no counting at all.  As we will show, these two mentioned methods of measuring parity, while ultimately yielding the same parity values, amounts to different kinds of measurements if used to perform, for example, state-projective measurements.  

\subsection{\label{sec:QND_AtomPar} QND measures of atomic parity}

\noindent Here we will discuss several methods of performing a QND measure of atomic parity through the use of coupling between the atomic system with an ancillary subsystem.  We begin by defining the even/odd atomic parity projection operators.  That is, we define the operators that project the atomic state into even/odd numbers of excitations.  These projectors are given by 

\begin{align}
\hat{\Pi}_{\text{even}} &= \sum_{m=-j}^{j}\cos^{2}\Big[\frac{\left(j+m\right)\pi}{2}\Big]\ket{j,m}\bra{j,m},\label{eqn:54a} \\
& \nonumber \\
\hat{\Pi}_{\text{odd}} &= \sum_{m=-j}^{j}\sin^{2}\Big[\frac{\left(j+m\right)\pi}{2}\Big]\ket{j,m}\bra{j,m},
\label{eqn:54}
\end{align}

\noindent satisfying the POVM condition $\sum_{i}\hat{\Pi}_{i}=\hat{I}$. Note that $j+m$ represents the total number of atoms found in the excited state while $j-m$ represents the total number of atoms found in the ground state. Consequently $j\pm m$ takes on only integer values such that for a given $j \pm m$, only one of the projectors will be nonzero.  These projection operators can be used to express the atomic parity operator as

\begin{align}
\hat{\Pi}&=\sum_{i}\lambda_{i}\;\hat{\Pi}_{i}=\hat{\Pi}_{\text{even}}-\hat{\Pi}_{\text{odd}} \nonumber \\
& = \sum_{m=-j}^{j}\cos\left[\left(j+m\right)\pi\right]\ket{j,m}\bra{j,m}
\label{eqn:55}
\end{align}

\noindent where $\lambda_{i}$ are the eigenvalues of the parity operator, respectively.  From Eqs.~\ref{eqn:54a} and \ref{eqn:54}, it follows that

\begin{align}
\braket{\hat{\Pi}_{\text{even}}}&=\frac{1}{2}\left(1+\braket{\hat{\Pi}}\right), \\
&\nonumber \\
\braket{\hat{\Pi}_{\text{odd}}}&=\frac{1}{2}\left(1-\braket{\hat{\Pi}}\right), \\
&\nonumber
\label{eqn:56}
\end{align}

\noindent where $\braket{\hat{\Pi}_{\text{odd}}}+\braket{\hat{\Pi}_{\text{even}}}=1$. We note here that 'parity' is defined with respect to the number of atoms found in the excited state.  That is, 'even' parity denotes an even number of atoms found in the excited state and 'odd' parity denotes and odd number of atoms found in the excited state.  Let us move on to consider a couple of different cases.

\subsubsection{\label{sec: Unphysical} Coupling to an ancillary atomic system}

\noindent Given an atomic system in which we wish to measure, denoted by the $a$-mode, we introduce an ancillary atomic system (occupying the $b$-mode) prepared in an atomic coherent state $\ket{\tau,j_{b}}_{b}$ as well as the coupling Hamiltonian and corresponding evolution operator:

\begin{equation}
\hat{H} = \hbar\chi\big(\hat{J}_{0}^{a}+\hat{J}_{z}^{a}\big)\big(\hat{J}_{0}^{b}+\hat{J}_{z}^{b}\big)\;\;\;\;\;\; \to \;\;\;\;\;\; \hat{U}=e^{-it\hat{H}/\hbar},
\label{eqn:57}
\end{equation}

\noindent where $\chi$ is the coupling strength.  This interaction Hamiltonian can be thought of as the atomic analog to the field coupling cross-Kerr interaction. We define the initial state as

\begin{align}
\ket{\text{initial}} &= \ket{\psi}_{a}\otimes\ket{\tau,j_{b}}_{b} =\sum_{m=-j_{a}}^{j_{a}}C_{m}^{\left(j_{a}\right)}\ket{j_{a},m}_{a}\otimes \nonumber \\
&\otimes \left(1+|\tau|^{2}\right)^{-j_{b}}\sum_{m'=-j_{b}}^{j_{b}}\binom{2j_{b}}{j_{b}+m'}\tau^{j_{b}+m'}\ket{j_{b},m'}_{b}.
\label{eqn:58}
\end{align}

\noindent Note that we wish to make a projective measurement on the ancillary atomic system ($b$-system) in order to determine the atomic parity of the target system ($a$-system). The final state is then

\begin{align}
\ket{\text{final}} &= e^{-it\hat{H}/\hbar}\ket{\text{initial}} \nonumber \\
&= \sum_{m=-j_{a}}^{j_{a}}C_{m}^{\left(j_{a}\right)}\ket{j_{a},m}_{a}\otimes \ket{\tau e^{-i t\chi\left(j_{a}+m\right)},j_{b}}_{b}.
\label{eqn:59}
\end{align}

\noindent For the choice of $\chi t \to \pi$, this becomes

\begin{align}
\ket{\text{final}} &= \ket{\psi_{\text{even}}}_{a}\ket{\tau,j_{b}}_{b}+\ket{\psi_{\text{odd}}}_{a}\ket{-\tau,j_{b}}_{b} \nonumber \\
&= \sum_{m=-j_{a}}^{j_{a}}\left[\cos^{2}\Big(\tfrac{\left(j_{a}+m\right)\pi}{2}\Big)\ket{\tau,j_{b}}_{b} \;\;+ \right. \nonumber \\
& \;\;\;\;\;\;\;\;\;\;\;\;\;\;\;\;\;\;\;\left. +\;\;\; \sin^{2}\Big(\tfrac{\left(j_{a}+m\right)\pi}{2}\Big)\ket{-\tau,j_{b}}_{b} \right]\;C_{m}^{\left(j_{a}\right)}\ket{j_{a},m} \nonumber \\
&= \sum_{m=-j_{a}}^{j_{a}}C_{m}^{\left(j_{a}\right)}\ket{j_{a},m}_{a}\ket{\Phi_{m}}_{b}.
\label{eqn:60}
\end{align}

\noindent  The system is entangled such that a projection onto the ancillary atomic system yielding the ACS characterized by parameter $-\tau$ will project out odd atomic states in the target system and projection onto the ACS characterized by parameter $+\tau$ will project out even atomic states in the target system.  What we require is a means of determining which state the ancillary atomic system is in. Let us assume the ancillary atomic coherent state is prepared such that $\tau=-1$;  this corresponds to a separable state in which all atoms of the ancillary atomic system are in the same superposition state

\begin{equation}
e^{-i\tfrac{\pi}{2}\hat{J_{y}}}\ket{j,-j}=\ket{-1,j}=\ket{\psi_{-}}^{\otimes N} =\frac{1}{2^{N/2}}\left(\ket{g}-\ket{e}\right)^{\otimes N}.
\label{eqn:61}
\end{equation}      

\noindent Similarly, we can also define the phase-rotated atomic coherent state

\begin{equation}
e^{-i\tfrac{\pi}{2}\hat{J_{y}}}\ket{j,j}=\ket{1,j}=\ket{\psi_{+}}^{\otimes N} =\frac{1}{2^{N/2}}\left(\ket{g}+\ket{e}\right)^{\otimes N}.
\label{eqn:62}
\end{equation}   

\noindent With this, we can rewrite the state $\ket{\Phi_{m}}_{b}$, assuming $\tau = -1$, as 

\begin{align}
\ket{\Phi_{m}}_{b} &= \cos^{2}\Big(\tfrac{\left(j_{a}+m\right)\pi}{2}\Big)\ket{-1,j_{b}}_{b} + \sin^{2}\Big(\tfrac{\left(j_{a}+m\right)\pi}{2}\Big)\ket{1,j_{b}}_{b} \nonumber \\
&= \cos^{2}\Big(\tfrac{\left(j_{a}+m\right)\pi}{2}\Big)e^{-i\tfrac{\pi}{2}\hat{J^{b}_{y}}}\ket{j_{b},-j_{b}}_{b} + \nonumber \\
& \;\;\;\;\;\;\;\;\;\;\;\;\;\;\;\;\;\;\;\;\;\;\;\;+  \sin^{2}\Big(\tfrac{\left(j_{a}+m\right)\pi}{2}\Big)e^{-i\tfrac{\pi}{2}\hat{J^{b}_{y}}}\ket{j_{b},j_{b}}_{b}.
\label{eqn:63}
\end{align}

\noindent It is important to note that only a single term in the superposition state Eq.~\ref{eqn:63} can be  present at a time. Performing a single $\pi/2$-pulse yields the ancillary atomic system in the state

\begin{equation}
e^{i\tfrac{\pi}{2}\hat{J}_{y}^{b}}\ket{\Phi_{m}}_{b}=
\begin{cases}
\cos^{2}\Big(\tfrac{\left(j_{a}+m\right)\pi}{2}\Big)\ket{j_{b},-j_{b}}_{b},  & j_{a}+m \;\;\text{even} \\
\\
\sin^{2}\Big(\tfrac{\left(j_{a}+m\right)\pi}{2}\Big)\ket{j_{b},j_{b}}_{b}, & j_{a}+m \;\;\text{odd} 
\end{cases}
\label{eqn:64}
\end{equation}

\noindent where $j_{a}+m$ is an integer corresponding to the number of excited atoms.  A state-reductive measurement performed on the ancillary atomic system would inform the experimenter of the parity of the target system \textit{without} providing explicit knowledge of the value of $j_{a}+m$.  Note that this works for an arbitrary value of $j_{b}$; the ancillary atomic system can be as small as a single atom.  Next we will discuss a method involving a known coupling Hamiltonian readily capable of being experimentally implemented. \\

\subsubsection{\label{sec: Physical} Coupling to a field state}

\noindent Some work has been done in developing a QND measure of photon number parity and a means of projecting out parity eigenstates in optical fields \cite{GerryQND}. Similarly, QND parity measurements have been performed in the context of error correction for a hardware-efficient protected quantum memory using Schr\"{o}dinger cat states \cite{PhotonJump} as well as a fault-tolerant detection of quantum error \cite{QuantumError}. Parity measurements have also been utilized in the entanglement of bosonic modes through the realization of the eSWAP operation \cite{eSWAP1} using bosonic qubits stored in two superconducting microwave cavities \cite{eSWAP2}.  Here will consider a similar scheme in which one couples the target atomic system to an ancillary field state.  The interaction Hamiltonian coupling the two subsystems is given by

\begin{equation}
\hat{H} = \hbar \chi \hat{b}^{\dagger}\hat{b}\hat{J}_{z}\;\;\;\; \to \;\;\;\; \hat{U} = e^{-i\frac{t}{\hbar}\hat{H}},
\label{eqn:65}
\end{equation}

\noindent where $\{\hat{b}$, $\hat{b}^{\dagger}\}$ are the usual boson annihilation and creation operators, respectively. Consider an ancillary field state, given by the usual coherent state

\begin{equation}
\ket{\alpha}_{b} = e^{-\tfrac{1}{2}|\alpha|^{2}}\sum_{n=0}^{\infty}\frac{\alpha^n}{\sqrt{n!}}\ket{n}_{b}.
\label{eqn:66}
\end{equation}

\noindent Following the same procedure as the previous section and setting $\chi t= \pi $ this coupling Hamiltonian yields the final state

\begin{equation}
\ket{\text{final}}=\hat{U}\ket{\text{initial}}=\sum_{m=-j}^{j}C_{m}^{\left(j\right)}\ket{j,m}_{a}\otimes\ket{\left(-1\right)^{m}\alpha}_{b}.
\label{eqn:67}
\end{equation}

\noindent The final entangled state will depend greatly on the value of $j$.  More specifically, noting $N=2j$, this simplifies to

\begin{widetext}
	\begin{equation}
	\ket{\text{final}}=
	\begin{cases}
	\ket{\psi_\text{odd}}_{a}\ket{-\alpha}_{b} + \ket{\psi_\text{even}}_{a}\ket{\alpha}_{b} & \;\;j=0,2,4...,\;\;\;\;N=0,4,8,... \\
	\\
	\ket{\psi_\text{even}}_{a}\ket{-\alpha}_{b} + \ket{\psi_\text{odd}}_{a}\ket{\alpha}_{b} & \;\;j=1,3,5...,\;\;\;\;N=2,6,10,... \\
	\\
	\ket{\psi_\text{even}}_{a}\ket{-i\alpha}_{b} + \ket{\psi_\text{odd}}_{a}\ket{i\alpha}_{b} & \;\;j=\tfrac{1}{2},\tfrac{5}{2},\tfrac{9}{2}...,\;\;\;\;N=1,5,9,...\\
	\\
	\ket{\psi_\text{odd}}_{a}\ket{-i\alpha}_{b} + \ket{\psi_\text{even}}_{a}\ket{i\alpha}_{b} & \;\;j=\tfrac{3}{2},\tfrac{7}{2},\tfrac{11}{2}...,\;\;\;\;N=3,7,11,... 
	\end{cases}
	\label{eqn:68}
	\end{equation}
\end{widetext}

\noindent In order for this procedure to be applicable, the total number of atoms involved must be known with certainty as the resulting superposition state and subsequent detection scheme will greatly depend on having this information.  As a proof of concept, let us consider the case where the number of atoms is $N=0,4,8,...\;$.  As per Eq.~\ref{eqn:68}, the final state is given by $\ket{\text{final}} = \ket{\psi_\text{odd}}_{a}\ket{-\alpha}_{b} + \ket{\psi_\text{even}}_{a}\ket{\alpha}_{b}$. Mixing the field at a 50:50 beam splitter with an equal-amplitude phase-adjusted coherent state $\ket{\alpha}_{c}$.  Assuming a $50:50$ $\hat{J}_{y}$-type beam splitter of angle $-\pi/2$ (see Appendix \ref{app:subsecA3}), the beam splitter (labeled BS 2 for reasons that will become clear) results in the transformation

\begin{align}
\ket{\alpha}_{b}\ket{\alpha}_{c} &\xrightarrow{\text{BS 2}} \ket{\sqrt{2}\alpha}_{b}\ket{0}_{c}, \nonumber \\ \label{eqn:69} \\
\ket{-\alpha}_{b}\ket{\alpha}_{c} &\xrightarrow{\text{BS 2}} \ket{0}_{b}\ket{\sqrt{2}\alpha}_{c}. \nonumber
\end{align}

\noindent The total state is now given by 

\begin{equation}
\ket{\text{final}}_{\text{BS2}} = \ket{\psi_\text{odd}}_{a}\ket{0}_{b}\ket{\sqrt{2}\alpha}_{c} + \ket{\psi_\text{even}}_{a}\ket{\sqrt{2}\alpha}_{b}\ket{0}_{c},
\label{eqn:70}
\end{equation}

\noindent in which all photons are in either one optical mode or the other.  A simple detection scheme informs the experimenter of the parity of the target atomic system.  Upon a state reductive measurement performed on the optical modes, the atomic system becomes

\begin{align}
\ket{\Psi}_{a} =
\begin{cases} 
\ket{\psi_{\text{even}}}_{a},  & b\text{-mode click} \\
\\
\ket{\psi_{\text{odd}}}_{a}.  & c\text{-mode click} 
\end{cases}
\label{eqn:71}
\end{align}

\noindent The optical fields $\ket{\alpha}_{b}$ and $\ket{\alpha}_{c}$ used in this procedure can be derived from the same beam if one starts with the state $\ket{\sqrt{2}\alpha}_{c}$ and uses a $50:50$ $\hat{J}_{y}$-type beam splitter of angle $+\pi/2$ such that

\begin{equation}
\ket{\sqrt{2}\alpha}_{b}\ket{0}_{c} \xrightarrow{\text{BS 1}} \ket{\alpha}_{b}\ket{\alpha}_{c}.
\label{eqn:72}
\end{equation} 

\noindent After beamsplitting, the $b$-mode coherent state can be coupled with the target atomic system before being photomixed with the coherent state $\ket{\alpha}_{c}$ at the second beam splitter. This method would allow one to determine the parity of the atomic system without explicit knowledge of the number of exicted atoms $j+m$.

\subsection{\label{sec:exp_opt} Detection of photon parity}

\begin{figure}
	\centering
	\includegraphics[width=1.0\linewidth,keepaspectratio]{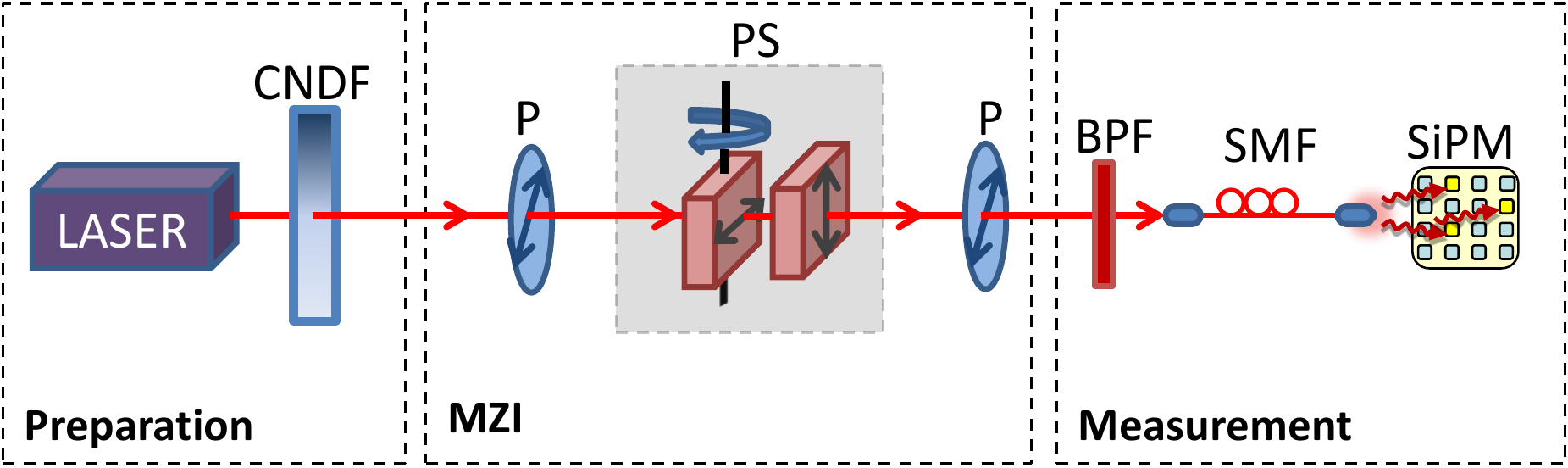}
	\caption{A schematic of the experimental setup used by Cohen \textit{et al.} The coherent states are produced by a Ti:Sapphire laser with average photon numbers controlled by a calibrated neutral density filer (NDF) and the MZI is composed from two polarizers $P$ at $45^{\circ}$ and a phase shifter. What is defined in the text as the output $a$-mode is filtered spectrally by $3\;nm$ band pass filter (BPF) and spatially by a single mode fiber (SMF).  This mode is detected by the silicon photomultiplier.  Figure taken from Cohen \textit{et al.} \cite{Cohen1} with permission from the authors.}
	\label{fig:Exp_Fig1}
\end{figure}

\noindent As mentioned above, one way to obtain photon number parity is perform photon number counts $n_{\text{count}}$ and raise $-1$ to that power:  $\Pi_{\text{measured}} = \left(-1\right)^{n_{\text{count}}}$. That raises the issue of the general lack of photon number counting techniques having resolution at the level of one photon.  Yet there are ways around this problem.  In fact, an experiment to detect a phase shift through optical interferometry using coherent light and parity measurements was performed a few years ago by Eisenberg's group at Hebrew University \cite{Cohen1}.  Recall from above, and from Gao \textit{et al.} \cite{SupRes}, that parity measurements performed in this context are predicted to result in phase-shift detections that are super-resolved even though the phase sensitivity is at the SQL.  The experiment of L. Cohen \textit{et al.} \cite{Cohen1} confirm this. \\

\noindent In the experiment reported, the phase shift to be detected was set to be $\pi$.  If we look at Eqs.~\ref{eqn:coh_6a} and \ref{eqn:coh_6}, and based on the discussion to follow, we see that with our labeling scheme, we should be performing parity measurements on the output $a$-mode.  That is, for $\phi$ near $\pi$, we have $\braket{\hat{\Pi}_{a}}=e^{-\bar{n}_{0}\left(1+\cos\phi\right)} \simeq e^{-\tfrac{\bar{n}_{0}}{2}\left(\phi-\pi\right)^{2}}$, where $\bar{n}_{0}$ is the total number of photons within the interferometer, which peaks at unity for $\phi \to \pi$.  In this limit $\Delta\phi \to 1/\sqrt{\bar{n}_{0}}$, the SQL.\\

\noindent These authors also measured a different kind of parity in which the outcomes are either no photons detected, as described by the projector $\ket{0}\bra{0}$, or any number of photons detected but without resolution, $\sum_{n=0}^{\infty}\ket{n}\bra{n}=\bm{I} - \ket{0}\bra{0}$.  Defining $\hat{Z} = \ket{0}_{a}\bra{0}$, from the output state given in Eq.~\ref{eqn:coh_3} we find the probability of there being no photon detections to be 

\begin{equation}
	P_{0}\left(\phi\right) = \braket{\hat{Z}}_{\text{out}} = e^{-\tfrac{\bar{n}_{0}}{2}\left(1+\cos\phi\right)}\stackrel{\phi\to\pi}{\simeq} e^{-\frac{\bar{n}_{0}}{4}\left(\phi-\pi\right)^{2}}. 
	\label{eqn:exp_1}
\end{equation}  

\noindent From the error propagation calculus one easily finds for $\phi\to\pi$, the phase uncertainty is given by $\Delta\phi_{a} \to 1/\sqrt{\bar{n}_{0}}$, which is also shot-noise limited.  Evidently the two observables give similar results for the case of input coherent light.  \\

\begin{figure}
	\centering
	\hspace*{-0.5cm}
	\includegraphics[width=1.02\linewidth,keepaspectratio]{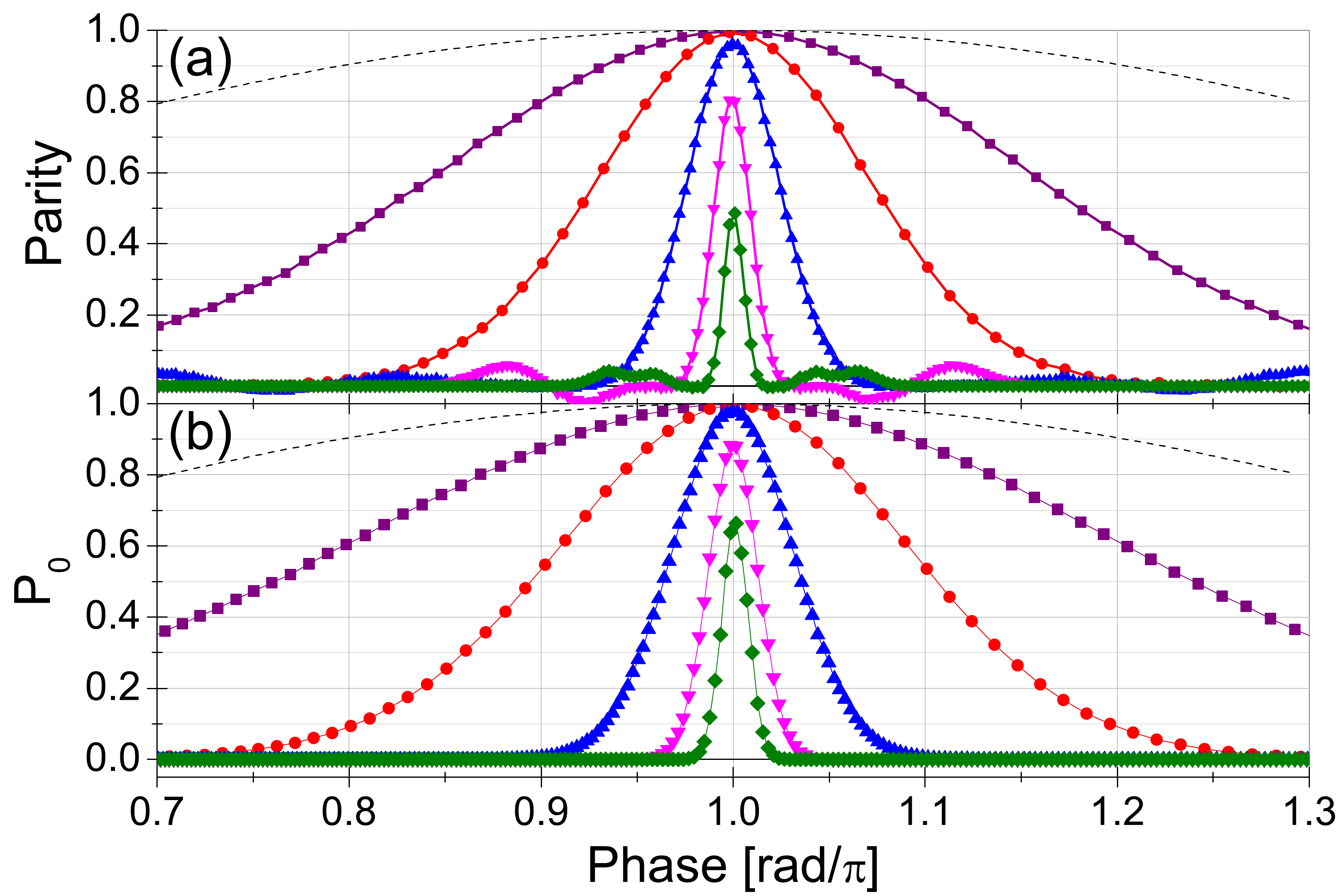}
	\caption{Expectation values of the parity operator as well as the no-photons-detected projector $\hat{Z}$. For each curve the average photon numbers are: $4.6 \pm 0.2$ (purple, square), $25 \pm 1$ (red, circle), $200 \pm 8$ (blue, triangle), $1,190 \pm 50$ (pink, inverted triangle) and $4,150 \pm 150$ (green, rhombus).  The classic interference curves are presented in black-dashed lines on each plot.  Errors are not shown as they are too small to be represented.  Figure taken from Cohen \textit{et al.} \cite{Cohen1} with permission from the authors.}
	\label{fig:Exp_Fig2}
\end{figure}

\noindent The experimental setup used by Cohen \textit{et al.} \cite{Cohen1} is detailed in Fig.~\ref{fig:Exp_Fig1}.  The coherent inputs are produced by a Ti:Sapphire laser with a calibrated variable neutral density filter (NDF) employed to control the average photon numbers.  The MZI consists of two polarizers at $45^{\circ}$ and the phase-shift is produced by tilting the calcite crystal.  The output mode to be measured is band pass filtered (BPF) and spatially filtered by a single mode fiber (SMF).  This mode was detected by a silicon photomultiplier consisting of an array of beam splitters and single-photon detectors.  Such an arrangement for photon counting with a resolution at the single-photon level is described in Kok and Lovett \cite{Kok3}.  See also the review of photon detection by Silberhorn \cite{Silberhorn}.  For the fine details of the experiment, we refer the reader to the paper by Cohen \textit{et al.} \cite{Cohen1}.  \\

\begin{figure}
	\centering
	\hspace*{-0.35cm}
	\includegraphics[width=1.06\linewidth,keepaspectratio]{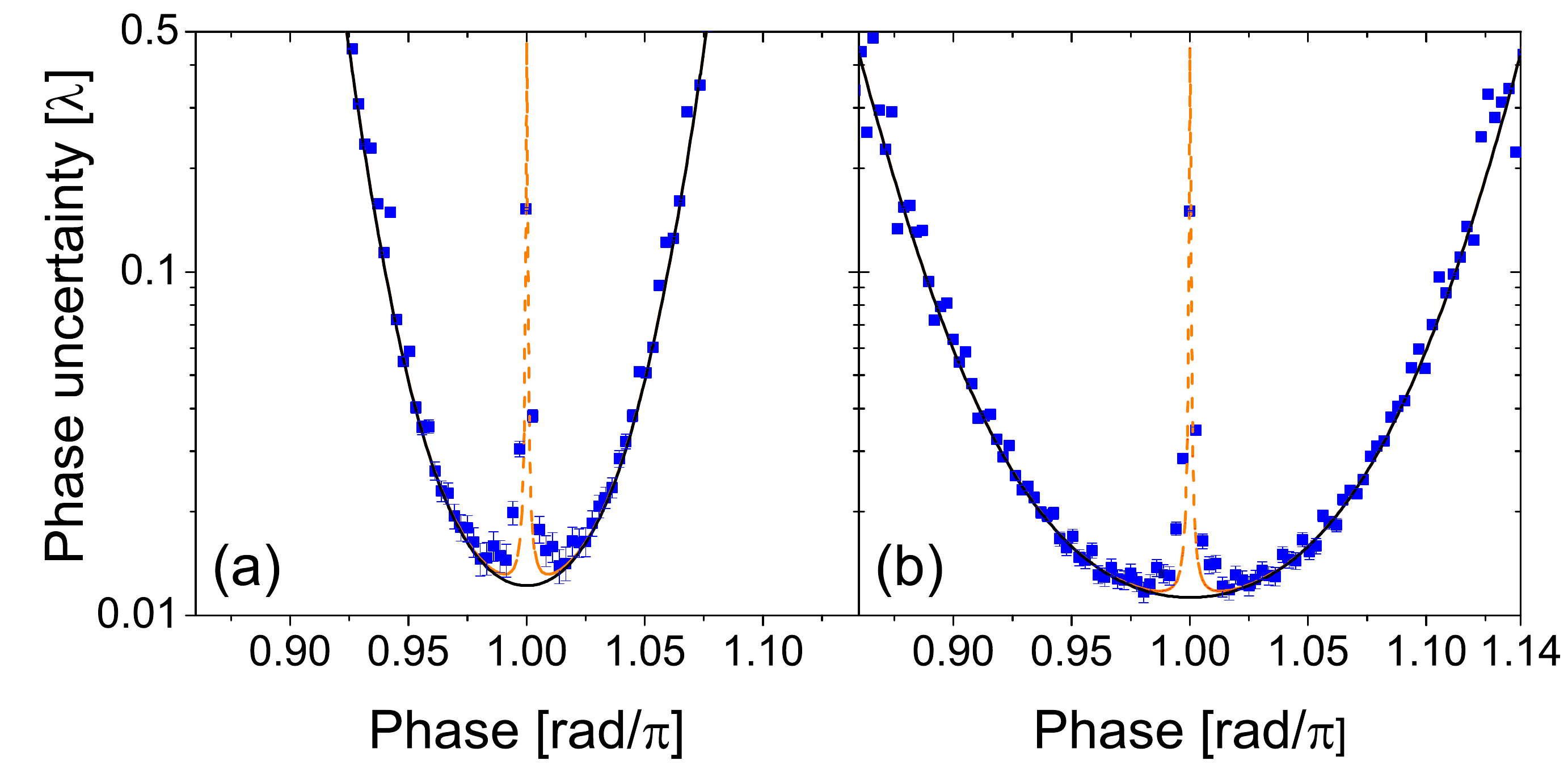}
	\caption{Phase uncertainties obtained through experimental data for $\bar{n}_{0}=200$ (blue, square) as a function of the phase $\phi$ for both parity (left-most) and no-photons-detected probability $P_{0}$ (right-most). Fits based on the theory are represented by solid black lines and a correction made to account for imperfect visibility (see Eq. 9 in Cohen \textit{et al.} \cite{Cohen1}) by orange-dashed lines. Errors representing the standard deviation in the phase uncertainty estimation process are shown when larger than their symbols. Figure taken from Cohen \textit{et al.} \cite{Cohen1} with permission from the authors.}
	\label{fig:Exp_Fig3}
\end{figure}

\begin{figure}
	\centering
	\hspace*{-0.3cm}
	\includegraphics[width=1.01\linewidth,keepaspectratio]{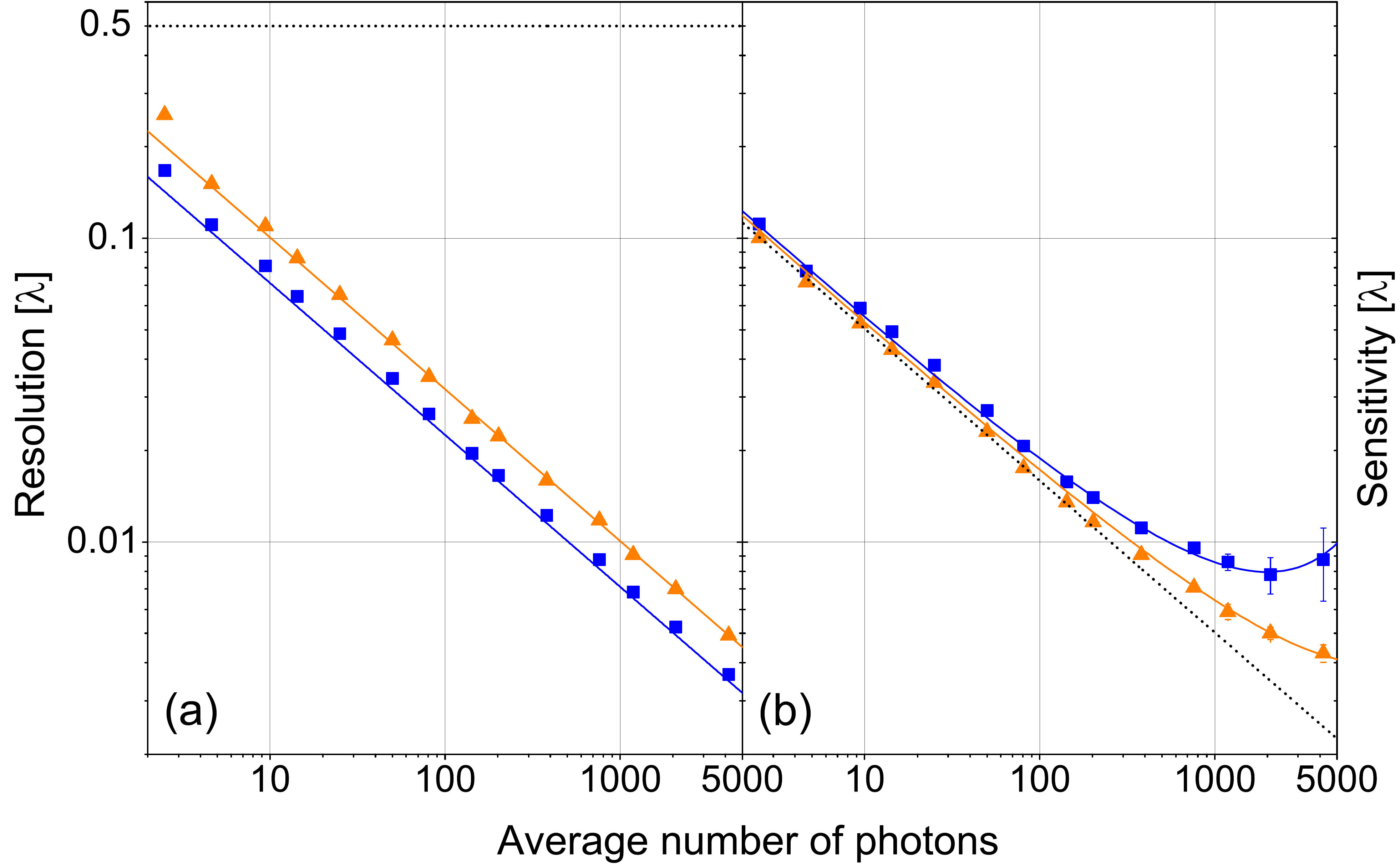}
	\caption{(a) The resolution and (b) the smallest uncertainty, plotted as a function of average photon number.  Parity results are represented by blue squares and $P_{0}$ with orange triangles.  Solid lines are the theoretical predictions.  For the resolution there are no free parameters and the predictions for sensitivity use the visibility values from the fits of Fig.~\ref{fig:Exp_Fig3} (the orange-dashed lines). Parity has better resolution but larger deviation from the SQL.  Errors were calculated as before and shown when larger than their symbol.  Figure taken from Cohen \textit{et al.} \cite{Cohen1} with permission from the authors.}
	\label{fig:Exp_Fig4}
\end{figure}

\noindent The expectation values of the parity operator $\hat{\Pi}_{a}$ (in our notation) and $\hat{Z}$ are plotted in Fig.~\ref{fig:Exp_Fig2} for several values of average photon number.  The predicted narrowing around the peak at $\phi=\pi$ is evident, though the degradation of the visibility for high average photon number is due to the imperfect visibility of the interferometer itself as well as dark counts.  Plots of the corresponding phase uncertainties for $\bar{n}_{0}=200$ are found in Fig.~\ref{fig:Exp_Fig3}.  Note that the peak at $\phi=\pi$ is wider for $P_{0}$ than for parity. Finally in Fig.~\ref{fig:Exp_Fig4} the results are summarized for resolution and sensitivty showing that, in part (a), parity yields greater resolution thatn $P_{0}$.  That is, for parity-based measurement, resolution reachs $\tfrac{\lambda}{288}$ where $\lambda = 780\;\text{nm}$ is the wavelength of the laser light, which is a factor of $1/144$ improvement (smaller) over the Rayleigh limit $\lambda/2$. The resolutions in the two cases differ by the expected amount $\sqrt{2}$,  On the other hand, as can be seen in part (b) of Fig.~\ref{fig:Exp_Fig4}, parity has a larger deviation from the SQL as compared to $P_{0}$.  In fact, up to 200 photons, sensitivity  of $P_{0}$ is maintained at the SQL.   \\

\noindent We mention that in recent years considerable effort has been directed towards the development of photon-number-resolving detectors.  These include superconducting transition edge detectors \cite{Detect1}, loop detectors \cite{Detect2}, detection by multiplexing \cite{Detect3}, an array of avalanche photodiodes (APDs) \cite{Detect4} and an array of single-photon detectors \cite{Detect5}.  This list by no means exhausts the literature on this topic.  In principle one could avoid photon counting entirely and instead perform a quantum non-demolition measurement of photonic parity in the same vain as discussed in Section \ref{sec:QND_AtomPar} pertaining to atomic parity.  This possibility has been discussed by Gerry \textit{et al.} \cite{GerryQND} as an extension of a technique proposed for the QND measurement of photon number \cite{QNDnum}.  The problem with these techniques is that they depend on a cross-Kerr interaction with a large third order nonlinearity which does not exist in optical materials. \\

\noindent But there is yet another possibility first mentioned by Campos \textit{et al.} \cite{Campos} in their analysis of interferometry with twin-Fock states.  Recall that the Wigner function is given by $W\left(\alpha\right) = \tfrac{2}{\pi}\braket{\hat{D}\left(\alpha\right)\hat{\Pi}\hat{D}^{\dagger}\left(\alpha\right)}$; it follows for $\alpha = 0$ that $\braket{\hat{\Pi}}\equiv \tfrac{\pi}{2}W\left(0\right)$, thus showing the expectation value of the parity operator can be found directly by value of the Wigner function at the origin of phase space.  Now the Wigner function can be constructed by the techniques of quantum state tomography which uses a balanced homodyning and the inverse Radon transformation to perform filtered back projections \cite{Tomography}.  However, one does not need the entire Wigner function: only its value at the origin of phase space is required.  Plick \textit{et al.} \cite{Plick} have examined this prospect in detail for Gaussian states of light.  

\subsection{\label{sec:exp_rannumgen} Quantum random number generator}

\noindent We close this section with a brief discussion of the prospect of using photon-number resolved detection as a basis for a quantum random number generator (QRNG) (for a discussion on QRNGs, see Herrero-Collantes \textit{et al.} \cite{QRNG} and references therein) with laser light.  As is well known, laser light shone on a $50:50$ beam splitter cannot be the basis of a QRNG as a matter of principle no matter the intensity of the light.  The ideal source of light is a single photon generated on demand, where the photon falls on the beam splitter and then appears randomly, i.e. is detected randomly, in one or the other output beams. Since on-demand single photon generation is problematic in practice, QRNG devices use weak (coherent) laser light; the idea being that on average only one photon is present at any given time.  This is a fiction, of course, as the probability of having two photons at a time is not zero due to the photon bunching effect \cite{GerryBook}.  Now consider light prepared in a coherent state $\ket{\alpha}$ by a well phase-stabilized laser.  The expectation value of the parity operator is 

\begin{equation}
	\braket{\alpha|\hat{\Pi}|\alpha} = e^{-|\alpha|^{2}}\sum_{n=0}^{\infty}\frac{|\alpha|^{2n}}{n!}\left(-1\right)^{n} = e^{-\bar{n}}\sum_{n=0}^{\infty}\frac{\left(-\bar{n}\right)^{n}}{n!} = e^{-2\bar{n}},
	\label{eqn:qrng_1}
\end{equation}

\noindent where $\bar{n} = |\alpha|^{2}$.  It is easy to show numerically that $\braket{\hat{\Pi}} \to 0$ as $\bar{n}$ becomes large.  Another way to put it is that the probabilities of getting even or odd photon-number parity must be equal to $1/2$ for sufficiently large $\bar{n}$.  For arbitrary $\alpha$ these probabilities are 

\begin{align}
	P_{e} &= \braket{\alpha|\hat{P}_{e}|\alpha} = e^{-|\alpha|^{2}}\sum_{m=0}^{\infty}\frac{|\alpha|^{4m}}{\left(2m\right)!} \nonumber \\
	& \;\;\;\;\;\;\;\;\;\;\;\;\;\;\;\;\;\;\;\;\;\;= e^{-\bar{n}}\sum_{m=0}^{\infty}\frac{\bar{n}^{2m}}{\left(2m\right)!} = e^{-\bar{n}}\cosh\bar{n} \label{eqn:qrng_2} \\
	&\nonumber \\
	P_{o} &= \braket{\alpha|\hat{P}_{o}|\alpha} = e^{-|\alpha|^{2}}\sum_{m=0}^{\infty}\frac{|\alpha|^{4m+2}}{\left(2m+1\right)!} \nonumber \\
	&\;\;\;\;\;\;\;\;\;\;\;\;\;\;\;\;\;\;\;\;\;\; = e^{-\bar{n}}\sum_{m=0}^{\infty}\frac{\bar{n}^{2m+1}}{\left(2m+1\right)!} = e^{-\bar{n}}\sinh\bar{n} \label{eqn:qrng_3}
\end{align}

\noindent where the even and odd photon number projectors are given, respectively, by $\hat{P}_{e} = \sum_{m=0}^{\infty}\ket{2m}\bra{2m}$ and $\hat{P}_{o} = \sum_{m=0}^{\infty}\ket{2m+1}\bra{2m+1}$ and where 

\begin{equation}
	\braket{\hat{\Pi}} = \braket{\hat{P}_{e}} - \braket{\hat{P}_{o}} = P_{e} - P_{o} = e^{-2\bar{n}}.
	\label{eqn:qrng_4}
\end{equation} 

\noindent The average photon number need not be very high for the probabilities in Eq.~\ref{eqn:qrng_2} and \ref{eqn:qrng_3} to equalize close to $1/2$.  In fact, for $\bar{n}=9$ one has $P_{e}=P_{o}=0.5$ after retaining seven decimal places.  For $\bar{n}=16$, this is true to thirteen decimal places. Thus a viable QRNG could be implemented with efficient parity detection in a regime where the average photon number need not be very high.  \\

\noindent Lastly we mention that laser light prepared in a pure coherent state is not necessary: a statistical mixture of coherent states of the same amplitude will suffice.  Consider the phase-averaged coherent states given by the density operator 

\begin{equation}
	\rho = \frac{1}{2\pi}\int_{0}^{2\pi} d\phi \ket{re^{i\phi}}\bra{re^{i\phi}},
	\label{eqn:qrng_5}
\end{equation}

\noindent where $r=|\alpha|=\sqrt{\bar{n}}$.  It is easily shown that 

\begin{align}
	\braket{\hat{\Pi}} = \text{Tr}\left[\rho\hat{\Pi}\right] &= \frac{1}{2\pi}\int_{0}^{2\pi}d\phi\sum_{n=0}^{\infty}\left(-1\right)^{n}|\braket{n|re^{i\phi}}|^{2} \nonumber \\
	& = e^{-\bar{n}}\sum_{n=0}^{\infty} \frac{\left(-\bar{n}\right)^{n}}{n!} = e^{-2\bar{n}},
	\label{eqn:qrng_6}
\end{align}

\noindent which is identical to Eq.~\ref{eqn:qrng_1}. \\

\noindent In the next section we will discuss parity-based measurement in the context of multi-atom spectroscopy with a particular interest in the construction of high-precision atomic clocks.  We will consider both the classical case in which the atoms are initially unentangled and prepared in the same initial state as well as the case in which the atoms are initially entangled and spin-squeezed. A comparison between parity and population difference is also made in the context of enhanced measurement resolution.

\section{\label{sec:Ramsey} Enhanced measurement resolution in multi-atom atomic clocks}

\noindent We have discussed the benefit of parity-based detection in the context of measurement resolution in two-mode optical systems as it pertains to interferometric measurements.  Enchanced measurement resolution may also be beneficial in the construction of highly precise multi-atom atomic clocks.  Collections of two level atoms have seen use in quantum information \cite{ColdAtoms} (in the context of cold atoms) and quantum metrology \cite{MetroAtoms}.  In this section we will endeavor to show how parity-based measurements yields finer resolution for increasingly large numbers of atoms whereas simply considering the population difference between ground and excited states does not.  We will initially restrict the discussion to unentangled ensembles of atoms, where each atom is initially prepared in the same atomic state, before considering the case of an initially entangled ensemble.  The latter case provides even greater resolution when parity measurements are performed.    

\subsection{\label{sec:Ramsey_ACS_resolution_pop} Population difference for an arbitrary atomic state}

\noindent Consider a general atomic state expressed in terms of a superposition of all Dicke states $\ket{j,m}$, where the Dicke states are defined relative to the individual atomic states in the previous section, given by 

\begin{equation}
\ket{\psi_{\pi/2}}= \sum_{m=-j}^{j}C_{m}^{\left(j\right)}\ket{j,m}.
\label{eqn:rs1}
\end{equation}

\noindent Let Eq.~\ref{eqn:rs1} describe the state \textit{after} the first $\pi/2$-pulse in a Ramsey interferometer, where $C_{m}^{\left(j\right)}$ are the probability amplitudes for obtaining each state $\ket{j,m}$.  The state after the time evolution and the final $\pi/2$-pulse is given by

\begin{align}
\ket{\Psi_{\text{F}}} &= e^{-i\tfrac{\pi}{2}\hat{J}_{y}}e^{-i\phi\hat{J}_{z}}\ket{\psi_{\pi/2}} \nonumber\\
&=\sum_{m'=-j}^{j}\Bigg[\sum_{m=-j}^{j}C_{m}^{\left(j\right)}e^{-i\phi 	m}d^{j}_{m',m}\left(\tfrac{\pi}{2}\right)\Bigg]\ket{j,m'} \nonumber \\
&=\sum_{m'=-j}^{j}\tilde{C}_{m'}^{\left(j\right)}\ket{j,m'},
\label{eqn:rs2}
\end{align}

\noindent where $\tilde{C}_{m'}^{\left(j\right)}$ are defined within the square brackets of Eq.~\ref{eqn:rs2} and $d_{m',m}^{j}\left(\beta\right)$ are the usual Wigner-$d$ matrix elements given by $\braket{j,m'|e^{-i\beta\hat{J}_{y}}|j,m}$.  We consider the measurement resolution obtained for this general case when using $\braket{\hat{J}_{z}}$ as our detection observable.  Physically, this corresponds to taking the difference between the number of atoms in the excited and ground states, respectively.  The optical analog would be taking the difference in intensity between the two output modes of an MZI.  The expectation value of $\hat{J}_{z}$ is given by 

\begin{align}
\braket{\hat{J}_{z}}&=\sum_{m=-j}^{j}\sum_{p=-j}^{j}C_{p}^{\left(j\right)*}C_{m}^{\left(j\right)}e^{i\phi\left(m-p\right)}\;\times \nonumber \\
&\;\;\;\;\;\;\;\;\;\;\;\;\;\;\;\;\;\;\;\;\;\;\;\;\;\;\times\Bigg[\sum_{m'=-j}^{j}m'd_{m',p}^{j}\left(\tfrac{\pi}{2}\right)d_{m',m}^{j}\left(\tfrac{\pi}{2}\right)\Bigg] \nonumber \\
&=\sum_{m=-j}^{j}\sum_{p=-j}^{j}C_{p}^{\left(j\right)*}C_{m}^{\left(j\right)}e^{i\phi\left(m-p\right)}\times \Gamma_{m,p}\left(\tfrac{\pi}{2}\right),
\label{eqn:rs3}
\end{align}

\noindent where the function $\Gamma_{m,p}\left(\lambda\right)$ is defined by Eq.~\ref{eqn:Ab20}.  Without loss of generality \footnote{For complex state coefficients,Eq.~\ref{eqn:rs5} can be written as $\braket{\hat{J}_{z}} = 2\text{Re}\left[z e^{i\phi}\right] + \alpha^{\left(j\right)}$, where $z$ is a complex number expressed in terms of the state coefficients. For real state coefficients, this simplifies to the form of Eq.~\ref{eqn:rs6a}}, we assume the coefficients $C_{m}^{\left(j\right)}$ are real and we obtain the result

\begin{equation}
\braket{\hat{J}_{z}} = \gamma^{\left(j\right)}\cos\phi + \alpha^{\left(j\right)},
\label{eqn:rs5}
\end{equation}

\noindent where 

\begin{align}
\gamma^{\left(j\right)} &= 2\sum_{m=-j}^{j}\sum_{m'=-j}^{j}C_{m+1}^{\left(j\right)}C_{m}^{\left(j\right)}m'd_{m',m+1}^{j}\left(\tfrac{\pi}{2}\right)d_{m',m}^{j}\left(\tfrac{\pi}{2}\right) ,
\nonumber \\
\label{eqn:rs6a} \\
\alpha^{\left(j\right)} &= \sum_{m=-j}^{j}\sum_{m'=-j}^{j}\Big|C_{m}^{\left(j\right)}d_{m',m}^{j}\left(\tfrac{\pi}{2}\right)\Big|^{2}m'. \nonumber 
\end{align}

\noindent For the case of an initial atomic coherent state, this yields the expected result of Eq.~\ref{eqn:ccg_9}

\begin{equation}
\braket{\hat{J}_{z}}_{\text{ACS}} = j\cos\phi,
\label{eqn:rs7}
\end{equation}

\begin{figure}
	\centering
	\includegraphics[width=0.95\linewidth,keepaspectratio]{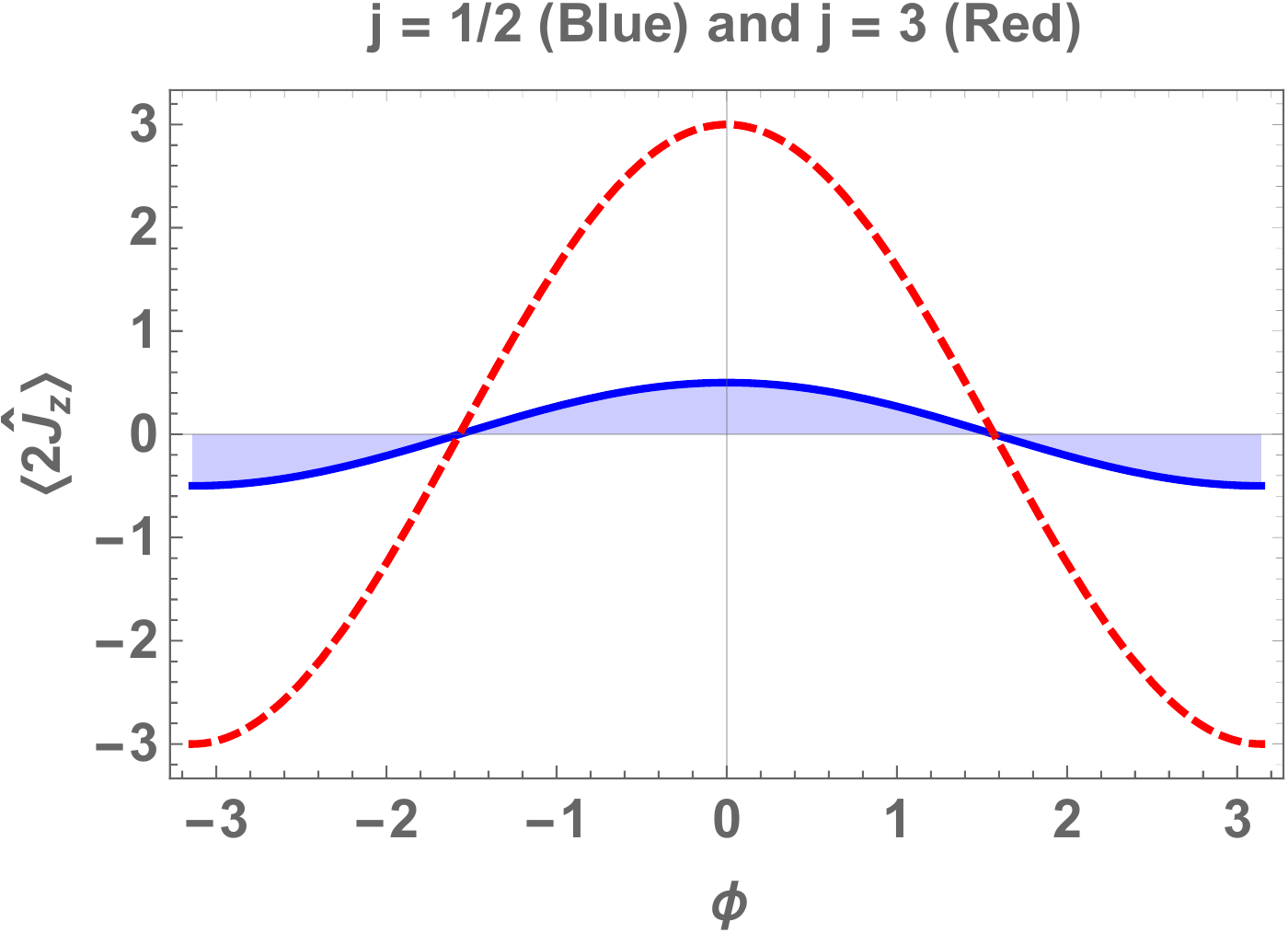}
	\caption{$\braket{\hat{J}_{z}}$ versus $\phi$ for a $j=1/2$ (blue, solid) and $j=3$ (red, dashed).   Note that we do not see greater resolution for larger numbers of atoms.}
	\label{fig:rs1}
\end{figure}

\noindent which we plot in Fig.~\ref{fig:rs1} for several values of $j$.  Similarly, an expression can be found for $\braket{\hat{J}_{z}^{2}}$ by using Eq.~\ref{eqn:Ab21} which yields

\begin{equation}
\braket{\hat{J}_{z}^{2}}_{\text{ACS}} = \Omega^{\left(j\right)}\cos 2\phi + \kappa^{\left(j\right)}\cos\phi +  \beta^{\left(j\right)},
\label{eqn:rs8}
\end{equation}

\noindent where

\begin{align}
\Omega^{\left(j\right)} &= 2\sum_{m=-j}^{j}\sum_{m'=-j}^{j}C_{m+2}^{\left(j\right)}C_{m}^{\left(j\right)}m^{\prime \;2}d_{m',m+2}^{j}\left(\tfrac{\pi}{2}\right)d_{m',m}^{\left(j\right)}\left(\tfrac{\pi}{2}\right), 
\nonumber \\
\kappa^{\left(j\right)} &= 2\sum_{m=-j}^{j}\sum_{m'=-j}^{j}C_{m+1}^{\left(j\right)}C_{m}^{\left(j\right)}m^{\prime\;2}d_{m',m+1}^{j}\left(\tfrac{\pi}{2}\right)d_{m',m}^{j}\left(\tfrac{\pi}{2}\right), \nonumber \\
\beta^{\left(j\right)} &= \sum_{m=-j}^{j}\sum_{m'=-j}^{j}\Big|C_{m}^{\left(j\right)}d_{m',m}^{j}\left(\tfrac{\pi}{2}\right)m'\Big|^{2}, \label{eqn:rs9a} 
\end{align}

\noindent Note that in neither case do we see a narrowing of the peak regardless of the state used.  The phase dependency goes as a cosine function (to the first power) regardless of the state coefficients and number of atoms used. This particular detection observable cannot yield enhanced measurement resolution regardless of the initial state under consideration.

\subsection{\label{sec:Ramsey_n00n} Parity-based resolution for an arbitrary atomic state}

\noindent Let us assume our state is initially prepared such that the state after the first $\pi/2$ pulse is once again given by Eq.~\ref{eqn:rs1}.  The action of the Ramsey interferometer is to transform this state to

\begin{align}
\ket{\Psi_{\text{F}}} &= e^{-i\tfrac{\pi}{2}\hat{J}_{y}}e^{-i\phi\hat{J}_{z}}\ket{\psi_{\pi/2}} \nonumber \\
& =\sum_{m=-j}^{j}\sum_{m'=-j}^{j}C_{m}^{\left(j\right)}d_{m',m}^{j}\left(\tfrac{\pi}{2}\right)e^{-i\phi m} \ket{j,m'},
\label{eqn:rs_new_1}
\end{align}

\noindent where the Wigner-$d$ matrix elements $d_{m',m}^{j}\left(\beta\right)$ are discussed in Appendix \ref{app:secB}.  The expectation value of the parity operator can be expressed as

\begin{align}
\braket{\hat{\Pi}} &= \sum_{m=-j}^{j}\sum_{p=-j}^{j}  C_{m}^{\left(j\right)}C_{p}^{\left(j\right)\;*}e^{-i\phi\left(m-p\right)} \times\nonumber\\ &\;\;\;\;\;\;\;\;\;\times \Bigg[\sum_{m'=-j}^{j}\left(-1\right)^{j-m'}d_{m',m}^{j}\left(\tfrac{\pi}{2}\right)d_{m',p}^{j}\left(\tfrac{\pi}{2}\right)\Bigg], \nonumber \\ 
& =\sum_{m=-j}^{j}\sum_{p=-j}^{j}C_{m}^{\left(j\right)}C_{p}^{\left(j\right)\;*}e^{-i\phi\left(m-p\right)}\;F_{m,p}\left(\tfrac{\pi}{2}\right),
\label{eqn:rs_new_2}
\end{align}

\noindent where the term in the square brackets is given by Eq.~\ref{eqn:Ab19}. Noting  that $F_{m,p}\left(\lambda \to \tfrac{\pi}{2}\right)=\left(-1\right)^{2j}\delta_{-m,p}$, we find for an arbitrary state 

\begin{equation}
\braket{\hat{\Pi}} = \left(-1\right)^{2j}\sum_{m=-j}^{j}C_{-m}^{\left(j\right)\;*}C_{m}^{\left(j\right)}e^{-i2\phi m}.
\label{eqn:rs_new_3}
\end{equation}

\noindent To aid in motivating the use of atomic parity as a detection observable let us first revisit the result for the most ideal of cases, that is, we analyze the measurement resolution obtained in a Ramsey interferometer when the state after the first $\pi/2$-pulse is the atomic N00N state, given by

\begin{equation}
\ket{\psi_{\pi/2}}=\ket{\psi_{\text{N00N}}}=\frac{1}{\sqrt{2}}\left(\ket{j,j}+e^{i\theta}\ket{j,-j}\right),
\label{eqn:rs-1}
\end{equation}

\noindent including an arbitrary relative phase $\theta$.  The representation of collections of two-level atoms in terms of the su(2) Lie alegbra, known as the Dicke model, is described in some detail in Appendix \ref{app:secA}.  This can be expressed as a finite sum over all Dicke states as 

\begin{equation}
\ket{\psi}_{\text{N00N}}= \sum_{m=-j}^{j}\Lambda_{m}^{\left(j\right)}\ket{j,m},
\label{eqn:rs-2}
\end{equation}

\noindent where the coefficients $\Lambda_{m}^{\left(j\right)}$ are given by

\begin{equation}
\Lambda_{m}^{\left(j\right)} = \frac{1}{\sqrt{2}}\left(\delta_{m,j} + e^{i\theta}\delta_{m,-j}\right).
\label{eqn:rs-3}
\end{equation}

\noindent This state is of interest as it is the atomic analog of the optical N00N state $\ket{\psi}_{\text{N00N}}=\tfrac{1}{\sqrt{2}}\big(\ket{N,0}+e^{i\theta}\ket{0,N}\big)$ (to be discussed in greater depth), which yields the lower bound on phase uncertainty in quantum optical interferometry: the Heisenberg limit. This follows from the heuristic uncertainty relation between photon number and phase $\Delta\phi\Delta N \geq 1$, where the N00N state maximizes the uncertainty in photon number, that is, $\Delta N = N\;\to\;\Delta\phi \simeq 1/\Delta N = 1/N$.  The action of the interferometer is once again given by Eq.~\ref{eqn:rs_new_1}  where we assume some means of experimentally generating the atomic N00N state is employed prior to free evolution.  Here we introduce the atomic parity operator as $\hat{\Pi}=\text{Exp}\big[i\pi\big(\hat{J}_{0}-\hat{J}_{3}\big)\big]$, with respect to the number of atoms found in the ground state.  We can likewise define atomic parity in terms of the number of atoms found in the excited state.  The relationship between the two being $\braket{\hat{\Pi}_{\text{Ground}}}=\left(-1\right)^{2j}\braket{\hat{\Pi}_{\text{Excited}}}$.  This is sensible, since if $N$ is even and $2n$ atoms are excited then $2n'$ atoms must also be in the ground state where $n,n' \in \mathbb{Z}^{+}$. A similar argument can be made for $2n+1$ exicted atoms.  Likewise, if $N$ were odd and $2n$ atoms are excited, then $2n'+1$ must be in the ground state, resulting in a sign difference in parity.  With this, we calculate the expectation value of the parity operator using Eq.~\ref{eqn:rs_new_3} to find

\begin{equation}
\braket{\hat{\Pi}}_{\text{N00N}} = \left(-1\right)^{N}\cos \left(N\phi+\theta\right),
\label{eqn:rs-5}
\end{equation}

\noindent where $N=2j$, the total number of atoms. Here, the relative phase $\theta$ acts as a phase shift of the interference pattern, which now oscillates at the enhanced rate $N\phi$.  Finer measurement resolution, characterized by a narrowing of the peak of $\braket{\hat{\Pi}}$ at $\phi=0$ for increasing values of $N$, tends towards enhancing phase sensitivity in interferometric measurements. This state would serve as the ideal, as it is maximally entangled and yields the greatest measurement resolution for a given detection observable.  The problem, however, is atomic N00N states are incredibly difficult to make.  We endeavor to find means of enhancing measurement resolution by considering parity-based measurements performed on both unentangled ensembles and initially entangled atomic systems.

\subsubsection{\label{sec:Ramsey_ACS_resolution} Unentangled ensemble of atoms}

\noindent Let us assume our state is initially prepared with all atoms in the ground state, that is $\ket{\text{in}}=\ket{j,-j}$.  The action of the Ramsey interferometer is to transform an initial state $\ket{\psi'}$ by

\begin{align}
\ket{\Psi_{\text{F}}} &= e^{-i\tfrac{\pi}{2}\hat{J}_{y}}e^{-i\phi\hat{J}_{z}}e^{-i\tfrac{\pi}{2}\hat{J}_{y}}\ket{\text{in}} \nonumber \\
&= e^{-i\tfrac{\pi}{2}\hat{J}_{y}}e^{-i\phi\hat{J}_{z}}\ket{\tau=-1,j},
\label{eqn:rs10}
\end{align}

\noindent where $e^{-i\tfrac{\pi}{2}\hat{J}_{y}}\ket{j,-j} = \ket{\tau=-1,j}$ is the atomic coherent state in which all of the atoms in the ensemble are in the superposition state $\tfrac{1}{\sqrt{2}}\big(\ket{g}-\ket{e}\big)$, that is,

\begin{align}
\ket{\tau=-1,j} &= \frac{1}{2^{N/2}}\left(\ket{\text{e}}-\ket{\text{g}}\right)^{\otimes N=2j}\; \nonumber \\
& = \; 2^{-j}\sum_{m=-j}^{j}\binom{2j}{j+m}^{1/2}\left(-1\right)^{j+m}\ket{j,m}.
\label{eqn:rs11}
\end{align}

\noindent Let us recover the known result for an initial atomic coherent state.  Taking the state coefficients of Eq.~\ref{eqn:rs11} and plugging them into Eq.~\ref{eqn:rs_new_3} we find

\begin{align}
\braket{\hat{\Pi}}_{\text{ACS}} &= 4^{-j}\sum_{m=-j}^{j}\binom{2j}{j+m}e^{-i2\phi m}\left(-1\right)^{4j} = d_{-j,-j}^{j}\left(2\phi\right) \nonumber \\
&= \cos^{2j}\phi = \cos^{N}\phi.
\label{eqn:rs14}
\end{align}

\begin{figure}
	\centering
	\includegraphics[width=1\linewidth,keepaspectratio]{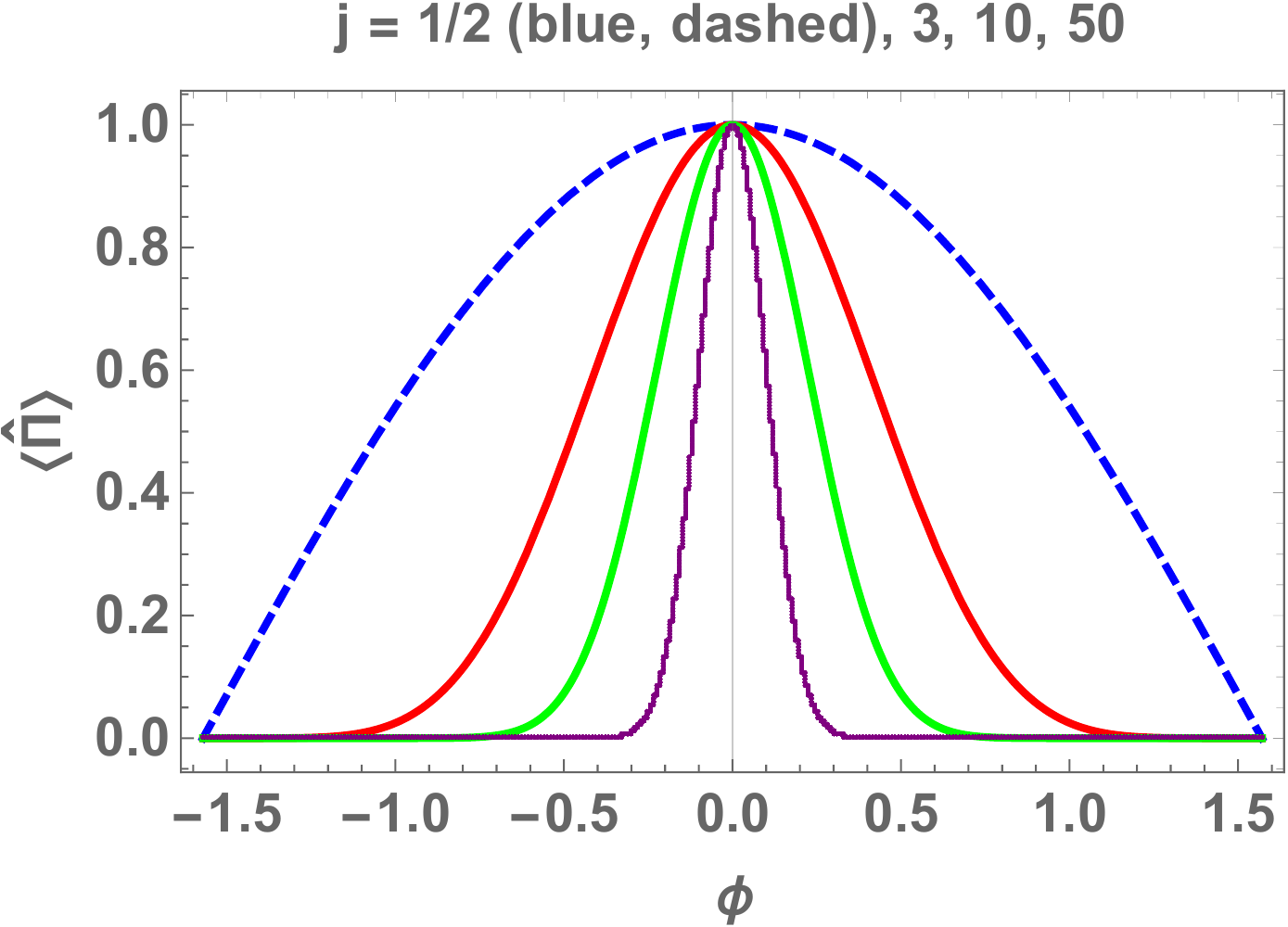}
	\caption{Expectation value of the parity operator for an initial atomic coherent state for $j=1/2,\;3,\;10,\;50$.  Note the peak becoming more narrow as the number of atoms is increased from $j=1/2$ (blue, dashed, broadest) to $j=50$ (purple, solid, narrowest).}
	\label{fig:rs2}
\end{figure}

\noindent This is plotted in Fig.~\ref{fig:rs2}. Unlike the case of simply taking the population difference, larger numbers of atoms yields greater measurement resolution, characterized by the narrowing of the peak about $\phi=0$. Comparing Eq.~\ref{eqn:rs14} result to Eq.~\ref{eqn:rs7}, we see that population difference measurement produces an interference cosine function raised to effectively $N=1$ photons, while the use of atomic parity instead has raised to the number of atoms $N$, leading to enhanced resolution at the origin as $\phi\to 0$. This can be demonstrated if we make the approximation $\cos^{N}\phi \sim e^{-N\phi^{2}/2}$ for small $\phi$ which is a Gaussian that narrows for increasing $N$.  We remind the reader that while parity-based measurement yields greater resolution, this does not translate to better phase sensitivity beyond the SQL as the initial state is classical.   

\subsubsection{\label{sec:Ramsey_Ent_resolution}  Entangled atomic ensemble}

\noindent Now let us next consider the case where we act with the $\hat{J}_{z}$ operator after the first $\pi/2$-pulse. The resulting state is both entangled at the atomic-state level and is known to exhibit spin-squeezing, which has been shown to reduce projection noise in high-precision population spectroscopy.  In this case the initial state is transformed by

\begin{align}
\ket{\Psi_{\text{F}}} &= e^{-i\tfrac{\pi}{2}\hat{J}_{y}}e^{-i\phi\hat{J}_{z}}\hat{J}_{z}e^{-i\tfrac{\pi}{2}\hat{J}_{y}}\ket{\text{in}} \nonumber \\
&=\ e^{-i\tfrac{\pi}{2}\hat{J}_{y}}e^{-i\phi\hat{J}_{z}}\ket{\tau=-1,j;1},
\label{eqn:16}
\end{align}

\noindent where the state

\begin{align}
&\ket{\tau=-1,j;1} \propto \hat{J}_{z}\ket{\tau=-1,j} \nonumber \\
&\;\;\;\;\;\;\;\;= \left(2^{2j-1}j\right)^{-1/2}\sum_{m=-j}^{j}\binom{2j}{j+m}^{1/2}\left(-1\right)^{j+m}m\ket{j,m}.
\label{eqn:17}
\end{align}

\noindent Once again, we use Eq.~\ref{eqn:rs_new_3} to find the expectation value of the parity operator

\begin{align}
&\braket{\hat{\Pi}}_{\hat{J}_{z}\text{-op ACS}} = -\left(2^{2j-1}j\right)^{-1}\sum_{m=-j}^{j}\binom{2j}{j+m}m^{2}e^{-i2\phi m} \nonumber \\
&= -\frac{2}{\pi}\frac{\Gamma\left(j+\tfrac{1}{2}\right)}{\Gamma\left(j+2\right)}e^{-i2\phi}\left[ {}_{3}F_{2}\left(2,2,1-j;1,j+2;-e^{-i2\phi}\right) + \right. \nonumber \\
&\;\;\;\;\;\;\;\;\;\;\;\;\;\;\;\;\;\;\;\;\;\;\; + \left. e^{i4\phi} {}_{3}F_{2}\left(2,2,1-j;1,j+2;-e^{i2\phi}\right)\right]
\label{eqn:18}
\end{align}

\noindent where ${}_{3}F_{2}$ is the hypergeometric function given generally by Eq.~\ref{eqn:Ab6}.  Simplifying for several different values of $j = N/2$, yields

\begin{align}
j=1/2 \;\;\;\;\to\;\;\;\; &\braket{\hat{\Pi}}_{\hat{J}_{z}\text{-op ACS}} = -\cos\phi, \nonumber\\
j=1 \;\;\;\;\to\;\;\;\; &\braket{\hat{\Pi}}_{\hat{J}_{z}\text{-op ACS}} = -\cos^{2}\phi\left(2-\cos^{-2}\phi\right) \nonumber\\
j=3/2 \;\;\;\;\to\;\;\;\; &\braket{\hat{\Pi}}_{\hat{J}_{z}\text{-op ACS}} = -\cos^{3}\phi\left(3-2\cos^{-2}\phi\right), \nonumber\\
j=2 \;\;\;\;\to\;\;\;\; &\braket{\hat{\Pi}}_{\hat{J}_{z}\text{-op ACS}} = -\cos^{4}\phi\left(4-3\cos^{-2}\phi\right), \nonumber\\
j=5/2 \;\;\;\;\to\;\;\;\; &\braket{\hat{\Pi}}_{\hat{J}_{z}\text{-op ACS}} = -\cos^{5}\phi\left(5-4\cos^{-2}\phi\right), \nonumber \\
j=3 \;\;\;\;\to\;\;\;\; &\braket{\hat{\Pi}}_{\hat{J}_{z}\text{-op ACS}} = -\cos^{6}\phi\left(6-5\cos^{-2}\phi\right),\nonumber \\
& \;\;\;\;\;\;\;\;\;\;\;\;\;\;\;\; .\nonumber \\
& \;\;\;\;\;\;\;\;\;\;\;\;\;\;\;\; .\nonumber \\
& \;\;\;\;\;\;\;\;\;\;\;\;\;\;\;\; .\label{eqn:20}
\end{align}

\noindent from which we can extrapolate for $N \geq 1$,

\begin{equation}
	\braket{\hat{\Pi}}^{\left(N\right)}_{\hat{J}_{z}\text{-op ACS}} = -\cos^{N}\phi\left[N-\left(N-1\right)\cos^{-2}\phi\right]
	\label{eqn:20a}
\end{equation}

\begin{figure}
	\centering
	\includegraphics[width=0.95\linewidth,keepaspectratio]{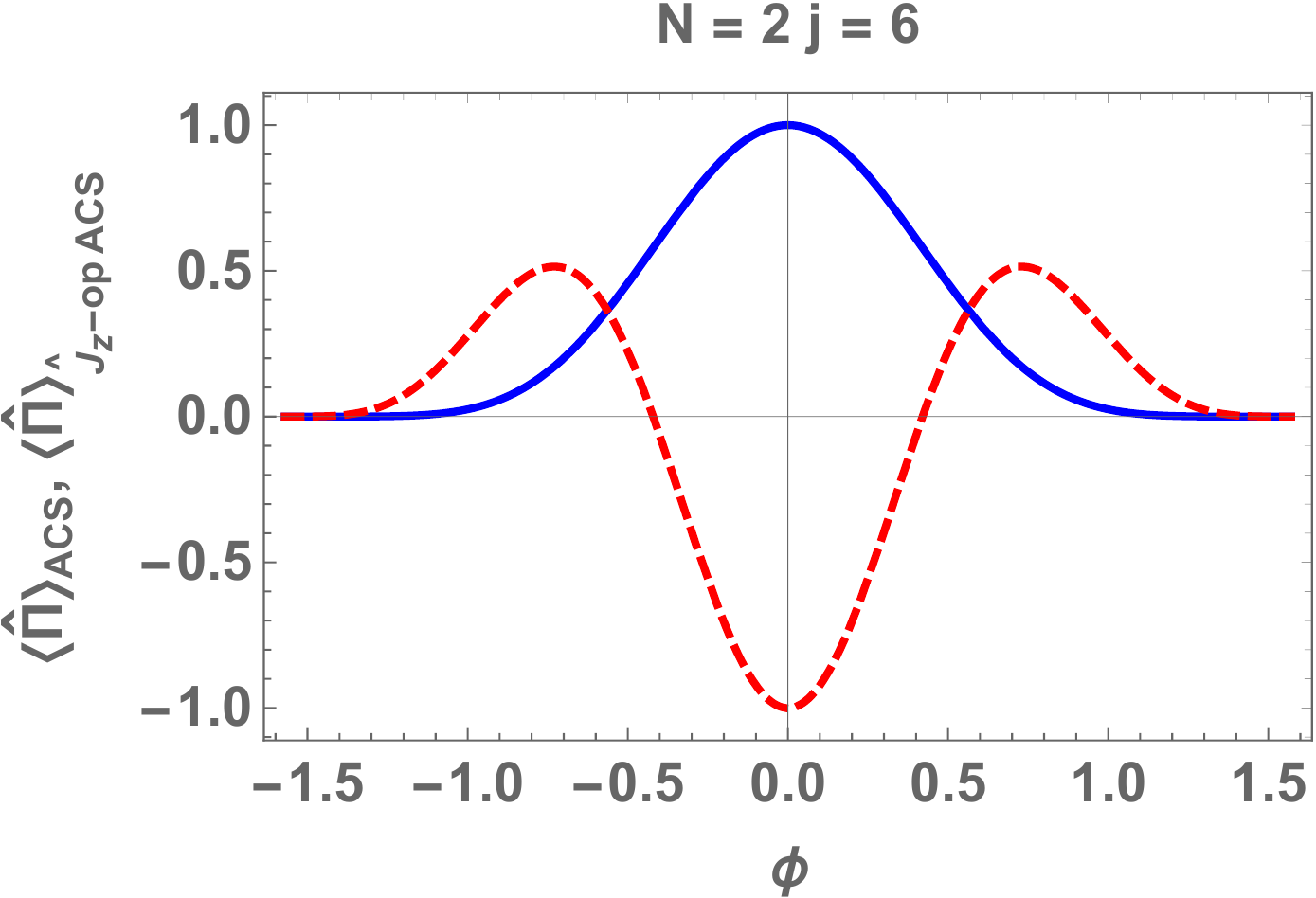}
	\caption{A comparison between expectation values of the parity operator for the case of input ACS (blue, solid) and input $\hat{J}_{z}$-op ACS (red,dashed) for the same value of $N=2j=6$. The resolution for the $\hat{J}_{z}$-op ACS is greater than the case of the ACS for all values of $N=1$.}
	\label{fig:rs3}
\end{figure}

\begin{figure}
	\centering
	\hspace{0.15cm}
	\includegraphics[width=0.93\linewidth,keepaspectratio]{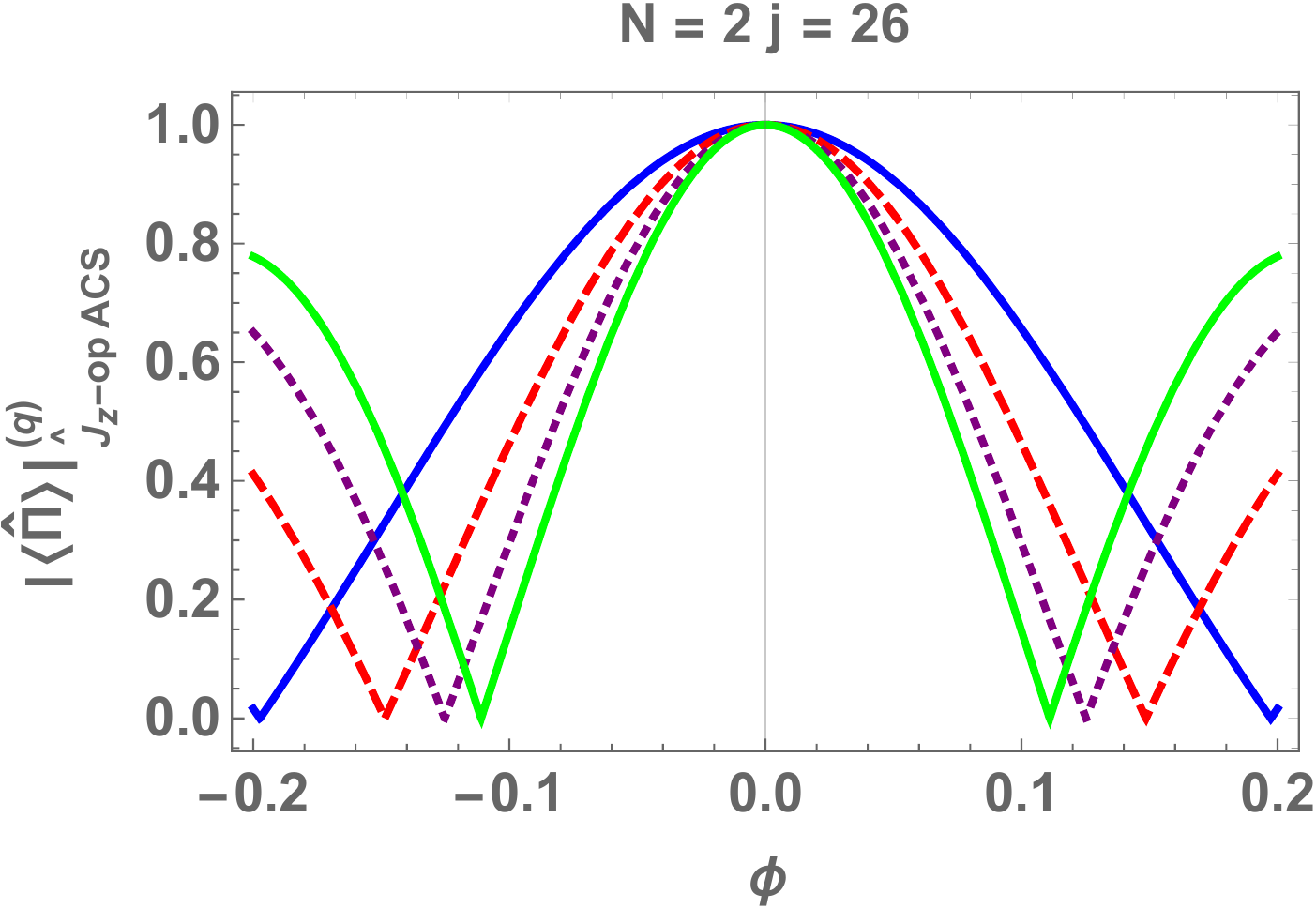}
	\caption{The expectation value of the parity operator $|\braket{\hat{\Pi}}|$ for the case of an initial $\hat{J}_{z}^{q}$-operated ACS, for $N=2j=26$.  The curves in the figure above correspond to $q=1$ (blue, outer-solid), $q=2$ (red,dashed), $q=3$ (purple, dotted), $q=4$ (green, inner-solid). Note the increased resolution as the initial ensemble becomes more entangled.}
	\label{fig:rs4}
\end{figure}

\noindent We see that for the case of an initially entangled atomic ensemble, there is greater measurement resolution with increasing number of atoms over the initially unentangled ACS case, except for the case of $j=1/2$ ($N=1$) in which case both initial states yield the same resolution. This is demonstrated explicitly in Fig.~\ref{fig:rs3} for the value of $N=2j=6$, where we can see a narrowing of the peak about the phase origin for the same number of atoms when considering the initially-entangled case. This trend of enhanced resolution persists for $q>1$ applications of $\hat{J}_{z}$ prior to free evolution, which further entangles the initial atomic ensemble. For the general case of $q$ applications of $\hat{J}_{z}$ prior to free evolution, or $\ket{\text{in}} \propto \hat{J}_{z}^{q}\ket{\zeta=-1,j}$, we can write the expectation value of the parity operator (unnormalized) as $\braket{\hat{\Pi}}^{\left(q\right)}_{\hat{J}_{z}\text{-op ACS}}\propto \sum_{m=-j}^{j}\binom{2j}{j+m}m^{2q}e^{-i2\phi m}$. Computations show greater resolutions obtained for the same (large) value of $j$ for increasing values of $q$, shown in Fig.~\ref{fig:rs4}.

\section{\label{sec:conclude} Conclusion}

\noindent In this review article we discussed the prospect of using parity-based measurements in phase estimation.  In the context of Ramsey spectroscopy, atomic parity is defined in terms of number of atoms found in the excited state while in the context of optical interferometry it is defined with regards to the average photon number in one output mode of the interferometer.  We provide a brief discussion on the basics of phase estimation and demonstrate how parity-based detection saturates the quantum Cram\'{e}r-Rao bound for path symmetric states in quantum optical interferometry, making parity the optimal observable for interferometric measurements.  We highlight the use of parity-measurements for several different initial states comprising of both quantum and classical light and show agreement between the phase uncertainty when considering parity-based measurements and the quantum Cram\'{e}r-Rao bound. For all quantum states considered, sub-SQL phase sensitivity is achieved when parity-based measurement is considered. \\

\noindent We also discuss the experimental efforts made towards measuring photon number parity in quantum optical interferometry using classical (coherent) light where the comparison is made to another dichotomic operator corresponding to "click"/"no click" detection.  QND measures of atomic parity are also considered through coupling to both an ancillary atomic system that can be as arbitrarily small as a single atom as well as coupling to a light field prepared in a coherent state.  We highlight that the latter case does not require a large third order non-linearity and in fact utilizes a readily available means of coupling.    \\ 

\noindent  We return to atomic spectroscopy in the closing section where the emphasis is placed on the enhanced measurement resolution obtained through parity-based measurement compared to the usual method of taking the population difference between the two internal degrees of freedom available to each atom.  We show in the latter case that for arbitrary $N$ atom initial states, the Ramsey technique cannot yield super-resolved measurements while the former shows greater resolution for increasingly large numbers of atoms $N$.  Of the cases discussed, we show an initially entangled atomic ensemble yields greater resolution for the same number of atoms than the case in which all atoms are initially unentangled and in the same atomic state.  

\begin{acknowledgments}
\noindent RJB acknowledges support from the National Research Council Research Associate Program (NRC RAP).
CCG acknowledges support under AFRL Summer Faculty Fellowship  Program (SFFP).
PMA and CCG acknowledge support from the Air Force Office of Scientific Research (AFOSR).  
Any opinions, findings and conclusions  or  recommendations  expressed  in  this  material are those of the author(s)
and do not necessarily reflect the views of the Air Force Research Laboratory (AFRL).
\end{acknowledgments}

\appendix

\section{\label{app:secA} A brief review of the SU(2) group}

\subsection{\label{app:subsecA1} The Dicke states}

\noindent In what follows, we prepare a brief review of the so-called Dicke states \cite{Dicke} which provides a representation of a collection of $N$ two-level atoms in terms of the SU(2) "angular momentum" states $\ket{j,m}$ where $j=N/2$, $m \in\left[-j,j\right]$ and where the internal basis states of the $i^{\text{th}}$ atom are $\{\ket{e}_{i},\ket{g}_{i}\}$.  The quantity $j$ in this context is sometimes called the cooperation parameter.  Only states symmetric with respect to the permutations of the $N$ atoms come into play in the cases considered here.  That is, the direct product of the spaces of the two-level atoms, as given by $2_{1}\otimes2_{2}\otimes2_{3}\otimes...\otimes2_{N}$, has one symmetric and generally several anti-symmetric decompositions, but the anti-symmetric states are dynamically decoupled from the symmetric states via the selection rules.  To demonstrate the construction of the symmetric states and connect them to the angular momentum states, we proceed by example where we use the ladder operators given in terms of the individual atoms as 

\begin{equation}
	\hat{J}_{+} = \sum_{i=1}^{N}\ket{e}_{i}\bra{g}, \;\;\;\ \hat{J}_{-}=\sum_{i=1}^{N}\ket{g}_{i}\bra{e},
	\label{eqn:Aa1a}
\end{equation}

\noindent and use the relations

\begin{equation}
	\hat{J}_{\pm}\ket{j,m} = \sqrt{j\left(j+1\right) -m\left(m\pm 1\right)}\ket{j,m\pm 1}
\end{equation}

\noindent acting on the collective angular momentum states. In actuality, multiple applications of the raising operator are used starting from the state in which all atoms are in their ground state. \\

\noindent For two atoms we have $j=N/2=1$; the state in which both atoms are in their ground state is state in which $m=-j=-1$ such that $\ket{j=1,m=-1}=\ket{g}_{1}\ket{g}_{2}$, or more simply 

\begin{equation}
	\ket{1,-1}=\ket{g}_{1}\ket{g}_{2}
	\label{eqn:Aa2a}
\end{equation}

\noindent Applying $\hat{J}_{+}$ to both sides and using the above relations we find 

\begin{equation}
	\ket{1,0} = \frac{1}{\sqrt{2}}\left(\ket{e}_{1}\ket{g}_{2}+\ket{g}_{1}\ket{e}_{2}\right),
	\label{eqn:Aa3a}
\end{equation}

\noindent and applying it again yields

\begin{equation}
	\ket{1,1} = \ket{e}_{1}\ket{e}_{2}.
	\label{Aa4a}
\end{equation}

\noindent We could have obtained the same set of states had we started with the state $\ket{1,1}=\ket{e}_{1}\ket{e}_{2}$ and laddered down to $\ket{1,-1}=\ket{g}_{1}\ket{g}_{2}$ through consecutive applications of the $\hat{J}_{-}$. These are the symmetric states, or Dicke states, for the case of two atoms.  Likewise one can see that for three atoms $j=N/2=3/2$ and the Dicke states are

\begin{align}
	&\ket{\tfrac{3}{2},-\tfrac{3}{2}} = \ket{g}_{1}\ket{g}_{2}\ket{g}_{3},\nonumber \\
	&\ket{\tfrac{3}{2},-\tfrac{1}{2}} = \frac{1}{\sqrt{3}}\left(\ket{e}_{1}\ket{g}_{2}\ket{g}_{3}+\ket{g}_{1}\ket{e}_{2}\ket{g}_{3}+\ket{g}_{1}\ket{g}_{2}\ket{e}_{3}\right),\nonumber \\
	&\ket{\tfrac{3}{2},\tfrac{1}{2}} = \frac{1}{\sqrt{3}}\left(\ket{e}_{1}\ket{e}_{2}\ket{g}_{3}+\ket{g}_{1}\ket{e}_{2}\ket{e}_{3}+\ket{e}_{1}\ket{g}_{2}\ket{e}_{3}\right),\nonumber \\
	&\ket{\tfrac{3}{2},\tfrac{3}{2}} = \ket{e}_{1}\ket{e}_{2}\ket{e}_{3}.\nonumber \\
	\label{eqn:Aa5a}
\end{align}

\noindent The process can be continued for any number of atoms.  Note that the extremal states $\ket{j,\pm j}$ are (separable) product states whereas all other states $m \neq \pm j$ are entangled at the atomic level.

\subsubsection{\label{app:ACS_A1} The atomic coherent state}

\noindent Two definitions of the atomic coherent state have been given in the literature: one by Radcliffe \cite{Radcliffe} and another by Arrechi \textit{et al.} \cite{Arrechi}  The definitions invlove the action of the SU(2) "displacement" operator $R\left(\theta,\varphi\right)=e^{\zeta\hat{J}_{+} - \zeta^{*}\hat{J}_{-}}$ on the extremal states $\hat{j,\pm j}$ where $\zeta=\tfrac{\theta}{2}e^{-i\varphi}$.  The angles $\theta,\;\varphi$ are the usual angles parameterizing the Bloch sphere with $0 \leq \theta \leq \pi$ and $0 \leq \varphi \leq 2\pi$ where we follow Arrechi \textit{et al.} in taking $\theta =0$ to be the south pole of the sphere.  This operator acts on the $\ket{j,j}$ state in the case of the Radcliffe definition and on the $\ket{j,-j}$ state in the case of Arrechi \textit{et al.}  For these two separate definitions, we require two different orderings of the "disentanglement" expressions of the operator $R\left(\theta,\varphi\right)$.  These are 

\begin{equation}
e^{\zeta\hat{J}_{+} - \zeta^{*}\hat{J}_{-}}
	=
	\begin{cases}
	e^{\tau\hat{J}_{+}}e^{\text{ln}\left(1+|\tau|^{2}\right)\hat{J}_{z}}e^{-\tau^{*}\hat{J}_{-}},  \\
	\\
	e^{-\tau^{*}\hat{J}_{-}}e^{-\text{ln}\left(1+|\tau|^{2}\right)\hat{J}_{z}}e^{\tau\hat{J}_{+}},
	\end{cases}
	\label{eqn:Aa6a}
\end{equation}

\noindent where $\tau = e^{-i\varphi}\tan\tfrac{\theta}{2}$.  In starting with the state $\ket{j,-j}$ we use the upper expression on the right hand side of Eq.~\ref{eqn:Aa6a} to arrive at 

\begin{align}
\ket{\tau,j} &= \left(1+|\tau|^{2}\right)^{-j}e^{\tau\hat{J}_{+}}\ket{j,-j} \nonumber \\
& =  \left(1+|\tau|^{2}\right)^{-j} \sum_{m=-j}^{j}\binom{2j}{j+m}^{1/2}\tau^{j+m}\ket{j,m},
\label{eqn:Aa7a}
\end{align}

\noindent and using the bottom line of the right hand side of Eq.~\ref{eqn:Aa6a} we obtain

\begin{align}
\ket{-\tau^{*},j} &= \left(1+|\tau|^{2}\right)^{-j}e^{-\tau^{*}\hat{J}_{+}}\ket{j,-j} \nonumber \\
& =  \left(1+|\tau|^{2}\right)^{-j} \sum_{m=-j}^{j}\binom{2j}{j+m}^{1/2}\left(-\tau^{*}\right)^{j+m}\ket{j,m}.
\label{eqn:Aa8a}
\end{align}

\noindent Note that under the replacement $\tau \leftrightarrow -\tau^{*}$, these states are the same. \\

\noindent For the special case used throughout Sections \ref{sec:Ramsey_1} and \ref{sec:Ramsey} where the rotation operator $e^{-i\tfrac{\pi}{2}\hat{J}_{y}}$ acts on the states $\ket{j,\pm j}$ states, we can set 

\begin{equation}
	R\left(\theta,\varphi\right) = e^{\zeta\hat{J}_{+} - \zeta^{*}\hat{J}_{-}} \equiv e^{-i\tfrac{\pi}{2}\hat{J}_{y}},
	\label{eqn:Aa9a}
\end{equation}

\noindent informing the value $\zeta=-\pi/4$ and consequently the angles $\theta=\pi/2$ and $\varphi=\pi$.  Thus we have $\tau = -1$ and $-\tau^{*} = 1$ which, for example, is demonstrated in Eq.~\ref{eqn:ccg_14}. More simply, one can insert a full set of states to find 

\begin{align}
	e^{-i\tfrac{\pi}{2}\hat{J}_{y}}\ket{j,\pm j} &= \sum_{m=-j}^{j}\ket{j,m}\braket{j,m|e^{-i\tfrac{\pi}{2}}|j,\pm j} \nonumber \\
	& = \sum_{m=-j}^{j} d_{m,\pm j}^{j}\left(\tfrac{\pi}{2}\right) \ket{j,m} \nonumber \\
	& = \left(1+|\tau|^{2}\right)^{-j}\sum_{m=-j}^{j}\binom{2j}{j+m}\left(\pm 1\right)^{j+m}\ket{j,m},
	\label{eqn:Aa10a}
\end{align}   

\noindent where the matrix elements $d_{m,\pm j}^{j}\left(\tfrac{\pi}{2}\right)$ are the Wigner-$d$ matrix elements found in Rose \cite{Rose} and given for this special case in Eq.~\ref{eqn:bb4}.  

\subsection{\label{app:subsecA2} The Schwinger realization of SU(2)}

\noindent Consider a two mode field with creation and annihilation operators satisfying the usual boson commutation relations $\left[\hat{a}_{i},\hat{a}_{j}\right] = \big[\hat{a}^{\dagger}_{i},\hat{a}^{\dagger}_{j}\big] = 0$ and $\big[\hat{a}_{i},\hat{a}^{\dagger}_{j}\big] = \delta_{i,j}$. One can introduce the Hermitian operators

\begin{align}
	\hat{J}_{x} &= \dfrac{1}{2}\big(\hat{a}_{1}^{\dagger}\hat{a}_{2} + \hat{a}^{\dagger}_{2}\hat{a}_{1}\big), \nonumber \\
	\hat{J}_{y} &= -\dfrac{i}{2}\big(\hat{a}_{1}^{\dagger}\hat{a}_{2} - \hat{a}^{\dagger}_{2}\hat{a}_{1}\big), \label{eqn:Aa1}\\
	\hat{J}_{z} &= \dfrac{1}{2}\big(\hat{a}_{1}^{\dagger}\hat{a}_{1} - \hat{a}^{\dagger}_{2}\hat{a}_{2}\big), \nonumber
\end{align}

\noindent and $\hat{N} = \tfrac{1}{2}\big(\hat{a}_{1}^{\dagger}\hat{a}_{1} + \hat{a}^{\dagger}_{2}\hat{a}_{2}\big)$, satisfying the commutation relations of the Lie algebra of SU(2):

\begin{equation}
	\big[\hat{J}_{i},\hat{J}_{j}\big] = i\hat{J}_{k}\epsilon_{i,j,k}.
	\label{eqn:Aa2}
\end{equation}

\noindent The Casimir invariant for the group can be found to be of the form $\hat{J}^{2} = \hat{J}_{x}^{2} + \hat{J}_{y}^{2} + \hat{J}_{z}^{2}=\frac{\hat{N}}{2}\left(\frac{\hat{N}}{2} + 1\right)$.  Note that $\hat{N}$ commutes will all operators in Eq.~\ref{eqn:Aa1}.  One can also define the operator $\hat{J}_{0}=\tfrac{1}{2}\hat{N}$ such that $\hat{J}_{0}\ket{j,m}=j\ket{j,m}$. It is also useful to recall the action of the angular momentum operatos $\hat{J}_{i}$ on the states $\ket{j,m}$:

\begin{align}
	\hat{J}^{2}\ket{j,m} &= j\left(j+1\right)\ket{j,m} \nonumber \\
	\hat{J}_{z}\ket{j,m} &= m\ket{j,m} \\
	\hat{J}_{\pm}\ket{j,m} &= \sqrt{j\left(j+1\right) -m\left(m\pm 1\right)}\ket{j,m\pm 1}, \nonumber 
	\label{eqn:Aa2.1}
\end{align}

\noindent where the ladder operators can be written as $\hat{J}_{\pm} = \hat{J}_{x} \pm i\hat{J}_{y}$. For bosons, we can translate between the Dicke (angular momentum) states $\ket{j,m}^{\left(a,b\right)}$ and Fock states

\noindent A beam splitter transforms the boson operators associated with each input mode according to the scattering matrix for the device, that is

\begin{equation}
	\vec{\hat{a}}_{\text{out}} = \hat{U}\;\vec{\hat{a}}_{\text{in}}
	\;\;\;\;\;\to\;\;\;\;\;
	\begin{pmatrix}
		\hat{a}_{1}  \\
		\hat{a}_{2} 
	\end{pmatrix}_{\text{out}} 
	=
	\begin{pmatrix}
		U_{11} & U_{12} \\
		U_{21} & U_{22}
	\end{pmatrix} 
	\begin{pmatrix}
		\hat{a}_{1}  \\
		\hat{a}_{2} 
	\end{pmatrix}_{\text{in}}.
	\label{eqn:Aa3}
\end{equation}

\noindent Note that since the boson creation and annihilation must satisfy the commutation relations both before and after beamsplitting, the matrix $\hat{U}$ must be unitary. Let us see how this transforms the operators of SU(2), $\vec{J} = \big(\hat{J}_{x},\hat{J}_{y},\hat{J}_{z}\big)$.  Consider the $\hat{J}_{x}$-type beam splitter scattering matrix, 

\begin{equation}
	\hat{U} = 
	\begin{pmatrix}
		\cos\tfrac{\theta}{2} & -i\sin\tfrac{\theta}{2} \\
	-i\sin\tfrac{\theta}{2} & \cos\tfrac{\theta}{2}
	\end{pmatrix} 
	=
	\begin{pmatrix}
		t & -ir \\
		-ir & t
	\end{pmatrix} ,
	\label{eqn:Aa4}
\end{equation}

\noindent which corresponds to a beam splitter with transmittance and reflectivity $T=\cos^{2}\tfrac{\theta}{2}$ and $R=\sin^{2}\tfrac{\theta}{2}$, respectively.  For this scattering matrix, $\vec{J}$ transforms to 

\begin{align} 
	\begin{pmatrix}
		\hat{J}_{x}  \\
		\hat{J}_{y}  \\
		\hat{J}_{z}
	\end{pmatrix}_{\text{out}}
	& =
	\begin{pmatrix}
		1 & 0 & 0  \\
		0 & \cos\theta & -\sin\theta  \\
		0 & \sin\theta & \cos\theta 
	\end{pmatrix}
	\begin{pmatrix}
		\hat{J}_{x}  \\
		\hat{J}_{y}  \\
		\hat{J}_{z}
	\end{pmatrix}_{\text{in}}\nonumber\\
	& = 
	e^{i\theta\hat{J}_{x}}
	\begin{pmatrix}
		\hat{J}_{x}  \\
		\hat{J}_{y}  \\
		\hat{J}_{z}
	\end{pmatrix}_{\text{in}}
	e^{-i\theta\hat{J}_{x}},
	\label{eqn:Aa5}
\end{align}

\noindent which amounts to a rotation about the fictitious $x$-axis.  Note that the last line of Eq.~\ref{eqn:Aa5} can be verified via the use of the Baker-Hausdorff identity

\begin{equation}
	e^{\tau\hat{A}}\hat{B}e^{-\tau\hat{A}} = \hat{B} + \tau\big[\hat{A},\hat{B}\big] + \frac{1}{2}\tau^{2}\big[\hat{A},\big[\hat{A},\hat{B}\big]\big] + ..\;.
	\label{eqn:Aa6}
\end{equation}

\noindent Working in the Schr\"{o}dinger picture, the action of the beam splitter corresponds to a transformation of the initial state given by  

\begin{equation}
	\ket{\text{out, BS}} = e^{-i\theta\hat{J}_{x}}\ket{\text{in}}.
	\label{eqn:Aa7}
\end{equation}

\noindent We can also express a two-mode state in the Fock basis in terms of the basis states of SU(2) (angular momentum states) using Eq.~\ref{eqn:Aa1}, yielding

\begin{align}
	\hat{J}_{z}\ket{j,m} = m\ket{j,m} \;\;\;\;\;&\to\;\;\;\;\; \hat{J}_{z}\ket{n,n'}_{1,2} = \frac{n-n'}{2}\ket{n,n'}_{1,2}, \nonumber \\
	\label{eqn:Aa8}\\
	\hat{J}^{2}\ket{j,m}=j\big(j+1\big)\ket{j,m}\;&\to\;\hat{J}^{2}\ket{n,n'}_{1,2} = \frac{n+n'}{2}\left(\frac{n+n'}{2} + 1\right), \nonumber
\end{align}

\noindent which informs us that $\ket{n,n'}_{1,2}\to\ket{j,m}$ where the values of $j$ and $m$ are given by $j=\tfrac{n+n'}{2}$ and $m=\tfrac{n-n'}{2}$. Inversely $\ket{j,m}\to\ket{n,n'}_{1,2}$ where $n=j+m$ and $n'=j-m$ with $n+n'=2j$ and $m \in \{-j,...,j\}$. Lastly, the action of the su(2) ladder operators in Fock basis is

\begin{align}
	&\hat{J}_{\pm}\ket{j,m}  = \sqrt{j\left(j+1\right)-m\left(m\pm 1\right)}\ket{j,m\pm 1} \to  \nonumber \\
	& \to \hat{J}_{\pm}\ket{n,n'}_{1,2} = \nonumber \\
	&  \tfrac{1}{2}\sqrt{\left(n+n'\right)\left(n+n'+2\right)-\left(n-n'\right)\left(n-n'\pm 2\right)}\times \nonumber \\
	& \;\;\;\;\;\;\;\;\;\;\;\;\;\;\;\;\;\;\;\;\;\;\;\;\;\;\;\;\;\;\;\;\;\;\;\;\;\;\;\;\;\;\;\;\;\;\;\;\;\;\;\;\;\;\;\;\;\;\;\;\;\;\;\;\;\;\;\;\;\;\;\;\times\ket{n\pm 1,n'\mp 1}_{1,2}.
	\label{eqn:Aa9}
\end{align}

\noindent The connection between two-mode boson fields and the "angular momentum" states of su(2) is now complete.  

\subsection{\label{app:subsecA3} Beam splitter transformations}

\noindent Throughout this paper, several different beam splitter transformations are utilized in conjunction with optical interferometry.  Here we endeavor to show how these particular transformations affect the corresponding boson mode operators.  First, let us reconsider the scattering matrix of Eq.~\ref{eqn:Aa4} corresponding to a $\hat{J}_{x}$-type beam splitter such that $\ket{\text{out, BS}}=e^{-i\theta\hat{J}_{x}}\ket{\text{in}}$ as per Eq.~\ref{eqn:Aa7}. For the choices of angle $\theta = \pm \pi/2$, both corresponding to a $50:50$ beam splitter but parameterized by different rotations about the fictitious $x$-axis, the mode operators transform according to

\begin{equation}
\begin{pmatrix}
\hat{a}_{0}'  \\
\hat{a}_{1}'
\end{pmatrix}
=
\frac{1}{\sqrt{2}}
\begin{pmatrix}
1 & \mp i \\
\mp i & 1
\end{pmatrix}
\begin{pmatrix}
\hat{a}_{0}  \\
\hat{a}_{1} 
\end{pmatrix}.
\label{eqn:Ac1}
\end{equation} 

\noindent Let us consider two modes of an optical field labeled $a$- and $b$- modes for which operators $\{\hat{a}_{0},\hat{a}_{0}'\}$ act on the $a$-mode and $\{\hat{a}_{1},\hat{a}_{1}'\}$ on the $b$-mode.  For $\theta= \pm\pi/2$ and using Eq.~\ref{eqn:Ac1} we can write

\begin{align}
	\hat{a}_{0}' &= \tfrac{1}{\sqrt{2}}\left(\hat{a}_{0} \mp i\hat{a}_{1}\right),\;\;\;\;\;\;\;\;    \hat{a}_{0} = \tfrac{1}{\sqrt{2}}\left(\hat{a}_{0}' \pm i\hat{a}_{1}'\right),\nonumber \\
	\hat{a}_{1}' &=  \tfrac{1}{\sqrt{2}}\left(\hat{a}_{1} \mp i\hat{a}_{0}\right),\;\;\;\;\;\;\;\;   \hat{a}_{1} =  \tfrac{1}{\sqrt{2}}\left(\hat{a}_{1}' \pm i\hat{a}_{0}'\right).
	\label{eqn:Ac3}
\end{align} 

\noindent Direct substitution of the mode operators in Eq.~\ref{eqn:Ac3} yields for the displacement operators 

\begin{align}
\hat{D}_{a}\left(\alpha\right) = e^{\alpha\hat{a}_{0}^{\dagger} - \alpha^{*}\hat{a}_{0}} 
&\to  e^{\tfrac{\alpha}{\sqrt{2}}\big(\hat{a}_{0}^{\dagger\prime} \mp i\hat{a}_{1}^{\dagger\prime}\big) - \tfrac{\alpha^{*}}{\sqrt{2}}\big(\hat{a}_{0}' \pm i \hat{a}_{1}'\big)} \nonumber \\
& \to e^{\tfrac{\alpha}{\sqrt{2}}\hat{a}_{0}^{\dagger\prime} - \tfrac{\alpha^{*}}{\sqrt{2}}\hat{a}_{0}'}\;\; e^{\mp\tfrac{i\alpha}{\sqrt{2}}\hat{a}_{1}^{\dagger\prime}  \mp\tfrac{i\alpha^{*}}{\sqrt{2}}\hat{a}_{1}'} \nonumber \\
& \to \hat{D}_{a}\big(\tfrac{\alpha}{\sqrt{2}}\big)\hat{D}_{b}\big(\mp\tfrac{i\alpha}{\sqrt{2}}\big),\label{eqn:Ac4} \\ \nonumber\\
\hat{D}_{b}\left(\beta\right) = e^{\beta\hat{a}_{1}^{\dagger} - \beta^{*}\hat{a}_{1}} 
&\to  e^{\tfrac{\beta}{\sqrt{2}}\big(\hat{a}_{1}^{\dagger\prime} \mp i\hat{a}_{0}^{\dagger\prime}\big) - \tfrac{\beta^{*}}{\sqrt{2}}\big(\hat{a}_{1}' \pm i \hat{a}_{0}'\big)} \nonumber \\
& \to e^{\mp\tfrac{i\beta}{\sqrt{2}}\hat{a}_{0}^{\dagger\prime} \mp \tfrac{i\beta^{*}}{\sqrt{2}}\hat{a}_{0}'}\;\; e^{\tfrac{\beta}{\sqrt{2}}\hat{a}_{1}^{\dagger\prime}  -\tfrac{\beta^{*}}{\sqrt{2}}\hat{a}_{1}'} \nonumber \\
& \to \hat{D}_{a}\big(\mp \tfrac{i\beta}{\sqrt{2}}\big)\hat{D}_{b}\big(\tfrac{\beta}{\sqrt{2}}\big), 
\label{eqn:Ac4b}
\end{align}

\noindent or noting the mode operators transform as $e^{\mp i\tfrac{\pi}{2}\hat{J}_{x}}\hat{a}_{0}e^{\pm i\tfrac{\pi}{2}\hat{J}_{x}} = \tfrac{1}{\sqrt{2}}\big(\hat{a}_{0} \pm i\hat{a}_{1}\big)$ and $e^{\mp i\tfrac{\pi}{2}\hat{J}_{x}}\hat{a}_{1}e^{\pm i\tfrac{\pi}{2}\hat{J}_{x}} = \tfrac{1}{\sqrt{2}}\big(\hat{a}_{1} \pm i\hat{a}_{0}\big) $, we can write more succinctly 

\begin{align}
	e^{\mp i\tfrac{\pi}{2}\hat{J}_{x}}\hat{D}_{a}\left(\alpha\right)e^{\pm i\tfrac{\pi}{2}\hat{J}_{x}} &= \hat{D}_{a}\big(\tfrac{\alpha}{\sqrt{2}}\big)\hat{D}_{b}\big(\mp\tfrac{i\alpha}{\sqrt{2}}\big) \nonumber \\
	\label{eqn:Ac4c} \\
	e^{\mp i\tfrac{\pi}{2}\hat{J}_{x}}\hat{D}_{b}\left(\beta\right)e^{\pm i\tfrac{\pi}{2}\hat{J}_{x}} &= \hat{D}_{a}\big(\mp \tfrac{i\beta}{\sqrt{2}}\big)\hat{D}_{b}\big(\tfrac{\beta}{\sqrt{2}}\big)  \nonumber 
\end{align}

\noindent For the case in which coherent light is incident on a beamsplitter from each input port, $\ket{\text{in}} = \ket{\alpha}_{a}\ket{\beta}_{b}$, the resulting output state is straight forward to work out from the discussion above \footnote{Note that $\hat{D}_{a}\left(\alpha\right)\hat{D}_{a}\left(\beta\right) = e^{\tfrac{1}{2}\left(\alpha\beta^{*} - \alpha^{*}\beta\right)}\hat{D}_{a}\left(\alpha+\beta\right)$. The product of displacement operators is the sum of the displacements up to an overall phase factor.}: 

\begin{align}
	\ket{\text{out, BS}} &= e^{-i\theta\hat{J}_{x}}\ket{\text{in}} \stackrel{\theta \to \pm \tfrac{\pi}{2}}{=} e^{\mp i\tfrac{\pi}{2}\hat{J}_{x}}\ket{\alpha}_{a}\ket{\beta}_{b} \nonumber \\
	& =\ket{\tfrac{1}{\sqrt{2}}\big(\alpha \mp i\beta\big)}_{a}\ket{\tfrac{1}{\sqrt{2}}\big(\beta \mp i\alpha\big)}_{b}. 
	\label{eqn:Ac5}
\end{align}

\noindent The resulting state is an equal-intensity distribution of light in each output mode, but where the reflected beam picks up a phase shift of $\pm \pi/2$. For the case in which $\beta \to 0$, the initial state is  $\ket{\text{in}} = \ket{\alpha}_{a}\ket{0}_{b}$ and the output state is simply $\ket{\text{out}} = e^{\mp i \tfrac{\pi}{2}\hat{J}_{x}}\ket{\text{in}} =  \ket{\tfrac{\alpha}{\sqrt{2}}}_{a}\ket{\mp\tfrac{i\alpha}{\sqrt{2}}}$. \\

\noindent For a $\hat{J}_{y}$-type beam splitter such that $\ket{\text{out}} = e^{-i\theta\hat{J}_{y}}\ket{\text{in}}$, the action of the scattering matrix transforms the mode operators according to 

\begin{align}
\begin{pmatrix}
\hat{a}_{0}'  \\
\hat{a}_{1}'
\end{pmatrix}
&=
\begin{pmatrix}
\cos\tfrac{\theta}{2} & - \sin\tfrac{\theta}{2}  \\
\sin\tfrac{\theta}{2} & \cos\tfrac{\theta}{2}
\end{pmatrix}
\begin{pmatrix}
\hat{a}_{0}  \\
\hat{a}_{1} 
\end{pmatrix} 
\stackrel{\theta \to \pm\tfrac{\pi}{2}}{=}
\frac{1}{\sqrt{2}}
\begin{pmatrix}
1 & \mp 1 \\
\pm 1 & 1
\end{pmatrix}
\begin{pmatrix}
\hat{a}_{0}  \\
\hat{a}_{1} 
\end{pmatrix}.
\label{eqn:Ac6}
\end{align} 

\noindent leading to 

\begin{align}
\hat{a}_{0}' &= \tfrac{1}{\sqrt{2}}\left(\hat{a}_{0} \mp \hat{a}_{1}\right),\;\;\;\;\;\;\;\;    \hat{a}_{0} = \tfrac{1}{\sqrt{2}}\left(\hat{a}_{0}' \pm\hat{a}_{1}'\right),\nonumber \\
\hat{a}_{1}' &=  \tfrac{1}{\sqrt{2}}\left(\hat{a}_{1} \pm \hat{a}_{0}\right),\;\;\;\;\;\;\;\;   \hat{a}_{1} =  \tfrac{1}{\sqrt{2}}\left(\hat{a}_{1}' \mp \hat{a}_{0}'\right).
\label{eqn:Ac7}
\end{align} 

\noindent Direct substitution into the single-mode displacement operators yields

\begin{align}
	\hat{D}_{a}\left(\alpha\right) = e^{\alpha\hat{a}_{0}^{\dagger} - \alpha^{*}\hat{a}_{0}} &\to e^{\tfrac{\alpha}{\sqrt{2}}\big(\hat{a}_{0}^{\dagger\prime} \pm \hat{a}_{1}^{\dagger\prime}\big) - \tfrac{\alpha^{*}}{\sqrt{2}}\big(\hat{a}_{0}' \pm \hat{a}_{1}'\big)} \nonumber \\
	&\to e^{\tfrac{\alpha}{\sqrt{2}}\hat{a}_{0}^{\dagger\prime}-\tfrac{\alpha^{*}}{\sqrt{2}}\hat{a}_{0}'}\;\;e^{\pm\tfrac{\alpha}{\sqrt{2}}\hat{a}_{1}^{\dagger\prime}\mp\tfrac{\alpha^{*}}{\sqrt{2}}\hat{a}_{1}'} \nonumber \\
	&\to \hat{D}_{a}\big(\tfrac{\alpha}{\sqrt{2}}\big)\hat{D}_{b}\big(\pm\tfrac{\alpha}{\sqrt{2}}\big), \label{eqn:Ac8} \\ \nonumber \\
	\hat{D}_{b}\left(\beta\right) = e^{\beta\hat{a}_{1}^{\dagger} - \beta^{*}\hat{a}_{1}} &\to e^{\tfrac{\beta}{\sqrt{2}}\big(\hat{a}_{1}^{\dagger\prime} \mp \hat{a}_{0}^{\dagger\prime}\big) - \tfrac{\beta^{*}}{\sqrt{2}}\big(\hat{a}_{1}' \mp \hat{a}_{0}'\big)} \nonumber \\
	&\to e^{\tfrac{\beta}{\sqrt{2}}\hat{a}_{1}^{\dagger\prime}-\tfrac{\beta^{*}}{\sqrt{2}}\hat{a}_{1}'}\;\;e^{\mp\tfrac{\beta}{\sqrt{2}}\hat{a}_{0}^{\dagger\prime}\pm\tfrac{\beta^{*}}{\sqrt{2}}\hat{a}_{0}'} \nonumber \\
	&\to \hat{D}_{a}\big(\mp\tfrac{\beta}{\sqrt{2}}\big)\hat{D}_{b}\big(\tfrac{\beta}{\sqrt{2}}\big), \label{eqn:Ac8b}
\end{align}

\noindent Once again assuming coherent light initially in each port incident on a $50:50$ beam splitter, the output state is given by 

\begin{align}
\ket{\text{out, BS}} &= e^{-i\theta\hat{J}_{y}}\ket{\text{in}} \stackrel{\theta \to \pm \tfrac{\pi}{2}}{=} e^{\mp i\tfrac{\pi}{2}\hat{J}_{y}}\ket{\alpha}_{a}\ket{\beta}_{b} \nonumber \\
& =\ket{\tfrac{1}{\sqrt{2}}\big(\alpha \mp \beta\big)}_{a}\ket{\tfrac{1}{\sqrt{2}}\big(\beta \pm \alpha\big)}_{b}.  
\label{eqn:Ac9}
\end{align}

\noindent where once again the result is an equal-intensity distribution in both output modes. For $\beta \to 0$, the output state is given by $\ket{\text{out}} = \ket{\tfrac{\alpha}{\sqrt{2}}}_{a}\ket{\pm \tfrac{\alpha}{\sqrt{2}}}$, where the choice of $\theta= -\pi/2$ results in a phase shift of $\pi$ in the reflected beam.  

\section{\label{app:secB} The Wigner-$d$ matrix elements}

\subsection{\label{app:subsecB1} Defined}

\noindent Here we provide a brief discussion on the matrix elements of an arbitrary rotation specified by an axis of rotation $\boldsymbol{\hat{n}}$ and angle of rotation $\phi$.  The matrix elements, with $\hbar \to 1$ for convenience, are 

\begin{equation}
	\mathcal{D}_{m',m}^{j}\left(R\right) = \braket{j,m'|e^{-i\phi\; \boldsymbol{J}\cdot\boldsymbol{\hat{n}}}|j,m}.
	\label{eqn:Ab1}
\end{equation}

\noindent Since the rotation operator commutes with the $\hat{J}^{2}$ operator, a rotation cannot change the $j$ value of a state. The $\big(2j+1\big)\times\big(2j+1\big)$ matrix formed by $\mathcal{D}_{m',m}^{j}\big(R\big)$ is referred to as the $\big(2j+1\big)$-dimensional irreducible representation of the rotation operator $\mathcal{D}\big(R\big)$ .  We now consider the matrix realization of the Euler Rotation,

\begin{align}
	\mathcal{D}_{m',m}^{j}\left(\alpha,\beta,\gamma\right) &= \braket{j,m'|R_{z_{f}}\left(\alpha\right)R_{y_{f}}\left(\beta\right)R_{z_{f}}\left(\gamma\right)|j,m} \nonumber\\
	&= \braket{j,m'|e^{-i\alpha\hat{J}_{z}}e^{-i\beta\hat{J}_{y}}e^{-i\gamma\hat{J}_{z}}|j,m}.
	\label{eqn:Ab2}
\end{align} 

\noindent These matrix elements are referred to as the Wigner-$D$ matrix elements. Notice that the first and last rotation only add a phase factor to the expression, thus making only the rotation about the fixed $y$-axis the only non-trival part of the matrix.  For this reason, the Wigner-$D$ matrix elements are written in terms of a new matrix

\begin{align}
	\mathcal{D}_{m',m}^{j}\left(\alpha,\beta,\gamma\right) &=\braket{j,m'|e^{-i\alpha\hat{J}_{z}}e^{-i\beta\hat{J}_{y}}e^{-i\gamma\hat{J}_{z}}|j,m} \nonumber \\
	&= e^{-i\left(m'\alpha + m\gamma\right)}\braket{j,m'|e^{-i\beta\hat{J}_{y}}|j,m} \nonumber\\
	&= e^{-i\left(m'\alpha + m\gamma\right)}\; d_{m',m}^{j}\left(\beta\right),
\label{eqn:Ab3}
\end{align}

\noindent where the matrix elements $d_{m',m}^{j}\left(\beta\right) = \braket{j,m'|e^{-i\beta\hat{J}_{y}}|j,m}$ are known as the Wigner-$d$ matrix elements and are given by

\begin{align}
	d_{m',m}^{j}\left(\beta\right) &= \Bigg(  \dfrac{\big(j-m\big)!\big(j+m'\big)!}{\big(j+m\big)!\big(j-m'\big)!}  \Bigg)^{1/2} \times\nonumber \\
	&\;\;\;\times\dfrac{\big(-1\big)^{m'-m}\cos^{2j+m-m'}\big(\tfrac{\beta}{2}\big)\sin^{m'-m}\big(\tfrac{\beta}{2}\big)}{\big(m'-m\big)!}\times \nonumber \\
	&\times _{2}F_{1}\left(m'-j,-m-j;m'-m+1;-\tan^{2}\big(\tfrac{\beta}{2}\big)\right),
	\label{eqn:Ab4}
\end{align}

\noindent with the property

\begin{equation}
d_{m'm}^{j}\left(\beta\right) = 
\begin{cases}
	d_{m',m}^{j}\left(\beta\right) & m' \geq m \\
	\\
	d_{m,m'}^{j}\left(-\beta\right) & m' < m 
	\end{cases}
	\label{eqn:Ab5}
\end{equation}

\noindent and where $_{2}F_{1}\big(a,b;c;z\big)$ is the hypergeometric function defined formally by
 
\begin{equation}
	{}_{p}F_{q}\left(a_{1},...,a_{p};b_{1},...,b_{q};z\right)= \sum_{n=0}^{\infty}\frac{\left(a_{1}\right)_{n}...\left(a_{p}\right)_{n}}{\left(b_{1}\right)_{n}...\left(b_{q}\right)_{n}}\frac{z^{n}}{n!}.
	\label{eqn:Ab6}
\end{equation} 

\noindent The Pochhammer symbol above is used to express $\left(x\right)_{n}=x\left(x+1\right)\left(x+2\right)...\left(x+n-1\right) = \Gamma\left(x+n\right)/\Gamma\left(x\right)$ for $n \geq 1$ with the Gamma (generalized factorial) function $\Gamma\left(x\right)=\left(x-1\right)!$.  It is worth noting that in our interferometric calculations, we naturally end up with an expression that depends on the Wigner-$d$ matrix elements.  However, when simply dealing with a single $\hat{J}_{x}$ beam splitter of angle $\theta$, one encounters the matrix elements $\braket{j,m'|e^{-i\theta\hat{J}_{x}}|j,m}$.  This can be simplified using the relations to

\begin{align}
	\braket{j,m'|e^{-i\theta\hat{J}_{x}}|j,m} &= \braket{j,m'|e^{i\tfrac{\pi}{2}\hat{J}_{z}}e^{-i\theta\hat{J}_{y}}e^{-i\tfrac{\pi}{2}\hat{J}_{z}}|j,m} \nonumber\\
	&= \mathcal{D}_{m',m}^{j}\left(-\tfrac{\pi}{2},\theta,\tfrac{\pi}{2}\right)\nonumber \\
	&= i^{m'-m}d_{m',m}^{j}\left(\theta\right).
	\label{eqn:Ab7}
\end{align}

\noindent In our closing section we state a handful of useful identities used throughout this paper.

\subsection{\label{app:subsecB2} Useful identities}

\noindent Within the body of this review article, several simplified forms of the Wigner-$d$ matrix are used. The cases relevant to the material herein are listed below:

\begin{align}
	d_{-j,-j}^{j}\left(\beta\right) = d_{j,j}^{j}\left(\beta\right) &=  \cos^{2j}\left(\tfrac{\beta}{2}\right), \label{eqn:bb1}\\
	\nonumber \\
	d_{j,-j}^{j}\left(\beta\right) = \left(-1\right)^{2j} d_{-j,j}^{j}\left(\beta\right) &= \left(-1\right)^{2j}\sin^{2j}\left(\tfrac{\beta}{2}\right), \label{eqn:bb2}
\end{align}

\noindent where we have used the identity

\begin{equation}
	d_{m',m}^{j}\left(\beta\right) = \left(-1\right)^{m'-m}d_{m,m'}^{j}\left(\beta\right).
	\label{eqn:bb3}
\end{equation}

\noindent We have also used the simplified form

\begin{equation}
	d_{m,-j}^{j}\left(\tfrac{\pi}{2}\right) = 2^{-j}\binom{2j}{j+m}^{1/2}\left(-1\right)^{j+m}
 = \left(-1\right)^{j+m}d_{m,j}^{j}\left(\tfrac{\pi}{2}\right), 
	\label{eqn:bb4}
\end{equation}

\noindent in our derivations pertaining to Ramsey spectroscopy. Through applications of the standard relations involving the Wigner-$d$ rotation elements \cite{WignerIdent}, it can be shown

\begin{equation}
	\sum_{m' = -j}^{j}\left(-1\right)^{j-m'}d_{j-n,m'}^{j}\left(-\beta\right)d_{m',j-n}^{j}\left(\beta\right) = \left(-1\right)^{n}d_{j-n,j-n}^{j}\left(2\beta\right),
	\label{eqn:Ab17}
\end{equation}

\noindent so that in particular for $n=0$ we find $d_{j,j}^{j}\left(2\phi\right)=\cos^{2j}\phi$.  This identity is used in Eq.~\ref{eqn:cohN_4} in the derivation of the expectation value of the parity operator $\braket{\hat{\Pi}_{b}}$ for the input state $\ket{\alpha}_{a}\otimes\ket{N}_{b}$. \\

\noindent The following functions are defined in conjunction with the Ramsey spectroscopy derivations of Eqs.~\ref{eqn:rs14}, \ref{eqn:rs7} and \ref{eqn:rs8}, respectively:

\begin{align}
F_{m,p}\left(\lambda\right) &= \sum_{m'=-j}^{j}\left(-1\right)^{j-m'}d_{m',m}^{j}\left(\lambda\right)d_{m',p}^{j}\left(\lambda\right) \nonumber \\
&= \left(-1\right)^{2j}\sum_{m'}d_{m',-m}^{j}\left(\pi-\lambda\right)d_{m',p}^{j}\left(\lambda\right)\nonumber\\
&= \left(-1\right)^{2j}\sum_{m'}d_{p,m'}^{j}\left(-\lambda\right)d_{m',-m}^{j}\left(\pi-\lambda\right) \nonumber \\
&= \left(-1\right)^{2j}d_{p,-m}^{j}\left(\pi-2\lambda\right).
\label{eqn:Ab19}
\end{align}

\begin{align}
	\Gamma_{m,p}&\left(\lambda\right) = \sum_{m'=-j}^{j} m'd^{j}_{m',p}\left(\lambda\right)d_{m',m}^{j}\left(\lambda\right) \nonumber \\ 
	 &= \sum_{m'=-j}^{j} m'd^{j}_{m',p}\left(\lambda\right)d_{m',m}^{j}\left(\lambda\right)\left(\delta_{p,m+1} + \delta_{p+1,m} + \delta_{p,m}\right)
	\label{eqn:Ab20}
\end{align}

\begin{align}
	\Lambda_{m,p}\left(\lambda\right) &= \sum_{m'=-j}^{j} m^{\prime\; 2}d^{j}_{m',p}\left(\lambda\right)d_{m',m}^{j}\left(\lambda\right) \nonumber \\ 
	&= \sum_{m'=-j}^{j} m^{\prime \;2}d^{j}_{m',p}\left(\lambda\right)d_{m',m}^{j}\left(\lambda\right) \times \nonumber \\
	&\;\;\;\;\;\;\;\times \left(\delta_{p,m+2} + \delta_{p+2,m}  + \delta_{p,m+1} + \delta_{p+1,m} + \delta_{p,m}\right)
	\label{eqn:Ab21}
\end{align}

\nocite{*}
\bibliography{ParityReviewArticle}% Produces the bibliography via BibTeX.

\providecommand{\noopsort}[1]{}\providecommand{\singleletter}[1]{#1}%
\begin{thebibliography}{103}%
\makeatletter
\providecommand \@ifxundefined [1]{%
 \@ifx{#1\undefined}
}%
\providecommand \@ifnum [1]{%
 \ifnum #1\expandafter \@firstoftwo
 \else \expandafter \@secondoftwo
 \fi
}%
\providecommand \@ifx [1]{%
 \ifx #1\expandafter \@firstoftwo
 \else \expandafter \@secondoftwo
 \fi
}%
\providecommand \natexlab [1]{#1}%
\providecommand \enquote  [1]{``#1''}%
\providecommand \bibnamefont  [1]{#1}%
\providecommand \bibfnamefont [1]{#1}%
\providecommand \citenamefont [1]{#1}%
\providecommand \href@noop [0]{\@secondoftwo}%
\providecommand \href [0]{\begingroup \@sanitize@url \@href}%
\providecommand \@href[1]{\@@startlink{#1}\@@href}%
\providecommand \@@href[1]{\endgroup#1\@@endlink}%
\providecommand \@sanitize@url [0]{\catcode `\\12\catcode `\$12\catcode
  `\&12\catcode `\#12\catcode `\^12\catcode `\_12\catcode `\%12\relax}%
\providecommand \@@startlink[1]{}%
\providecommand \@@endlink[0]{}%
\providecommand \url  [0]{\begingroup\@sanitize@url \@url }%
\providecommand \@url [1]{\endgroup\@href {#1}{\urlprefix }}%
\providecommand \urlprefix  [0]{URL }%
\providecommand \Eprint [0]{\href }%
\providecommand \doibase [0]{http://dx.doi.org/}%
\providecommand \selectlanguage [0]{\@gobble}%
\providecommand \bibinfo  [0]{\@secondoftwo}%
\providecommand \bibfield  [0]{\@secondoftwo}%
\providecommand \translation [1]{[#1]}%
\providecommand \BibitemOpen [0]{}%
\providecommand \bibitemStop [0]{}%
\providecommand \bibitemNoStop [0]{.\EOS\space}%
\providecommand \EOS [0]{\spacefactor3000\relax}%
\providecommand \BibitemShut  [1]{\csname bibitem#1\endcsname}%
\let\auto@bib@innerbib\@empty
%</preamble>
\bibitem [{\citenamefont {Wigner}(1932)}]{Wigner1}%
  \BibitemOpen
  \bibfield  {author} {\bibinfo {author} {\bibfnamefont {E.}~\bibnamefont
  {Wigner}},\ }\bibfield  {title} {\enquote {\bibinfo {title} {On the quantum
  correction for thermodynamic equilibrium},}\ }\href@noop {} {\bibfield
  {journal} {\bibinfo  {journal} {Phys.\ Rev.}\ }\textbf {\bibinfo {volume}
  {40}},\ \bibinfo {pages} {749--759} (\bibinfo {year} {1932})}\BibitemShut
  {NoStop}%
\bibitem [{\citenamefont {Chaill}\ and\ \citenamefont
  {Glauber}(1969)}]{Wigner3}%
  \BibitemOpen
  \bibfield  {author} {\bibinfo {author} {\bibfnamefont {K.~E.}\ \bibnamefont
  {Chaill}}\ and\ \bibinfo {author} {\bibfnamefont {R.~J.}\ \bibnamefont
  {Glauber}},\ }\bibfield  {title} {\enquote {\bibinfo {title} {Density
  operators and quasiprobability distributions},}\ }\href@noop {} {\bibfield
  {journal} {\bibinfo  {journal} {Phys.\ Rev.}\ }\textbf {\bibinfo {volume}
  {177}},\ \bibinfo {pages} {5} (\bibinfo {year} {1969})}\BibitemShut {NoStop}%
\bibitem [{\citenamefont {Royer}(1977)}]{Wigner2}%
  \BibitemOpen
  \bibfield  {author} {\bibinfo {author} {\bibfnamefont {A.}~\bibnamefont
  {Royer}},\ }\bibfield  {title} {\enquote {\bibinfo {title} {Wigner function
  as the expectation value of a parity operator},}\ }\href@noop {} {\bibfield
  {journal} {\bibinfo  {journal} {Phys.\ Rev. A}\ }\textbf {\bibinfo {volume}
  {15}},\ \bibinfo {pages} {2} (\bibinfo {year} {1977})}\BibitemShut {NoStop}%
\bibitem [{\citenamefont {Leonhardt}(1997)}]{Tomography}%
  \BibitemOpen
  \bibfield  {author} {\bibinfo {author} {\bibfnamefont {U.}~\bibnamefont
  {Leonhardt}},\ }\href@noop {} {\emph {\bibinfo {title} {Measuring the Quantum
  State of Light}}}\ (\bibinfo  {publisher} {Cambridge University Press,
  Cambridge, UK.},\ \bibinfo {year} {1997})\BibitemShut {NoStop}%
\bibitem [{\citenamefont {Banaszek}\ and\ \citenamefont
  {W\'{o}dkiewicz}(1998)}]{Bell}%
  \BibitemOpen
  \bibfield  {author} {\bibinfo {author} {\bibfnamefont {K.}~\bibnamefont
  {Banaszek}}\ and\ \bibinfo {author} {\bibfnamefont {K.}~\bibnamefont
  {W\'{o}dkiewicz}},\ }\bibfield  {title} {\enquote {\bibinfo {title}
  {Nonlocality of the {E}instein-{P}odolsky-{R}osen state in the {W}igner
  representation},}\ }\href@noop {} {\bibfield  {journal} {\bibinfo  {journal}
  {Phys.\ Rev. A}\ }\textbf {\bibinfo {volume} {58}},\ \bibinfo {pages} {6}
  (\bibinfo {year} {1998})}\BibitemShut {NoStop}%
\bibitem [{\citenamefont {Barsotti}, \citenamefont {Harms},\ and\ \citenamefont
  {Schnabel}(2019)}]{LIGO}%
  \BibitemOpen
  \bibfield  {author} {\bibinfo {author} {\bibfnamefont {L.}~\bibnamefont
  {Barsotti}}, \bibinfo {author} {\bibfnamefont {J.}~\bibnamefont {Harms}}, \
  and\ \bibinfo {author} {\bibfnamefont {R.}~\bibnamefont {Schnabel}},\
  }\bibfield  {title} {\enquote {\bibinfo {title} {Squeezed vacuum states of
  light for gravitational wave detectors},}\ }\href@noop {} {\bibfield
  {journal} {\bibinfo  {journal} {Rep. Prog. Phys.}\ }\textbf {\bibinfo
  {volume} {82}},\ \bibinfo {pages} {016905} (\bibinfo {year}
  {2019})}\BibitemShut {NoStop}%
\bibitem [{\citenamefont {Bollinger}\ \emph {et~al.}(1996)\citenamefont
  {Bollinger}, \citenamefont {Itano}, \citenamefont {Wineland},\ and\
  \citenamefont {Heinzen}}]{Bollinger}%
  \BibitemOpen
  \bibfield  {author} {\bibinfo {author} {\bibfnamefont {J.~J.}\ \bibnamefont
  {Bollinger}}, \bibinfo {author} {\bibfnamefont {W.~M.}\ \bibnamefont
  {Itano}}, \bibinfo {author} {\bibfnamefont {D.~J.}\ \bibnamefont {Wineland}},
  \ and\ \bibinfo {author} {\bibfnamefont {D.~J.}\ \bibnamefont {Heinzen}},\
  }\bibfield  {title} {\enquote {\bibinfo {title} {Optimal frequency
  measurements with maximally correlated states},}\ }\href@noop {} {\bibfield
  {journal} {\bibinfo  {journal} {Phys.\ Lett. A}\ }\textbf {\bibinfo {volume}
  {54}},\ \bibinfo {pages} {{R}4649} (\bibinfo {year} {1996})}\BibitemShut
  {NoStop}%
\bibitem [{\citenamefont {Yurke}, \citenamefont {McCall},\ and\ \citenamefont
  {Klauder}(1986)}]{Yurke}%
  \BibitemOpen
  \bibfield  {author} {\bibinfo {author} {\bibfnamefont {B.}~\bibnamefont
  {Yurke}}, \bibinfo {author} {\bibfnamefont {S.~L.}\ \bibnamefont {McCall}}, \
  and\ \bibinfo {author} {\bibfnamefont {J.~R.}\ \bibnamefont {Klauder}},\
  }\bibfield  {title} {\enquote {\bibinfo {title} {{SU(2)} and {SU(1,1)}
  interferometers},}\ }\href@noop {} {\bibfield  {journal} {\bibinfo  {journal}
  {Phys.\ Rev. A}\ }\textbf {\bibinfo {volume} {33}},\ \bibinfo {pages} {4033}
  (\bibinfo {year} {1986})}\BibitemShut {NoStop}%
\bibitem [{\citenamefont {Ramsey}(1956)}]{RamseyBook}%
  \BibitemOpen
  \bibfield  {author} {\bibinfo {author} {\bibfnamefont {N.}~\bibnamefont
  {Ramsey}},\ }\href@noop {} {\emph {\bibinfo {title} {Molecular Beams}}}\
  (\bibinfo  {publisher} {Oxford University Press, Oxford},\ \bibinfo {year}
  {1956})\BibitemShut {NoStop}%
\bibitem [{\citenamefont {Arrechi}\ \emph {et~al.}(1972)\citenamefont
  {Arrechi}, \citenamefont {Courtnes}, \citenamefont {Gilmore},\ and\
  \citenamefont {Thomas}}]{Arrechi}%
  \BibitemOpen
  \bibfield  {author} {\bibinfo {author} {\bibfnamefont {F.~T.}\ \bibnamefont
  {Arrechi}}, \bibinfo {author} {\bibfnamefont {E.}~\bibnamefont {Courtnes}},
  \bibinfo {author} {\bibfnamefont {R.}~\bibnamefont {Gilmore}}, \ and\
  \bibinfo {author} {\bibfnamefont {H.}~\bibnamefont {Thomas}},\ }\bibfield
  {title} {\enquote {\bibinfo {title} {Atomic coherent states in quantum
  optics},}\ }\href@noop {} {\bibfield  {journal} {\bibinfo  {journal} {Phys.\
  Rev. A}\ }\textbf {\bibinfo {volume} {6}},\ \bibinfo {pages} {6} (\bibinfo
  {year} {1972})}\BibitemShut {NoStop}%
\bibitem [{\citenamefont {Steinbach}\ and\ \citenamefont
  {Gerry}(1998)}]{Steinbach}%
  \BibitemOpen
  \bibfield  {author} {\bibinfo {author} {\bibfnamefont {J.}~\bibnamefont
  {Steinbach}}\ and\ \bibinfo {author} {\bibfnamefont {C.~C.}\ \bibnamefont
  {Gerry}},\ }\bibfield  {title} {\enquote {\bibinfo {title} {Efficient scheme
  for the deterministic maximal entanglement of {N} trapped ions},}\
  }\href@noop {} {\bibfield  {journal} {\bibinfo  {journal} {Phys.\ Rev. A}\
  }\textbf {\bibinfo {volume} {81}},\ \bibinfo {pages} {5528} (\bibinfo {year}
  {1998})}\BibitemShut {NoStop}%
\bibitem [{\citenamefont {Leibfried}\ \emph {et~al.}(2004)\citenamefont
  {Leibfried}, \citenamefont {Barrett}, \citenamefont {Schaetz}, \citenamefont
  {Briton}, \citenamefont {Chiaverini}, \citenamefont {Itano}, \citenamefont
  {Jost}, \citenamefont {Langer},\ and\ \citenamefont {Wineland}}]{Leibfried}%
  \BibitemOpen
  \bibfield  {author} {\bibinfo {author} {\bibfnamefont {D.}~\bibnamefont
  {Leibfried}}, \bibinfo {author} {\bibfnamefont {M.~D.}\ \bibnamefont
  {Barrett}}, \bibinfo {author} {\bibfnamefont {T.}~\bibnamefont {Schaetz}},
  \bibinfo {author} {\bibfnamefont {J.}~\bibnamefont {Briton}}, \bibinfo
  {author} {\bibfnamefont {J.}~\bibnamefont {Chiaverini}}, \bibinfo {author}
  {\bibfnamefont {W.~M.}\ \bibnamefont {Itano}}, \bibinfo {author}
  {\bibfnamefont {J.~D.}\ \bibnamefont {Jost}}, \bibinfo {author}
  {\bibfnamefont {C.}~\bibnamefont {Langer}}, \ and\ \bibinfo {author}
  {\bibfnamefont {D.~J.}\ \bibnamefont {Wineland}},\ }\bibfield  {title}
  {\enquote {\bibinfo {title} {Toward {H}eisenberg-limited spectroscopy with
  multiparticle entangled states},}\ }\href@noop {} {\bibfield  {journal}
  {\bibinfo  {journal} {Science}\ }\textbf {\bibinfo {volume} {304}},\ \bibinfo
  {pages} {1476} (\bibinfo {year} {2004})}\BibitemShut {NoStop}%
\bibitem [{\citenamefont {Pegg}\ and\ \citenamefont
  {Barnett}(1989)}]{PeggBarn}%
  \BibitemOpen
  \bibfield  {author} {\bibinfo {author} {\bibfnamefont {D.~T.}\ \bibnamefont
  {Pegg}}\ and\ \bibinfo {author} {\bibfnamefont {S.~M.}\ \bibnamefont
  {Barnett}},\ }\bibfield  {title} {\enquote {\bibinfo {title} {Phase
  properties of the quantized single-mode electromagnetic field},}\ }\href@noop
  {} {\bibfield  {journal} {\bibinfo  {journal} {Phys.\ Rev. A}\ }\textbf
  {\bibinfo {volume} {39}},\ \bibinfo {pages} {1665} (\bibinfo {year}
  {1989})}\BibitemShut {NoStop}%
\bibitem [{\citenamefont {Pezz\'{e}}\ and\ \citenamefont
  {Smerzi}(2014)}]{Pezze}%
  \BibitemOpen
  \bibfield  {author} {\bibinfo {author} {\bibfnamefont {L.}~\bibnamefont
  {Pezz\'{e}}}\ and\ \bibinfo {author} {\bibfnamefont {A.}~\bibnamefont
  {Smerzi}},\ }\href@noop {} {\emph {\bibinfo {title} {Quantum theory of phase
  estimation}}}\ (\bibinfo  {publisher} {Atom {I}nterferometry: {P}roceedings
  of the {I}nternational {S}chool of {Physics} {E}nrico {F}ermi, edited by {G}.
  {M}. {T}ino and {M}. {A}. {K}asevich, IOS, Amsterdam, p.691},\ \bibinfo
  {year} {2014})\BibitemShut {NoStop}%
\bibitem [{Note1()}]{Note1}%
  \BibitemOpen
  \bibinfo {note} {The derivation of the CRB is straightforward. First, we have
  $\protect \genfrac {}{}{}1{\partial \mathinner {\delimiter "426830A {\Phi
  }\delimiter "526930B }_{\varphi }}{\partial \varphi } = \DOTSB \sum@
  \slimits@ _{\epsilon }\partial _{\varphi }\protect \tmspace +\thinmuskip
  {.1667em}P(\epsilon |\varphi ) \Phi (\epsilon )= \mathinner {\delimiter
  "426830A {\Phi \protect \genfrac {}{}{}1{\partial L}{\partial \varphi
  }}\delimiter "526930B }$, where $L(\epsilon |\varphi ) \equiv \protect
  \qopname \relax o{ln}P(\epsilon |\varphi )$. Noting that $\DOTSB \sum@
  \slimits@ _{\epsilon }\partial _{\varphi }\protect \tmspace +\thinmuskip
  {.1667em}P(\epsilon |\varphi ) = \mathinner {\delimiter "426830A {\protect
  \genfrac {}{}{}1{\partial L}{\partial \varphi }}\delimiter "526930B } = 0$ we
  have $\left ( \protect \genfrac {}{}{}1{\partial \mathinner {\delimiter
  "426830A {\Phi }\delimiter "526930B }_{\varphi }}{\partial \varphi } \right
  )^2 = \mathinner {\delimiter "426830A {{\setbox \z@ \hbox {\frozen@everymath
  \@emptytoks \mathsurround \z@ $\nulldelimiterspace \z@ \left (\vcenter to\@ne
  \big@size {}\right .$}\box \z@ } \Phi - \mathinner {\delimiter "426830A {\Phi
  }\delimiter "526930B }_\varphi {\setbox \z@ \hbox {\frozen@everymath
  \@emptytoks \mathsurround \z@ $\nulldelimiterspace \z@ \left )\vcenter to\@ne
  \big@size {}\right .$}\box \z@ } \protect \genfrac {}{}{}1{\partial
  L}{\partial \varphi }}\delimiter "526930B }^2_\varphi $ $\le \mathinner
  {\delimiter "426830A {{\setbox \z@ \hbox {\frozen@everymath \@emptytoks
  \mathsurround \z@ $\nulldelimiterspace \z@ \left (\vcenter to\@ne \big@size
  {}\right .$}\box \z@ } \Phi - \mathinner {\delimiter "426830A {\Phi
  }\delimiter "526930B }_\varphi {\setbox \z@ \hbox {\frozen@everymath
  \@emptytoks \mathsurround \z@ $\nulldelimiterspace \z@ \left )\vcenter to\@ne
  \big@size {}\right .$}\box \z@ }^2}\delimiter "526930B }_\varphi \mathinner
  {\delimiter "426830A { \left (\protect \genfrac {}{}{}1{\partial L}{\partial
  \varphi } \right )^2}\delimiter "526930B }_\varphi $ $= \left (\Delta \Phi
  \right )^2_\varphi \protect \tmspace +\thinmuskip {.1667em}F(\varphi )$,
  where we have invoked the Cauchy-Schwarz inequality $\mathinner {\delimiter
  "426830A {A B}\delimiter "526930B }^2_\varphi \le \mathinner {\delimiter
  "426830A {A^2}\delimiter "526930B }_\varphi \protect \tmspace +\thinmuskip
  {.1667em}\mathinner {\delimiter "426830A {B^2}\delimiter "526930B }_\varphi
  $. Dividing by $F(\varphi )$ yields the CRB in Eq~\ref
  {eqn:2.13}.}\BibitemShut {Stop}%
\bibitem [{Note2()}]{Note2}%
  \BibitemOpen
  \bibinfo {note} {We follow the language found in the literature with regards
  to defining the upper and lower bounds on phase estimation; i.e. the CRB is
  the upper bound and the qCRB is the lower bound.}\BibitemShut {Stop}%
\bibitem [{\citenamefont {Braunstein}\ and\ \citenamefont {Caves}(1994)}]{QFI}%
  \BibitemOpen
  \bibfield  {author} {\bibinfo {author} {\bibfnamefont {S.~L.}\ \bibnamefont
  {Braunstein}}\ and\ \bibinfo {author} {\bibfnamefont {C.~M.}\ \bibnamefont
  {Caves}},\ }\bibfield  {title} {\enquote {\bibinfo {title} {Statistical
  distance and the geometry of quantum states},}\ }\href@noop {} {\bibfield
  {journal} {\bibinfo  {journal} {Phys.\ Rev. Lett.}\ }\textbf {\bibinfo
  {volume} {72}},\ \bibinfo {pages} {3439} (\bibinfo {year}
  {1994})}\BibitemShut {NoStop}%
\bibitem [{\citenamefont {Braunstein}, \citenamefont {Milburn},\ and\
  \citenamefont {Caves}(1996)}]{QFI2}%
  \BibitemOpen
  \bibfield  {author} {\bibinfo {author} {\bibfnamefont {S.~L.}\ \bibnamefont
  {Braunstein}}, \bibinfo {author} {\bibfnamefont {G.~J.}\ \bibnamefont
  {Milburn}}, \ and\ \bibinfo {author} {\bibfnamefont {C.~M.}\ \bibnamefont
  {Caves}},\ }\bibfield  {title} {\enquote {\bibinfo {title} {Generalized
  uncertainty relations: theory, examples and {L}orentz invariants},}\
  }\href@noop {} {\bibfield  {journal} {\bibinfo  {journal} {Ann. of Phys.}\
  }\textbf {\bibinfo {volume} {247}} (\bibinfo {year} {1996})}\BibitemShut
  {NoStop}%
\bibitem [{Note3()}]{Note3}%
  \BibitemOpen
  \bibinfo {note} {The QFI in Eq.~\ref {eqn:2.17} has a pleasing geometrical
  interpretation: it is the infinitesimal version of the quantum fidelity
  $\protect \mathcal {F}_Q(\protect \mathaccentV {hat}05E{\rho }_1,\protect
  \mathaccentV {hat}05E{\rho }_2) = \protect \textrm {Tr}\left [\protect \sqrt
  {\protect \sqrt {\protect \mathaccentV {hat}05E{\rho }_2}\protect \tmspace
  +\thinmuskip {.1667em} \protect \mathaccentV {hat}05E{\rho }_1\protect \sqrt
  {\protect \mathaccentV {hat}05E{\rho }_2}}\right ] $ between two density
  matrices in the sense that $\DOTSI \intop \ilimits@ _0^\zeta \protect \sqrt
  {F_Q}(\zeta ') d\zeta ' = \protect \sqrt {\protect \mathcal {F}_Q}(\zeta )$
  along the geodesic curve connecting $\protect \mathaccentV {hat}05E{\rho }_1$
  and $\protect \mathaccentV {hat}05E{\rho }_2$ parameterized by $\zeta $. This
  can be seen a follows: $\protect \mathcal {F}_Q(\protect \mathaccentV
  {hat}05E{\rho }_1,\protect \mathaccentV {hat}05E{\rho }_2) = \protect \textrm
  {max}_U \protect \textrm {Tr}[\protect \sqrt {\protect \mathaccentV
  {hat}05E{\rho }_1} \protect \sqrt {\protect \mathaccentV {hat}05E{\rho }_2}]
  $ $ =\protect \textrm {Tr}[|\protect \sqrt {\protect \mathaccentV
  {hat}05E{\rho }_1} \protect \sqrt {\protect \mathaccentV {hat}05E{\rho
  }_2}\protect \tmspace +\thinmuskip {.1667em}|] $ = (where $|A| = \protect
  \sqrt {A A^\dagger }$) over all purifications of the density matrices. A
  purification of a density matrix $\protect \mathaccentV {hat}05E{\rho }_i$ is
  a pure state $\mathinner {|{\psi _i}\delimiter "526930B } =(U^{(f)}\otimes
  \protect \sqrt {\protect \mathaccentV {hat}05E{\rho }_i}) \mathinner
  {|{\Gamma }\delimiter "526930B }$ where $\mathinner {|{\Gamma }\delimiter
  "526930B }=\DOTSB \sum@ \slimits@ _k \mathinner {|{k}\delimiter "526930B
  }\otimes \mathinner {|{k}\delimiter "526930B }$ \cite {Wilde:2017} such that
  $\protect \textrm {Tr}_f[\mathinner {|{\psi _i}\delimiter "526930B
  }\mathinner {\delimiter "426830A {\psi _i}|} ]= \protect \mathaccentV
  {hat}05E{\rho }_i$. We can think of this as a fibre bundle where the base
  space is the space of all positive Hermitian operators, not necessarily of
  unit trace (the positive cone), and sitting above each (un-normalized)
  quantum state $\protect \mathaccentV {hat}05E{\rho }_i$ is the vector space
  (fibre, $f$) of its purifications, which as operators can be represented the
  vector $A_i=\protect \sqrt {\protect \mathaccentV {hat}05E{\rho }_i}$. The
  arbitrary unitary $U^{(f)}$ is the freedom to move the vector $A_i$ around in
  the fibre. Now the fidelity is given as $\protect \sqrt {\protect \mathcal
  {F}_Q}(\zeta ) = \protect \textrm {max}|\protect \textrm {Tr}[A_1 A^\dagger
  _2]|$ over all purifications (i.e. over all $U^{(f)}_1 U^{\dagger (f)}_2)$.
  The \protect \textit {Bures angle} $d_B$ is given by $\protect \qopname
  \relax o{cos}\left [d_B(\protect \mathaccentV {hat}05E{\rho }_1,\protect
  \mathaccentV {hat}05E{\rho }_2)\right ] = \protect \sqrt {\protect \mathcal
  {F}_Q}(\protect \mathaccentV {hat}05E{\rho }_1,\protect \mathaccentV
  {hat}05E{\rho }_2)$ is the length of the geodesic curve within the subspace
  of (unit trace) density matrices connecting $\protect \mathaccentV
  {hat}05E{\rho }_1$ and $\protect \mathaccentV {hat}05E{\rho }_2$. The
  infinitesimal version of this is given by the \protect \textit {Bures metric}
  \cite {GeomQStates:2006} $ds^2_B = \protect \textrm {Tr}[dA_1 dA^\dagger _2]
  = \protect \genfrac {}{}{}1{1}{4} \protect \textrm {Tr}[d\protect
  \mathaccentV {hat}05E{\rho }(\zeta ) L_\zeta ] = \protect \genfrac
  {}{}{}1{1}{4} \protect \textrm {Tr}[\protect \mathaccentV {hat}05E{\rho
  }(\zeta ) L^2_\zeta ] = F_Q(\zeta ) d\zeta ^2$, where the last expression is
  just Eq.~\ref {eqn:2.17}. This last expression could also be interpreted as
  the speed $ds_B/d\zeta = \protect \sqrt {F_Q}(\zeta )$ along the geodesic
  connecting the two quantum states $\protect \mathaccentV {hat}05E{\rho }_1$
  and $\protect \mathaccentV {hat}05E{\rho }_2$ (our \protect \textit {input}
  and \protect \textit {output} states along which $\zeta $ varies) is governed
  by the (square root) of the QFI. Note further that for pure states $\protect
  \genfrac {}{}{}1{1}{4}F_Q(\zeta ) =\left [ \mathinner {\delimiter "426830A
  {\partial _\zeta \psi |\partial _\zeta \psi }\delimiter "526930B } -
  |\mathinner {\delimiter "426830A {\psi |\partial _\zeta \psi }\delimiter
  "526930B }|^2\right ] = \protect \textrm {Tr}[\mathinner {|{\partial _\zeta
  \psi }\delimiter "526930B }\mathinner {\delimiter "426830A {\partial _\zeta
  \psi }|}\protect \tmspace +\thinmuskip {.1667em} P_\perp ]$ where $P_\perp =
  I - \mathinner {|{\psi }\delimiter "526930B }\mathinner {\delimiter "426830A
  {\psi }|}$ is the projector onto states perpendicular to $\mathinner {|{\psi
  }\delimiter "526930B }$. Thus $\mathinner {|{\nabla _\zeta \protect \tmspace
  +\thinmuskip {.1667em}\psi }\delimiter "526930B }\equiv P_\perp \protect
  \tmspace +\thinmuskip {.1667em}\mathinner {|{\partial _\zeta \psi }\delimiter
  "526930B }$ $=\mathinner {|{\partial _\zeta \psi }\delimiter "526930B } -
  \mathinner {|{\psi }\delimiter "526930B } \mathinner {\delimiter "426830A
  {\psi |\partial _\zeta \psi }\delimiter "526930B } $ is the intrinsic
  \protect \emph {covariant derivative} \cite {Provost_Vallee:1980} pointing
  across fibres that is horizontal (tangent) to the base (parameter, $\zeta $)
  space, with $\mathinner {\delimiter "426830A {\psi |\partial _\zeta \psi
  }\delimiter "526930B }$ the U(1) connection (in the complex Hermitian line
  bundle) \cite {Ben-Aryeh:2004,GeomQStates:2006,Frankel_3rdEd:2012}. The QFI
  is just the norm of this covariant derivative, $\protect \genfrac
  {}{}{}1{1}{4}F_Q(\zeta ) = ||\mathinner {|{\nabla _\zeta \protect \tmspace
  +\thinmuskip {.1667em}\psi }\delimiter "526930B }||^2 = \mathinner
  {\delimiter "426830A {\nabla _\zeta \protect \tmspace +\thinmuskip
  {.1667em}\psi |\nabla _\zeta \protect \tmspace +\thinmuskip {.1667em}\psi
  }\delimiter "526930B }$. From the discussion after Eq.~\ref {eqn:2.50} with
  the parity operator considered as a unitary evolution operator $\mathinner
  {|{\psi (\phi )}\delimiter "526930B }=\protect \mathaccentV {hat}05E{\Pi
  }_b^{(\phi )}\mathinner {|{\psi (0)}\delimiter "526930B } = e^{-i\phi
  \protect \tmspace +\thinmuskip {.1667em}\protect \mathaccentV
  {hat}05E{n}_b}\mathinner {|{\psi (0)}\delimiter "526930B }$ we obtain
  $\mathinner {|{\nabla _\phi \psi }\delimiter "526930B } = -i(\protect
  \mathaccentV {hat}05E{n}_b - \protect \mathaccentV {bar}016{n}_b)\mathinner
  {|{\psi (\phi )}\delimiter "526930B }$, with $ \protect \mathaccentV
  {bar}016{n}_b = \mathinner {\delimiter "426830A {\psi (\phi )}|} \protect
  \mathaccentV {hat}05E{n}_b\mathinner {|{\psi (\phi )}\delimiter "526930B }$,
  and $F_{\Pi _b}(\phi ) = 4\protect \tmspace +\thinmuskip
  {.1667em}||\mathinner {|{\nabla _\phi \protect \tmspace +\thinmuskip
  {.1667em}\psi }\delimiter "526930B }||^2=4\protect \tmspace +\thinmuskip
  {.1667em}\mathinner {\delimiter "426830A {(\Delta \protect \mathaccentV
  {hat}05E{n}_b)^2}\delimiter "526930B }_\phi $ as before. Note, these
  geometric concepts can be extended to multiparameter estimation \cite
  {Guo:2016}, where $H = H(\zeta _1,\zeta _2,\protect \ldots )$ and the Quantum
  Geometric Tensor $Q_{ij} = \mathinner {\delimiter "426830A {\partial _i \psi
  |P_\perp |\partial _i \psi }\delimiter "526930B }$ $=\mathinner {\delimiter
  "426830A {\partial _i H\protect \tmspace +\thinmuskip {.1667em}\partial _j
  H}\delimiter "526930B } - \mathinner {\delimiter "426830A {\partial _i
  H}\delimiter "526930B }\mathinner {\delimiter "426830A {\partial _j
  H}\delimiter "526930B }$ (where $\partial _i = \partial _{\zeta _i}$) takes
  central role \cite {QFI}, with the unifying properties that $\protect \textrm
  {Re}(Q_{ij}) = \protect \genfrac {}{}{}1{1}{4} \protect \mathcal {F}_{QFI}=
  Cov(H_i,H_j) = \mathinner {\delimiter "426830A {\protect \genfrac
  {}{}{}1{1}{2}\protect \{H_i\protect \tmspace +\thinmuskip
  {.1667em}H_j\protect \}}\delimiter "526930B } - \mathinner {\delimiter
  "426830A {H_i}\delimiter "526930B }\mathinner {\delimiter "426830A
  {H_j}\delimiter "526930B }$ are the elements of the QFI Matrix \cite {QFI},
  and $\protect \textrm {Im}(Q_{ij})=-\protect \genfrac {}{}{}1{1}{2} \Omega
  _{ij} = -\protect \genfrac {}{}{}1{1}{4} \left (\mathinner {\delimiter
  "426830A {\partial _i \psi | \partial _j \psi }\delimiter "526930B } -
  \mathinner {\delimiter "426830A { \partial _j \psi | \partial _i \psi
  }\delimiter "526930B }\right )$ are the Berry (Phase) Curvatures \cite
  {Samuel_Bhandari:1988,BerryBook,Ben-Aryeh:2004}.}\BibitemShut {Stop}%
\bibitem [{\citenamefont {Drummond}\ and\ \citenamefont {Hillery}(2009)}]{SLD}%
  \BibitemOpen
  \bibfield  {author} {\bibinfo {author} {\bibfnamefont {P.~D.}\ \bibnamefont
  {Drummond}}\ and\ \bibinfo {author} {\bibfnamefont {M.}~\bibnamefont
  {Hillery}},\ }\bibfield  {title} {\enquote {\bibinfo {title} {The quantum
  theory of nonlinear optics},}\ }\href@noop {} {\bibfield  {journal} {\bibinfo
   {journal} {Acta Physica Slovaca}\ }\textbf {\bibinfo {volume} {59}}
  (\bibinfo {year} {2009})}\BibitemShut {NoStop}%
\bibitem [{\citenamefont {Helstrom}(1967)}]{Helstrom}%
  \BibitemOpen
  \bibfield  {author} {\bibinfo {author} {\bibfnamefont {C.~W.}\ \bibnamefont
  {Helstrom}},\ }\bibfield  {title} {\enquote {\bibinfo {title} {Minimum
  mean-squared error of estimates in quantum statistics},}\ }\href@noop {}
  {\bibfield  {journal} {\bibinfo  {journal} {Phys.\ Lett. A}\ }\textbf
  {\bibinfo {volume} {25}},\ \bibinfo {pages} {101--102} (\bibinfo {year}
  {1967})}\BibitemShut {NoStop}%
\bibitem [{\citenamefont {Hubner}(1992)}]{Hubner}%
  \BibitemOpen
  \bibfield  {author} {\bibinfo {author} {\bibfnamefont {M.}~\bibnamefont
  {Hubner}},\ }\bibfield  {title} {\enquote {\bibinfo {title} {Explicit
  computation of the {B}ures distance for density matricies},}\ }\href@noop {}
  {\bibfield  {journal} {\bibinfo  {journal} {Phys.\ Lett. A}\ }\textbf
  {\bibinfo {volume} {163}},\ \bibinfo {pages} {239--242} (\bibinfo {year}
  {1992})}\BibitemShut {NoStop}%
\bibitem [{\citenamefont {Gerry}(2000)}]{GerryParityAgain}%
  \BibitemOpen
  \bibfield  {author} {\bibinfo {author} {\bibfnamefont {C.~C.}\ \bibnamefont
  {Gerry}},\ }\bibfield  {title} {\enquote {\bibinfo {title} {Heisenberg-limit
  interferometry with four-wave mixers operating in a nonlinear regime},}\
  }\href@noop {} {\bibfield  {journal} {\bibinfo  {journal} {Phys.\ Rev. A}\
  }\textbf {\bibinfo {volume} {61}},\ \bibinfo {pages} {043811} (\bibinfo
  {year} {2000})}\BibitemShut {NoStop}%
\bibitem [{\citenamefont {Gerry}\ and\ \citenamefont
  {Mimih}(2010{\natexlab{a}})}]{GerryParity2}%
  \BibitemOpen
  \bibfield  {author} {\bibinfo {author} {\bibfnamefont {C.}~\bibnamefont
  {Gerry}}\ and\ \bibinfo {author} {\bibfnamefont {J.}~\bibnamefont {Mimih}},\
  }\bibfield  {title} {\enquote {\bibinfo {title} {The parity operator in
  quantum optical metrology},}\ }\href@noop {} {\bibfield  {journal} {\bibinfo
  {journal} {Contemp. Phys.}\ }\textbf {\bibinfo {volume} {51}},\ \bibinfo
  {pages} {6} (\bibinfo {year} {2010}{\natexlab{a}})}\BibitemShut {NoStop}%
\bibitem [{\citenamefont {Hofmann}(2009)}]{PathSym1}%
  \BibitemOpen
  \bibfield  {author} {\bibinfo {author} {\bibfnamefont {H.~F.}\ \bibnamefont
  {Hofmann}},\ }\bibfield  {title} {\enquote {\bibinfo {title} {All
  path-symmetric pure states achieve their maximal phase sensitivity in
  conventional two-path interferometry},}\ }\href@noop {} {\bibfield  {journal}
  {\bibinfo  {journal} {Phys.\ Lett. A}\ }\textbf {\bibinfo {volume} {79}},\
  \bibinfo {pages} {033822} (\bibinfo {year} {2009})}\BibitemShut {NoStop}%
\bibitem [{\citenamefont {Kim}\ \emph {et~al.}(2012)\citenamefont {Kim},
  \citenamefont {Seshadreesan}, \citenamefont {Dowling},\ and\ \citenamefont
  {Lee}}]{PathSym2}%
  \BibitemOpen
  \bibfield  {author} {\bibinfo {author} {\bibfnamefont {S.}~\bibnamefont
  {Kim}}, \bibinfo {author} {\bibfnamefont {K.~P.}\ \bibnamefont
  {Seshadreesan}}, \bibinfo {author} {\bibfnamefont {J.~P.}\ \bibnamefont
  {Dowling}}, \ and\ \bibinfo {author} {\bibfnamefont {H.}~\bibnamefont
  {Lee}},\ }\bibfield  {title} {\enquote {\bibinfo {title} {Parity measurement
  is sufficient for phase estimation at the quantum cram\'{e}r-rao bound for
  path symmetric states},}\ }\href@noop {} {\bibfield  {journal} {\bibinfo
  {journal} {ArXiv:}\ }\textbf {\bibinfo {volume} {1207.049}} (\bibinfo {year}
  {2012})}\BibitemShut {NoStop}%
\bibitem [{\citenamefont {Seshadreesan}\ \emph {et~al.}(2013)\citenamefont
  {Seshadreesan}, \citenamefont {Kim}, \citenamefont {Dowling},\ and\
  \citenamefont {Lee}}]{PathSym}%
  \BibitemOpen
  \bibfield  {author} {\bibinfo {author} {\bibfnamefont {K.~P.}\ \bibnamefont
  {Seshadreesan}}, \bibinfo {author} {\bibfnamefont {S.}~\bibnamefont {Kim}},
  \bibinfo {author} {\bibfnamefont {J.~P.}\ \bibnamefont {Dowling}}, \ and\
  \bibinfo {author} {\bibfnamefont {H.}~\bibnamefont {Lee}},\ }\bibfield
  {title} {\enquote {\bibinfo {title} {Phase estimation and the quantum
  {C}ram\'{e}r-{R}ao bound via parity detection},}\ }\href@noop {} {\bibfield
  {journal} {\bibinfo  {journal} {Phys.\ Lett. A}\ }\textbf {\bibinfo {volume}
  {87}},\ \bibinfo {pages} {043833} (\bibinfo {year} {2013})}\BibitemShut
  {NoStop}%
\bibitem [{\citenamefont {Uys}\ and\ \citenamefont {Meystre}(2007)}]{Uys}%
  \BibitemOpen
  \bibfield  {author} {\bibinfo {author} {\bibfnamefont {H.}~\bibnamefont
  {Uys}}\ and\ \bibinfo {author} {\bibfnamefont {P.}~\bibnamefont {Meystre}},\
  }\bibfield  {title} {\enquote {\bibinfo {title} {Quantum states for
  {H}eisenberg-limited interferometry},}\ }\href@noop {} {\bibfield  {journal}
  {\bibinfo  {journal} {Phys.\ Lett. A}\ }\textbf {\bibinfo {volume} {76}},\
  \bibinfo {pages} {013804} (\bibinfo {year} {2007})}\BibitemShut {NoStop}%
\bibitem [{Note4()}]{Note4}%
  \BibitemOpen
  \bibinfo {note} {The exception being the optical $N00N$ state, for which
  parity is optimal for all values of the phase.}\BibitemShut {Stop}%
\bibitem [{\citenamefont {Barndorff-Nielsen}\ and\ \citenamefont
  {Gill}(2000)}]{QFI3}%
  \BibitemOpen
  \bibfield  {author} {\bibinfo {author} {\bibfnamefont {O.~E.}\ \bibnamefont
  {Barndorff-Nielsen}}\ and\ \bibinfo {author} {\bibfnamefont {R.~D.}\
  \bibnamefont {Gill}},\ }\bibfield  {title} {\enquote {\bibinfo {title}
  {Fisher information in quantum statistics},}\ }\href@noop {} {\bibfield
  {journal} {\bibinfo  {journal} {J. Phys. A}\ }\textbf {\bibinfo {volume}
  {33}},\ \bibinfo {pages} {24} (\bibinfo {year} {2000})}\BibitemShut {NoStop}%
\bibitem [{\citenamefont {Anisimov}\ \emph {et~al.}(2010)\citenamefont
  {Anisimov}, \citenamefont {Raterman}, \citenamefont {Chiruvelli},
  \citenamefont {Plick},\ and\ \citenamefont {Huver}}]{AnisimovTMSVS}%
  \BibitemOpen
  \bibfield  {author} {\bibinfo {author} {\bibfnamefont {P.~M.}\ \bibnamefont
  {Anisimov}}, \bibinfo {author} {\bibfnamefont {G.~M.}\ \bibnamefont
  {Raterman}}, \bibinfo {author} {\bibfnamefont {A.}~\bibnamefont
  {Chiruvelli}}, \bibinfo {author} {\bibfnamefont {W.~N.}\ \bibnamefont
  {Plick}}, \ and\ \bibinfo {author} {\bibfnamefont {S.~D.}\ \bibnamefont
  {Huver}},\ }\bibfield  {title} {\enquote {\bibinfo {title} {Quantum metrology
  with two-mode squeezed vacuum: parity detection beats the {H}eisenberg
  limit},}\ }\href@noop {} {\bibfield  {journal} {\bibinfo  {journal} {Phys.\
  Rev. Lett.}\ }\textbf {\bibinfo {volume} {104}},\ \bibinfo {pages} {103602}
  (\bibinfo {year} {2010})}\BibitemShut {NoStop}%
\bibitem [{Note5()}]{Note5}%
  \BibitemOpen
  \bibinfo {note} {Generally speaking, the expectation value of the parity
  operator for each output mode is related by a phase shift. Often the output
  mode for which the parity expectation value peaks at $\phi =0$ is considered.
  However, either output mode is suitable with the right choice of
  phase-shifter.}\BibitemShut {Stop}%
\bibitem [{\citenamefont {Dowling}(1998)}]{Dowling1}%
  \BibitemOpen
  \bibfield  {author} {\bibinfo {author} {\bibfnamefont {J.~P.}\ \bibnamefont
  {Dowling}},\ }\bibfield  {title} {\enquote {\bibinfo {title} {Correlated
  input-port matter-wave interferometer: quantum-noise limits to the atom-laser
  gyroscope},}\ }\href@noop {} {\bibfield  {journal} {\bibinfo  {journal}
  {Phys.\ Rev. A}\ }\textbf {\bibinfo {volume} {57}},\ \bibinfo {pages} {4736}
  (\bibinfo {year} {1998})}\BibitemShut {NoStop}%
\bibitem [{\citenamefont {Lee}, \citenamefont {Kok},\ and\ \citenamefont
  {Dowling}(2002)}]{Dowling2}%
  \BibitemOpen
  \bibfield  {author} {\bibinfo {author} {\bibfnamefont {H.}~\bibnamefont
  {Lee}}, \bibinfo {author} {\bibfnamefont {P.}~\bibnamefont {Kok}}, \ and\
  \bibinfo {author} {\bibfnamefont {J.~P.}\ \bibnamefont {Dowling}},\
  }\bibfield  {title} {\enquote {\bibinfo {title} {A quantum {R}osetta stone
  for interferometry},}\ }\href@noop {} {\bibfield  {journal} {\bibinfo
  {journal} {J. Mod. Opt.}\ }\textbf {\bibinfo {volume} {49}},\ \bibinfo
  {pages} {2325} (\bibinfo {year} {2002})}\BibitemShut {NoStop}%
\bibitem [{\citenamefont {Kok}, \citenamefont {Braunstein},\ and\ \citenamefont
  {Dowling}(2004)}]{Dowling3}%
  \BibitemOpen
  \bibfield  {author} {\bibinfo {author} {\bibfnamefont {P.}~\bibnamefont
  {Kok}}, \bibinfo {author} {\bibfnamefont {S.~L.}\ \bibnamefont {Braunstein}},
  \ and\ \bibinfo {author} {\bibfnamefont {J.~P.}\ \bibnamefont {Dowling}},\
  }\bibfield  {title} {\enquote {\bibinfo {title} {Quantum lithography,
  entanglement and {H}eisenberg-limited parameter estimation},}\ }\href@noop {}
  {\bibfield  {journal} {\bibinfo  {journal} {J. Opt. B}\ }\textbf {\bibinfo
  {volume} {6}},\ \bibinfo {pages} {8} (\bibinfo {year} {2004})}\BibitemShut
  {NoStop}%
\bibitem [{\citenamefont {Joo}, \citenamefont {Munro},\ and\ \citenamefont
  {Spiller}(2011)}]{ECS3}%
  \BibitemOpen
  \bibfield  {author} {\bibinfo {author} {\bibfnamefont {J.}~\bibnamefont
  {Joo}}, \bibinfo {author} {\bibfnamefont {W.~J.}\ \bibnamefont {Munro}}, \
  and\ \bibinfo {author} {\bibfnamefont {T.~P.}\ \bibnamefont {Spiller}},\
  }\bibfield  {title} {\enquote {\bibinfo {title} {Quantum metrology with
  entangled coherent states},}\ }\href@noop {} {\bibfield  {journal} {\bibinfo
  {journal} {Phys.\ Rev. Lett.}\ }\textbf {\bibinfo {volume} {107}},\ \bibinfo
  {pages} {083601} (\bibinfo {year} {2011})}\BibitemShut {NoStop}%
\bibitem [{\citenamefont {Israel}\ \emph {et~al.}(2019)\citenamefont {Israel},
  \citenamefont {Cohen}, \citenamefont {Song}, \citenamefont {Joo},\ and\
  \citenamefont {Eisenberg}}]{Israel}%
  \BibitemOpen
  \bibfield  {author} {\bibinfo {author} {\bibfnamefont {Y.}~\bibnamefont
  {Israel}}, \bibinfo {author} {\bibfnamefont {L.}~\bibnamefont {Cohen}},
  \bibinfo {author} {\bibfnamefont {X.}~\bibnamefont {Song}}, \bibinfo {author}
  {\bibfnamefont {J.}~\bibnamefont {Joo}}, \ and\ \bibinfo {author}
  {\bibfnamefont {H.~S.}\ \bibnamefont {Eisenberg}},\ }\bibfield  {title}
  {\enquote {\bibinfo {title} {Entangled coherent states created by mixing
  squeezed vacuum and coherent light},}\ }\href@noop {} {\bibfield  {journal}
  {\bibinfo  {journal} {Optica}\ }\textbf {\bibinfo {volume} {6}},\ \bibinfo
  {pages} {6} (\bibinfo {year} {2019})}\BibitemShut {NoStop}%
\bibitem [{\citenamefont {Yurke}\ and\ \citenamefont {Stoler}(1986)}]{Yurke2}%
  \BibitemOpen
  \bibfield  {author} {\bibinfo {author} {\bibfnamefont {B.}~\bibnamefont
  {Yurke}}\ and\ \bibinfo {author} {\bibfnamefont {D.}~\bibnamefont {Stoler}},\
  }\bibfield  {title} {\enquote {\bibinfo {title} {Generating quantum
  mechanical superpositions of macroscopically distinguishable states via
  amplitude dispersion},}\ }\href@noop {} {\bibfield  {journal} {\bibinfo
  {journal} {Phys.\ Rev. Lett.}\ }\textbf {\bibinfo {volume} {57}},\ \bibinfo
  {pages} {13} (\bibinfo {year} {1986})}\BibitemShut {NoStop}%
\bibitem [{\citenamefont {Luis}(2001)}]{ECS1}%
  \BibitemOpen
  \bibfield  {author} {\bibinfo {author} {\bibfnamefont {A.}~\bibnamefont
  {Luis}},\ }\bibfield  {title} {\enquote {\bibinfo {title} {Equivalence
  between macroscopic quantum superpositions and maxmially entangled states:
  application to phase-shift detection},}\ }\href@noop {} {\bibfield  {journal}
  {\bibinfo  {journal} {Phys.\ Rev. A}\ }\textbf {\bibinfo {volume} {64}},\
  \bibinfo {pages} {054102} (\bibinfo {year} {2001})}\BibitemShut {NoStop}%
\bibitem [{\citenamefont {Gerry}, \citenamefont {Benmoussa},\ and\
  \citenamefont {Campos}(2002)}]{ECS}%
  \BibitemOpen
  \bibfield  {author} {\bibinfo {author} {\bibfnamefont {C.~C.}\ \bibnamefont
  {Gerry}}, \bibinfo {author} {\bibfnamefont {A.}~\bibnamefont {Benmoussa}}, \
  and\ \bibinfo {author} {\bibfnamefont {R.~A.}\ \bibnamefont {Campos}},\
  }\bibfield  {title} {\enquote {\bibinfo {title} {Nonlinear interferometer as
  a resource for maximally entangled photonic states: application to
  interferometry},}\ }\href@noop {} {\bibfield  {journal} {\bibinfo  {journal}
  {Phys.\ Rev. A}\ }\textbf {\bibinfo {volume} {66}},\ \bibinfo {pages}
  {013804} (\bibinfo {year} {2002})}\BibitemShut {NoStop}%
\bibitem [{\citenamefont {Holland}\ and\ \citenamefont
  {Burnett}(1993)}]{TwinFock}%
  \BibitemOpen
  \bibfield  {author} {\bibinfo {author} {\bibfnamefont {M.~J.}\ \bibnamefont
  {Holland}}\ and\ \bibinfo {author} {\bibfnamefont {K.}~\bibnamefont
  {Burnett}},\ }\bibfield  {title} {\enquote {\bibinfo {title} {Interferometric
  detection of optical phase shifts at the {H}eisenberg limit},}\ }\href@noop
  {} {\bibfield  {journal} {\bibinfo  {journal} {Phys.\ Rev. Lett.}\ }\textbf
  {\bibinfo {volume} {71}},\ \bibinfo {pages} {1355} (\bibinfo {year}
  {1993})}\BibitemShut {NoStop}%
\bibitem [{Note6()}]{Note6}%
  \BibitemOpen
  \bibinfo {note} {Their discussion is in the context of spectroscopy using
  maximally entangled states of a system of $N$ two-level trapped
  ions.}\BibitemShut {Stop}%
\bibitem [{\citenamefont {Gerry}\ and\ \citenamefont {Campos}(2001)}]{Campos1}%
  \BibitemOpen
  \bibfield  {author} {\bibinfo {author} {\bibfnamefont {C.~C.}\ \bibnamefont
  {Gerry}}\ and\ \bibinfo {author} {\bibfnamefont {R.~A.}\ \bibnamefont
  {Campos}},\ }\bibfield  {title} {\enquote {\bibinfo {title} {Generation of
  maximally entangled photonic states with a quantum-optical {F}redkin gate},}\
  }\href@noop {} {\bibfield  {journal} {\bibinfo  {journal} {Phys.\ Rev. A}\
  }\textbf {\bibinfo {volume} {64}},\ \bibinfo {pages} {063814} (\bibinfo
  {year} {2001})}\BibitemShut {NoStop}%
\bibitem [{\citenamefont {Gerry}\ and\ \citenamefont
  {Benmoussa}(2001)}]{Benmoussa}%
  \BibitemOpen
  \bibfield  {author} {\bibinfo {author} {\bibfnamefont {C.~C.}\ \bibnamefont
  {Gerry}}\ and\ \bibinfo {author} {\bibfnamefont {A.}~\bibnamefont
  {Benmoussa}},\ }\bibfield  {title} {\enquote {\bibinfo {title}
  {{H}eisenberg-limited interferometry and photolithograpy with nonlinear
  four-wave mixing},}\ }\href@noop {} {\bibfield  {journal} {\bibinfo
  {journal} {Phys.\ Rev. A}\ }\textbf {\bibinfo {volume} {65}},\ \bibinfo
  {pages} {033822} (\bibinfo {year} {2001})}\BibitemShut {NoStop}%
\bibitem [{\citenamefont {Kok}, \citenamefont {Lee},\ and\ \citenamefont
  {Dowling}(2002)}]{Kok}%
  \BibitemOpen
  \bibfield  {author} {\bibinfo {author} {\bibfnamefont {P.}~\bibnamefont
  {Kok}}, \bibinfo {author} {\bibfnamefont {H.}~\bibnamefont {Lee}}, \ and\
  \bibinfo {author} {\bibfnamefont {J.~P.}\ \bibnamefont {Dowling}},\
  }\bibfield  {title} {\enquote {\bibinfo {title} {Creation of
  large-photon-number path entanglement conditioned on photodetection},}\
  }\href@noop {} {\bibfield  {journal} {\bibinfo  {journal} {Phys.\ Rev. A}\
  }\textbf {\bibinfo {volume} {65}},\ \bibinfo {pages} {052104} (\bibinfo
  {year} {2002})}\BibitemShut {NoStop}%
\bibitem [{\citenamefont {Lee}, \citenamefont {Kok},\ and\ \citenamefont {amd
  J.~P.~Dowling}(2002)}]{Kok2}%
  \BibitemOpen
  \bibfield  {author} {\bibinfo {author} {\bibfnamefont {H.}~\bibnamefont
  {Lee}}, \bibinfo {author} {\bibfnamefont {P.}~\bibnamefont {Kok}}, \ and\
  \bibinfo {author} {\bibfnamefont {N.~J.~C.}\ \bibnamefont {amd
  J.~P.~Dowling}},\ }\bibfield  {title} {\enquote {\bibinfo {title} {Linear
  optics and projective measurements alone suffice to create
  large-photon-number path entanglement},}\ }\href@noop {} {\bibfield
  {journal} {\bibinfo  {journal} {Phys.\ Rev. A}\ }\textbf {\bibinfo {volume}
  {65}},\ \bibinfo {pages} {030101} (\bibinfo {year} {2002})}\BibitemShut
  {NoStop}%
\bibitem [{\citenamefont {Fiur\'{a}\u{s}ek}(2002)}]{Fiurasek}%
  \BibitemOpen
  \bibfield  {author} {\bibinfo {author} {\bibfnamefont {J.}~\bibnamefont
  {Fiur\'{a}\u{s}ek}},\ }\bibfield  {title} {\enquote {\bibinfo {title}
  {Conditional generation of {N}-photon entangled states of light},}\
  }\href@noop {} {\bibfield  {journal} {\bibinfo  {journal} {Phys.\ Rev. A}\
  }\textbf {\bibinfo {volume} {65}},\ \bibinfo {pages} {053818} (\bibinfo
  {year} {2002})}\BibitemShut {NoStop}%
\bibitem [{\citenamefont {Gerry}\ and\ \citenamefont
  {Benmoussa}(2002)}]{Benmoussa2}%
  \BibitemOpen
  \bibfield  {author} {\bibinfo {author} {\bibfnamefont {C.~C.}\ \bibnamefont
  {Gerry}}\ and\ \bibinfo {author} {\bibfnamefont {A.}~\bibnamefont
  {Benmoussa}},\ }\bibfield  {title} {\enquote {\bibinfo {title} {Nonlinear
  interferometer as a resource for maximally entangled photonic states:
  application to interferometry},}\ }\href@noop {} {\bibfield  {journal}
  {\bibinfo  {journal} {Phys.\ Rev. A}\ }\textbf {\bibinfo {volume} {66}},\
  \bibinfo {pages} {013804} (\bibinfo {year} {2002})}\BibitemShut {NoStop}%
\bibitem [{\citenamefont {Campos}, \citenamefont {Gerry},\ and\ \citenamefont
  {Benmoussa}(2003)}]{Campos}%
  \BibitemOpen
  \bibfield  {author} {\bibinfo {author} {\bibfnamefont {R.~A.}\ \bibnamefont
  {Campos}}, \bibinfo {author} {\bibfnamefont {C.~C.}\ \bibnamefont {Gerry}}, \
  and\ \bibinfo {author} {\bibfnamefont {A.}~\bibnamefont {Benmoussa}},\
  }\bibfield  {title} {\enquote {\bibinfo {title} {Optical interferometry at
  the {H}eisenberg limit with twin {F}ock states and parity measurements},}\
  }\href@noop {} {\bibfield  {journal} {\bibinfo  {journal} {Phys.\ Rev. A}\
  }\textbf {\bibinfo {volume} {68}},\ \bibinfo {pages} {023810} (\bibinfo
  {year} {2003})}\BibitemShut {NoStop}%
\bibitem [{\citenamefont {Hong}, \citenamefont {Ou},\ and\ \citenamefont
  {Mandel}(1987)}]{HOM}%
  \BibitemOpen
  \bibfield  {author} {\bibinfo {author} {\bibfnamefont {C.~K.}\ \bibnamefont
  {Hong}}, \bibinfo {author} {\bibfnamefont {Z.~Y.}\ \bibnamefont {Ou}}, \ and\
  \bibinfo {author} {\bibfnamefont {L.}~\bibnamefont {Mandel}},\ }\bibfield
  {title} {\enquote {\bibinfo {title} {Measurement of sub-picosecond time
  interverals between two photons by interference},}\ }\href@noop {} {\bibfield
   {journal} {\bibinfo  {journal} {Phys.\ Rev. Lett.}\ }\textbf {\bibinfo
  {volume} {59}},\ \bibinfo {pages} {2044} (\bibinfo {year}
  {1987})}\BibitemShut {NoStop}%
\bibitem [{\citenamefont {Feller}(1968)}]{MathBook}%
  \BibitemOpen
  \bibfield  {author} {\bibinfo {author} {\bibfnamefont {W.}~\bibnamefont
  {Feller}},\ }\href@noop {} {\emph {\bibinfo {title} {An Introduction to
  Probability Theory and its Applications, {3rd Edition}}}}\ (\bibinfo
  {publisher} {Wiley},\ \bibinfo {year} {1968})\BibitemShut {NoStop}%
\bibitem [{\citenamefont {Boyd}(1992)}]{BoydBook}%
  \BibitemOpen
  \bibfield  {author} {\bibinfo {author} {\bibfnamefont {R.~W.}\ \bibnamefont
  {Boyd}},\ }\href@noop {} {\emph {\bibinfo {title} {Nonlinear Optics}}}\
  (\bibinfo  {publisher} {Academic Press},\ \bibinfo {year} {1992})\BibitemShut
  {NoStop}%
\bibitem [{\citenamefont {Gerry}\ and\ \citenamefont
  {Mimih}(2010{\natexlab{b}})}]{GerryPCS}%
  \BibitemOpen
  \bibfield  {author} {\bibinfo {author} {\bibfnamefont {C.~C.}\ \bibnamefont
  {Gerry}}\ and\ \bibinfo {author} {\bibfnamefont {J.}~\bibnamefont {Mimih}},\
  }\bibfield  {title} {\enquote {\bibinfo {title} {Heisenberg-limited
  interferometry with pair coherent states and parity measurements},}\
  }\href@noop {} {\bibfield  {journal} {\bibinfo  {journal} {Phys.\ Rev. A}\
  }\textbf {\bibinfo {volume} {82}},\ \bibinfo {pages} {013831} (\bibinfo
  {year} {2010}{\natexlab{b}})}\BibitemShut {NoStop}%
\bibitem [{\citenamefont {Birrittella}, \citenamefont {Gura},\ and\
  \citenamefont {Gerry}(2015)}]{TMSCS}%
  \BibitemOpen
  \bibfield  {author} {\bibinfo {author} {\bibfnamefont {R.}~\bibnamefont
  {Birrittella}}, \bibinfo {author} {\bibfnamefont {A.}~\bibnamefont {Gura}}, \
  and\ \bibinfo {author} {\bibfnamefont {C.~C.}\ \bibnamefont {Gerry}},\
  }\bibfield  {title} {\enquote {\bibinfo {title} {Coherently stimulated
  parametric down-conversion, phase effects, and quantum optical
  interferometry},}\ }\href@noop {} {\bibfield  {journal} {\bibinfo  {journal}
  {Phys.\ Rev. A}\ }\textbf {\bibinfo {volume} {91}},\ \bibinfo {pages}
  {053801} (\bibinfo {year} {2015})}\BibitemShut {NoStop}%
\bibitem [{\citenamefont {Birrittella}, \citenamefont {Alsing},\ and\
  \citenamefont {Gerry}(2019)}]{Birrittella3}%
  \BibitemOpen
  \bibfield  {author} {\bibinfo {author} {\bibfnamefont {R.}~\bibnamefont
  {Birrittella}}, \bibinfo {author} {\bibfnamefont {P.~M.}\ \bibnamefont
  {Alsing}}, \ and\ \bibinfo {author} {\bibfnamefont {C.~C.}\ \bibnamefont
  {Gerry}},\ }\bibfield  {title} {\enquote {\bibinfo {title} {Phase effects in
  coherently stimulated down conversion with a quantized pump field},}\
  }\href@noop {} {\bibfield  {journal} {\bibinfo  {journal} {Phys.\ Rev. A}\
  }\textbf {\bibinfo {volume} {101}},\ \bibinfo {pages} {013813} (\bibinfo
  {year} {2019})}\BibitemShut {NoStop}%
\bibitem [{\citenamefont {Agarwal}(1986)}]{PCS}%
  \BibitemOpen
  \bibfield  {author} {\bibinfo {author} {\bibfnamefont {G.~S.}\ \bibnamefont
  {Agarwal}},\ }\bibfield  {title} {\enquote {\bibinfo {title} {Generation of
  pair coherent states and squeezing via the competition of four-wave mixing
  and amplified spontaneous emission},}\ }\href@noop {} {\bibfield  {journal}
  {\bibinfo  {journal} {Phys.\ Rev. Lett.}\ }\textbf {\bibinfo {volume} {57}},\
  \bibinfo {pages} {827} (\bibinfo {year} {1986})}\BibitemShut {NoStop}%
\bibitem [{Note7()}]{Note7}%
  \BibitemOpen
  \bibinfo {note} {The generalized pair coherent state is defined such that
  $\left (\protect \mathaccentV {hat}05E{a}^{\dagger }\protect \mathaccentV
  {hat}05E{a}-\protect \mathaccentV {hat}05E{b}^{\dagger }\protect \mathaccentV
  {hat}05E{b}\right )\mathinner {|{\zeta ,q}\delimiter "526930B }=q\mathinner
  {|{\zeta ,q}\delimiter "526930B }$. For the purposes of this review article,
  we assume $q=0$ without loss of generality.}\BibitemShut {Stop}%
\bibitem [{\citenamefont {Gerry}, \citenamefont {Mimih},\ and\ \citenamefont
  {Birrittella}(2011)}]{PCSgen}%
  \BibitemOpen
  \bibfield  {author} {\bibinfo {author} {\bibfnamefont {C.~C.}\ \bibnamefont
  {Gerry}}, \bibinfo {author} {\bibfnamefont {J.}~\bibnamefont {Mimih}}, \ and\
  \bibinfo {author} {\bibfnamefont {R.}~\bibnamefont {Birrittella}},\
  }\bibfield  {title} {\enquote {\bibinfo {title} {State-projective scheme for
  generating pair coherent states in traveling-wave optical fields},}\
  }\href@noop {} {\bibfield  {journal} {\bibinfo  {journal} {Phys.\ Rev. A}\
  }\textbf {\bibinfo {volume} {84}},\ \bibinfo {pages} {023810} (\bibinfo
  {year} {2011})}\BibitemShut {NoStop}%
\bibitem [{\citenamefont {Caves}(1981)}]{Caves}%
  \BibitemOpen
  \bibfield  {author} {\bibinfo {author} {\bibfnamefont {C.~M.}\ \bibnamefont
  {Caves}},\ }\bibfield  {title} {\enquote {\bibinfo {title}
  {Quantum-mechanical noise in an interferometer},}\ }\href@noop {} {\bibfield
  {journal} {\bibinfo  {journal} {Phys.\ Rev. D}\ }\textbf {\bibinfo {volume}
  {23}},\ \bibinfo {pages} {9} (\bibinfo {year} {1981})}\BibitemShut {NoStop}%
\bibitem [{\citenamefont {Buz\u{e}k}\ and\ \citenamefont
  {Hillery}(1995)}]{Buzek}%
  \BibitemOpen
  \bibfield  {author} {\bibinfo {author} {\bibfnamefont {V.}~\bibnamefont
  {Buz\u{e}k}}\ and\ \bibinfo {author} {\bibfnamefont {M.}~\bibnamefont
  {Hillery}},\ }\bibfield  {title} {\enquote {\bibinfo {title} {Quantum
  disentanglement and phase measurement},}\ }\href@noop {} {\bibfield
  {journal} {\bibinfo  {journal} {Czech. J. Phys. B}\ }\textbf {\bibinfo
  {volume} {45}},\ \bibinfo {pages} {711--726} (\bibinfo {year}
  {1995})}\BibitemShut {NoStop}%
\bibitem [{\citenamefont {Pezz\'{e}}\ and\ \citenamefont
  {Smerzi}(2008)}]{Pezze2}%
  \BibitemOpen
  \bibfield  {author} {\bibinfo {author} {\bibfnamefont {L.}~\bibnamefont
  {Pezz\'{e}}}\ and\ \bibinfo {author} {\bibfnamefont {A.}~\bibnamefont
  {Smerzi}},\ }\bibfield  {title} {\enquote {\bibinfo {title} {Mach-{Z}ehnder
  interferometry at the {H}eisenberg limit with coherent and squeezed-vacuum
  light},}\ }\href@noop {} {\bibfield  {journal} {\bibinfo  {journal} {Phys.\
  Rev. Lett.}\ }\textbf {\bibinfo {volume} {100}},\ \bibinfo {pages} {073601}
  (\bibinfo {year} {2008})}\BibitemShut {NoStop}%
\bibitem [{\citenamefont {Birrittella}, \citenamefont {Mimih},\ and\
  \citenamefont {Gerry}(2012)}]{Birrittella1}%
  \BibitemOpen
  \bibfield  {author} {\bibinfo {author} {\bibfnamefont {R.}~\bibnamefont
  {Birrittella}}, \bibinfo {author} {\bibfnamefont {J.}~\bibnamefont {Mimih}},
  \ and\ \bibinfo {author} {\bibfnamefont {C.~C.}\ \bibnamefont {Gerry}},\
  }\bibfield  {title} {\enquote {\bibinfo {title} {Multiphoton quantum
  interference at a beam splitter and the approach to {H}eisenberg-limited
  interferometry},}\ }\href@noop {} {\bibfield  {journal} {\bibinfo  {journal}
  {Phys.\ Rev. A}\ }\textbf {\bibinfo {volume} {86}},\ \bibinfo {pages}
  {063828} (\bibinfo {year} {2012})}\BibitemShut {NoStop}%
\bibitem [{\citenamefont {Afek}, \citenamefont {Ambar},\ and\ \citenamefont
  {Silberberg}(2010)}]{Afek}%
  \BibitemOpen
  \bibfield  {author} {\bibinfo {author} {\bibfnamefont {I.}~\bibnamefont
  {Afek}}, \bibinfo {author} {\bibfnamefont {O.}~\bibnamefont {Ambar}}, \ and\
  \bibinfo {author} {\bibfnamefont {Y.}~\bibnamefont {Silberberg}},\ }\bibfield
   {title} {\enquote {\bibinfo {title} {High {N00N} states by mixing quantum
  and classical light},}\ }\href@noop {} {\bibfield  {journal} {\bibinfo
  {journal} {Science}\ }\textbf {\bibinfo {volume} {328}},\ \bibinfo {pages}
  {{(5980)} 879--881} (\bibinfo {year} {2010})}\BibitemShut {NoStop}%
\bibitem [{\citenamefont {Seshadreesan}\ \emph {et~al.}(2011)\citenamefont
  {Seshadreesan}, \citenamefont {Anisimov}, \citenamefont {Lee},\ and\
  \citenamefont {Dowling}}]{Seshadreesan}%
  \BibitemOpen
  \bibfield  {author} {\bibinfo {author} {\bibfnamefont {K.~P.}\ \bibnamefont
  {Seshadreesan}}, \bibinfo {author} {\bibfnamefont {P.~M.}\ \bibnamefont
  {Anisimov}}, \bibinfo {author} {\bibfnamefont {H.}~\bibnamefont {Lee}}, \
  and\ \bibinfo {author} {\bibfnamefont {J.~P.}\ \bibnamefont {Dowling}},\
  }\bibfield  {title} {\enquote {\bibinfo {title} {Parity detection achieves
  the {H}eisenberg limit in interferometry with coherent mixed with squeezed
  vacuum light},}\ }\href@noop {} {\bibfield  {journal} {\bibinfo  {journal}
  {New J. Phys.}\ }\textbf {\bibinfo {volume} {13}},\ \bibinfo {pages} {083026}
  (\bibinfo {year} {2011})}\BibitemShut {NoStop}%
\bibitem [{\citenamefont {Birrittella}\ and\ \citenamefont
  {Gerry}(2014)}]{Birrittella2}%
  \BibitemOpen
  \bibfield  {author} {\bibinfo {author} {\bibfnamefont {R.}~\bibnamefont
  {Birrittella}}\ and\ \bibinfo {author} {\bibfnamefont {C.~C.}\ \bibnamefont
  {Gerry}},\ }\bibfield  {title} {\enquote {\bibinfo {title} {Quantum optical
  interferometry via the mixing of coherent and photon-subtracted squeezed
  vacuum states of light},}\ }\href@noop {} {\bibfield  {journal} {\bibinfo
  {journal} {J. Opt. Soc. Am. B}\ }\textbf {\bibinfo {volume} {31}},\ \bibinfo
  {pages} {3} (\bibinfo {year} {2014})}\BibitemShut {NoStop}%
\bibitem [{\citenamefont {Chiruvelli}\ and\ \citenamefont
  {Lee}(2011)}]{Chiruvelli}%
  \BibitemOpen
  \bibfield  {author} {\bibinfo {author} {\bibfnamefont {A.}~\bibnamefont
  {Chiruvelli}}\ and\ \bibinfo {author} {\bibfnamefont {H.}~\bibnamefont
  {Lee}},\ }\bibfield  {title} {\enquote {\bibinfo {title} {Parity measurements
  in quantum optical metrology},}\ }\href@noop {} {\bibfield  {journal}
  {\bibinfo  {journal} {J. Mod. Opt.}\ }\textbf {\bibinfo {volume} {58}},\
  \bibinfo {pages} {945--953} (\bibinfo {year} {2011})}\BibitemShut {NoStop}%
\bibitem [{\citenamefont {Gao}\ \emph {et~al.}(2010)\citenamefont {Gao},
  \citenamefont {Anisimov}, \citenamefont {Wildfeuer}, \citenamefont {Luine},
  \citenamefont {Lee},\ and\ \citenamefont {Dowling}}]{SupRes}%
  \BibitemOpen
  \bibfield  {author} {\bibinfo {author} {\bibfnamefont {Y.}~\bibnamefont
  {Gao}}, \bibinfo {author} {\bibfnamefont {P.}~\bibnamefont {Anisimov}},
  \bibinfo {author} {\bibfnamefont {C.~F.}\ \bibnamefont {Wildfeuer}}, \bibinfo
  {author} {\bibfnamefont {J.}~\bibnamefont {Luine}}, \bibinfo {author}
  {\bibfnamefont {H.}~\bibnamefont {Lee}}, \ and\ \bibinfo {author}
  {\bibfnamefont {J.~P.}\ \bibnamefont {Dowling}},\ }\bibfield  {title}
  {\enquote {\bibinfo {title} {Super-resolution at the shot-noise limit with
  coherent states and photon-number-resolving detectors},}\ }\href@noop {}
  {\bibfield  {journal} {\bibinfo  {journal} {J. Opt. Soc. Am. B}\ }\textbf
  {\bibinfo {volume} {27}},\ \bibinfo {pages} {6} (\bibinfo {year}
  {2010})}\BibitemShut {NoStop}%
\bibitem [{\citenamefont {Gerry}\ and\ \citenamefont
  {Knight}(2005)}]{GerryBook}%
  \BibitemOpen
  \bibfield  {author} {\bibinfo {author} {\bibfnamefont {C.~C.}\ \bibnamefont
  {Gerry}}\ and\ \bibinfo {author} {\bibfnamefont {P.~L.}\ \bibnamefont
  {Knight}},\ }\href@noop {} {\emph {\bibinfo {title} {Introductory Quantum
  Optics}}}\ (\bibinfo  {publisher} {Cambridge University Press, Cambridge,
  UK.},\ \bibinfo {year} {2005})\BibitemShut {NoStop}%
\bibitem [{\citenamefont {Resch}\ \emph {et~al.}(2007)\citenamefont {Resch},
  \citenamefont {Pregnell}, \citenamefont {Prevedel}, \citenamefont
  {Gilchrist}, \citenamefont {Pryde}, \citenamefont {O'Brien},\ and\
  \citenamefont {White}}]{Resch}%
  \BibitemOpen
  \bibfield  {author} {\bibinfo {author} {\bibfnamefont {K.~J.}\ \bibnamefont
  {Resch}}, \bibinfo {author} {\bibfnamefont {K.~L.}\ \bibnamefont {Pregnell}},
  \bibinfo {author} {\bibfnamefont {R.}~\bibnamefont {Prevedel}}, \bibinfo
  {author} {\bibfnamefont {A.}~\bibnamefont {Gilchrist}}, \bibinfo {author}
  {\bibfnamefont {G.~J.}\ \bibnamefont {Pryde}}, \bibinfo {author}
  {\bibfnamefont {J.~L.}\ \bibnamefont {O'Brien}}, \ and\ \bibinfo {author}
  {\bibfnamefont {A.~G.}\ \bibnamefont {White}},\ }\bibfield  {title} {\enquote
  {\bibinfo {title} {Time-reversal and super-resolving phase measurements},}\
  }\href@noop {} {\bibfield  {journal} {\bibinfo  {journal} {Phys.\ Rev.
  Lett.}\ }\textbf {\bibinfo {volume} {98}},\ \bibinfo {pages} {223601}
  (\bibinfo {year} {2007})}\BibitemShut {NoStop}%
\bibitem [{\citenamefont {Li}\ \emph {et~al.}(2016)\citenamefont {Li},
  \citenamefont {Gard}, \citenamefont {Gao}, \citenamefont {Yuan},
  \citenamefont {Zhang}, \citenamefont {Lee},\ and\ \citenamefont
  {Dowling}}]{Li}%
  \BibitemOpen
  \bibfield  {author} {\bibinfo {author} {\bibfnamefont {D.}~\bibnamefont
  {Li}}, \bibinfo {author} {\bibfnamefont {B.~T.}\ \bibnamefont {Gard}},
  \bibinfo {author} {\bibfnamefont {Y.}~\bibnamefont {Gao}}, \bibinfo {author}
  {\bibfnamefont {C.}~\bibnamefont {Yuan}}, \bibinfo {author} {\bibfnamefont
  {W.}~\bibnamefont {Zhang}}, \bibinfo {author} {\bibfnamefont
  {H.}~\bibnamefont {Lee}}, \ and\ \bibinfo {author} {\bibfnamefont {J.~P.}\
  \bibnamefont {Dowling}},\ }\bibfield  {title} {\enquote {\bibinfo {title}
  {Phase sensitivity at the {H}eisenberg limit in an {SU(1,1)} interferometer
  via parity detection},}\ }\href@noop {} {\bibfield  {journal} {\bibinfo
  {journal} {Phys.\ Rev. A}\ }\textbf {\bibinfo {volume} {94}},\ \bibinfo
  {pages} {063840} (\bibinfo {year} {2016})}\BibitemShut {NoStop}%
\bibitem [{\citenamefont {Hume}\ \emph {et~al.}(2013)\citenamefont {Hume},
  \citenamefont {Stroescu}, \citenamefont {Joos}, \citenamefont {Muessel},
  \citenamefont {Strobel},\ and\ \citenamefont {Oberthaler}}]{Hume}%
  \BibitemOpen
  \bibfield  {author} {\bibinfo {author} {\bibfnamefont {D.~B.}\ \bibnamefont
  {Hume}}, \bibinfo {author} {\bibfnamefont {I.}~\bibnamefont {Stroescu}},
  \bibinfo {author} {\bibfnamefont {M.}~\bibnamefont {Joos}}, \bibinfo {author}
  {\bibfnamefont {W.}~\bibnamefont {Muessel}}, \bibinfo {author} {\bibfnamefont
  {H.}~\bibnamefont {Strobel}}, \ and\ \bibinfo {author} {\bibfnamefont
  {M.~K.}\ \bibnamefont {Oberthaler}},\ }\bibfield  {title} {\enquote {\bibinfo
  {title} {Accurate atom counting in mesoscopic ensembles},}\ }\href@noop {}
  {\bibfield  {journal} {\bibinfo  {journal} {Phys.\ Rev. Lett.}\ }\textbf
  {\bibinfo {volume} {111}},\ \bibinfo {pages} {253001} (\bibinfo {year}
  {2013})}\BibitemShut {NoStop}%
\bibitem [{\citenamefont {Gerry}, \citenamefont {Benmoussa},\ and\
  \citenamefont {Campos}(2005)}]{GerryQND}%
  \BibitemOpen
  \bibfield  {author} {\bibinfo {author} {\bibfnamefont {C.~C.}\ \bibnamefont
  {Gerry}}, \bibinfo {author} {\bibfnamefont {A.}~\bibnamefont {Benmoussa}}, \
  and\ \bibinfo {author} {\bibfnamefont {R.~A.}\ \bibnamefont {Campos}},\
  }\bibfield  {title} {\enquote {\bibinfo {title} {Quantum nondemolition
  measurement of parity and generation of parity eigenstates in optical
  fields},}\ }\href@noop {} {\bibfield  {journal} {\bibinfo  {journal} {Phys.\
  Rev. A}\ }\textbf {\bibinfo {volume} {72}},\ \bibinfo {pages} {053818}
  (\bibinfo {year} {2005})}\BibitemShut {NoStop}%
\bibitem [{\citenamefont {Sun}\ \emph {et~al.}(2014)\citenamefont {Sun},
  \citenamefont {Petrenko}, \citenamefont {Leghtas}, \citenamefont {Vlastakis},
  \citenamefont {Kirchmair}, \citenamefont {Sliwa}, \citenamefont {Narla},
  \citenamefont {Hatridge}, \citenamefont {Shankar}, \citenamefont {Blumoff},
  \citenamefont {Frunzio}, \citenamefont {Mirrahimi}, \citenamefont {Devoret},\
  and\ \citenamefont {Schoelkopf}}]{PhotonJump}%
  \BibitemOpen
  \bibfield  {author} {\bibinfo {author} {\bibfnamefont {L.}~\bibnamefont
  {Sun}}, \bibinfo {author} {\bibfnamefont {A.}~\bibnamefont {Petrenko}},
  \bibinfo {author} {\bibfnamefont {Z.}~\bibnamefont {Leghtas}}, \bibinfo
  {author} {\bibfnamefont {B.}~\bibnamefont {Vlastakis}}, \bibinfo {author}
  {\bibfnamefont {G.}~\bibnamefont {Kirchmair}}, \bibinfo {author}
  {\bibfnamefont {K.~M.}\ \bibnamefont {Sliwa}}, \bibinfo {author}
  {\bibfnamefont {A.}~\bibnamefont {Narla}}, \bibinfo {author} {\bibfnamefont
  {M.}~\bibnamefont {Hatridge}}, \bibinfo {author} {\bibfnamefont
  {S.}~\bibnamefont {Shankar}}, \bibinfo {author} {\bibfnamefont
  {J.}~\bibnamefont {Blumoff}}, \bibinfo {author} {\bibfnamefont
  {L.}~\bibnamefont {Frunzio}}, \bibinfo {author} {\bibfnamefont
  {M.}~\bibnamefont {Mirrahimi}}, \bibinfo {author} {\bibfnamefont {M.~H.}\
  \bibnamefont {Devoret}}, \ and\ \bibinfo {author} {\bibfnamefont {R.~J.}\
  \bibnamefont {Schoelkopf}},\ }\bibfield  {title} {\enquote {\bibinfo {title}
  {Tracking photon jumps with repeated quantum non-demolition parity
  measurements},}\ }\href@noop {} {\bibfield  {journal} {\bibinfo  {journal}
  {Nature}\ }\textbf {\bibinfo {volume} {511}},\ \bibinfo {pages} {444--448}
  (\bibinfo {year} {2014})}\BibitemShut {NoStop}%
\bibitem [{\citenamefont {Rosenblum}\ \emph {et~al.}(2018)\citenamefont
  {Rosenblum}, \citenamefont {Reinhold}, \citenamefont {Mirrahimi},
  \citenamefont {Jiang}, \citenamefont {Frunzio},\ and\ \citenamefont
  {Schoelkopf}}]{QuantumError}%
  \BibitemOpen
  \bibfield  {author} {\bibinfo {author} {\bibfnamefont {S.}~\bibnamefont
  {Rosenblum}}, \bibinfo {author} {\bibfnamefont {P.}~\bibnamefont {Reinhold}},
  \bibinfo {author} {\bibfnamefont {M.}~\bibnamefont {Mirrahimi}}, \bibinfo
  {author} {\bibfnamefont {L.}~\bibnamefont {Jiang}}, \bibinfo {author}
  {\bibfnamefont {L.}~\bibnamefont {Frunzio}}, \ and\ \bibinfo {author}
  {\bibfnamefont {R.~J.}\ \bibnamefont {Schoelkopf}},\ }\bibfield  {title}
  {\enquote {\bibinfo {title} {Fault-tolerant detection of a quantum error},}\
  }\href@noop {} {\bibfield  {journal} {\bibinfo  {journal} {Science}\ }\textbf
  {\bibinfo {volume} {361}},\ \bibinfo {pages} {6399 pp. 266--270} (\bibinfo
  {year} {2018})}\BibitemShut {NoStop}%
\bibitem [{\citenamefont {Lao}\ and\ \citenamefont {Plenio}(2016)}]{eSWAP1}%
  \BibitemOpen
  \bibfield  {author} {\bibinfo {author} {\bibfnamefont {H.~K.}\ \bibnamefont
  {Lao}}\ and\ \bibinfo {author} {\bibfnamefont {M.~B.}\ \bibnamefont
  {Plenio}},\ }\bibfield  {title} {\enquote {\bibinfo {title} {Universal
  quantum computing with arbitrary continuous-variable encoding},}\ }\href@noop
  {} {\bibfield  {journal} {\bibinfo  {journal} {Phys.\ Rev. Lett.}\ }\textbf
  {\bibinfo {volume} {117}},\ \bibinfo {pages} {100501} (\bibinfo {year}
  {2016})}\BibitemShut {NoStop}%
\bibitem [{\citenamefont {Gao}\ \emph {et~al.}(2019)\citenamefont {Gao},
  \citenamefont {Lester}, \citenamefont {Chou}, \citenamefont {Frunzio},
  \citenamefont {Devoret}, \citenamefont {Jiang}, \citenamefont {Girvin},\ and\
  \citenamefont {Schoelkopf}}]{eSWAP2}%
  \BibitemOpen
  \bibfield  {author} {\bibinfo {author} {\bibfnamefont {Y.~Y.}\ \bibnamefont
  {Gao}}, \bibinfo {author} {\bibfnamefont {B.~J.}\ \bibnamefont {Lester}},
  \bibinfo {author} {\bibfnamefont {K.~S.}\ \bibnamefont {Chou}}, \bibinfo
  {author} {\bibfnamefont {L.}~\bibnamefont {Frunzio}}, \bibinfo {author}
  {\bibfnamefont {M.~H.}\ \bibnamefont {Devoret}}, \bibinfo {author}
  {\bibfnamefont {L.}~\bibnamefont {Jiang}}, \bibinfo {author} {\bibfnamefont
  {S.~M.}\ \bibnamefont {Girvin}}, \ and\ \bibinfo {author} {\bibfnamefont
  {R.~J.}\ \bibnamefont {Schoelkopf}},\ }\bibfield  {title} {\enquote {\bibinfo
  {title} {Entanglement of bosonic modes through an engineered exchange
  interaction},}\ }\href@noop {} {\bibfield  {journal} {\bibinfo  {journal}
  {Nature}\ }\textbf {\bibinfo {volume} {566}},\ \bibinfo {pages} {509--512}
  (\bibinfo {year} {2019})}\BibitemShut {NoStop}%
\bibitem [{\citenamefont {Cohen}\ \emph {et~al.}(2014)\citenamefont {Cohen},
  \citenamefont {Istrati}, \citenamefont {Dovratm},\ and\ \citenamefont
  {Eisenberg}}]{Cohen1}%
  \BibitemOpen
  \bibfield  {author} {\bibinfo {author} {\bibfnamefont {L.}~\bibnamefont
  {Cohen}}, \bibinfo {author} {\bibfnamefont {D.}~\bibnamefont {Istrati}},
  \bibinfo {author} {\bibfnamefont {L.}~\bibnamefont {Dovratm}}, \ and\
  \bibinfo {author} {\bibfnamefont {H.~S.}\ \bibnamefont {Eisenberg}},\
  }\bibfield  {title} {\enquote {\bibinfo {title} {Super-resolved phase
  measurements at the shot noise limit by parity measurements},}\ }\href@noop
  {} {\bibfield  {journal} {\bibinfo  {journal} {Optics Express}\ }\textbf
  {\bibinfo {volume} {22}},\ \bibinfo {pages} {11945} (\bibinfo {year}
  {2014})}\BibitemShut {NoStop}%
\bibitem [{\citenamefont {Kok}\ and\ \citenamefont {Lovett}(2010)}]{Kok3}%
  \BibitemOpen
  \bibfield  {author} {\bibinfo {author} {\bibfnamefont {P.}~\bibnamefont
  {Kok}}\ and\ \bibinfo {author} {\bibfnamefont {B.~W.}\ \bibnamefont
  {Lovett}},\ }\href@noop {} {\emph {\bibinfo {title} {Introduction to Optical
  Quantum Information Processing}}}\ (\bibinfo  {publisher} {Cambridge
  University Press, Cambridge, UK., p. 124},\ \bibinfo {year}
  {2010})\BibitemShut {NoStop}%
\bibitem [{\citenamefont {Silberhorn}(2007)}]{Silberhorn}%
  \BibitemOpen
  \bibfield  {author} {\bibinfo {author} {\bibfnamefont {C.}~\bibnamefont
  {Silberhorn}},\ }\bibfield  {title} {\enquote {\bibinfo {title} {Detecting
  quantum light},}\ }\href@noop {} {\bibfield  {journal} {\bibinfo  {journal}
  {Contemp. Phys.}\ }\textbf {\bibinfo {volume} {48}},\ \bibinfo {pages} {143}
  (\bibinfo {year} {2007})}\BibitemShut {NoStop}%
\bibitem [{\citenamefont {Gerrits}\ \emph {et~al.}(2010)\citenamefont
  {Gerrits}, \citenamefont {Clancy}, \citenamefont {Clement}, \citenamefont
  {Calkins}, \citenamefont {Lita}, \citenamefont {Miller}, \citenamefont
  {Migdall}, \citenamefont {Nam}, \citenamefont {Mirin},\ and\ \citenamefont
  {Knill}}]{Detect1}%
  \BibitemOpen
  \bibfield  {author} {\bibinfo {author} {\bibfnamefont {T.}~\bibnamefont
  {Gerrits}}, \bibinfo {author} {\bibfnamefont {S.}~\bibnamefont {Clancy}},
  \bibinfo {author} {\bibfnamefont {T.~S.}\ \bibnamefont {Clement}}, \bibinfo
  {author} {\bibfnamefont {B.}~\bibnamefont {Calkins}}, \bibinfo {author}
  {\bibfnamefont {A.~E.}\ \bibnamefont {Lita}}, \bibinfo {author}
  {\bibfnamefont {A.~J.}\ \bibnamefont {Miller}}, \bibinfo {author}
  {\bibfnamefont {A.~L.}\ \bibnamefont {Migdall}}, \bibinfo {author}
  {\bibfnamefont {S.~W.}\ \bibnamefont {Nam}}, \bibinfo {author} {\bibfnamefont
  {R.~P.}\ \bibnamefont {Mirin}}, \ and\ \bibinfo {author} {\bibfnamefont
  {E.}~\bibnamefont {Knill}},\ }\bibfield  {title} {\enquote {\bibinfo {title}
  {Generation of optical coherent-state superpositions by number-resolving
  photon subtraction from the squeezed vacuum},}\ }\href@noop {} {\bibfield
  {journal} {\bibinfo  {journal} {Phys.\ Rev. A}\ }\textbf {\bibinfo {volume}
  {82}},\ \bibinfo {pages} {031802{(R)}} (\bibinfo {year} {2010})}\BibitemShut
  {NoStop}%
\bibitem [{\citenamefont {Banaszek}\ and\ \citenamefont
  {Walmsley}(2003)}]{Detect2}%
  \BibitemOpen
  \bibfield  {author} {\bibinfo {author} {\bibfnamefont {K.}~\bibnamefont
  {Banaszek}}\ and\ \bibinfo {author} {\bibfnamefont {I.~A.}\ \bibnamefont
  {Walmsley}},\ }\bibfield  {title} {\enquote {\bibinfo {title} {Photon
  counting with a loop detector},}\ }\href@noop {} {\bibfield  {journal}
  {\bibinfo  {journal} {Opt. Lett.}\ }\textbf {\bibinfo {volume} {28}},\
  \bibinfo {pages} {52} (\bibinfo {year} {2003})}\BibitemShut {NoStop}%
\bibitem [{\citenamefont {Achilles}\ \emph {et~al.}(2004)\citenamefont
  {Achilles}, \citenamefont {Silberhorn}, \citenamefont {\'{S}liwa},
  \citenamefont {Banaszek}, \citenamefont {Walmsley}, \citenamefont {Finch},
  \citenamefont {Jacobs}, \citenamefont {Pittman},\ and\ \citenamefont
  {Franson}}]{Detect3}%
  \BibitemOpen
  \bibfield  {author} {\bibinfo {author} {\bibfnamefont {D.}~\bibnamefont
  {Achilles}}, \bibinfo {author} {\bibfnamefont {C.}~\bibnamefont
  {Silberhorn}}, \bibinfo {author} {\bibfnamefont {C.}~\bibnamefont
  {\'{S}liwa}}, \bibinfo {author} {\bibfnamefont {K.}~\bibnamefont {Banaszek}},
  \bibinfo {author} {\bibfnamefont {I.~A.}\ \bibnamefont {Walmsley}}, \bibinfo
  {author} {\bibfnamefont {M.~J.}\ \bibnamefont {Finch}}, \bibinfo {author}
  {\bibfnamefont {B.~C.}\ \bibnamefont {Jacobs}}, \bibinfo {author}
  {\bibfnamefont {T.~B.}\ \bibnamefont {Pittman}}, \ and\ \bibinfo {author}
  {\bibfnamefont {J.~D.}\ \bibnamefont {Franson}},\ }\bibfield  {title}
  {\enquote {\bibinfo {title} {Photon-number resolving detection using
  time-multiplexing},}\ }\href@noop {} {\bibfield  {journal} {\bibinfo
  {journal} {J. Mod. Opt.}\ }\textbf {\bibinfo {volume} {51}},\ \bibinfo
  {pages} {1499} (\bibinfo {year} {2004})}\BibitemShut {NoStop}%
\bibitem [{\citenamefont {Jiang}, \citenamefont {Dauler},\ and\ \citenamefont
  {Chang}(2007)}]{Detect4}%
  \BibitemOpen
  \bibfield  {author} {\bibinfo {author} {\bibfnamefont {L.~A.}\ \bibnamefont
  {Jiang}}, \bibinfo {author} {\bibfnamefont {E.~A.}\ \bibnamefont {Dauler}}, \
  and\ \bibinfo {author} {\bibfnamefont {J.~T.}\ \bibnamefont {Chang}},\
  }\bibfield  {title} {\enquote {\bibinfo {title} {Photon-number-resolving
  detector with 10 bits of resolution},}\ }\href@noop {} {\bibfield  {journal}
  {\bibinfo  {journal} {Phys.\ Rev. A}\ }\textbf {\bibinfo {volume} {75}},\
  \bibinfo {pages} {062325} (\bibinfo {year} {2007})}\BibitemShut {NoStop}%
\bibitem [{\citenamefont {Cohen}\ \emph {et~al.}(2018)\citenamefont {Cohen},
  \citenamefont {Pilnyak}, \citenamefont {Istrali}, \citenamefont {Studer},
  \citenamefont {Dowling},\ and\ \citenamefont {Eisenberg}}]{Detect5}%
  \BibitemOpen
  \bibfield  {author} {\bibinfo {author} {\bibfnamefont {L.}~\bibnamefont
  {Cohen}}, \bibinfo {author} {\bibfnamefont {Y.}~\bibnamefont {Pilnyak}},
  \bibinfo {author} {\bibfnamefont {D.}~\bibnamefont {Istrali}}, \bibinfo
  {author} {\bibfnamefont {N.~M.}\ \bibnamefont {Studer}}, \bibinfo {author}
  {\bibfnamefont {J.~P.}\ \bibnamefont {Dowling}}, \ and\ \bibinfo {author}
  {\bibfnamefont {H.}~\bibnamefont {Eisenberg}},\ }\bibfield  {title} {\enquote
  {\bibinfo {title} {Absolute calibration of single-photon and multiplexed
  photon-number-resolving detectors},}\ }\href@noop {} {\bibfield  {journal}
  {\bibinfo  {journal} {Phys.\ Rev. A}\ }\textbf {\bibinfo {volume} {98}},\
  \bibinfo {pages} {013811} (\bibinfo {year} {2018})}\BibitemShut {NoStop}%
\bibitem [{\citenamefont {Imoto}, \citenamefont {Haus},\ and\ \citenamefont
  {Yamamoto}(1985)}]{QNDnum}%
  \BibitemOpen
  \bibfield  {author} {\bibinfo {author} {\bibfnamefont {N.}~\bibnamefont
  {Imoto}}, \bibinfo {author} {\bibfnamefont {H.~A.}\ \bibnamefont {Haus}}, \
  and\ \bibinfo {author} {\bibfnamefont {Y.}~\bibnamefont {Yamamoto}},\
  }\bibfield  {title} {\enquote {\bibinfo {title} {Quantum nondemolition
  measurements of the photon number via the optical {K}err effect},}\
  }\href@noop {} {\bibfield  {journal} {\bibinfo  {journal} {Phys.\ Rev. A}\
  }\textbf {\bibinfo {volume} {32}},\ \bibinfo {pages} {2287} (\bibinfo {year}
  {1985})}\BibitemShut {NoStop}%
\bibitem [{\citenamefont {Plick}\ \emph {et~al.}(2010)\citenamefont {Plick},
  \citenamefont {Anisimov}, \citenamefont {Dowling}, \citenamefont {Lee},\ and\
  \citenamefont {Agarwal}}]{Plick}%
  \BibitemOpen
  \bibfield  {author} {\bibinfo {author} {\bibfnamefont {W.~N.}\ \bibnamefont
  {Plick}}, \bibinfo {author} {\bibfnamefont {P.~M.}\ \bibnamefont {Anisimov}},
  \bibinfo {author} {\bibfnamefont {J.~P.}\ \bibnamefont {Dowling}}, \bibinfo
  {author} {\bibfnamefont {H.}~\bibnamefont {Lee}}, \ and\ \bibinfo {author}
  {\bibfnamefont {G.~S.}\ \bibnamefont {Agarwal}},\ }\bibfield  {title}
  {\enquote {\bibinfo {title} {Parity detection in quantum optical metrology
  without number resolving detectors},}\ }\href@noop {} {\bibfield  {journal}
  {\bibinfo  {journal} {N. J. Phys.}\ }\textbf {\bibinfo {volume} {12}},\
  \bibinfo {pages} {113025} (\bibinfo {year} {2010})}\BibitemShut {NoStop}%
\bibitem [{\citenamefont {Herrero-Collantes}\ and\ \citenamefont
  {Garcia-Escartin}(2017)}]{QRNG}%
  \BibitemOpen
  \bibfield  {author} {\bibinfo {author} {\bibfnamefont {M.}~\bibnamefont
  {Herrero-Collantes}}\ and\ \bibinfo {author} {\bibfnamefont {J.~C.}\
  \bibnamefont {Garcia-Escartin}},\ }\bibfield  {title} {\enquote {\bibinfo
  {title} {Quantum random number generators},}\ }\href@noop {} {\bibfield
  {journal} {\bibinfo  {journal} {Rev. Mod. Phys.}\ }\textbf {\bibinfo {volume}
  {89}},\ \bibinfo {pages} {015004} (\bibinfo {year} {2017})}\BibitemShut
  {NoStop}%
\bibitem [{\citenamefont {Huang}\ \emph {et~al.}(2014)\citenamefont {Huang},
  \citenamefont {Wu}, \citenamefont {Zhong},\ and\ \citenamefont
  {Lee}}]{ColdAtoms}%
  \BibitemOpen
  \bibfield  {author} {\bibinfo {author} {\bibfnamefont {J.}~\bibnamefont
  {Huang}}, \bibinfo {author} {\bibfnamefont {S.}~\bibnamefont {Wu}}, \bibinfo
  {author} {\bibfnamefont {H.}~\bibnamefont {Zhong}}, \ and\ \bibinfo {author}
  {\bibfnamefont {C.}~\bibnamefont {Lee}},\ }\href@noop {} {\emph {\bibinfo
  {title} {Annual Review of Cold Atoms anf Molecules, {Chapter 7: Quantum
  Metrology with Cold Atoms, pp. 365-415}}}}\ (\bibinfo  {publisher} {World
  Scientific},\ \bibinfo {year} {2014})\BibitemShut {NoStop}%
\bibitem [{\citenamefont {Pezz\'{e}}\ \emph {et~al.}(2018)\citenamefont
  {Pezz\'{e}}, \citenamefont {Smerzi}, \citenamefont {Overthaler},
  \citenamefont {Schmied},\ and\ \citenamefont {Treutlein}}]{MetroAtoms}%
  \BibitemOpen
  \bibfield  {author} {\bibinfo {author} {\bibfnamefont {L.}~\bibnamefont
  {Pezz\'{e}}}, \bibinfo {author} {\bibfnamefont {A.}~\bibnamefont {Smerzi}},
  \bibinfo {author} {\bibfnamefont {M.~K.}\ \bibnamefont {Overthaler}},
  \bibinfo {author} {\bibfnamefont {R.}~\bibnamefont {Schmied}}, \ and\
  \bibinfo {author} {\bibfnamefont {P.}~\bibnamefont {Treutlein}},\ }\bibfield
  {title} {\enquote {\bibinfo {title} {Quantum metrology with nonclassical
  states of atomic ensembles},}\ }\href@noop {} {\bibfield  {journal} {\bibinfo
   {journal} {Rev. Mod. Phys.}\ }\textbf {\bibinfo {volume} {90}},\ \bibinfo
  {pages} {035005} (\bibinfo {year} {2018})}\BibitemShut {NoStop}%
\bibitem [{Note8()}]{Note8}%
  \BibitemOpen
  \bibinfo {note} {For complex state coefficients,Eq.~\ref {eqn:rs5} can be
  written as $\mathinner {\delimiter "426830A {\protect \mathaccentV
  {hat}05E{J}_{z}}\delimiter "526930B } = 2\protect \text {Re}\left [z e^{i\phi
  }\right ] + \alpha ^{\left (j\right )}$, where $z$ is a complex number
  expressed in terms of the state coefficients. For real state coefficients,
  this simplifies to the form of Eq.~\ref {eqn:rs6a}}\BibitemShut {NoStop}%
\bibitem [{\citenamefont {Dicke}(1954)}]{Dicke}%
  \BibitemOpen
  \bibfield  {author} {\bibinfo {author} {\bibfnamefont {R.~H.}\ \bibnamefont
  {Dicke}},\ }\bibfield  {title} {\enquote {\bibinfo {title} {Coherent in
  spontaneous radiation processes},}\ }\href@noop {} {\bibfield  {journal}
  {\bibinfo  {journal} {Phys.\ Rev.}\ }\textbf {\bibinfo {volume} {93}},\
  \bibinfo {pages} {99} (\bibinfo {year} {1954})}\BibitemShut {NoStop}%
\bibitem [{\citenamefont {Radcliffe}(1971)}]{Radcliffe}%
  \BibitemOpen
  \bibfield  {author} {\bibinfo {author} {\bibfnamefont {J.~M.}\ \bibnamefont
  {Radcliffe}},\ }\bibfield  {title} {\enquote {\bibinfo {title} {Some
  properties of coherent spin states},}\ }\href@noop {} {\bibfield  {journal}
  {\bibinfo  {journal} {J. Phys. A: Gen. Phys.}\ }\textbf {\bibinfo {volume}
  {4}},\ \bibinfo {pages} {313} (\bibinfo {year} {1971})}\BibitemShut {NoStop}%
\bibitem [{\citenamefont {Rose}(1995)}]{Rose}%
  \BibitemOpen
  \bibfield  {author} {\bibinfo {author} {\bibfnamefont {M.~E.}\ \bibnamefont
  {Rose}},\ }\href@noop {} {\emph {\bibinfo {title} {Elementary Theory of
  Angular Momentum}}}\ (\bibinfo  {publisher} {Dover Publications, Inc., New
  York},\ \bibinfo {year} {1995})\BibitemShut {NoStop}%
\bibitem [{Note9()}]{Note9}%
  \BibitemOpen
  \bibinfo {note} {Note that $\protect \mathaccentV {hat}05E{D}_{a}\left
  (\alpha \right )\protect \mathaccentV {hat}05E{D}_{a}\left (\beta \right ) =
  e^{\protect \genfrac {}{}{}1{1}{2}\left (\alpha \beta ^{*} - \alpha ^{*}\beta
  \right )}\protect \mathaccentV {hat}05E{D}_{a}\left (\alpha +\beta \right )$.
  The product of displacement operators is the sum of the displacements up to
  an overall phase factor.}\BibitemShut {Stop}%
\bibitem [{\citenamefont {Tajima}(2015)}]{WignerIdent}%
  \BibitemOpen
  \bibfield  {author} {\bibinfo {author} {\bibfnamefont {N.}~\bibnamefont
  {Tajima}},\ }\bibfield  {title} {\enquote {\bibinfo {title} {Analytical
  formula for numerical evaluations of the {W}igner rotation matricies at high
  spins},}\ }\href@noop {} {\bibfield  {journal} {\bibinfo  {journal} {Phys.\
  Rev. C}\ }\textbf {\bibinfo {volume} {91}},\ \bibinfo {pages} {014320}
  (\bibinfo {year} {2015})}\BibitemShut {NoStop}%
\bibitem [{\citenamefont {Bengtsson}\ and\ \citenamefont
  {Zyczkowski}(2006)}]{GeomQStates:2006}%
  \BibitemOpen
  \bibfield  {author} {\bibinfo {author} {\bibfnamefont {I.}~\bibnamefont
  {Bengtsson}}\ and\ \bibinfo {author} {\bibfnamefont {K.}~\bibnamefont
  {Zyczkowski}},\ }\href@noop {} {\emph {\bibinfo {title} {Geometry of Quantum
  States}}}\ (\bibinfo  {publisher} {Cambridge University Press, Cambridge,
  UK.},\ \bibinfo {year} {2006})\BibitemShut {NoStop}%
\bibitem [{\citenamefont {Wilde}(2017)}]{Wilde:2017}%
  \BibitemOpen
  \bibfield  {author} {\bibinfo {author} {\bibfnamefont {M.}~\bibnamefont
  {Wilde}},\ }\href@noop {} {\emph {\bibinfo {title} {Quantum Information, 2nd
  Ed.}}}\ (\bibinfo  {publisher} {Cambridge University Press, Cambridge, UK.},\
  \bibinfo {year} {2017})\BibitemShut {NoStop}%
\bibitem [{\citenamefont {W.~Guo}\ and\ \citenamefont {Wang}(2019)}]{Guo:2016}%
  \BibitemOpen
  \bibfield  {author} {\bibinfo {author} {\bibfnamefont {X.~J. L.~F.}\
  \bibnamefont {W.~Guo}, \bibfnamefont {W.~Zhong}}\ and\ \bibinfo {author}
  {\bibfnamefont {X.}~\bibnamefont {Wang}},\ }\bibfield  {title} {\enquote
  {\bibinfo {title} {Berry curvature as a lower bound for multiparameter
  estimation},}\ }\href@noop {} {\bibfield  {journal} {\bibinfo  {journal}
  {Phys. Rev. A}\ }\textbf {\bibinfo {volume} {93}},\ \bibinfo {pages} {042115}
  (\bibinfo {year} {2019})}\BibitemShut {NoStop}%
\bibitem [{\citenamefont {Samuel}\ and\ \citenamefont
  {Bhandari}(1988)}]{Samuel_Bhandari:1988}%
  \BibitemOpen
  \bibfield  {author} {\bibinfo {author} {\bibfnamefont {J.}~\bibnamefont
  {Samuel}}\ and\ \bibinfo {author} {\bibfnamefont {R.}~\bibnamefont
  {Bhandari}},\ }\bibfield  {title} {\enquote {\bibinfo {title} {General
  setting for berry's phase},}\ }\href@noop {} {\bibfield  {journal} {\bibinfo
  {journal} {Phys. Rev. Lett.}\ }\textbf {\bibinfo {volume} {60}},\ \bibinfo
  {pages} {2339} (\bibinfo {year} {1988})}\BibitemShut {NoStop}%
\bibitem [{\citenamefont {Shapere}\ and\ \citenamefont
  {Wilczek}(1989)}]{BerryBook}%
  \BibitemOpen
  \bibfield  {author} {\bibinfo {author} {\bibfnamefont {A.}~\bibnamefont
  {Shapere}}\ and\ \bibinfo {author} {\bibfnamefont {F.}~\bibnamefont
  {Wilczek}},\ }\href@noop {} {\emph {\bibinfo {title} {Geometric Phases in
  Physics}}}\ (\bibinfo  {publisher} {World Scientific, Singapore},\ \bibinfo
  {year} {1989})\ pp.\ \bibinfo {pages} {113--123}\BibitemShut {NoStop}%
\bibitem [{\citenamefont {Provost}\ and\ \citenamefont
  {Vallee}(1980)}]{Provost_Vallee:1980}%
  \BibitemOpen
  \bibfield  {author} {\bibinfo {author} {\bibfnamefont {J.}~\bibnamefont
  {Provost}}\ and\ \bibinfo {author} {\bibfnamefont {G.}~\bibnamefont
  {Vallee}},\ }\bibfield  {title} {\enquote {\bibinfo {title} {Riemann
  structure on manifolds of quantum states},}\ }\href@noop {} {\bibfield
  {journal} {\bibinfo  {journal} {Comm. Math. Phys.}\ }\textbf {\bibinfo
  {volume} {76}},\ \bibinfo {pages} {289} (\bibinfo {year} {1980})}\BibitemShut
  {NoStop}%
\bibitem [{\citenamefont {Ben-Aryeh}(2004)}]{Ben-Aryeh:2004}%
  \BibitemOpen
  \bibfield  {author} {\bibinfo {author} {\bibfnamefont {Y.}~\bibnamefont
  {Ben-Aryeh}},\ }\bibfield  {title} {\enquote {\bibinfo {title} {Berry and
  {P}ancharatnam topological phases of atomic and optical systems},}\
  }\href@noop {} {\bibfield  {journal} {\bibinfo  {journal} {J. Opt. B: Quantum
  Semiclass. Opt.}\ }\textbf {\bibinfo {volume} {6}},\ \bibinfo {pages}
  {R1--R18} (\bibinfo {year} {2004})}\BibitemShut {NoStop}%
\bibitem [{\citenamefont {Frankel}(2012)}]{Frankel_3rdEd:2012}%
  \BibitemOpen
  \bibfield  {author} {\bibinfo {author} {\bibfnamefont {T.}~\bibnamefont
  {Frankel}},\ }\href@noop {} {\emph {\bibinfo {title} {The Geometry of
  Physics, 3rd Ed., Chap. 17}}}\ (\bibinfo  {publisher} {Cambridge University
  Press, Cambridge, UK.},\ \bibinfo {year} {2012})\ pp.\ \bibinfo {pages}
  {451--474}\BibitemShut {NoStop}%
\end{thebibliography}%


\providecommand{\noopsort}[1]{}\providecommand{\singleletter}[1]{#1}%
%

\end{document}